# Tumor-associated CD19+ macrophages induce immunosuppressive microenvironment in hepatocellular carcinoma

Junli Wang[1,2,#], Wanyue Cao[1,2,#], Jinyan Huang[2,7], Yu Zhou[2], Rujia Zheng[2,7], Yu Lou[1,2], Jiaqi Yang[1,2], Jianghui Tang[1,2], Mao Ye[1,2], Zhengtao Hong[1,2], Jiangchao Wu[1,2], Haonan Ding[1,2], Yuquan Zhang[1,2], Jianpeng Sheng[2,5,6], Xinjiang Lu[2,8], Pinglong Xu[2,5,9], Xiongbin Lu[1,2], Xueli Bai[1,2,3,4,5,6], Tingbo Liang[1,2,3,4,5,6], Qi Zhang[1,2,3,4,5,6,*]

[1] Department of Hepatobiliary and Pancreatic Surgery, the First Affiliated Hospital, Zhejiang University School of Medicine, Hangzhou 310003, China.

[2] Zhejiang Provincial Key Laboratory of Pancreatic Disease, the First Affiliated Hospital, Zhejiang University School of Medicine, Hangzhou 310003, China.

[3] Clinical Research Center of Hepatobiliary and Pancreatic Diseases, Zhejiang Province, Hangzhou 31003, China.

[4] The Innovation Center for the Study of Pancreatic Diseases of Zhejiang Province, Hangzhou 310003, China.

[5] Zhejiang University Cancer Center, Hangzhou 310063, China.

[6] MOE Joint International Research Laboratory of Pancreatic Diseases, Hangzhou 310003, China.

[7] Biomedical Big Data Center, the First Affiliated Hospital, Zhejiang University School of Medicine, Hangzhou 31003, China.

[8] Department of Physiology, Zhejiang University School of Medicine, Hangzhou 310063, China.

[9] Life Sciences Institute, Zhejiang University, Hangzhou 310063, China

[#] These authors contributed equally.

[*]Correspondence: qi.zhang@zju.edu.cn (Q.Z.).

**Abstract**

Tumor-associated macrophages are a key component that contributes to the immunosuppressive microenvironment in human cancers. However, therapeutic targeting of macrophages has been a challenge in clinic due to the limited understanding of their heterogeneous subpopulations and distinct functions. Here, we identify a unique and clinically relevant CD19$^+$ subpopulation of macrophages that is enriched in many types of cancer, particularly in hepatocellular carcinoma (HCC). The CD19$^+$ macrophages exhibit increased levels of PD-L1 and CD73, enhanced mitochondrial oxidation, and compromised phagocytosis, indicating their immunosuppressive functions. Targeting CD19$^+$ macrophages with anti-CD19 chimeric antigen receptor T (CAR-T) cells inhibited HCC tumor growth. We identify PAX5 as a primary driver of up-regulated mitochondrial biogenesis in CD19$^+$ macrophages, which depletes cytoplasmic Ca$^{2+}$, leading to lysosomal deficiency and consequent accumulation of CD73 and PD-L1. Inhibiting CD73 or mitochondrial oxidation enhanced the efficacy of immune checkpoint blockade therapy in treating HCC, suggesting great promise for CD19$^+$ macrophage-targeting therapeutics.

**Significance**

Hepatocellular carcinoma (HCC) remains a lethal malignancy with limited therapeutic options. Our study reveals the critical role of CD19$^+$ tumor-associated macrophages in driving HCC progression through enhanced mitochondrial oxidation, and lysosomal dysfunction mediated by the Pax5-[Ca$^{2+}$]-TFEB axis with PD-L1 and CD73 overexpression, . Importantly, targeting CD19+ macrophages via CD19 CAR-T cells or CD73 blockade significantly enhances antitumor immunity, offering a promising combinatorial immunotherapy strategy. Our findings provide novel insights into the immunosuppressive tumor microenvironment and present a transformative approach to improve outcomes for HCC patients.

**Introduction**

Liver cancer has emerged as the third leading cause of cancer-related death worldwide, with hepatocellular carcinoma (HCC) being the predominant pathological subtype[1]. Current therapies for treating hepatocellular carcinoma (HCC) comprise surgical resection, liver transplantation, locoregional therapies, and systemic treatments including targeted therapies and immunotherapy[2]. Surgical resection and liver transplantation offer offer the best chance for a cure, but are only options for a small subset of patients diagnosed at an early stage. Locoregional therapies, including radiofrequency ablation and transarterial chemoembolization, are used for intermediate stages of HCC or when surgery is not feasible. Systemic treatments, like tyrosine kinase inhibitors (e.g., sorafenib and lenvatinib) and immune checkpoint inhibitors (e.g., atezolizumab, nivolumab and pembrolizumab), have expanded treatment options for the majority of patients with advanced HCC that misses the window for surgical intervention[3]. Despite these advancements, the 5-year overall survival rates for HCC remain below 20%[4]. These therapies face significant challenges in the clinic, including tumor heterogeneity, which leads to variable treatment responses, and the immunosuppressive tumor microenvironment (TME), which hampers the efficacy of immunotherapies[5, 6].

The immunosuppressive microenvironment of HCC is characterized by a complex interplay of various cell types that collectively inhibit effective anti-tumor immune responses. Key players include myeloid-derived suppressor cells (MDSCs), which accumulate and inhibit T cell activation[7]. Regulatory T cells (Tregs) suppress cytotoxic T cell function, while cancer-associated fibroblasts (CAFs) produce extracellular matrix components and cytokines that further support an immunosuppressive milieu[8]. Particularly, tumor-associated macrophages (TAMs) are the most abundant in HCC that promotes tumor growth and suppresses inflammation[9]. These various types of stroma cells interact through direct contact and the release of soluble factors, creating a TME that hinders immune-mediated tumor eradication[10].

Macrophages exhibit different features in the presence of different inducers, such as lipopolysaccharide and interleukin-4, which lead to the two classic polarizations of macrophages, defining as M1 and M2 macrophages[11]. TAMs are believed to promote tumor progression by reprogramming local immunosuppression, inducing angiogenesis and drug resistance, and promoting tumor cell invasion. Phenotypically, despite being considered to have an M2-like polarization[12], TAMs are noncanonical and dynamically altered in response to stimuli within the TME. Consequently, the binary classification of macrophages is inadequate for describing them within the highly heterogeneous TME, both spatially and temporally[13]. Studies have identified specific TAM subgroups, such as PD-1$^+$ TAMs in colon cancer, TREM1$^+$ TAMs in hepatocellular carcinoma, and CD169$^+$ TAMs in glioblastoma[14, 15, 16]. These subgroups have diverse or even contradictory effects on tumor progression, suggesting that precise interventions targeting certain TAM subgroups, rather than the wholesale removal of M1, M2 or all TAMs in the TME, are likely to achieve better anti-tumor efficacy. Therefore, understanding the full spectrum of TAMs is crucial for developing therapeutic approaches that target macrophages in HCC.

In the present study, we identified a unique subgroup of CD19$^+$ TAMs that are enriched in a number of solid tumor types, particularly in HCC. As a transmembrane glycoprotein, CD19 is considered as a biomarker and functionally crucial for the regulation of B cell receptor signaling in both the early and late stages of B cell development. Interestingly, a recent analysis of single-cell RNA sequencing (scRNA-seq) data defined a transdifferentiated cell types in the ischemic brain that was characterized by co-expression of B-cell markers and several macrophage markers[17], suggesting a potential role of CD19 in macrophages. Here, we elucidated the clinical relevance, cellular functions, and molecular features of CD19$^+$ TAMs and determined the underlying regulatory mechanisms of their immunosuppressive characteristics. Based on these findings, we propose that targeting CD19$^+$ TAMs or mitigating their pro-tumoral effects could be a promising strategy for treating HCC.

## Methods

A detailed description of the methods used in the study can be found in the Supplementary information.

## Clinical specimens

The collection and application of clinical samples were approved by the ethics committee of the First Affiliated Hospital, Zhejiang University School of Medicine. All patients' informed consent was obtained.

## Cell culture

THP-1 cells, Hepa1-6 and Plat-E cells were obtained from the American Type Culture Collection (ATCC, Manassas, VA), immortalized murine bone marrow-derived macrophages (iBMDMs) were provided by Jingying Zhang at Zhejiang University. THP-1 cells were cultured in Roswell Park Memorial Institute medium (RPMI 1640, Gibco, Carlsbad, CA). Hepa1-6 and Plat-E cells were cultured in high-glucose Dulbecco's modified Eagle's medium (DMEM, Gibco). Media were supplemented with 10% fetal bovine serum (FBS, Gibco) and 1% penicillin/streptomycin (Sigma-Aldrich, Saint Louis, MO), and maintained at 37.0 ± 0.2 °C in a humidified incubator with 5.0% $CO_2$.

BMDMs were isolated from femurs and tibia of adult C57BL/6 mice (The Model Animal Research Center of Nanjing University, Nanjing, China) and cultured in RPMI 1640 medium containing 50 ng/ml recombinant murine macrophage colony-stimulating factor (M-CSF, Peprotech, Rocky Hill, NJ) for 5-day differentiation.

## Mouse models

All mice were bred under pathogen-free conditions. Animal experiments were approved by the ethics committee of The First Affiliated Hospital, Zhejiang University School of Medicine. Lyz-Cre mice were obtained from The Jackson Laboratory (Bar Harbor, ME). Homozygous muMT mice were provided by Prof. Chao Wang at Zhejiang University. Pax5flox/flox and CD19flox/flox mice were purchased from Cyagen Biosciences (Suzhou,

China). C57BL/6 (CD45.2) mice were purchased from The Model Animal Research Center of Nanjing University (Nanjing, China). Mice at 6-12 weeks of age were used for animal experiments.

For in situ mice model, a 20 μl suspension consisting of 50% Matrigel Basement Membrane Matrix (Corning, Teterboro, NJ) and 50% phosphate buffered saline (PBS; Cienry, Hangzhou, China) with Hepa1-6 cells in combination with macrophages were injected into the liver of mice. For syngeneic model, Hepa1-6 cells with 100 μl PBS were inoculated subcutaneously into the left or right flanks of mice. Sacrificed tumors were processed for further experiments.

**Transfection of lentiviral vectors and siRNA**

THP-1 cells and iBMDMs were seeded in 6-well plates at a density of $2 \times 10^5$ cells/ml and infected with lenti-*PAX5* or lenti-*TFEB*, lenti-*NC* (Jikai, Shanghai, China). Stably transduced cells were purified using fluorescence activated cell sorting for further experiments. For siRNA transfection, $1.5 \times 10^5$ per well THP-1 cells were seeded in 24-well plates and transfected with human *TFEB* siRNA (Santa Cruz Biotechnology, Santa Cruz, CA) using lipofectamine RNAimax (ThermoFisher Scientific, Waltham, MA) for 48 hours according to manufacturer's protocols.

**Targeted anti-CD19 CAR-T cell therapy**

$CD19^+/CD19^-$ macrophages were isolated from the spleens of mice ($CD45^+CD11b^+F4/80^+CD19^+$), and were mixed with Hepa1-6 cells to establish subcutaneous xenograft model at the left ($CD19^+$ macrophages) or right ($CD19^-$ macrophages) flank of C57BL/6 or *muMT* mice. Anti-CD19 CAR-T cells were prepared and expanded. Mice were randomly assigned into groups for T cell infusion ($1 \times 10^7$ cells in 100 μL PBS), and were sacrificed after 14 days. Harvested tumors were processed for subsequent experiments.

**ImageStream analysis**

Single-cell suspensions were stained and observed using an ImageStream II system (Amnis Corp, Seattle, WA) to identify CD19+ macrophages, CD19- macrophages, and B cells. Totally $1 \times 10^5$ single and focus cells were collected and analyzed by IDEAS™ 6.2 software (Amnis Corp).

**Multiplex immunohistochemistry (mIHC)**

HCC and para-tumor were prepared into slides with 4-μm thickness. mIHC was performed using Opal™ 6-Plex Detection Kits (Akoya Biosciences, Marlborough, MA) according to the manufacturer's instructions.

**Single-cell RNA sequencing (scRNA-seq) and RNA-seq**

ScRNA-seq libraries were constructed using a Chromium Single Cell 3' Library and Gel Bead Kit v3.1 (10 x Genomics) according to the manufacturer's protocol. RNA-seq was performed using the Illumina Hiseq XTEN platform (Illumina, San Diego, CA, USA) at Novogene Co. Ltd (Beijing, China).

**Statistical analysis**

Statistical analysis was conducted using GraphPad Prism 8 (GraphPad Inc., La Jolla, CA).

## Results

### CD19$^+$ macrophages are enriched in HCC

A group of tumor tissue samples (n=30) from patients with HCC were analyzed with mass cytometry (CyTOF) for profiling their tumor immune microenvironment [6]. Unexpectedly, we identified a subgroup of TAMs expressing CD19 at a level comparable to that in B cells in the tumors (**Figure 1A**). To validate the presence of CD19$^+$ TAMs, we further conducted immune profiling analysis of HCC samples in a 41-patient cohort. Approximately a half (29 out of 41) of the HCC samples had a considerable proportion (>20% of total TAMs) of CD19$^+$ TAMs. The existence of a CD19$^+$ subpopulation in the TAMs was further confirmed using immunofluorescence (**Figure 1B**), ImageStream (**Figures 1C and S1A**) and flow cytometry (**Figures 1D and S1B**). Additionally, we applied multiplex immunohistochemical (mIHC) staining on tumors and their adjacent normal tissues using CD14, CD68, CD19, and CD20 markers to demonstrate CD19$^+$ TAMs at the single-cell level (**Figure 1E**). Conventional macrophage markers, CD11b and CD68/CD14, were used to confirm their macrophage identity. Notably, the proportion of CD19$^+$ TAMs was much higher in HCC than in adjacent normal tissue or peripheral blood (**Figures 1B and1D**). In terms of cell size, the CD19$^+$ TAMs had greater FSC/SSC (forward scatter/side scatter) values in flow cytometry than those of B cells, suggesting their macrophage cell lineage rather than a B cell lineage (**Figure S1C**).

Next, we wanted to determine whether CD19$^+$ TAMs exist in other types of solid tumors. In a total of six types of human solid tumors, including HCC, CD19$^+$ TAMs are enriched (**Figure 1F**) compared with those in peripheral blood and normal tissues. In particular, HCC, renal carcinoma, colorectal cancer, pancreatic cancer, and gastric cancer had a median proportion of CD19$^+$ TAMs over 20% of total TAMs. Similarly, we detected CD19$^+$ macrophages in various normal tissues and peripheral blood in mice

(**Figure S1D**). Moreover, the proportion of CD19$^+$ macrophages did not correlate with that of B cells in these tissues (**Figures S1E and S1F**), further suggesting that they are not derived from B cells. The results collectively suggest that CD19$^+$ macrophages naturally exist in the body but are enriched in solid tumors.

**CD19$^+$ TAMs are associated with poor clinical outcomes in HCC**

As no previous study reported this subgroup of TAMs, we wanted to examine the clinical relevance of CD19$^+$ macrophages in HCC. A human HCC tissue microarray was co-immunostained with anti-CD19 and anti-CD68 antibodies (**Figure S2A**). In a total of 191 tissue samples with valid co-immunostaining, we found that the abundance (shown as density) of CD19$^+$ TAMs was positively correlated with tumor size (**Figure S2B**) and tumor cell differentiation (**Figure S2C**), but not with disease stages (**Figure S2D**), suggesting CD19$^+$ TAMs may be recruited at early stage of tumor initiation for creating immunosuppressive TME. Importantly, HCC patients with greater density of CD19$^+$ TAMs had worse overall survival (hazard ratio 3.12, 95% confidential interval 1.23–7.89) and disease-free survival (**Figures 2A and S2E**). By contrast, the density of total TAMs had no correlation with overall survival of these patients (**Figure 2B**). Consistent with its clinical relevance, the density of CD19$^+$ TAMs was positively correlated with tumor cell proliferation, as indicated by Ki-67 staining in human HCC samples (**Figure 2C**). TME analysis of the HCC tissue microarray indicated that a high density of CD19$^+$ TAMs was associated with increased infiltration of Tregs but reduced infiltration of CD8$^+$ T cells (**Figure S2F**). Furthermore, in a prospective cohort of HCC patients treated with anti-PD-1 antibodies (NCT03732547), we observed a notable correlation between a favorable tumor response (partial response or stable disease) and elevated CD19$^+$ TAMs levels, rather than total TAMs levels (**Figures 2D, and 2E**), suggesting that CD19$^+$ TAM-associated immunosuppression can be overcome, at least partially, by immune checkpoint blockade therapy.

**CD19+ TAMs possess immunosuppressive abilities and promote HCC growth**

To study the role of CD19+ TAMs in shaping the TME, we further analyzed the CyTOF data from the 41 patients with HCC (the same cohort analyzed in **Figure 1A**). CD19+ TAMs represent a major subgroup of TAMs, accounting for approximately 30% of TAMs and 5% of CD45+ cells in the tumors (**Figure S2G**). The CD19+ and CD19- subsets of TAMs showed a sharp difference at the level of the M2 macrophage marker CD163 (**Figure S2H**). The proportion of CD19+ TAMs was also negatively correlated with the abundance of total T, CD4+ and CD8+ T cells (**Figure S2I**). In-depth analysis showed that the proportion of CD19+ TAMs was positively correlated with that of PD-1+ effector memory T cells (TEMs), but negatively correlated with that of CD127+ TEMs (**Figure S2I**), consistent with the clinical data in (**Figures 2D and 2E**. These data strongly suggest that CD19+ TAMs possess immunosuppressive abilities.

To determine the role of CD19+ macrophages in tumor growth in vivo, we isolated CD19+ and CD19- macrophages from mouse spleen and mixed them with mouse HCC cells (Hepa 1-6) to generate orthotopic syngeneic mouse models (**Figure S3A**). The presence of CD19+ macrophages significantly promoted tumor growth in comparison to CD19- macrophages (**Figure 2F**). Meanwhile, the number of infiltrating CD8+ T cells was much lower, but Gr-1+ cells were more abundant in tumors with CD19+ macrophages (**Figures S3B and S3C**). To test the clinical potential of targeting CD19+ macrophages in HCC, we developed mouse chimeric antigen receptor T (CAR-T) cells specifically targeting mouse CD19 (see details in **Supplementary Methods**) for adoptive T cell therapy (**Figure S3D**). The growth of the tumors with CD19+ macrophages, but not with CD19- macrophages, was significantly inhibited when the tumor-bearing mice received

anti-CD19 CAR-T cells. This inhibition was not observed in the mice treated with control CAR-T cells for tumors containing either CD19$^+$ or CD19$^-$ macrophages (**Figures 2G and S2E**). To exclude the possibility that the inhibition is associated with B cells that are also positive for CD19, we tested the anti-tumor effect of the anti-CD19 CAR-T in *muMT* mice that lack mature B cells[18]. Low levels of B cells in *mu*MT mice were verified by flow cytometry (**Figure S3F**). GFP$^+$ anti-CD19 CAR-T cells were infiltrated into tumors in *muMT* mice (**Figure S3G**) and CD19$^+$ macrophages were effectively depleted by the anti-CD19 CAR-T treatment (**Figure S3H**) The similar anti-tumor effect of anti-CD19 CAR-T cells was observed in the tumors with CD19$^+$ macrophages (**Figure 2H**), independent of the B cell levels in the mice. These results highlight the role of CD19$^+$ macrophages in HCC progression and suggested CD19$^+$ macrophages are a potential clinical target for HCC treatment.

**CD19$^+$ TAMs exhibit a distinct gene expression profile**

To understand the molecular features of CD19$^+$ macrophages, we isolated CD19$^+$ TAMs, CD19$^-$ TAMs and B cells from three HCC patients, and conducted scRNA-seq analysis for a total of 52,749 cells (**Figures 3A and S4A**). TAMs were annotated by a high expression level of *CD11b, CD14* and *CD68*, while B cells were identified by *CD45, CD19* and *CD79A.* CD19$^+$ TAMs, CD19$^-$ TAMs and B cells were clustered (**Figure 3B**). We next characterized the transcriptional profiles of the three types of cells (**Figure 3C**). CD19$^+$ TAMs exhibited a distinctive gene signature, among which ribonuclease 1 (*RNASE1)*, selenoprotein P (*SELENOP)* and apolipoprotein E *(APOE)* are closely associated with cell metabolism. CD19$^-$ TAMs expressed high levels of genes encoding small calcium-binding proteins in the S100 family including *S100A9*, *S100A8* and *S100A4*. Interestingly, CD19$^+$ TAMs and B cells both express high levels of genes (*IGHG1*, *IGLC1* and *IGKC*) encoding immunoglobin proteins. Furthermore, CD19$^+$ TAMs appeared to be in proximity to M2-like macrophages, while CD19⁻ TAMs were close to M1 macrophages to some extent, although no definite associations were found (**Figures 3D and S4B**). Genes in metabolic pathways and lysosome were significantly enriched in CD19⁺ TAMs (**Figure 3E**). Specifically, CD19⁺ TAMs had higher expression levels of genes associated with mitochondrial functions and higher mitochondria scores (**Figures 3F and 3G**). The gene expression profiling analysis demonstrated that CD19⁺ macrophages are a subgroup of cells distinct from CD19⁻ macrophages and B cells.

Based on the scRNA-seq data analysis, we identified a 10-gene signature capable of distinguishing between CD19⁺ and CD19⁻ macrophages. This signature includes five genes specific to CD19⁺ TAMs (*RNASE1, APOE, APOA2, APOC1, SELENOP*) and five genes specific to CD19⁻ TAMs (*S100A9, S100A8, S100A4, FCN1, VCAN*) (**Figure S4C**). The mRNA expression profiles of these 10 genes effectively differentiate CD19⁺ TAMs from CD19⁻ TAMs, aligning well with their protein levels determined by flow cytometry (**Figures S4D and S4E**). This 10-gene signature was also applied to analyze two public scRNA-seq datasets (Bioproject: PRJCA010606 and PRJCA007744) (**Figures S5A, S5B, S5C and S6A, S6B, S6C**). The defined CD19⁺ TAMs are highly enriched in HCC compared to normal adjacent tissues and show a higher level of mitochondrial activity than CD19⁻ TAMs (**Figures S5D, S5E and S6D, S6E**).

**CD19⁺ TAMs are highly proliferative and immunosuppressive**

Given the high mitochondria score of CD19⁺ TAMs in HCC, we next analyzed the metabolic activity of mitochondria based on the expression levels of oxidative phosphorylation (OXPHOS)-related genes, including NADH dehydrogenase members, cytochrome c oxidase subunit members, and ATP synthase membrane subunit members in CD19$^+$ TAMs and CD19$^-$ TAMs. Overall, a majority of these genes was significantly up-regulated in CD19$^+$ TAMs (**Figure 4A**). Seahorse analysis of cell metabolism showed remarkably enhanced mitochondrial respiration of CD19$^+$ macrophages, including both basal and maximal respiratory capacities (**Figure 4B**). The energetic feature of CD19$^+$ macrophages may lead to the alterations of their functions and cell activities. First, CD19$^+$ TAMs showed much compromised phagocytotic activity (**Figure 4C**). Second, we found that the proportion of CD19$^+$ TAMs was significantly higher in large tumors than in small ones in an HCC syngeneic mouse model (**Figure S7A**), consistent with their clinical relevance in human HCC. CD19$^+$ TAMs showed higher levels of Ki-67 expression than CD19$^-$ TAMs from the same tumor, both in mice and patients (**Figures S7B and S7C**). The enhanced proliferation of CD19$^+$ TAMs was confirmed using an orthotopic mouse model and 5-ethynyl-2′-deoxyuridine (EdU) assays (**Figure S7D**). Therefore, we reasoned that abundance of CD19$^+$ TAMs in tumors, especially large ones, is likely due to their high proliferation rate.

As CD19$^+$ TAMs are a biomarker for poor clinical outcomes of anti-PD-1 immunotherapy in patients with HCC (**Figures 2D and 2E**), it was not of much surprise to see increased PD-L1 levels in CD19$^+$ TAMs compared to CD19$^-$ TAMs (**Figure 4D**). During the HCC metabolite analysis[6], we were particularly interested in adenosine, a primary immunosuppressive metabolite present in high levels in the TME of most solid tumors[19]. In the metabolite data from the analysis using liquid chromatography with tandem mass spectrometry (LC-MS/MS), we observed significant associations between the concentrations of adenosine diphosphate (ADP), adenosine monophosphate (AMP), and adenosine and the abundance of CD19$^+$ TAMs in human HCC tumors (**Figure S7E**). Factors that contribute to adenosine production include hypoxia, high cell turnover, and particularly, the expression of CD73, a 5′-nucleotidase (NT5E) that converts autocrine and paracrine danger signals of ATP to anti-inflammatory adenosine[20]. Given its pivotal role in regulating the ADP/AMP/adenosine conversion[21], CD73 has been an emerging therapeutic target for immune checkpoint blockade[22, 23], and numerous CD73-targeted antibodies and small-molecule inhibitors are currently undergoing clinical testing for cancer therapy[24]. As expected, we observed significantly elevated levels of CD73 in CD19$^+$ TAMs (**Figure 4D**). Biochemical assay verified a high 5′-nucleotidase activity of CD73 in the CD19$^+$ TAMs (**Figure 4E**). As an indicator of their immunosuppressive functions, CD19$^+$ macrophages significantly inhibited the proliferation of total T cells, CD4$^+$ T cells, and CD8$^+$ T cells when these cells were isolated from the clinical tumor samples and co-cultured in vitro in mixed lymphocyte reaction (MLR) assays (**Figure 4F**).

**PAX5 is a master regulator for CD19$^+$ macrophages**

CD19 is important for B cell development and differentiation[25]. Therefore, it is intriguing to determine whether CD19 expression is essential for the immunosuppressive activity of CD19$^+$ TAMs. To this end, *CD19$^{\triangle M\varphi}$* (*CD19$^{flox/flox}$;Lyz-Cre*) mice were generated to establish orthotopic mouse HCC tumor models with macrophage-specific knockout of CD19. Unexpectedly, no differences in orthotopic tumor growth were observed between *CD19$^{\triangle M\varphi}$* mice and their littermate controls (**Figure 5A**), although CD19 expression was depleted in literally all the macrophages in the *CD19$^{\triangle M\varphi}$* mice (**Figure 5B**). These results suggest that CD19 can only be used as a biomarker for the CD19$^+$ macrophages, but functionally dispensable for this subgroup of macrophages in the tumor. Among upstream regulators of CD19, PAX5 is the key transcription factor that governs the expression of CD19[26, 27]. We found that CD19$^+$ TAMs indeed had up-regulated expression of PAX5 at both mRNA and protein levels (**Figures 5C, 5D, and 5E**). Thus, we reasoned that PAX5 may play a role in the CD19$^+$ macrophages. Exogenous overexpression of *PAX5* in human THP-1 cells (a model for macrophage-related cell activities) induced the expression of CD19, CD73 and PD-L1 (**Figures 5F and5G**). Similar results were observed in mouse immortalized bone marrow-derived macrophages (iBMDMs) (**Figure 5H**). Functionally, overexpression of *PAX5* impaired phagocytosis in the mouse iBMDMs (**Figure S8A**). Collectively, PAX5 appears to be a master regulator for the essential features of CD19$^+$ macrophages in the tumor.

In the human THP-1 cells, overexpression of *PAX5* also profoundly enhanced mitochondrial respiration in terms of basal respiration, maximal respiration and ATP production (**Figure 5I**). Electron microscopy analysis demonstrated that the number of mitochondria was much higher in the THP-1 cells with *PAX5* overexpression (**Figure 5J**). The increased number of mitochondria and decreased mitochondrial superoxide levels were detected by flow cytometry in these cells (**Figure 5K**). Importantly, PAX5 significantly enhanced the expression of PGC-1α, a potent inducer of mitochondrial biogenesis (**Figure S8B**)[28]. Additionally, *PAX5*-overexpressing THP-1 cells presented a higher level of adenosine in the culture medium than control cells, consistent with increased CD73 activity (**Figures S8C and S8D**). Since cellular metabolism is tightly related to macrophage functions[15], we asked whether regulation of mitochondrial metabolism affected CD73 activity in CD19$^+$ macrophages. Antimycin, an OXPHOS inhibitor, dramatically impaired CD73 activity (**Figure S8D**). Induction of reactive oxide species by diamide, rather than their removal by N-acetyl-cysteine (NAC), also inhibited CD73 activity. By contrast, blocking CD19 with neutralizing antibodies or inhibiting glycolysis with 2-deoxy-d-glucose (2-DG) had no effect on the CD73 activity (**Figure S8D**). Therefore, mitochondrial metabolism, but not CD19-associated cell signaling regulate the functions of CD19$^+$ macrophages in the tumor. Consistent with the cell studies, conditional knocked out of *Pax5* in mouse macrophages (*Pax5$^{flox/flox}$;Lyz-Cre*, named as *Pax5$^{\triangle M\varphi}$*) drastically inhibited the growth of orthotopically injected

Hepa1-6-derived tumors (**Figures 5L,S8E, S8F, and S8G**). As expected, depletion of PAX5 efficiently decreased the protein levels of CD19, CD73, and PD-L1 in the macrophages of these mice (**Figure S8H**). Collectively, these results suggested that PAX5, instead of CD19, plays a central role in the regulation of CD19$^+$ macrophages in the tumor.

**PAX5 induces PD-L1 and CD73 in a post-transcriptional manner**

Interestingly, *PAX5* overexpression in human THP1 cells did not significantly induce the mRNA levels of *PD-L1* and *CD73* (**Figure 6A**), indicating post-transcriptional regulation of PAX5 in CD19$^+$macrophages. We performed CUT&Tag (**C**leavage **U**nder **T**argets and **Tag**mentation) and RNA-seq assays on human THP-1 cells with or without *PAX5* overexpression and identified global expression changes of genes that are transcriptionally regulated by PAX5 (**Figure 6B**). Among these changes, genes associated with lysosome pathways were the second most significantly enriched, behind the genes involved in phagocytosis (**Figure 6C**), suggesting that PAX5 may control target protein degradation via lysosome. Heat shock cognate 71-kDa protein (HSC70) is a chaperon protein that brings target proteins to lysosomes for degradation in cells[29]. We identified three KFERQ-like motifs, which are potential binding sites to HSC70, in both PD-L1 and CD73 (**Figure 6D**). Coimmunoprecipitation assays validated protein interactions between HSC70 and PD-L1 or CD73 (**Figure 6E**). In lysosome function assays, we observed that lysosomes were significantly inhibited upon *PAX5* overexpression (**Figures 6F and S6G**). Meanwhile, both CD73 and PD-L1 were induced when *PAX5*-overexpresing THP-1 cells were treated with inhibitors of lysosomal proteolysis (NH$_4$Cl and leupeptin) (**Figure 6H**).

Because decreased number of lysosomes were seen in the *PAX5*-overexpressing

THP-1 cells, we speculated that PAX5 may also regulate the activity of TFEB (Transcription Factor EB), a master regulator of lysosomal biogenesis (**Figure S9A**)[30], in the nucleus. We found that nuclear TFEB levels were dramatically diminished in *PAX5*-overexpressing cells (**Figures 6I and S9B**). Knockdown of TFEB inhibited lysosomal functions and increased PD-L1 and CD73 protein levels, similar to the effects of *PAX5* overexpression (**Figures S9C, S9D, and S9E**). Overexpression of TFEB offset the effects of *PAX5* overexpression, leading to reduced levels of PD-L1 and CD73 in the *PAX5*-overexpressing cells (**Figures S9F and S9G**). We sought to delineate the mechanism by which PAX5 regulates the nuclear translocation of TFEB. TFEB subcellular translocation is regulated primarily by $Ca^{2+}$ signaling or mTOR-mediated phosphorylation[31, 32]. While no change in mTOR expression was observed in the *PAX5*-overexpressing cells (data not shown), we reasoned that increasing calcium uptake by mitochondria, due to their up-regulated biogenesis in CD19$^+$ TAMs, may impact $Ca^{2+}$ signaling in the cytosol[32]. Using a fluorescent probe, we found an increased level of mitochondrial $Ca^{2+}$ and, accordingly, a decreased level of cytosolic $Ca^{2+}$ in *PAX5*-overexpressing THP-1 cells (**Figure 6J**). Treatment with IACS-010759 (an OXPHOS inhibitor) or ionomycin (a $Ca^{2+}$ ionophore) significantly increased the cytosolic level of $Ca^{2+}$ and thus promoted the nuclear level of TFEB (**Figures S9H and S9I**). These results suggested that PAX5 modulates the $Ca^{2+}$ shuttling between mitochondria and the cytosol, consequently inhibiting the nuclear translocation of TFEB and lysosomal activity, which eventually increases the levels of CD73 and PD-L1 by inhibiting their lysosomal degradation.

**Blocking CD19$^+$ TAMs enhances the efficacy of immune checkpoint blockade therapy**

Given the role of CD19$^+$ TAMs in HCC progression, we wanted to determine whether they could be a therapeutic target in treating HCC. In a syngeneic HCC tumor model, we first depleted macrophages in mice using clodronate liposomes every other day for four doses (**Figures S10A, S10B, and S10C**) prior to mouse Hepa1-6-derived tumor models establishment **(Figure S10D)**[33, 34]. Control or *Pax5*-overexpressing iBMDMs were then infused into these mice to test their effects on tumor growth (**Figure S10D**). As expected, *Pax5*-overexpressing iBMDMs promoted greater tumor growth than wild-type iBMDMs (**Figure S10E**). In another orthotopic mouse model without systemic depletion of macrophages, *Pax5*-overexpressing iBMDMs also imposed a profound pro-tumor effect (**Figure 7A**). Immunohistochemical analysis revealed that *Pax5*-overexpressing iBMDMs limited the tumor infiltration of $CD8^+$ T cells (**Figure 7B**), indicating an immunosuppressive TME.

CD19$^+$ TAMs create an immunosuppressive tumor microenvironment with increased levels of PD-L1 and CD73. We next tested whether the combination of PD-L1 and CD73 blockade had a synergistic effect on tumor control (**Figure 7C**). Anti-CD73 neutralizing antibodies or a small-molecule inhibitor of CD73 significantly enhanced the anti-tumor effect of anti-PD-L1 antibodies in an orthotopic HCC mouse model (**Figure 7D**). These effects are solely dependent on PAX5 in the CD19$^+$ macrophages, as PD-L1 and CD73 blockade did not affect tumor progression in the $Pax5^{\triangle M\varphi}$ mice (**Figure S10F**). In the presence of anti-PD-L1 antibodies, CD73 blockade significantly increased the tumor infiltration of $CD8^+$ T cells, as well as $CD4^+$ T cells and natural killer cells. Tumor cell proliferation was greatly suppressed, as indicated by reduced Ki-67 staining (**Figures S11**).

Because induced mitochondrial respiration is critical for the functions of CD19$^+$ TAMs, we tested whether inhibiting mitochondrial respiration of CD19$^+$ TAMs had a similar anti-tumor effect. The OXPHOS inhibitor IACS-010759 exhibited an improved anti-tumor effect when combined with anti-PD-L1 antibodies in the orthotopic mouse HCC tumor model (**Figure 7E**). The combination of PD-L1 blockade and OXPHOS inhibition also significantly promoted tumor infiltration of immune cells, including CD4+ T cells and CD8$^+$ T cells (**Figures S12**). Collectively, therapeutic approaches targeting CD73 or OXPHOS in CD19$^+$ TAMs show great promise for improving the anti-tumor efficacy of immunotherapy for HCC.

**Discussion**

TAMs play a crucial role in shaping the immunosuppressive tumor microenvironment. However, the heterogeneity of TAMs and the mechanisms by which they contribute to tumor progression remain elusive. In this study, we identified CD19$^+$ macrophages enriched in a number of human tumor tissues. A high density of CD19$^+$ macrophages within tumors is associated with poor clinical outcomes and reduced response to immunotherapy in patients with HCC. It was previously known that B cell progenitors, and even mature B cells, can be induced to adopt the gene expression patterns, morphology, and functions of macrophages[35]. Overexpression of transcription factor C/EBP in differentiated B cells leads to their rapid reprogramming into macrophages by inhibiting the B cell commitment transcription factor PAX5 and downregulation of its target CD19[35]. Interestingly, CD19$^+$ macrophages retain high expression levels of PAX5 and CD19. Despite its essential role in B cell differentiation and signaling, CD19 itself appears to be dispensable for the pro-tumor functions of CD19$^+$ TAMs, at least in HCC. Depletion of CD19 in our preclinical mouse models had no notable effect on tumor growth (**Figure 5A**). The abundance of CD19$^+$ macrophages was also not correlated with the numbers of B cells in the tissues (**Figure S1F**). In addition, cell morphology and size analysis suggest that CD19$^+$ TMAs is not likely to originate from a B cell lineage (**Figures 1C and S1C**). Further studies are warranted to determine whether CD19+ TAMs are recruited from circulating non-resident macrophages or functionally switched from tissue-resident macrophages.

PAX5 is a deciding regulator for B-lineage commitment[36], and is usually undetectable in macrophages. However, PAX5 remains as a key player that drives immunosuppressive functions in the CD19$^+$ TAMs. *PAX5*-overexpressing macrophages phenocopied CD19$^+$ macrophages with all known characteristics, including enhanced oxidative phosphorylation, increased proliferative capacity, compromised phagocytic activity, and up-regulated levels of PD-L1 and CD73. However, CD19 does not seem to be transactivated in CD19$^+$ macrophages as it does in B cells. Instead, PAX5 induces the CD19 protein level in the macrophages by inhibiting its lysosomal degradation. This differential regulation may distinguish signaling networks between B cells and macrophages. It is now unclear how PAX5 is highly expressed in CD19$^+$ macrophages. This study is the first report showing a potential lineage crossing between B cells and macrophages in the tumor microenvironment. We observed varying levels of CD19$^+$ macrophages in the circulation and organs of both humans and mice, indicating the possible existence of these cells since embryonic development. This suggests that lineage restriction during hematopoietic stem cell differentiation may not be as stringent as previously thought, allowing for the generation of a small population of "hybrid cells."

CD19$^+$ TAMs are immunosuppressive due to their high levels of PD-L1 and CD73. CD73 expression in solid tumors has been identified as an independent biomarker of poor prognosis in clinic[37, 38]. Recent evidence has indicated that CD73 is selectively expressed in a diverse range of immune cell types, including T cells, natural killer cells, and macrophages[39]. The expression of CD73 on macrophages is closely associated with M2 polarization and resistance to anti-PD-1 antibodies[21, 40]. Our study demonstrated that PAX5 increases the levels of CD73 and PD-L1 by compromising lysosomal activity due to decreased nuclear translocation of TFEB in CD19$^+$ macrophages. The elevated activity of CD73 in these macrophages leads to increased adenosine production, which supports cancer cell proliferation in the tumor microenvironment. Given the hyperactivity of OXPHOS in CD19$^+$ TAMs, the combined inhibition of CD73 and OXPHOS, along with immune checkpoint blockade, presents a promising therapeutic approach for treating HCC and potentially other tumors enriched with CD19$^+$ TAMs.

In this study, we explored potential therapeutic strategies to deplete or inhibit CD19+ TAMs, including CD19-targeted CAR-T cells, CD73 blockade, and OXPHOS inhibitors. CD19-targeted CAR-T therapies have consistently demonstrated high antitumor efficacy in both pediatric and adult patients with relapsed B-cell acute lymphoblastic leukemia, chronic lymphocytic leukemia, and B-cell non-Hodgkin lymphoma[41]. However, their clinical application in non-hematologic solid tumors remains to be established[42, 43]. Our results from a preclinical HCC model indicate significant translational potential for CD19-targeted CAR-T therapy in treating solid tumors with high levels of CD19+ TAMs. CD73 blockade has demonstrated antitumor effects in preclinical experiments by eliciting an effective immune response, including enhanced NK cell activity, improved CD4+ and CD8+ T cell function, and elevated levels of proinflammatory cytokines[22]. Beyond macrophages, CD73 expression on tumor cells and Tregs also suppresses antitumor immunity[21]. Thus, targeting CD73 may exert multifaceted effects in cancer treatment. Finally, OXPHOS inhibitors have become a focus in cancer therapeutics, enhancing treatment responses in various cancers, including melanomas, lymphomas, colorectal cancers, leukemias, and pancreatic ductal adenocarcinoma[44, 45]. However, the clinical efficacy of OXPHOS inhibitors is influenced by the complex TME. Our study suggests that OXPHOS inhibition may enhance clinical efficacy by targeting and suppressing CD19+ TAMs in cancer immunotherapy.

In summary, we identified a new subgroup of CD19+ TAMs, which show immunosuppressive capacity and promoted HCC progression. Mechanistically, CD19+ TAMs are primarily driven by PAX5 and have up-regulated PD-L1 and CD73 levels. Elimination of these macrophages by CD19-specific CAR-T cells or functional inhibition of them by inhibiting CD73 or OXPHOS sensitized HCC to immunotherapy.

**Acknowledgements**

This work was supported by National Key Research & Development Program (No. 2020YFA0804300), National Natural Science Foundation of China (Nos. 82071865, 82403723, 81871320, 32321002, 82188102, 92359304), Zhejiang Provincial Natural Science Funds (Nos. LR20H160002, HDMD22H319373), Zhejiang Provincial Key Research & Development Program (No. 2021C03063), Zhejiang Provincial Medical and Health Technology Project (No. WKJ-ZJ-2403), and Zhejiang Provincial Traditional Chinese Medicine Science and Technology Project (GZY-ZJ-KJ-23025). We thank Jianfeng Wang from the Zhejiang Provincial Key Laboratory of Pancreatic Disease for sample collection. Dr. Qi Zhang also gratefully acknowledges the support of K.C.Wong Education Foundation.

**Author contributions**

T.L. and Q.Z. conceived the project. Q.Z., J.W. and J.S. designed the experiments. Q.Z., J.W., Y.Z., J.Y., M.Y., J.Z., J.W., J.W., and Z.H. performed most of the experiments under the supervision of T.L., X.B, X.L and P.X. R.Z., Y.L, X.L., and J.H. performed the bioinformatic analysis. Q.Z. and J.W. wrote the manuscript and the other authors made critical revisions.

**Competing Interests statement**

The authors declare no competing interests.

**Lead contact**

Further information and requests for resources and reagents should be addressed to Qi Zhang (qi.zhang@zju.edu.cn).

**Materials availability**

Plasmids, cell lines and mice generated in this study will be made available with a completed Materials Transfer Agreement by the lead contact upon request.

**Data and code availability**

All data generated or analyzed during this study are included in this manuscript (and its supplementary information files). Human sequencing data were deposited in Genome Sequence Archive (GSA) (GSA: HRA008143). All processed sequencing data are available via https://www.scidb.cn/en/s/b63uAr. Any additional information required to reanalyze the data reported in this paper is available from the lead contact upon request.

**Reference**

1. Sung H, Ferlay J, Siegel RL, Laversanne M, Soerjomataram I, Jemal A, Bray F: (2021).Global Cancer Statistics 2020: GLOBOCAN Estimates of Incidence and Mortality Worldwide for 36 Cancers in 185 Countries. *CA: a cancer journal for clinicians* 71(3):209-249.

2. Forner A, Llovet JM, Bruix J: (2012).Hepatocellular carcinoma. *Lancet (London, England)* 379(9822):1245-1255.

3. Gordan JD, Kennedy EB, Abou-Alfa GK, Beg MS, Brower ST, Gade TP, Goff L, Gupta S, Guy J, Harris WP *et al*: (2020).Systemic Therapy for Advanced Hepatocellular Carcinoma: ASCO Guideline. *Journal of Clinical Oncology* 38(36):4317-4345.

4. Vogel A, Meyer T, Sapisochin G, Salem R, Saborowski A: (2022).Hepatocellular carcinoma. *Lancet (London, England)* 400(10360):1345-1362.

5. Marra F, Tacke F: (2014).Roles for chemokines in liver disease. *Gastroenterology* 147(3):577-594.e571.

6. Zhang Q, Lou Y, Yang J, Wang J, Feng J, Zhao Y, Wang L, Huang X, Fu Q, Ye M *et al*: (2019).Integrated multiomic analysis reveals comprehensive tumour heterogeneity and novel immunophenotypic classification in hepatocellular carcinomas. *Gut* 68(11):2019-2031.

7. Kalathil S, Lugade AA, Miller A, Iyer R, Thanavala Y: (2013).Higher frequencies of GARP(+)CTLA-4(+)Foxp3(+) T regulatory cells and myeloid-derived suppressor cells in hepatocellular carcinoma patients are associated with impaired T-cell functionality. *Cancer research* 73(8):2435-2444.

8. Shen KY, Zhu Y, Xie SZ, Qin LX: (2024).Immunosuppressive tumor microenvironment and immunotherapy of hepatocellular carcinoma: current status and prospectives. *Journal of hematology & oncology* 17(1):25.

9. Noy R, Pollard JW: (2014).Tumor-associated macrophages: from mechanisms to therapy. *Immunity* 41(1):49-61.

10. Binnewies M, Roberts EW, Kersten K, Chan V, Fearon DF, Merad M, Coussens LM, Gabrilovich DI, Ostrand-Rosenberg S, Hedrick CC *et al*: (2018).Understanding the tumor immune microenvironment (TIME) for effective therapy. *Nature medicine* 24(5):541-550.

11. Murray PJ: (2017).Macrophage Polarization. *Annu Rev Physiol* 79:541-566.

12. Mantovani A, Sozzani S, Locati M, Allavena P, Sica A: (2002).Macrophage polarization: tumor-associated macrophages as a paradigm for polarized M2 mononuclear phagocytes. *Trends in immunology* 23(11):549-555.

13. DeNardo DG, Ruffell B: (2019).Macrophages as regulators of tumour immunity and immunotherapy. *Nat Rev Immunol* 19(6):369-382.

14. Wu Q, Zhou W, Yin S, Zhou Y, Chen T, Qian J, Su R, Hong L, Lu H, Zhang F *et al*: (2019).Blocking Triggering Receptor Expressed on Myeloid Cells-1-Positive Tumor-Associated Macrophages Induced by Hypoxia Reverses Immunosuppression and Anti-Programmed Cell Death Ligand 1 Resistance in Liver Cancer. *Hepatology (Baltimore, Md)* 70(1):198-214.

15. Gordon SR, Maute RL, Dulken BW, Hutter G, George BM, McCracken MN, Gupta R, Tsai JM, Sinha R, Corey D *et al*: (2017).PD-1 expression by tumour-associated macrophages inhibits phagocytosis and tumour immunity. *Nature* 545(7655):495-499.

16. Kim HJ, Park JH, Kim HC, Kim CW, Kang I, Lee HK: (2022).Blood monocyte-derived CD169(+) macrophages contribute to antitumor immunity against glioblastoma. *Nature communications* 13(1):6211.

17. Wang R, Li H, Ling C, Zhang X, Lu J, Luan W, Zhang J, Shi L: (2023).A novel phenotype of B cells associated with enhanced phagocytic capability and chemotactic function after ischemic stroke. *Neural regeneration research* 18(11):2413-2423.

18. Kitamura D, Roes J, Kühn R, Rajewsky K: (1991).A B cell-deficient mouse by targeted disruption of the membrane exon of the immunoglobulin mu chain gene. *Nature* 350(6317):423-426.

19. Allard B, Allard D, Buisseret L, Stagg J: (2020).The adenosine pathway in immuno-oncology. *Nature reviews Clinical oncology* 17(10):611-629.

20. Junger WG: (2011).Immune cell regulation by autocrine purinergic signalling. *Nat Rev Immunol* 11(3):201-212.

21. Antonioli L, Pacher P, Vizi ES, Haskó G: (2013).CD39 and CD73 in immunity and inflammation. *Trends in molecular medicine* 19(6):355-367.

22. Zhang B: (2010).CD73: a novel target for cancer immunotherapy. *Cancer research* 70(16):6407-6411.

23. Vijayan D, Young A, Teng MWL, Smyth MJ: (2017).Targeting immunosuppressive adenosine in cancer. *Nature reviews Cancer* 17(12):765.

24. Alcedo KP, Bowser JL, Snider NT: (2021).The elegant complexity of mammalian ecto-5'-nucleotidase (CD73). *Trends in cell biology* 31(10):829-842.

25. Depoil D, Fleire S, Treanor BL, Weber M, Harwood NE, Marchbank KL, Tybulewicz VL, Batista FD: (2008).CD19 is essential for B cell activation by promoting B cell receptor-antigen microcluster formation in response to membrane-bound ligand. *Nature immunology* 9(1):63-72.

26. Tedder TF: (2009).CD19: a promising B cell target for rheumatoid arthritis. *Nature reviews Rheumatology* 5(10):572-577.

27. Chung EY, Psathas JN, Yu D, Li Y, Weiss MJ, Thomas-Tikhonenko A: (2012).CD19 is a major B cell receptor-independent activator of MYC-driven B-lymphomagenesis. *The Journal of clinical investigation* 122(6):2257-2266.

28. Finck BN, Kelly DP: (2006).PGC-1 coactivators: inducible regulators of energy metabolism in health and disease. *The Journal of clinical investigation*

116(3):615-622.

29. Cuervo AM, Mann L, Bonten EJ, d'Azzo A, Dice JF: (2003).Cathepsin A regulates chaperone-mediated autophagy through cleavage of the lysosomal receptor. *The EMBO journal* 22(1):47-59.

30. Settembre C, Di Malta C, Polito VA, Garcia Arencibia M, Vetrini F, Erdin S, Erdin SU, Huynh T, Medina D, Colella P *et al*: (2011).TFEB links autophagy to lysosomal biogenesis. *Science (New York, NY)* 332(6036):1429-1433.

31. Roczniak-Ferguson A, Petit CS, Froehlich F, Qian S, Ky J, Angarola B, Walther TC, Ferguson SM: (2012).The transcription factor TFEB links mTORC1 signaling to transcriptional control of lysosome homeostasis. *Science signaling* 5(228):ra42.

32. Medina DL, Di Paola S, Peluso I, Armani A, De Stefani D, Venditti R, Montefusco S, Scotto-Rosato A, Prezioso C, Forrester A *et al*: (2015).Lysosomal calcium signalling regulates autophagy through calcineurin and TFEB. *Nature cell biology* 17(3):288-299.

33. Van Rooijen N, Sanders A: (1994).Liposome mediated depletion of macrophages: mechanism of action, preparation of liposomes and applications. *Journal of immunological methods* 174(1-2):83-93.

34. Moreno SG: (2018).Depleting Macrophages In Vivo with Clodronate-Liposomes. *Methods in molecular biology (Clifton, NJ)* 1784:259-262.

35. Xie H, Ye M, Feng R, Graf T: (2004).Stepwise reprogramming of B cells into macrophages. *Cell* 117(5):663-676.

36. Hodawadekar S, Yu D, Cozma D, Freedman B, Sunyer O, Atchison ML, Thomas-Tikhonenko A: (2007).B-Lymphoma cells with epigenetic silencing of Pax5 trans-differentiate into macrophages, but not other hematopoietic lineages. *Experimental cell research* 313(2):331-340.

37. Ma XL, Shen MN, Hu B, Wang BL, Yang WJ, Lv LH, Wang H, Zhou Y, Jin AL, Sun YF

*et al*: (2019).CD73 promotes hepatocellular carcinoma progression and metastasis via activating PI3K/AKT signaling by inducing Rap1-mediated membrane localization of P110β and predicts poor prognosis. *Journal of hematology & oncology* 12(1):37.

38. Supernat A, Markiewicz A, Welnicka-Jaskiewicz M, Seroczynska B, Skokowski J, Sejda A, Szade J, Czapiewski P, Biernat W, Zaczek A: (2012).CD73 expression as a potential marker of good prognosis in breast carcinoma. *Applied immunohistochemistry & molecular morphology : AIMM* 20(2):103-107.

39. Beavis PA, Stagg J, Darcy PK, Smyth MJ: (2012).CD73: a potent suppressor of antitumor immune responses. *Trends in immunology* 33(5):231-237.

40. Eichin D, Laurila JP, Jalkanen S, Salmi M: (2015).CD73 Activity is Dispensable for the Polarization of M2 Macrophages. *PloS one* 10(8):e0134721.

41. Cappell KM, Kochenderfer JN: (2023).Long-term outcomes following CAR T cell therapy: what we know so far. *Nature reviews Clinical oncology* 20(6):359-371.

42. The Lancet O: (2021).CAR T-cell therapy for solid tumours. *The Lancet Oncology* 22(7):893.

43. Albelda SM: (2024).CAR T cell therapy for patients with solid tumours: key lessons to learn and unlearn. *Nature reviews Clinical oncology* 21(1):47-66.

44. Ashton TM, McKenna WG, Kunz-Schughart LA, Higgins GS: (2018).Oxidative Phosphorylation as an Emerging Target in Cancer Therapy. *Clinical cancer research : an official journal of the American Association for Cancer Research* 24(11):2482-2490.

45. Yap TA, Daver N, Mahendra M, Zhang J, Kamiya-Matsuoka C, Meric-Bernstam F, Kantarjian HM, Ravandi F, Collins ME, Francesco MED *et al*: (2023).Complex I inhibitor of oxidative phosphorylation in advanced solid tumors and acute myeloid leukemia: phase I trials. *Nature medicine* 29(1):115-126.

**Figure legends**

**Figure 1. CD19⁺ macrophages are enriched in HCC (See also Figure S1).**

(**A**) Visualized tSNE representation of the immune cell subtypes in human HCC tissues, including CD19⁺ TAMs, CD19⁻ TAMs, and B cells (left) and CD19 protein expression levels in these three clusters (right). Clusters are differently color-coded and annotated in the figure. Data were generated from CyTOF. (**B**) Representative immunofluorescence images of CD19⁺ TAMs in HCC tissue and adjacent normal tissue. Scale bar, 50 μm. White arrows indicate CD19⁺ TAMs. (**C**) ImageStream showing CD19⁺ TAMs, CD19⁻ TAMs and B cells in HCC tissue at a single-cell level. ×40 magnification. (**D**) Representative flow cytometry plots showing CD19 expression on TAMs in HCC tissue, adjacent normal tissue, and peripheral blood. (**E**) Multiplex IHC assay was used to identify CD19⁺ macrophages and B cells in formalin-fixed and paraffin-embedded HCC and adjacent normal tissues. Different colors indicate CD14 (yellow), CD68 (green), CD20 (orange), CD19 (red), and DAPI (blue). CD19⁺ macrophages are identified as CD14⁺CD68⁺CD20⁻CD19⁺, whereas B cells are identified as CD14⁻CD68⁻CD20⁺CD19⁺. Scale bars, 50 μm. (**F**) Flow cytometry quantification of CD19⁺ TAMs proportion in different types of cancer. n (hepatocellular carcinoma) = 28, n (renal cancer) =11, n (colorectal cancer) = 11, n (pancreatic ductal adenocarcinoma) = 13, n (breast cancer) = 12, n (gastric cancer) = 6. Data are presented as the mean ± SEM. A paired two-tailed *t*-test was used in **F**. *$P$ < 0.05; ***$P$ < 0.001.****$P$ < 0.0001.

**Figure 2. CD19⁺ TAMs are associated with poor clinical outcome and immunotherapy response (See also Figure S2 and S3).**

(**A**) Kaplan-Meier curves showing overall survival of HCC patients with a high (> median) or low (< median) ratio of CD19⁺ TAMs. Data are derived from immunofluorescence co-staining of human HCC tissue microarray. n = 149. (**B**) Kaplan-Meier curves show overall survival of HCC patients with a high or low TAMs ratio. Data are derived from immunofluorescence co-staining of human HCC tissue microarray. n = 149. (**C**) Correlation analysis between Ki-67⁺ cells and CD19⁺ TAMs density in HCC tissues. Data are derived from immunofluorescence co-staining of human HCC tissue microarray. n = 156. Scale bar, 50 μm; 100 μm. (**D**) Representative images of immunofluorescence co-staining of CD19 and CD68 in human HCC tissues from the responders and non-responders in a cohort of patients treated with anti-PD-1 antibody. Scale bar, 25 μm. (**E**) Representative magnetic resonance images showing a responder and a non-responder before and after anti-PD-1 antibody treatment from (**D**) (top panel). Quantification of total TAMs and CD19⁺ TAMs density in all the responders and non-responders from (**D**) (bottom panel). Tumors are indicated by red circles or arrows. n = 17. PR, partial response; SD, stable disease; PD, progressive disease. (**F**) Splenic CD19⁺ and CD19⁻ macrophages were sorted and separately mixed with 5 × 10⁵ Hepa1-6 cells at the ratio of 1:5, and then injected into the liver of mice. Images and weight of the tumors were showed at 14 days post-inoculation. n = 5 per group. Tumors are indicated by red circles. (**G**) Splenic CD19⁺ macrophages and CD19⁻ macrophages were separately mixed with 5× 10⁵ Hepa1-6 cells at the ratio of 1:5, and then inoculated subcutaneously (*s.c.*) into mice. When tumors reached 35 to 45 mm², mice were received intravenous injection (*i.v.*) with GFP⁺ control CAR-T cells or GFP⁺ anti-CD19 CAR-T cells (1 × 10⁷ cells in 100 μL of PBS). The harvested tumors were analyzed. n = 5 per group. (**H**) Splenic CD19⁺ macrophages and CD19⁻ macrophages were mixed with 5 × 10⁵ Hepa1-6 cells at the ratio of 1:5, and then inoculated subcutaneously in the left and right flanks of *muMT* (lacking mature B cells) or wild-type (WT) mice. When tumors reached 35 to 45 mm$^2$, mice were received intravenous injection with GFP$^+$ control CAR-T cells or GFP$^+$ anti-CD19 CAR-T cells (1 × 10$^7$ cells in 100 μL of PBS). The harvested tumors were analyzed. n = 5 per group. Data are presented as the mean ± SEM. Log-rank test (**A, B**), Pearson correlation test (**C**), unpaired two-tailed *t*-test (**E, F, H**), or paired two-tailed *t* test (**G**) were used for statistical analysis. *$P$ < 0.05; **$P$ < 0.01; ***$P$ < 0.001, NS, not significant.

**Figure 3. CD19$^+$ TAMs display a distinct gene expression profile (See also Figure S4, S5 and S6).**

(**A**) Schematic diagram of scRNA-seq experimental strategy for human HCC tissues. n = 3. (**B**) UMAP plot analysis showing HCC samples (top panel) and cell types in the samples (bottom) as indicated. (**C**) Heatmap showing the top 10 upregulated genes in CD19$^+$ TAMs, CD19$^-$ TAMs and B cells, respectively. (**D**) mRNA expression levels of classical M1 and M2 macrophage marker genes in CD19$^+$ TAMs, compared with CD19$^-$ TAMs. Bubble size and color represent the percentages and relative levels, respectively, of upregulated or downregulated genes as indicated at the x axis. (**E**) Bar chart showing the top-ranked biological pathways in CD19$^+$ TAMs from KEGG enrichment analysis of the scRNA-seq data in (**A**). (**F**) t-SNE plots showing the expression levels of mitochondrion-related genes in CD19$^+$ and CD19$^-$ TAMs. (**G**) Violin plot analysis of mitochondrion-related gene expression levels in CD19$^+$ and CD19$^-$ TAMs. Data are mean ± SEM; An unpaired two-tailed *t*-test was used in (**G**). ****$P$ < 0.0001.

**Figure 4. CD19+ TAMs are highly proliferative and immunosuppressive (See also Figure S7).**

(**A**) Relative expression levels of mitochondrial metabolism-related genes in CD19+ TAMs and CD19- TAMs in human HCC tissue. (**B**) Mitochondrial OCR measurement in CD19+ TAMs and CD19- TAMs (5× 10$^5$ cells per well). n = 3 independent experiments. (**C**) Flow cytometry histograms and quantification showing the phagocytic ability in CD19+ TAMs and CD19- TAMs. n = 3 independent experiments. (**D**) Flow cytometry histograms and quantification showing the levels of PD-L1 and CD73 in CD19+ TAMs and CD19- TAMs. n = 5 independent experiments. (**E**) Normalized 5'-nucleotidase (CD73) specific activity of CD19+ TAMs and CD19- TAMs lysate. n = 3 independent experiments. (**F**) Co-culture of CD19+ TAMs or CD19- TAMs with CD3+ T, CD4+ T, CD8+ T cells at different ratios for 72 h. n = 3 independent experiments. Data are mean ± SEM.; Unpaired two-tailed *t*-test (**B**, **E**), paired two-tailed *t* test (**C**, **D**), or two-way ANOVA test (**F**) were used. *$P < 0.05$; **$P < 0.01$; ***$P < 0.001$.

**Figure 5. PAX5 is a master regulator for CD19+ macrophages (See also Figure S8).**

(**A**) Tumor images and weights of orthotopic HCC model in $CD19^{\triangle M\varphi}$ ($CD19^{flox/flox}$;*Lyz-Cre*) and littermate wildtype mice. n = 4 per group. (**B**) Representative flow cytometrical images and quantification showing the proportion of CD19+ TAMs in the tumors from (**A**). (**C**) *PAX5* mRNA expression level in CD19+ TAMs and CD19- TAMs. (**D**) Flow cytometry histograms and quantification showing protein expression levels of PAX5 in CD19+ TAMs and CD19- TAMs. n = 5 independent experiments. (**E**) Western blotting analysis of PAX5 expression level in CD19+ TAMs and CD19- TAMs. Relative PAX5 blot intensities were shown. (**F**) Representative flow cytometry histograms showing the levels CD19, PD-L1, and CD73 in THP-1 cells with or without *PAX5* overexpression. (**G**) Western blotting analysis of CD19, PD-L1, and CD73 levels in THP-1 cells with or without *PAX5* overexpression. (**H**) Representative flow cytometry histograms showing the levels of CD19, PD-L1, and CD73 in iBMDM with or without *Pax5* overexpression. (**I**) Mitochondrial OCR measurement in THP-1 cells with or without *PAX5* overexpression (5× $10^5$ cells per well). n = 3 independent experiments. (**J**) Electron microscopy images and quantification showing mitochondrias in THP-1 cells with or without *PAX5* overexpression. scale bar: 0.5 μm or 0.2 μm. n = 5 independent experiments. (**K**) Representative flow cytometry histograms showing mitochondrial number (mitotracker), and mitochondrial ROS level (MitoSOX) in THP-1 cells with or without *PAX5* overexpression. (**L**) Tumor images and sizes of orthotopic HCC model in $Pax5^{\triangle M\varphi}$ ($Pax59^{flox/flox}$;*Lyz-Cre*) and littermate control mice. n= 5 per group. Data are mean ± SEM; One-way ANOVA with Welch's correction (**A**), unpaired two-tailed *t*-test (**B,I, J, L**), or paired two-tailed *t* test (**C, D**) were used. *$P < 0.05$; **$P < 0.01$; ***$P < 0.001$; ****$P < 0.0001$; NS, not significant.

**Figure 6. PAX5 induces PD-L1 and CD73 in a post-transcriptional manner (See also Figure S9).**

(**A**) mRNA expression levels of *CD73,* and *PD-L1* in THP-1 cells with or without *PAX5* overexpression. n = 3 independent experiments. (**B**) Venn diagram showing the number and percentage of share genes in Cut&Tag and RNA-seq analyses of THP-1 cells with *PAX5* overexpression. (**C**) WebGestalt analysis of enriched biological pathways in the shared genes in (**B**). (**D**) KFERQ finder V0.8 identified KFERQ-like motifs in PD-L1 and CD73. (**E**) Immunoprecipitation (IP) of HSC70 followed by immunoblotting (IB) analyses for HSC70, PD-L1, and CD73 in THP-1 cells with or without *PAX5* overexpression. (**F**) Representative immunofluorescence images and quantification of Lysotracker (red) in THP-1 cells with or without *PAX5* overexpression or not. Blue color indicates DAPI staining; scale bar, 25 µm. n = 3 independent experiments. (**G**) Representative flow cytometry histograms and quantification showing the lysosome quantity assessed by LysoSensor fluorescence expression in THP-1 cells with or without *PAX5* overexpression. n = 3 independent experiments. (**H**) Western blotting detection of PD-L1 and CD73 in THP-1 cells with *PAX5* overexpression, or treated with inhibitors of lysosomal proteolysis ($NH_4Cl$, 20mM and leupeptin, 100µM), or inhibitor of proteasome (MG132, 5µM). LC3B is used as a marker of lysosomal proteolysis inhibition. (**I**) Western blotting detection of cytosol and nuclear TFEB in THP-1 cells with or without *PAX5* overexpression. (**J**) Fluorescent probe indicates relative $Ca^{2+}$ levels in mitochondria (Rhod2) and cytoplasm (Calbryte™ 630$^{AM}$) in THP-1 cells with or without *PAX5* overexpression. Fluorescence was measured every 30s for 30 minutes. n = 3 repeats. Data are mean ± SEM; An unpaired two-tailed *t*-test (**A, F, G**), or two-way ANOVA test (**J**) were used. $^*P < 0.05$; $^{**}P < 0.01$; $^{***}P < 0.001$; $^{****}P < 0.0001$; NS, not significant.

**Figure 7. Blocking CD19$^+$ TAMs enhances the efficacy of immune checkpoint blockade therapy (See also Figures S10, S11, and S12).**

(**A**) Hepa1-6 cells (5× 10$^5$) were mixed with iBMDMs (with or without *Pax5* overexpression) at a ratio of 1:1, and orthotopically injected into the livers of C57BL/6 mice. Tumor images and sizes were showed at 15 days post-inoculation. Tumor sites were indicated by red dotted line in liver. n = 5 per group. (**B**) Representative immunohistochemistrical images and quantification showing the infiltrated CD8$^+$ T cells in the tumors of (**A**). Scale bar, 50 μm. (**C**) Schematic diagram illustrating targeted therapy of CD73 and PD-L1 for CD19$^+$ TAMs in HCC. Tumors were established by inoculating 5× 10$^5$ Hepa1-6 cells and iBMDMs with *Pax5* overexpression at the ratio of 1:1. Seven days post-inoculation, mice were treated as indicated. i.p, intraperitoneal injections; i.g, oral gavage. (**D**) Tumor images and sizes of (**C**). n = 5 per group. Tumor sites were indicated by red dotted line in the liver. (**E**) Tumor images and sizes in the in vivo experiment targeting OXPHOS and PD-L1 in CD19$^+$ TAM. Tumors were established by inoculating 5× 10$^5$ Hepa1-6 cells and iBMDMs with *Pax5* overexpression at the ratio of 1:1. Seven days post-inoculation, mice were treated as indicated. n = 5 per group. Tumor sites were indicated by red line in the liver. Data are mean ± SEM; Unpaired two-tailed *t*-test (**A, B**), or unpaired one-way ANOVA with Welch's correction (**D, E**) were used. $^*P < 0.05$; $^{**}P < 0.01$.

**Inventory of Supplemental Information**

・**Supplementary Methods**

・**Figures S1-S12 and Figure S legends.**

・**Supplementary Table S1-8**.

**Supplementary Table S1**. Patients' clinical information, related to **Figure 1**, **Figure 2**, and **Figure S2**.

**Supplementary Table S2**. Sequences of primer for transgenic mice genotyping, related to **Figure 5**. **and Figure S8**.

**Supplementary Table S3**. CyTOF antibody panel, related to **Figure 1** and **Figure S2**.

**Supplementary Table S4**. CD19 CAR sequences, related to **Figure 2** and **Figure S3**.

**Supplementary Table S5**. qRT-PCR Primer sequences, related to **Figure 4** and **Figure 6**.

**Supplementary Table S6**. Differentially expressed genes (DEGs) in CD19$^+$ TAMs, CD19$^-$TAMs, and B cells (data from sc-RNA seq), related to **Figure 3**.

**Supplementary Table S7**. Classical M1 and M2 macrophage marker genes expression in CD19+ TAMs and CD19- TAMs, related to **Figure 3**.

**Supplementary table 8.** Key resources table.

Fig. 1

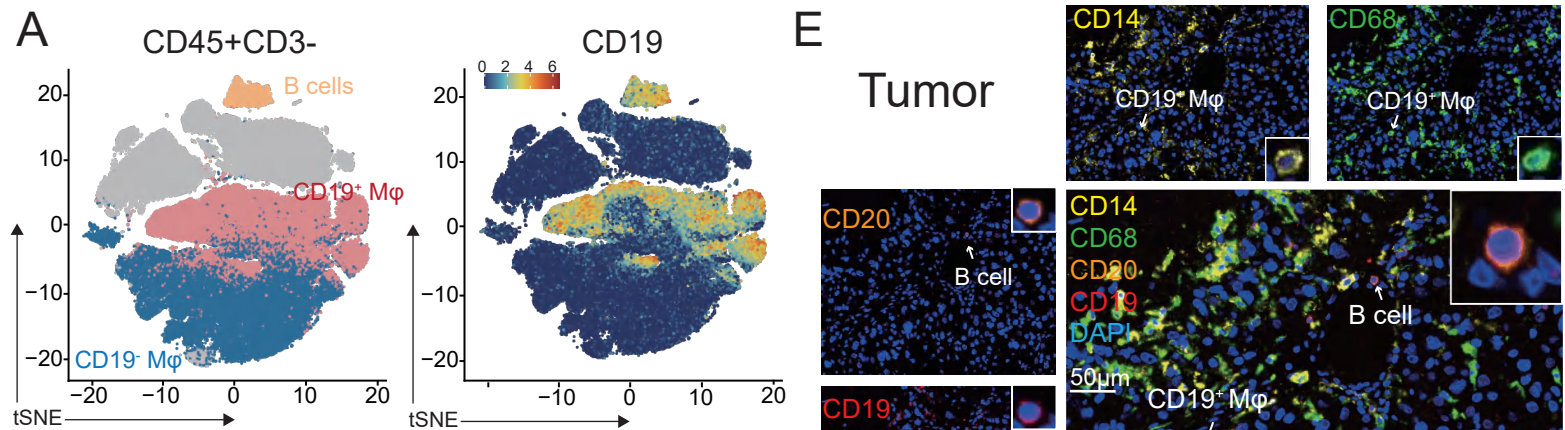
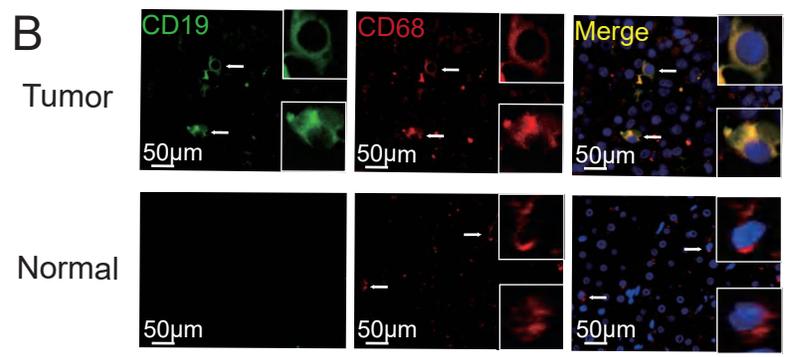
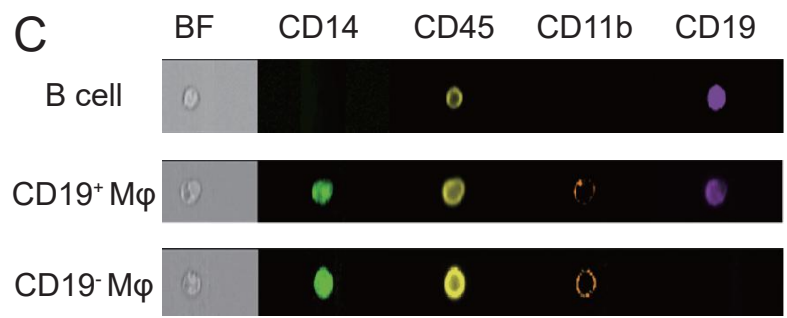
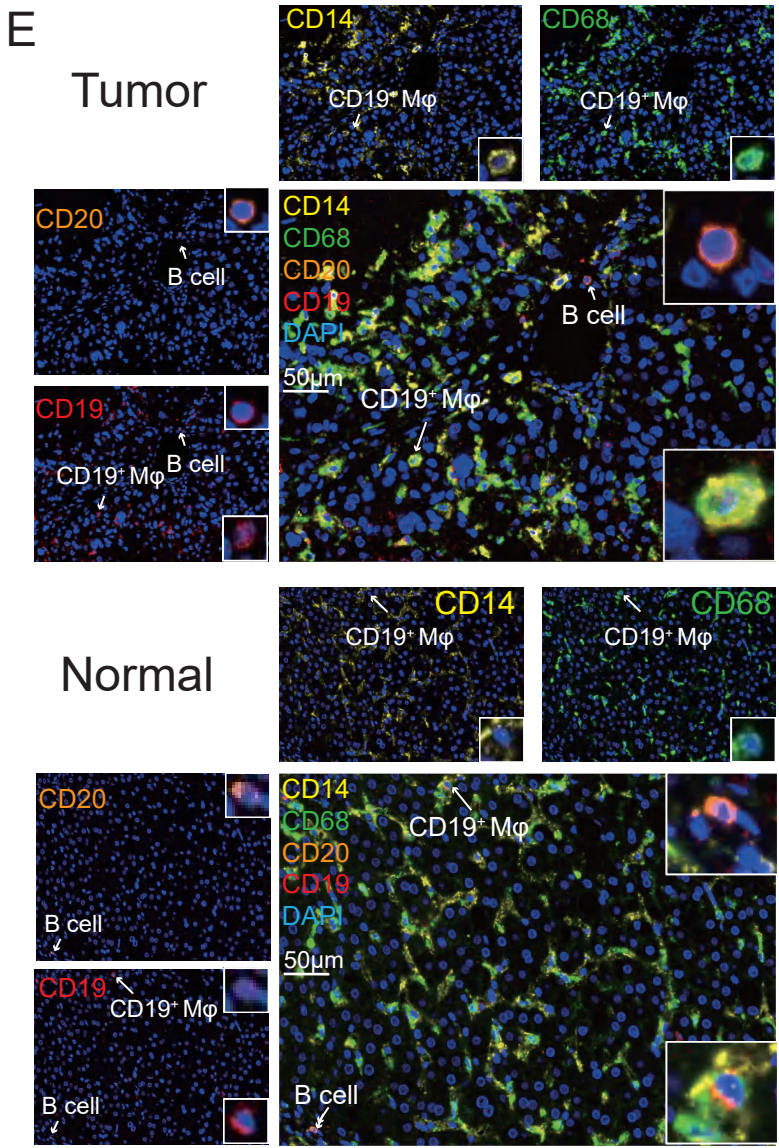
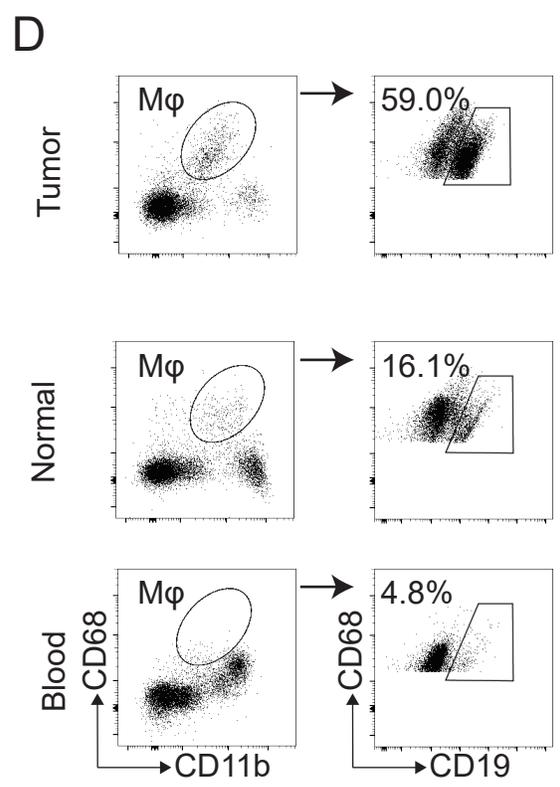
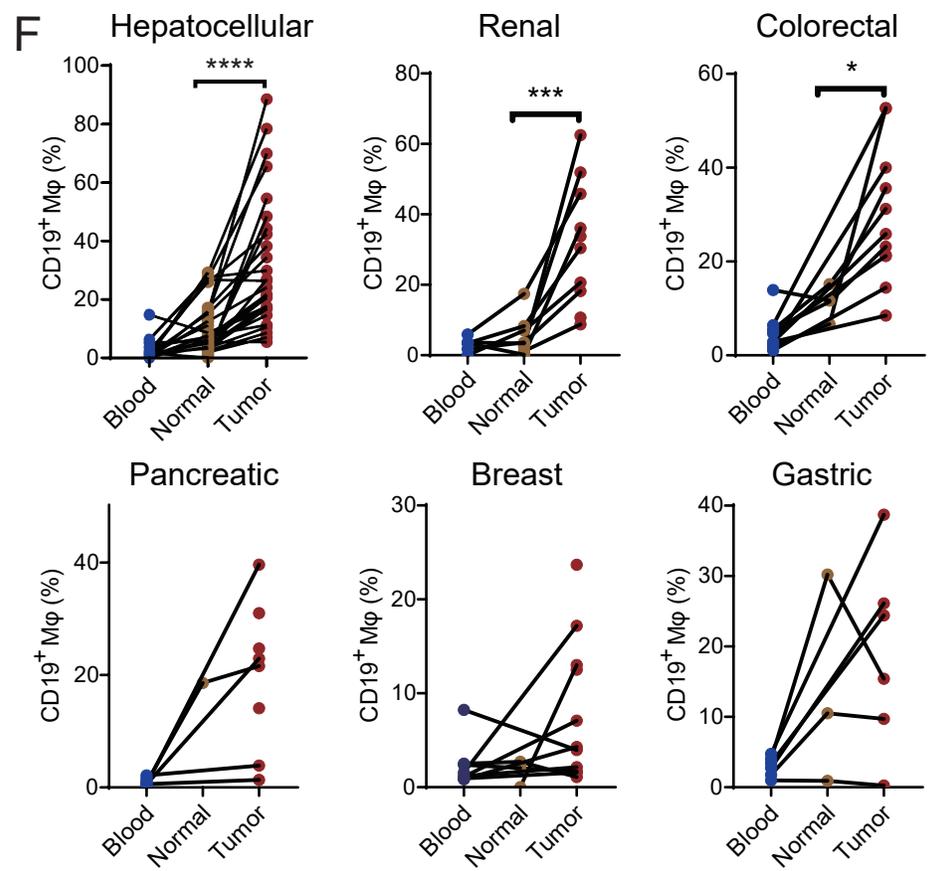

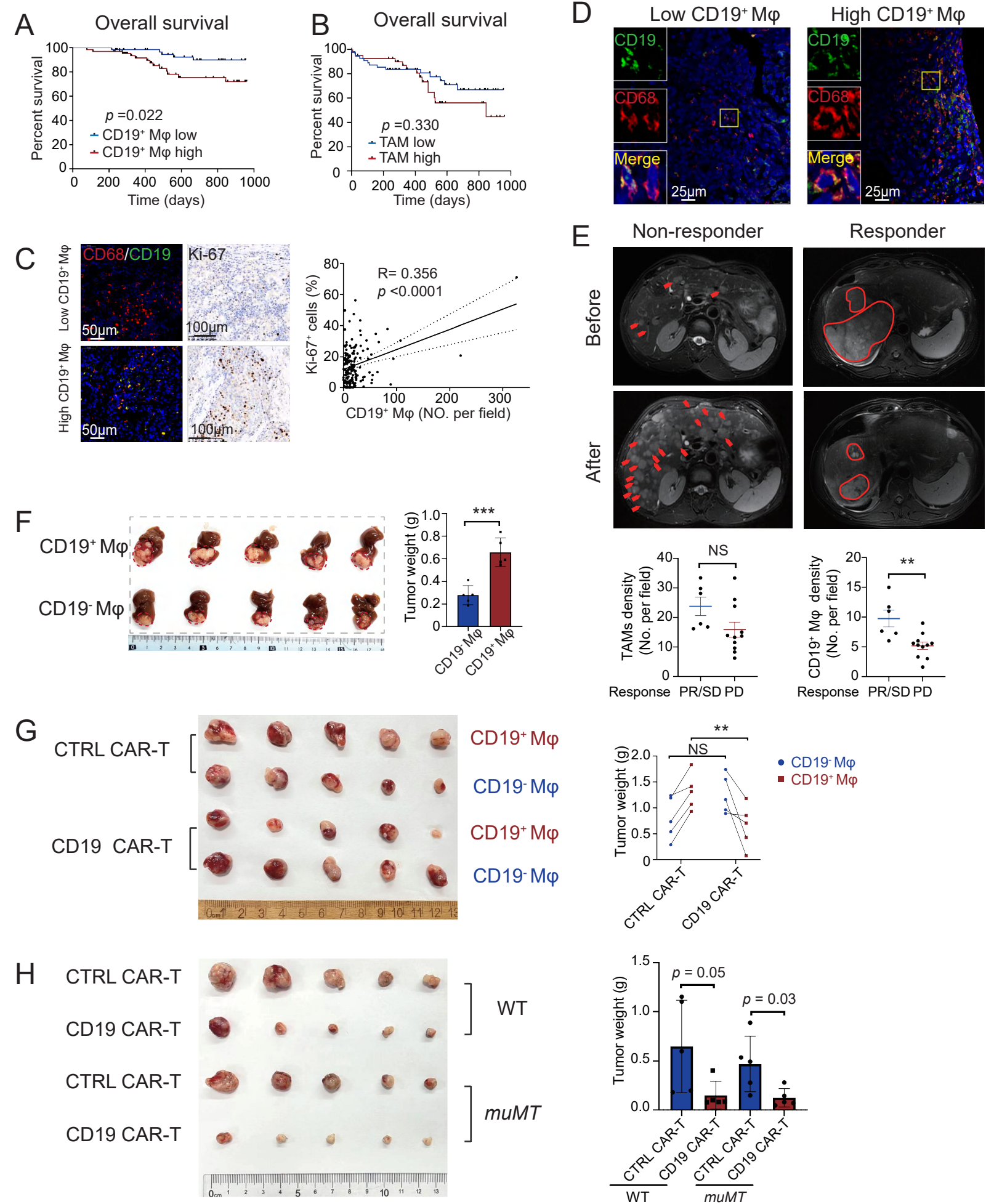

Fig. 2

Fig.3

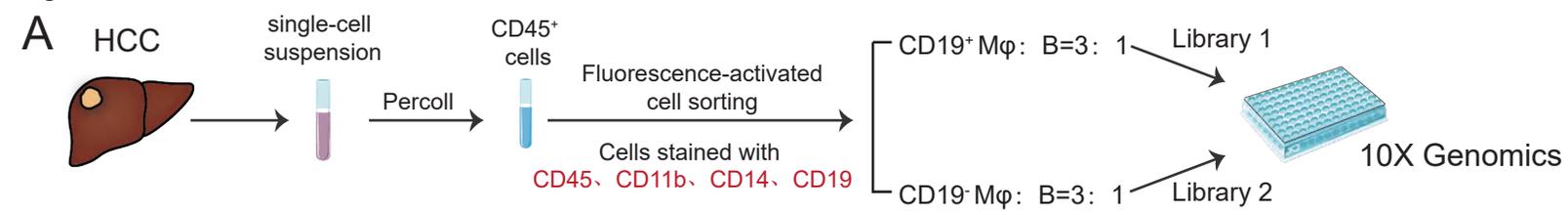
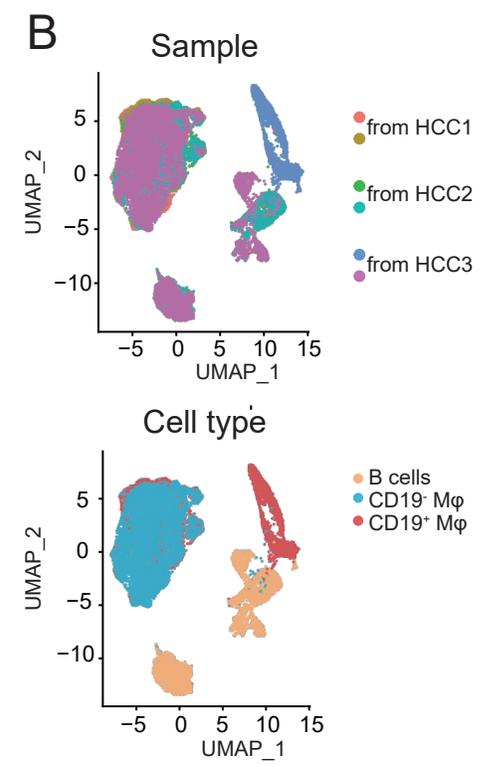
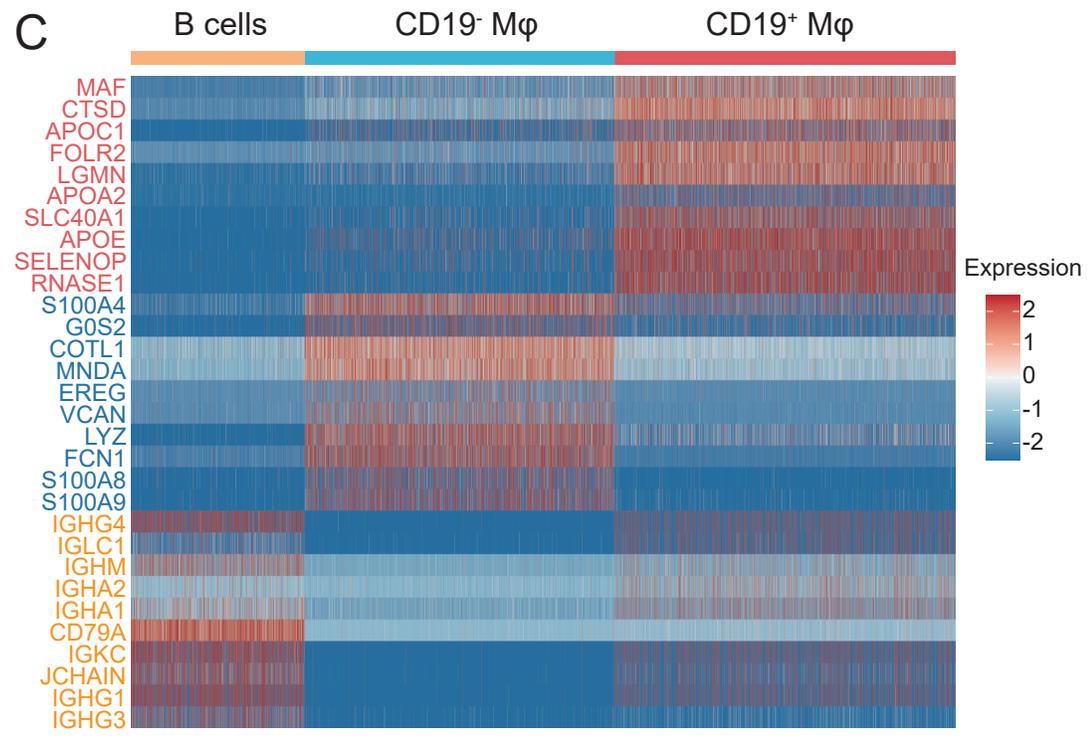
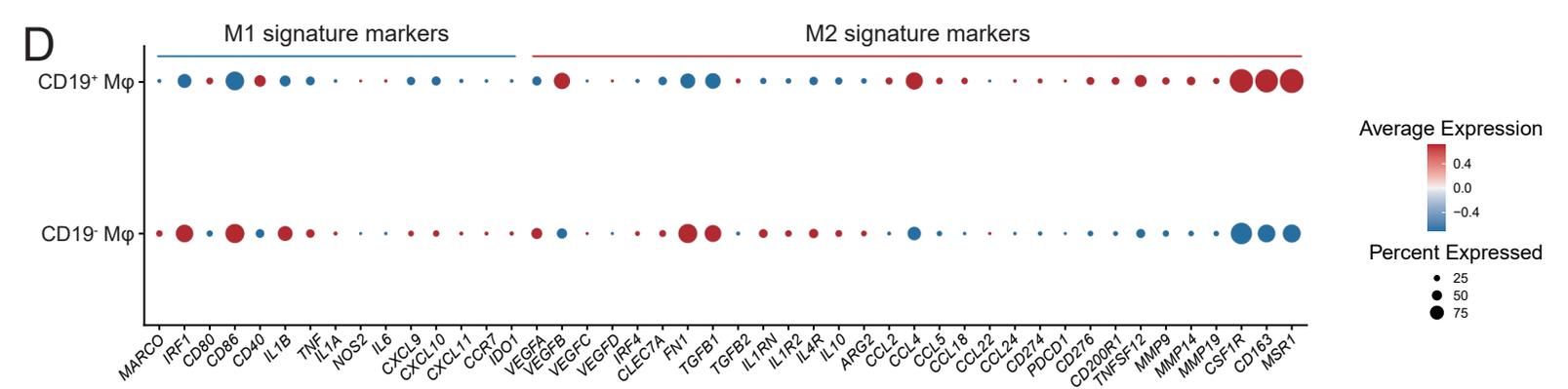
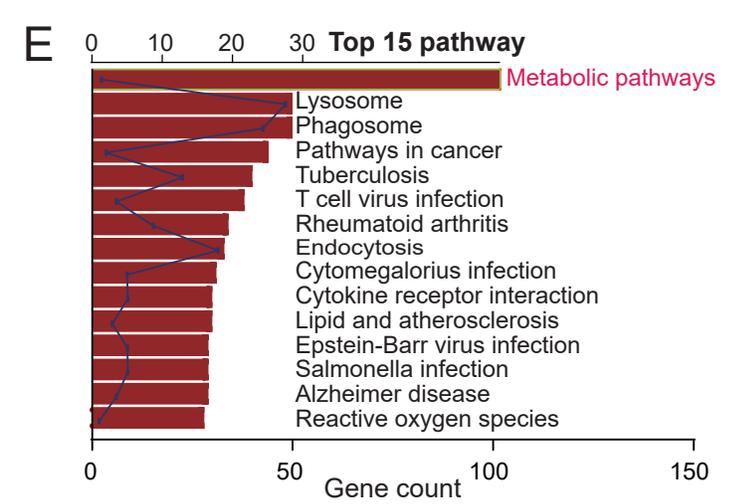
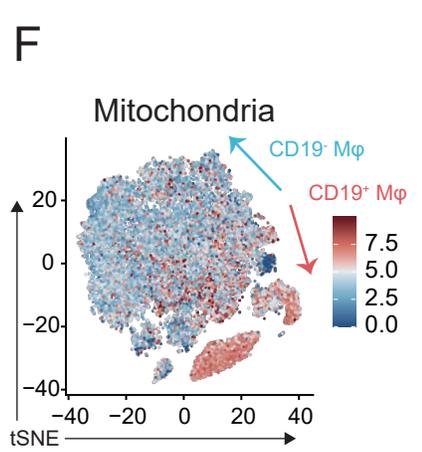
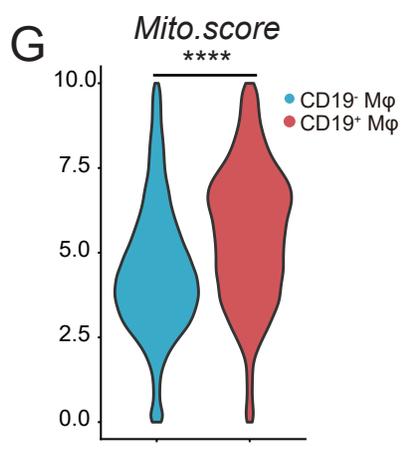

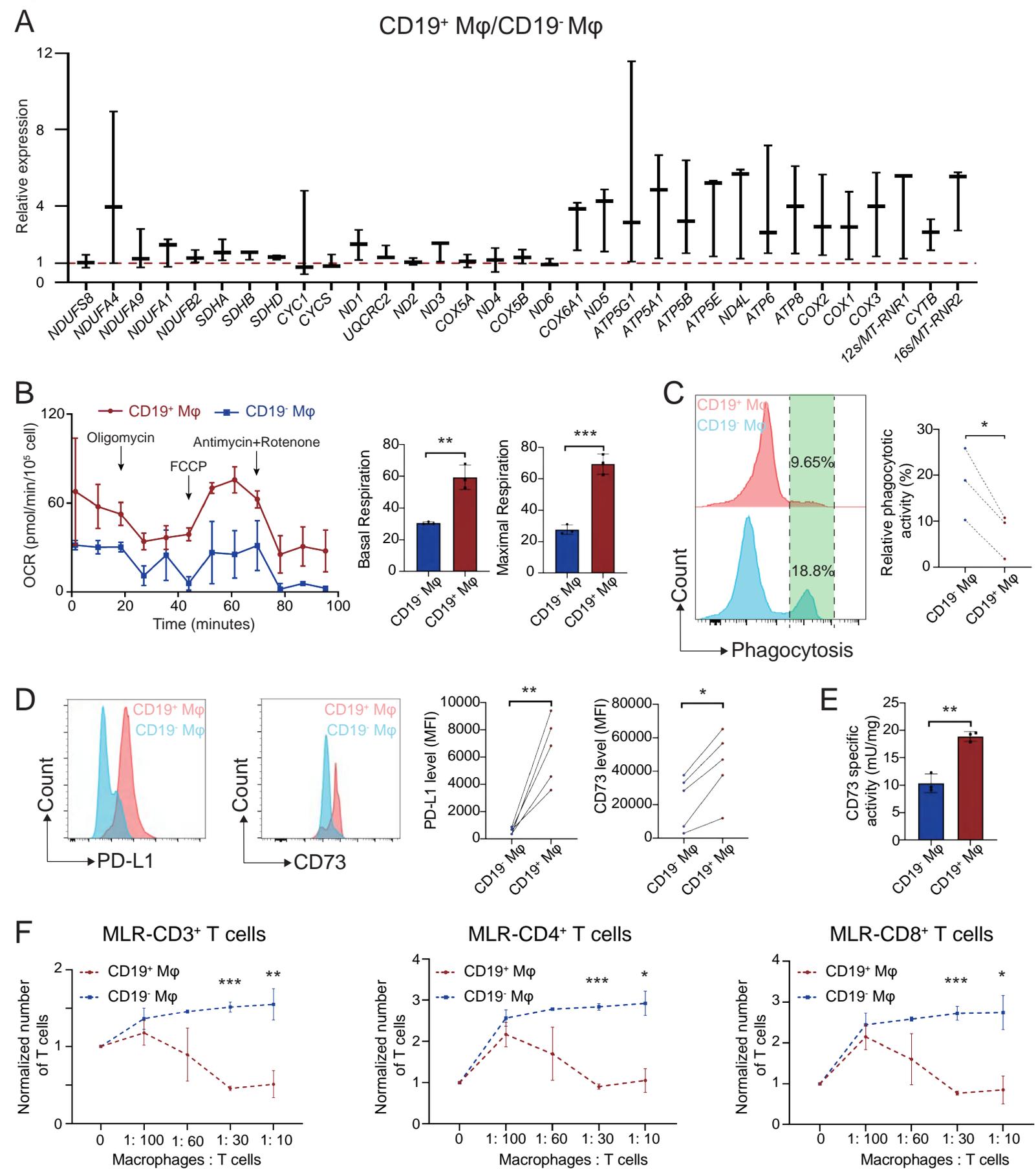

Fig. 4

Fig. 5

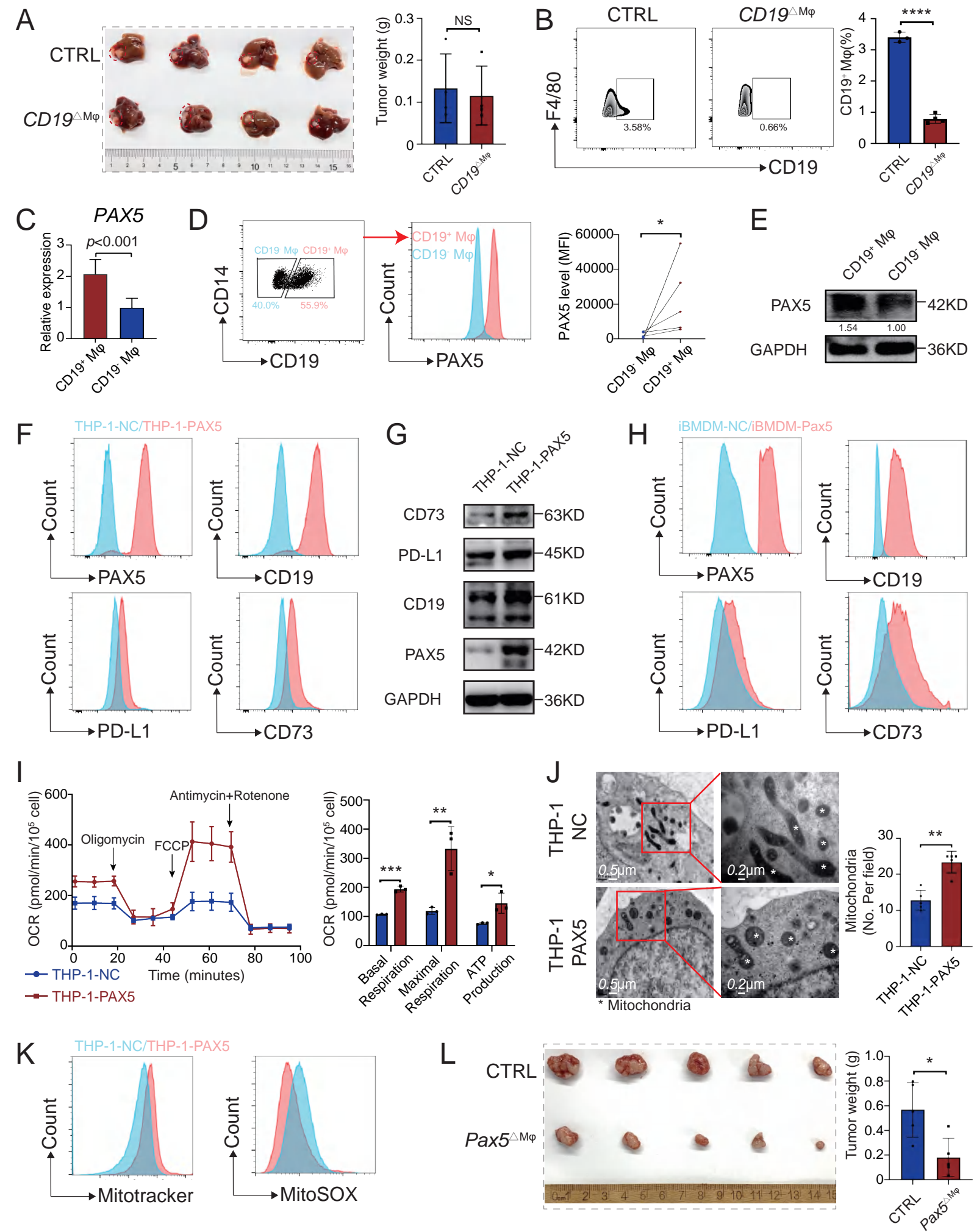

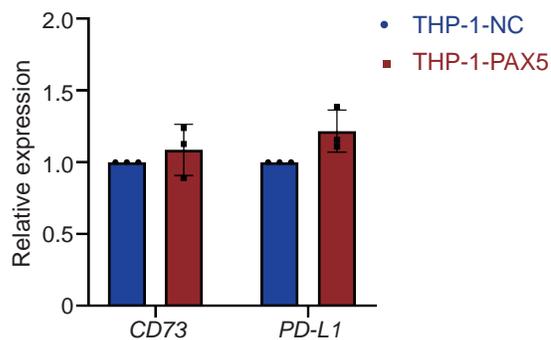
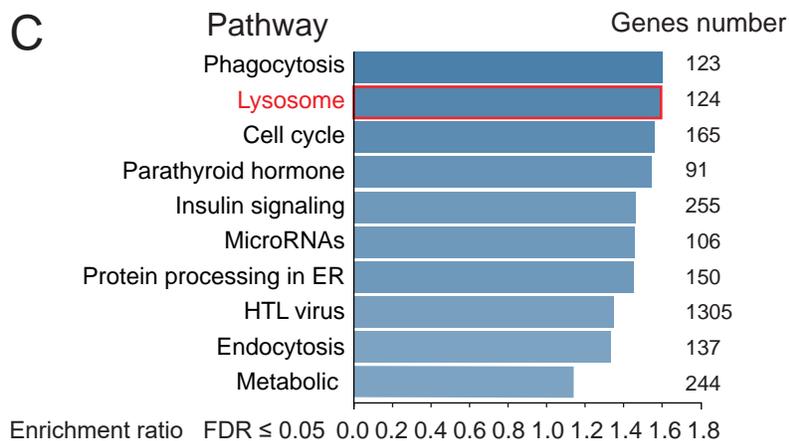
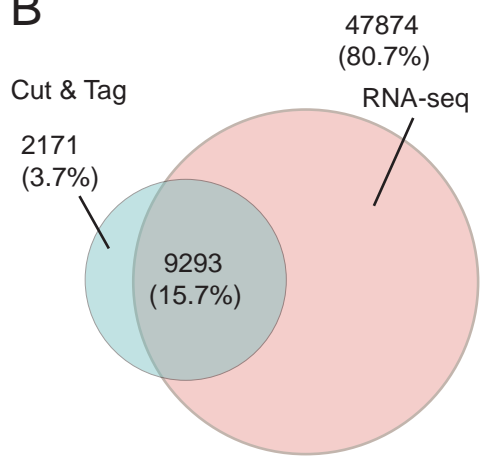
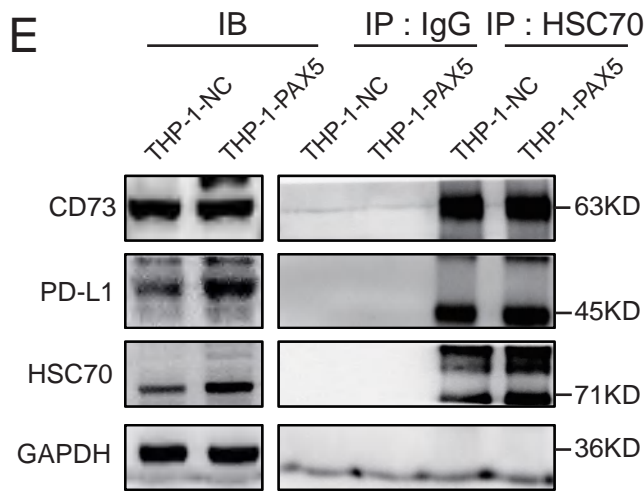
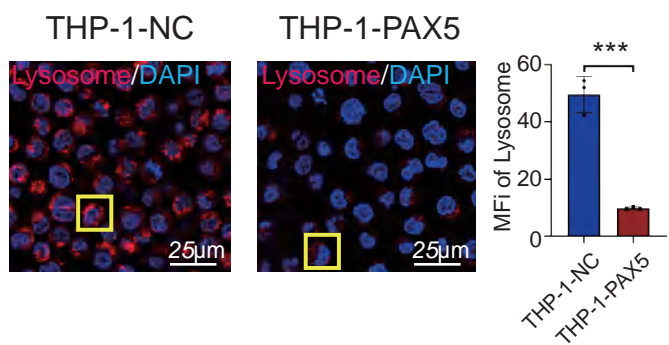
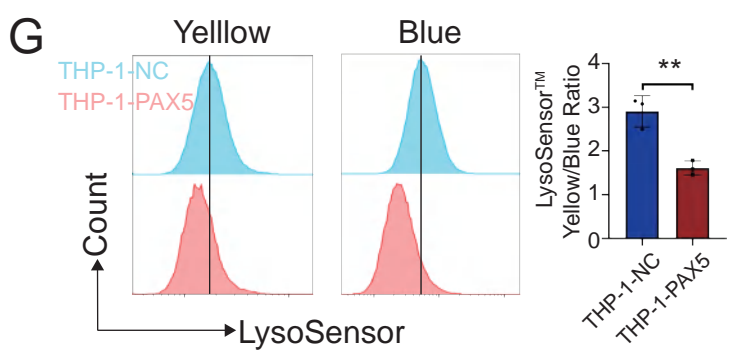
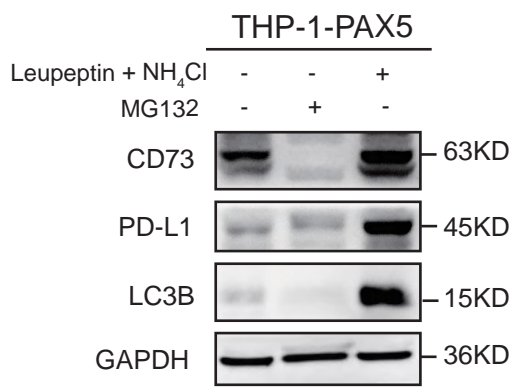
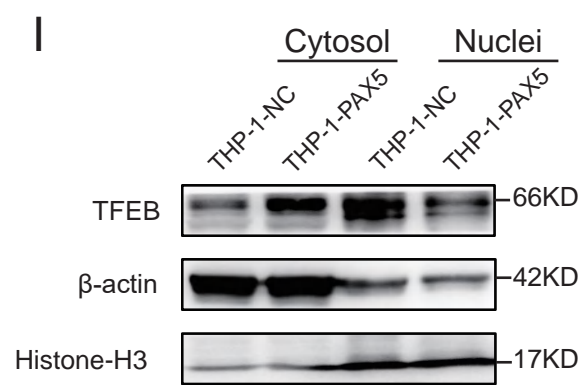
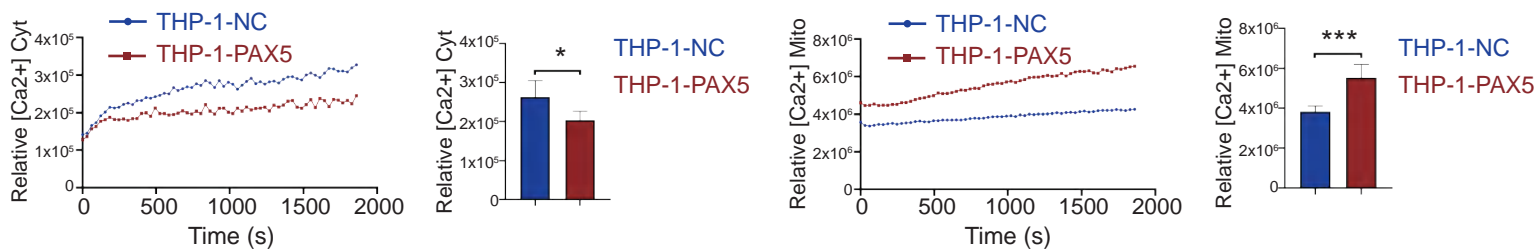

Fig. 7

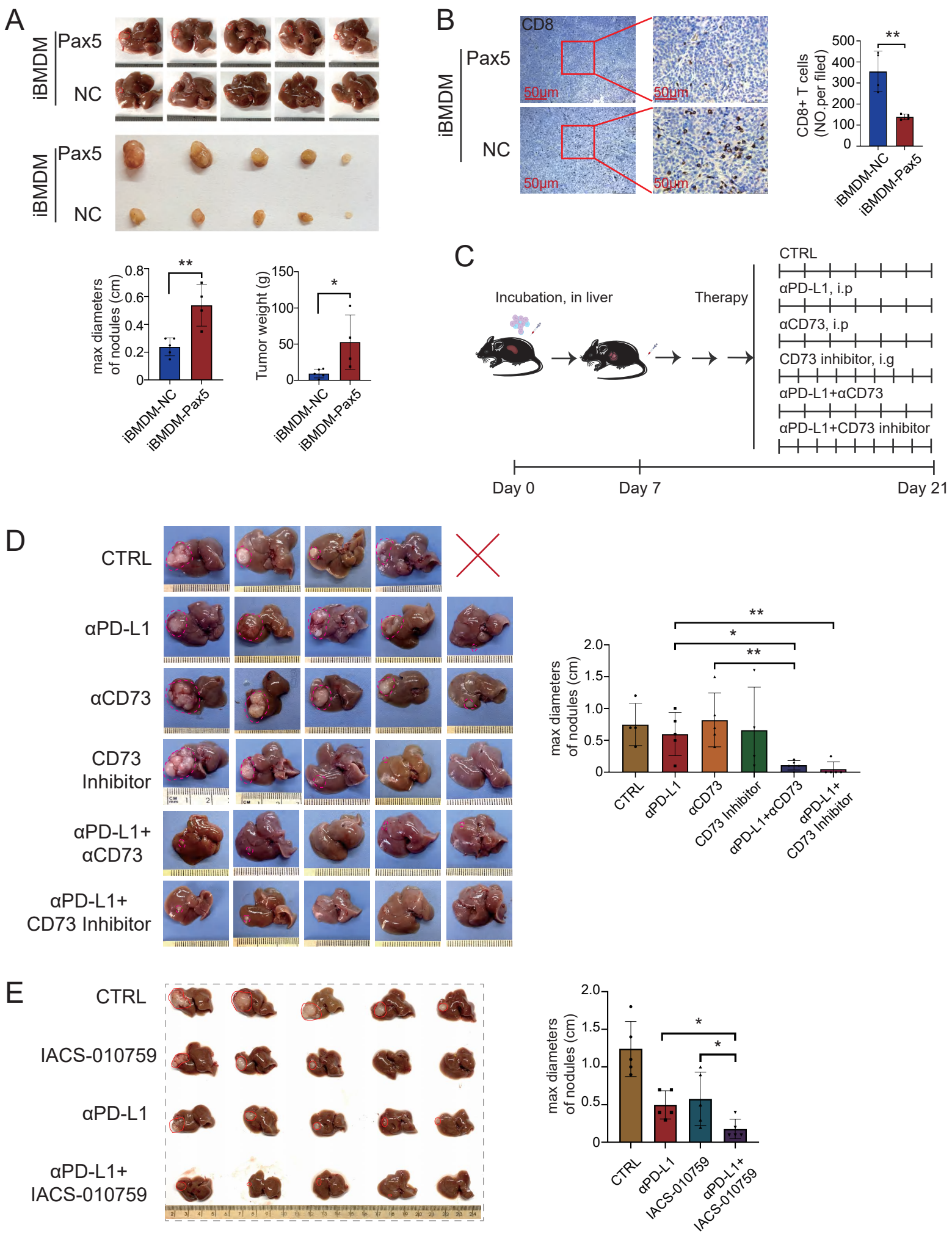

# Tumor-associated CD19⁺ macrophages induce immunosuppressive microenvironment in hepatocellular carcinoma

**Inventory of Supplemental Information**

- **Supplementary Methods**
- **Figures S1-S12 and Figure S legends.**
- **Supplementary Table S1-8**.

**Supplementary Table S1**. Patients' clinical information, related to **Figure 1**, **Figure 2**, and **Figure S2**.

**Supplementary Table S2**. Sequences of primer for transgenic mice genotyping, related to **Figure 5**. and **Figure S8**.

**Supplementary Table S3**. CyTOF antibody panel, related to **Figure 1** and **Figure S2**.

**Supplementary Table S4**. CD19 CAR sequences, related to **Figure 2** and **Figure S3**.

**Supplementary Table S5**. qRT-PCR Primer sequences, related to **Figure 4** and **Figure 6**.

**Supplementary Table S6**. Differentially expressed genes (DEGs) in CD19⁺ TAMs, CD19⁻TAMs, and B cells (data from sc-RNA seq), related to **Figure 3**.

**Supplementary Table S7**. Classical M1 and M2 macrophage marker genes expression in CD19+ TAMs and CD19- TAMs, related to **Figure 3**.

**Supplementary table 8.** Key resources table.

**Supplementary Methods**

**Human specimens**

Tumors (without necrotic foci) and matched adjacent normal tissues, along with peripheral and portal vein blood, were obtained from patients who were diagnosed with HCC or other specific cancers by pathology and underwent curative surgical resection between 2019 and 2024. Normal tissues were resected from a macroscopically normal part that was at least 2 cm away from the tumor tissue. Specifically, for flow cytometry analyses in Figure 1F, tumor samples with paired non-tumor tissues and peripheral blood of 81 patients were collected, including 28 patients with HCC, 13 patients with pancreatic ductal adenocarcinoma, 12 patients with breast carcinoma, 6 patients with gastric carcinoma, 11patients with renal carcinoma, 11 patients with colorectal carcinoma (**Supplementary Table 1**). The HCC tissue microarray contains 289 paired paraffin-embedded tumor and para-tumor tissue in our department between 2017 to 2021, by Wuha Servicebio Technology (Wuhan, China) (**Supplementary Table 1**). Clinical stages were classified according to the guidelines of the American Joint Committee on Cancer (AJCC), $^{8}$th edition. For survival analysis, patients of HCC received follow up every month for the first year and every 3–6 months thereafter. Survival time was calculated from the date of surgery to the date when death was confirmed. The prospective cohort of HCC patients with anti-PD-1 antibody treatment (**Figure 2E**) was a phase II clinical trial, patients were collected from our department and all diagnosed with HCC through pathology. Patients had not received any other treatment before enrollment (NCT03732547).

**Sample processing and single cell harvest**

Fresh tumor and adjacent normal tissues from surgical resections or mice were washed three times with PBS and cut into 1-3 mm³ pieces using sterile scalpel blades. The minced pieces were incubated with a digestion solution comprising DMEM, 2% fetal bovine serum (FBS; Gibco), type IV collagenase (1 mg/ml, Worthington, Lakewood, NJ, USA), DNase (15 μg/ml, Sigma-Aldrich, Saint Louis, MO, USA), and hyaluronidase (0.002 mg/ml, Sigma-Aldrich) at 37 °C for 1 hour under constant shaking. The digested mixture was filtered through a 70 μm nylon mesh (Corning, Teterboro, NJ, USA) and washed with PBS supplemented with 2% FBS. Immune cells were enriched using a 36% Percoll (GE Healthcare, Logan, UT, USA) density gradient medium (Gibco). Peripheral and portal vein blood lymphocytes were isolated using standard Ficoll gradient centrifugation with Lymphocyte Separation Medium (Solarbio, Beijing, China), according to the manufacturer's instructions, to harvest immune cells. Mouse spleens were ground with the flat end of a syringe in RPMI (Gibco) to collect single-cell suspensions. Red blood cells were removed using ACK Lysing Buffer (Sigma-Aldrich). The dissociated cells were used for subsequent experiments.

**Transgenic mouse genotyping**

To determine the genotype of transgenic mice, genomic DNA was isolated from tail tissues[1] and subjected to PCR using the corresponding primers (**Supplementary Table 2**). The PCR conditions for the *Pax5-floxP* fragment were as follows: 95°C for 10 min, 40 cycles of 95°C for 10s, 60°C for 60s, followed by 95°C for 10s, 60°C for 60s, and 95°C for 15s. The *Cre$^{Lyz}$* fragment was amplified using PCR conditions of 94°C for 10 min, 38 cycles of 94°C for 10s, 68°C for 30s, 72°C for 35s, followed by 72°C for 10 min. The PCR conditions for the *CD19-floxP* fragment were 94°C for 3 min, 35 cycles of 94°C for 30s, 62°C for 35s, 72°C for 35s, followed by 72°C for 5 min. PCR products were analyzed by electrophoresis in a 1% agarose gel. The *Lyz-Cre (+)* and *Lyz-Cre (-)* loci were identified as 700 bp and 350 bp, respectively. The *Pax5* FloxP-flanking exon (MT) and null loci (WT) were confirmed by two paired primers: primer 4 identified at 219 bp and 152 bp, and primer 5 identified at 212 bp and 156 bp. The *CD19* FloxP-flanking exon (MT) and null loci (WT) were recognized at 208 bp and 140 bp, respectively. Mice certified as *Floxp+/+-Lyz-Cre (+)* phenotypes were considered *Pax5$^{\triangle M\varphi}$* or *CD19$^{\triangle M\varphi}$* mice for subsequent experiments.

**Macrophage depletion and iBMDMs adoptive transfer**

To deplete macrophages, C57BL/6 mice were injected intraperitoneally with a standard suspension of clodronate liposomes (10 ml/kg, Vrije Universiteit, Amsterdam, Netherlands) every other day for four times[2]. Control animals received a control liposomal solution. iBMDMs with or without *Pax5* overexpression (1 × $10^7$ cells in 100 μL PBS) were injected into the tail veins of C57BL/6 recipient mice7 days after the in situ model formed.

**Transfection of lentiviral vectors and siRNA**

THP-1 cells and iBMDMs were seeded in 6-well plates at a density of 2 × $10^5$ cells/ml and infected with lenti-*PAX5* or or lenti-*TFEB*, lenti-*NC* (Jikai, Shanghai, China). The medium containing lentiviral vectors was removed and replaced with a maintenance medium after 16 hours. Infection efficiency was observed using a fluorescence microscope (Leica, Wetzlar, Germany) and verified by immunoblotting for PAX5 after 72 hours of culture. Stably transduced cells were purified using fluorescence activated cell sorting for further experiments. For siRNA transfection, 1.5 × $10^5$ per well THP-1 cells were seeded in 24- well plates and transfected with human *TFEB siRNA* (Santa Cruz Biotechnology, Santa Cruz, CA) using lipofectamine RNAimax (ThermoFisher Scientific, Waltham, MA) for 48 hours according to manufacturer's protocols[3].

**Mass cytometry staining**

Mass cytometry was performed by PLTTech Inc. (Hangzhou, China). The protocol has been described previously . Cells isolated from human liver tissues were blocked and stained for 30 min with a surface antibody mix panel (**Supplementary Table 3**) developed in-house, followed by fixation overnight. Permeabilization buffer was applied, and the cells were incubated in an intracellular antibody mix. The immuno-labeled samples were finally "barcoded" with a unique barcode isotope combination for 30 min, re-suspended in deionized (DI) water, and run through a CyTOF instrument (Helios, FLUIDIGM, Shanghai, China).

**Flow cytometry and fluorescence-activated cell sorting (FACS)**

Single-cell suspensions were generated and cells were stained with specific antibodies (**Supplementary Table 8**) for flow cytometry or FACS according to the manufacturer's instructions. To exclude dead cells, the cells were firstly stained using Live/Dead Fixable Viability Stain 700 or Fixable Viability Stain 780 or 7-AAD (BD Bioscience) in PBS for 30 min at 4 °C. For surface marker staining, cells were blocked with Fc block for 30 min and stained for 30 min at 4 °C with fluorescence-conjugated antibodies. For intracellular staining of transcription factors, we followed the manufacturer's protocol using a Transcription Factor Staining Buffer set (Thermo Fisher Scientific, Waltham, MA, USA). For staining of secreted cytokines, we fixed and permeabilized the cells using a Fixation/Permeabilization Solution Kit (BD Bioscience, Franklin Lakes, NJ, USA) after activating the cells with a Leukocyte Activation Cocktail with BD GolgiPlugⅢ (BD Bioscience) for 5 h. Flow cytometry acquisition was carried out on a BD LSRFortessa (BD Biosciences), and analyses were performed using FlowJo V.10.0 software (Tree Star, Ashland, OR, USA). Using specific antibodies (Supplementary Table 8), FACS was performed on a flow cell sorter (Beckman Coulter, Indianapolis, IN, USA) at 4 °C in 15 ml RNAse and DNAse free tubes (Simport, Saint-Mathieu-de-Beloeil, Canada) pre-filled with 5 ml of PBS with 5% FBS. The purity of the sorted populations was verified (98% pure) and cells were immediately used for further experiments.

**CD19-CAR plasmid construction**

The chimeric antigen receptor (CAR) was pieced together using murine CD8α signal peptide, anti-mouse-CD19 scFv, the hinge and transmembrane domains of murine CD8α, and the cytoplasmic domains of murine CD28 and CD3ζ. Then CAR was cloned into the pMSCV-IRES-GFP II (pMIGII; Addgene, Watertown, MA, USA) retroviral plasmid backbone through the EcoRI and BamHI site. Nucleotide sequence and amino acid sequence of CD19 CAR were showed in **Supplementary Table 4**.

**Mouse primary T cell activation and isolation**

Spleen cells from 8-week-old C57BL/6 mice were proceed into suspensions and activated using anti-CD3ε (1µg/ml; Biolegend, Waltham, MA, USA) and anti-CD28 (0.5 ug/ml; Biolegend) antibodies for 24 hours. Activated T cells were enriched by Percoll (GE Healthcare) gradient centrifugation. The residual B cells were removed by using biotin anti-mouse-CD19 antibodies and MojoSort™ Streptavidin Nanobeads (Biolegend). Briefly, the enriched activated spleen cells were resuspended to 1 x $10^8$/ml in complete T cell medium with 10µg/ml biotin anti-mouse-CD19 antibodies (Biolegend) and incubated at 4 °C for 30 min, followed by centrifugation at 500g for 5 min and resuspension in complete T cell medium with 20µl/ml MojoSort™ Streptavidin Nanobeads. After another round of incubation at 4 °C for 30 min, the CD19+ B cells were labeled with Nanobeads and then removed by MojoSort™ Magnet (Biolegend).

**Retrovirus production, transduction and anti-CD19-CAR-T/control CAR-T cell expansion**

Retroviral supernatant was produced in the Plat-E cell line. Briefly, the pMIGII plasmids containing CD19-CAR were transfected into Plat-E cells using calcium phosphate- mediated transfection protocol[3]. The retroviral supernatant was harvest at 48 and 72 hours after transfection respectively, and concentrated by ultrafiltration using Amicon Ultra-15 (Millipore, Billerica, MA, USA). Retrovirus transduction of anti-CD19 CAR-T/control CAR- T cells was performed in 6-well plates with polybrene (1:100). Transduced cells were resuspended after 4 hours and transferred to 10-cm plates for expansion. Mouse primary T cells were maintained incomplete T cell medium, which was RPMI 1640 Medium (Gibco) supplemented with 10% FBS (NEWZERUM, Christchurch, New Zealand), 100 IU/ml IL-2 (Novoprotein, Suzhou, China) and 0.05 mM 2-mercaptoethanol (Gibco).

**Targeted anti-CD19 CAR-T cell therapy**

$CD19^+/CD19^-$ macrophages were isolated from the spleens of mice ($CD45^+CD11b^+$ $F4/80^+CD19^+$), and were mixed with Hepa1-6 cells to establish subcutaneous xenograft model at the left ($CD19^+$ macrophages) or right ($CD19^-$ macrophages) flank of C57BL/6 or muMT mice. Anti-CD19 CAR-T cells were prepared and expanded. Mice were randomly assigned into groups for T cell infusion (1 × $10^7$ cells in 100 µL PBS), and were sacrificed after 14 days. Harvested tumors were processed for subsequent experiments.

**ImageStream analysis**

Single-cell suspensions were blocked with Fc Block for 30 min on ice, then incubated with fluorescence-conjugated antibodies for 30 min in the dark on ice, including anti-CD45 (Apc-cy7, Biolegend), anti-CD11b (PE-cy7, Biolegend), anti-CD14 (FITC, Biolegend), anti- CD19 (BV421, Biolegend). After washed with PBS-2% FBS, cells were resuspended in 50 μl PBS and analyzed with ImageStream II system (Amnis Corp, Seattle, WA, USA) for single cell image of $CD19^+$ TAMs, $CD19^-$ TAMs and B cells. A total of $1 \times 10^5$ cells were imaged at 40 × magnification at low speed for receiving high–quality images with an acquisition time of 30 min per sample. Data were analyzed using IDEAS software version 6.2 (Amnis Corp). The optimal compensation matrix between individual fluorescence channels was established using the same cells, stained with each of the above–mentioned antibodies separately or without stained. The settings for acquisition and analyses were used for all samples.

**Multiplex immunohistochemistry (mIHC) image acquisition**

We used Opal ™ 6-Plex Detection Kits (Akoya Biosciences, MA, USA) to perform mIHC according to the manufacturer's protocol. In brief, HCC and para-tumor tissues were processed to slides with 4-μm thickness. Tissue slides were baked at 65 °C for 3 hours to remove paraffin, then washed with xylene for 10 min three times，rehydrated in graded ethanol (100%-95%-70%-50%-30%), subsequently washed three with deionized distilled $H_2O$. The epitope retrieval was performed in microwave heating using AR buffer, and cooled to room temperature at least 15 min, following by washing slides with Tris Buffered Saline with Tween® 20 (TBST) buffer. PAP pen was used to mark around the tissue on the slide. After incubating in blocking solution at room temperature for 15 min, slides were covered with primary antibody for 60 min, corresponding horseradish peroxidase-conjugated secondary antibody for 30 min, opal fluorophore for 10 min, following by 2 minutes TBST wash for three times individually. Repeated epitope retrieval, blocking, primary antibody, secondary antibody, and opal fluorophore for each of the primary antibodies. Finally, tissue slides were counterstained with spectral DAPI for 5 min and coverslipped with fluoromount medium (SouthernBiotech, Birmingham, USA). Acquisition of the multi-color images was conducted using the Vectra Polaris System (Akoya Biosciences). Spectral unmixing was done in the inForm® software (Akoya Biosciences). The primary antibodies used for identifying CD19$^+$ TAMs and B cells, along with their respective fluorescent dyes, were as follows: CD68 (Abcam, Cat. no. ab192874, 1:200)/opal 520, CD14 (CST, Cat. no. 75181S, 1:200)/opal 570, CD19 (Abcam, Cat. no. ab134114, 1:50)/opal 690, CD20 (CST, Cat. no. 48750S, 1:200)/opal 620.

**Immunoblotting and immunoprecipitation (IP) analysis**

For immunoprecipitation, cells were washed with ice-cold PBS and harvested in a suitable volume of cell lysis buffer (25 mmol/L HEPES pH 7.5, 150 mmol/L NaCl, 0.25% Triton X-100, 0.25% NP-40, 0.25% CHAPS, 10% glycerol, and 1 × protease inhibitor cocktail) on ice for 30 min, and then centrifuged at 13,000 × g for 15 min. The protein concentration was evaluated using the bicinchoninic acid (BCA) method (Pierce, Rockford, IL, USA). For immunoprecipitation analysis, the supernatants were incubated with the indicated primary antibodies or isotype control IgG overnight at 4 °C and then incubated with 40 μl protein A/G-coupled agarose (Santa Cruz) for another 2 h at 4 °C. After three washes with lysis buffer, the immunoprecipitations were boiled in 2 × loading buffer (Thermo Fisher Scientific) for western blotting analysis. In some experiments, the nuclear and cytoplasmic proteins were extracted and isolated using a Nuclear and Cytoplasmic Protein Extraction Kit (Beyotime, Shanghai, China).

Proteins (40 μg per lane for a 10–lane comb and 20 μg for a 15–lane comb) were separated by 10 or 12% sodium dodecyl sulfate–polyacrylamide gel electrophoresis (SDS-PAGE), according to the molecular weight of the arget proteins, which were then transferred to polyvinylidene difluoride (PVDF) membranes (Millipore). Membranes were blocked with 10% skim milk (Bio-Rad Laboratories, Hercules, CA, USA) in tris-buffered saline/0.1% Tween 20 (TBST, Applygen, Beijing, China) for 1.5 hours and subsequently incubated overnight at 4 °C with specific primary antibodies (**Supplementary Table 8**) according to the manufacturer's recommendation. The next day, the membranes were washed three times with Tris-buffered saline with 0.1%

Tween® 20 Detergent (TBST) and further incubated with appropriate horseradish peroxidase (HRP)–labeled secondary antibodies for 2h at room temperature. Membranes were washed and the immunoreactive protein bands were visualized using a ChemiDoc XRS System (Bio-Rad). The photodensity of the immunoreactive protein bands was assessed using IPP software (Media Cybernetics, Rockville, MD, USA).

**RNA extraction and quantitative real-Time reverse transcription polymerase chain reaction (qRT-PCR)**

Tissue and cell RNA was extracted using the Trizol (Thermo Fisher Scientific) reagent[4], and reverse transcribed into cDNA using a Prime ScriptⅢ RT reagent Kit (Takara Biotechnology Co., Dalian, China) following the manufacturer's protocol. Then, the cDNA was subjected to quantitative real-time PCR (qPCR), which was performed in duplicate for each sample on a Prism 7900HT instruments (Applied Biosystems) with TB green premix EX TaqⅢ reagent Kit (Takara Biotechnology Co.). The qPCR conditions were: 50 °C for 2 min, 95 °C for 10 min, followed by 40 cycles at 95 °C for 15 s, and 60 °C for 1 min. Amplification of specific transcripts was confirmed by melting curve profiles generated at the end of the PCR program. The expression levels of target genes were normalized to the expression of glyceraldehyde-3-phosphate dehydrogenase (GAPDH, internal control) or corresponding house-keeping protein, and calculated based on the comparative cycle threshold (CT) method ($2^{-\Delta\Delta Ct}$) (PMID: 11846609). All primers were synthesized by Sunya (Hangzhou, China) and listed in **Supplementary Table 5**.

**Immunofluorescence and immunohistochemistry**

For cell immunofluorescence, cultured cells were fixed using 4% paraformaldehyde (PFA) for 20 min, and then treated with 0.1% Triton X-100 (Sigma-Aldrich) for 20 min. For tissue immunofluorescence, formalin-fixed paraffin-embedded samples were cut into 5 μm thick serial sections and heat-induced antigen retrieval was performed by microwaving in sodium citrate (pH 6.0, 98 °C, 10 min). The slides or cells were blocked in 3% bovine serum albumin (BSA, Sigma-Aldrich) for 1 h at room temperature and then incubated with the indicated primary antibodies

(**Supplementary Table 8**) diluted according to the manufacturer's recommendation in a buffer (PBS plus 1% BSA, 0.3% Triton X-100 at pH 7.4) overnight at 4 °C. After washing three times with PBS, slides or cells were incubated with fluorescent-dye-conjugated secondary antibodies and DAPI (Life Technologies, Invitrogen, and Carlsbad, CA). Immunofluorescent microscopic images were obtained and viewed using a confocal microscope (TSC SP8, Leica).

For tissue immunohistochemistry, formalin-fixed paraffin-embedded samples were cut into 5 μm thick serial sections. After dewaxing, heat-induced antigen retrieval was performed by microwaving in sodium citrate (pH 6.0, 98 °C, 10 min). The slides were blocked in 3% BSA for 1 h at room temperature and then incubated with the indicated primary antibodies (**Supplementary Table 8**), diluted according to the manufacturer's recommendation, overnight at 4 °C. The next day, slides were washed three times with PBS and then incubated with the corresponding horseradish peroxidase-conjugated secondary antibodies. Subsequently, the slides were stained using a Histostain-Plus Kit (ZSGB-BIO, Beijing, China) and then further counterstained with hematoxylin. Images were obtained and viewed using an inverted microscope (DMI8, Leica).

**Single-cell RNA sequencing (scRNA-seq) and analysis**

Based on FACS analysis, we first sorted CD19+ TAMs, CD19- TAMs and B cells from HCC samples. Single cells were counted and assessed for viability with Trypan blue using a Countess II automated counter (Thermo Fisher Scientific), subsequently processed for 10x Genomics when viability of > 90%. Individually, CD19+ TAMs and CD19- TAMs were mixed with B cells at a ratio of 3:1, numbers of $1 \times 10^4$ cells for each sample were then encapsulated into emulsion droplets using Chromium Single Cell Controller (10 x Genomics, Pleasanton, CA, USA). Reverse transcription and library preparation was performed on aVeriti Thermal Cycler with 96-Deep Well Reaction Module (Thermo Fisher Scientific). Amplified cDNA was purified using SPRIselect beads (Beckman Coulter) and sheared to 200-300 bp. ScRNA-seq libraries were constructed using a Chromium Single Cell 3' Library and Gel Bead Kit v3.1 (10 x Genomics) according to the manufacturer's protocol. Qualification was performed using a Qubit 3.0 Fluorometer (Thermo Fisher Scientific). Libraries were sequenced on an Illumina Hiseq PE150 system (Illumina). Raw reads were demultiplexed and mapped to the reference genome using the 10× Genomics Cell Ranger pipeline using default parameters. All downstream single-cell analysis were performed using Cell Ranger[5], Seurat[6], and Moncole[7], unless mentioned specifically. In brief, for each gene and each cell barcode (filtered by Cell Ranger), unique molecule identifiers were counted to construct digital expression matrices. For secondary filtration in Seurat analysis, a gene with expression in more than three cells was considered as expressed and each cell was required to have at least 200 expressed genes, but below 5000. Cells with unique molecular identifier (UMI) numbers < 1000, or with > 10% mitochondrial-derived UMI counts were considered low-quality cells and removed.

Principal component analysis (PCA) was performed to reduce the dimensionality of scRNA-seq data through the prcomp package of the R software (Version 4.0.5)[8]. The number of principal components was determined from a scree plot. Uniform Manifold Approximation and Projection (UMAP) or t-distributed stochastic neighbor embedding (t-SNE) analysis was then performed on the principal components with default parameters to visualize cells in a two-dimensional space. Conventional markers were used to categorize cells into a known biological cell type, including macrophages: CD11b, CD14, CD68; B cells: CD45, CD19, CD79a. Combined FACS and PCA analyses, we recognized three clusters of all samples, the differentially expressed genes (DEGs, Supplementary Table 6) were calculated in the one cluster in comparison with other two clusters. The false-positive result was corrected by Benjamini–Hochberg adjustment (adjusted p value; adj.p value). The "adj. p value < 0.05" and "|logFC| > 0.25" were considered as statistically significant. Particularly, we used pheatmap R package (Version:1.0.12) to construct the DEGs, and ComplexHeatmap R package (Version: 2.13.1) to picture the M1 and M2 macrophage markers gene expression heatmap with the scaled gen expression value (**Supplementary Table 7**). Furthermore, DEGs were applied to pathway enrichment analysis through Web-based Gene Set Analysis Tool Kit

(WebGestalt), with KEGG gene sets (c2.cp.kegg.v7.5.1). To define mitochondria score in CD19$^+$ TAMs and CD19$^-$ TAMs, the mean expression of "MT-ND1", "MT-ND2", "MT-CO1", "MT-CO2", "MT-ATP8", "MT-ATP6", "MT-CO3", "MT-ND3", "MT-ND4L", "MT-ND4", "MT-ND5", "MT-ND6", "MT-CYB" were individually calculated through Percentage FeatureSet Function (Seurat package).

**RNA-seq and gene expression analysis**

RNA-seq was performed using the Illumina Hiseq XTEN platform (Illumina, San Diego, CA, USA) at Novogene Co. Ltd (Beijing, China). In brief, total RNA was extracted and its purity were checked using a NanoPhotometer® spectrophotometer (IMPLEN, Westlake Village, CA, USA). mRNA was isolated using Oligo Magnetic Beads and cut into small fragments for cDNA synthesis. Libraries were generated using the NEBNext UltraⅢ RNA Library Prep Kit for the Illumina system (New England Biolabs, Ipswich, MA, USA) following the manufacturer's instructions, and index codes were added to attribute sequences to each sample. After cluster generation, the library preparations were sequenced on an Illumina HiseqXTEN platform. Raw data were processed through quality control steps to obtain clean data with high quality. After reads mapping, HTSeq was used to count the reads numbers mapped to each gene[9]. Then, the Fragments Per Kilobase of transcript per Million mapped reads (FPKM) value was calculated based on the gene length and reads count mapped to this gene. The list of ranked genes was based on the normalized enrichment score (NES).

**Untargeted metabolomics by Liquid chromatography tandem mass spectrometry (LC-MS/MS)**

Tumor tissues were individually grounded with liquid nitrogen and 100 μL of the homogenates was resuspend with precooled100% methanol (-20 °C) followed by vortexing, and then incubated at -20 °C for 60 min, centrifuged at 14,000 g, 4°C for 15 min. The supernatants were then transferred to a fresh microcentrifuge tube and dried under vacuum in a centrifugal evaporator. The dried metabolite pellets were redissolved by 80% methanol and analyzed by LC-MS/MS. LC-MS/MS analysis was performed using a Vanquish UHPLC system (Thermo Fisher Scientific) coupled with an Orbitrap Q Exactive HF-X mass spectrometer (Thermo Fisher Scientific) by Novogene Co. Ltd. Samples were injected onto an Accucore HILIC column (100×2.1 mm, 2.6 µm) using a 20-min linear gradient at a flow rate of 0.3 mL/min. Eluent A (95% ACN, 10 mM ammonium acetate, pH 9.0) and eluent B (50% ACN, 10 mM ammonium acetate, pH 9.0) were mixed for negative polarity mode. The solvent gradient was set as follows: 2% B, 1 min; 2-50% B, 16.5 min; 50-2% B, 2.5 min. Q-Exactive HF-X mass spectrometer was operated in positive/negative polarity mode with spray voltage of 3.2 kV, capillary temperature of 320°C, sheath gas flow rate of 35 arb and aux gas flow rate of 10 arb. Identification and quantification of metabolites were performed using the mzCloud database by the search engines: Compound Discoverer 3.0 (Thermo Fisher Scientific). Differentially expressed metabolites were screened based on VIP value from PLS-DA analysis and P value from t-test (VIP > 1 and P value < 0.05).

**Metabolic extracellular flux analysis (Seahorse XF Cell Mito Stress Test)**

Real-time bioenergetic profiles of $CD19^+$ macrophages, $CD19^-$ macrophages and THP-1 cells were obtained by measuring oxygen consumption rate (OCR) using a Seahorse XF24 Flux Analyzer (Seahorse Bioscience, Inc, North Billerica, MA, USA). Briefly, cells were seeded into a Cell-Tak (Corning) coated Seahorse 24-well plate (Agilent Technologies, Santa Clara, CA, USA) at a density of $5 \times 10^5$ cells per well and allowed to adhere for 2h. Cells were then washed and the medium was replaced with Seahorse XF RPMI Medium supplemented with 10 mM glucose, 2 mM glutamine, and 1 mM pyruvate (Agilent). Following incubation in an incubator without $CO_2$ at 37 °C for 60 min, the basal OCR and ECAR were recorded. A Mito Stress assay was performed by sequential addition of 1 µM oligomycin (Cayman, Ann Arbor, MI, USA), 1 µM carbonyl cyanide 4- (trifluoromethoxy) phenylhydrazone (FCCP, Cayman), and 0.5 µM rotenone/antimycin A (Cayman). Parameters such as ATP-linked OCR, maximal OCR, reserve capacity, and proton leak were calculated from Mito Stress assays of three independent experiments as previously described[10].

Phagocytosis assay

Cells were seeded into 96-well plates and allowed to adhere for 2 h at a density of 5 × 10⁵, then media were removed and replaced with 200 μg/ml pHrodoTM Red SE or pHrodo red Escherichia coli bioparticles (Thermo Fisher Scientific) labeled dead cells as per the manufacturer's instruction for 0.5, 1, 2, or 24 h before flow cytometry analyses.

**In vitro co-culture experiments**

T cells (CD3, CD4, CD8) generated by sorting (CD45$^+$CD3$^+$/CD45$^+$CD3$^+$CD4$^+$/CD45$^+$CD3$^+$CD8$^+$) were collected and co-cultured with target macrophages (1 × 10$^4$) for 72 h at different ratios in a 96-well plate. The plate was centrifuged at 500 × g for 1 min at 4°C to initiate cell-cell contact and transferred to a 37 °C water bath immediately. Thereafter, T cells were analyzed using flow cytometry.

**Electronic microscopy**

Cells were fixed with 2.5% glutaraldehyde in phosphate buffer (0.1 M, PH 7.0) for 4 h at 4 °C. The samples were then washed three times for 15 min each in 0.1 M phosphate buffer (0.1 M, PH 7.0), postfixed with ice-cold 1% osmium tetroxide in 0.1 M phosphate buffer for 1 h, and washed three times for 15 min each in phosphate buffer (0.1 M, PH 7.0). Then, all samples were dehydrated for 15 min using a graded ethanol series (30%, 50%, 70%, 80%, 90%, 95% and 100%) and then transferred to absolute acetone for 20 min. Subsequently, the sample was placed in a mixture of 1:1 absolute acetone and the final Spurr resin, left at room temperature for 1 h, transferred to 1:3 absolute acetone and final resin mixture for 3 h, transferred to the final Spur resin mixture, and left overnight. Next, the specimen was placed in Eppendorf container sprayed with resin and heated for more than 9 h at 70°C. The specimens were sliced on a Leica EM UC7 slicing machine and stained with uranyl acetate and basic lead citrate for 5-10 min before observation on a Hitachi Model H-7650 TEM (Tokyo, Japan).

**Mitochondrion assessment**

Cells ($5 \times 10^5$) were incubated with 50 nM MitotrackerⅢ probe (Thermo Fisher Scientific) for 30 min at 37 °C in the indicated dilutions before washing away the probe. Then cells were used for flow cytometry analyses.

**MitoSox assay**

To determine mitochondrial ROS production, cells ($5 \times 10^5$) were stained with 2 μm of Mitosox (Thermo Fisher Scientific) for 10 min at the end of the experiment, followed by flow cytometry analyses.

**Adenosine measurements**

Cells ($5 \times 10^5$) were seeded in 6-well plates in triplicate. When culture plates reached 8090% confluence, the media were replaced with 2 ml of serum-free phenol red free RPMI (Gibco) with inhibitors (erythro-9-(2-hydroxy-3-nonyl) adenine (EHNA) 100 μmol/L, Nitrobenzylthioinosine (NBMPR) 100 μmol/L, Dipyridamole 40μmol/L). Conditioned media were collected and centrifuged at 10,000 × g for 10 min at 4 °C to remove insoluble particles after 16 h incubation. Cells were harvested and cell counts were recorded for back calculations. For adenosine measurements, conditioned media were added in 96- well fluorescence black microtiter plates using an adenosine assay kit (Cell Biolabs, Inc. USA) according to a modified manufacturer's protocol. The plates were read using a SpectraMax i3x multi-Model Microplate Reader (Molecular Devices LLC.) for excitation in the 530-570 nm range and for emission in the 590–600 nm range.

**CD73 activity assay**

CD73 activity was measured using a 5′ nucleotidase (CD73) activity assay kit (Abcam, Cambridge, MA, USA). Briefly, samples and a standard curve were prepared, and then added into 96 wells plates along with reaction mix at 37 °C for 20 min. Next, stop solution added, and then developer solution I and II. After incubating at room temperature for 20 min, the absorbance was recorded at 670 nm by a SpectraMax i3x multi-Model Microplate Reader (Molecular Devices, LLC., San Jose, CA, USA).

**Cut & Tag (cleavage under targets and tagmentation) sequencing**

THP-1 cells with or without *PAX5* overexpression were harvested and dead cells were removed. After incubation with the indicated primary PAX5 antibodies (Abcam, Cambridge, MA, USA), the corresponding secondary antibody, and hyperactive PG/PA-Tn5 transposase, cells were fragmented and DNA was extracted for further library construction. Then, paired-end Illumina sequencing was performed on the barcoded libraries by OE BioLab Co. Ltd. (Shanghai, China), following the manufacturer's instructions.

**Proliferation analysis**

For short-term in vivo proliferation assay, mice were intraperitoneally injected with 50 mg/kg EdU (Beyotime), cells were obtained at 6 h post-injection and detected by using the Click-iTTM Plus EdU Flow Cytometry Assay Kits, then analyzed using flow cytometry.

**Lysosome labeling**

For Lysotracker staining, cells ($5 \times 10^5$) were incubated with 50 nM of the LysotrackerⅢ probe (Beyotime) diluted in serum-free RPMI for 5 min at 37 °C before washing away the probe. For Lysosensor staining, cells ($5 \times 10^5$) were incubated with 1 μM LysotrackerⅢ probe (Invitrogen) diluted in serum-free RPMI before washing away the probe. The cells were then used for confocal imaging or flow cytometrocal analyses.

**Measurements of [$Ca^{2+}$]**

For the measurement of [$Ca^{2+}$]cyt and [$Ca^{2+}$]mito, $5 \times 10^4$ THP-1 cells with or without *Pax5* overexpressing were collected and incubated with Calbryte 630$^{AM}$ (2 μM, 37 °C, 60 min, ABD Bioquest, Sunnyvale, CA) or Rhod2 (2 μM, 37 °C, 60 min, specific expression in mitochondria with a low concentration, ABD Bioquest) in HBSS (Cienry, supplemented with CaCl2 and MgCl2). Cells were detected at indicated excitation and emission wavelengths with Microplate Reader every 30s for 30 mins. In some experiments, cells were pretreated with IASC-010759 (100 nM) and ionomycin (1 μM) for 20 min at 37 °C for next calcium fluorophore procedure.

**Statistical analyses**

Statistical tests were selected based on the appropriate assumptions with respect to data distribution and variance characteristics using Prism8 (Graphpad). Student's t-test or the Mann–Whitney test was used to compare the data from two groups. For comparison of data from more than two groups, one-way analysis of variance (ANOVA) was performed. Two-way ANOVA with the Bonferroni multiple comparison test was used for co-culture analysis. Categorical variables were compared using a chi-squared test or Fisher's exact test. Survival analysis was conducted using the Kaplan–Meier method and compared using the log-rank test. Details were indicated in Figure legends. The number of animals used per group is indicated in the figure legends as 'n' . The data are shown as mean ± standard error of mean (SEM). NS, not significant, *$P < 0.05$; **$P < 0.01$; ***$P < 0.001$; ****$P < 0.0001$.

**Reference**

1. Truett GE, Heeger P, Mynatt RL, Truett AA, Walker JA, Warman ML: (2000).Preparation of PCR-quality mouse genomic DNA with hot sodium hydroxide and tris (HotSHOT). BioTechniques 29(1):52, 54.

2. Toda G, Yamauchi T, Kadowaki T, Ueki K: (2021).Preparation and culture of bone marrow-derived macrophages from mice for functional analysis. STAR protocols 2(1):100246.

3. Kumar P, NagarajanA, Uchil PD: (2019).Calcium Phosphate-Mediated Transfection of Adherent Cells or Cells Growing in Suspension: Variations on the Basic Method. Cold Spring Harbor protocols 2019(10).

4. Rio DC, Ares M, Jr., Hannon GJ, Nilsen TW: (2010).Purification of RNA using TRIzol (TRI reagent). Cold Spring Harbor protocols 2010(6):pdb.prot5439.

5. Zheng GX, Terry JM, Belgrader P, Ryvkin P, Bent ZW, Wilson R, Ziraldo SB, Wheeler TD, McDermott GP, Zhu J et al: (2017).Massively parallel digital transcriptional profiling of single cells. Nature communications 8:14049.

6. Butler A, Hoffman P, Smibert P, Papalexi E, Satija R: (2018).Integrating single-cell transcriptomic data across different conditions, technologies, and species. Nature biotechnology 36(5):411-420.

7. Qiu X, Mao Q, Tang Y, Wang L, Chawla R, Pliner HA, Trapnell C: (2017).Reversed graph embedding resolves complex single-cell trajectories. Nature methods 14(10):979-982.

8. Dessau RB, Pipper CB: (2008).["R"--project for statistical computing]. Ugeskrift for laeger 170(5):328-330.

9. Putri GH, Anders S, Pyl PT, Pimanda JE, Zanini F: (2022).Analysing high-throughput sequencing data in Python with HTSeq 2.0. Bioinformatics (Oxford, England) 38(10):2943-2945.

10. Leung DTH, Chu S: (2018).Measurement of Oxidative Stress: Mitochondrial Function Using the Seahorse System. Methods in molecular biology (Clifton, NJ) 1710:285- 293.

Fig. S1

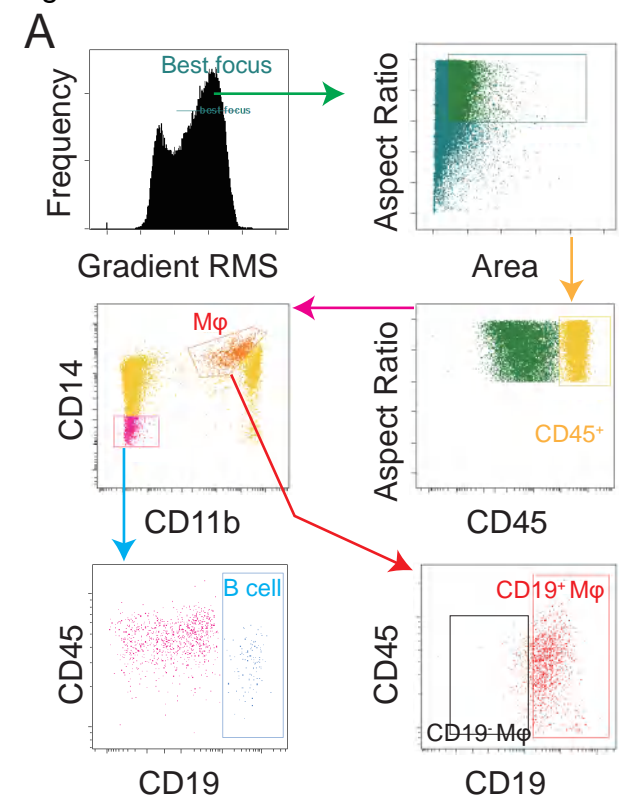
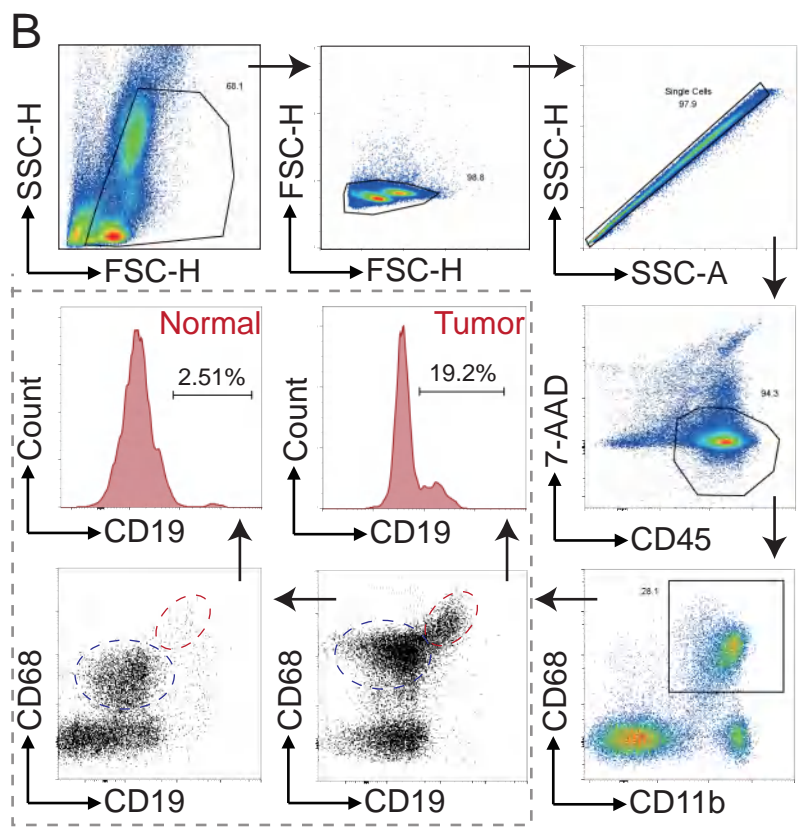
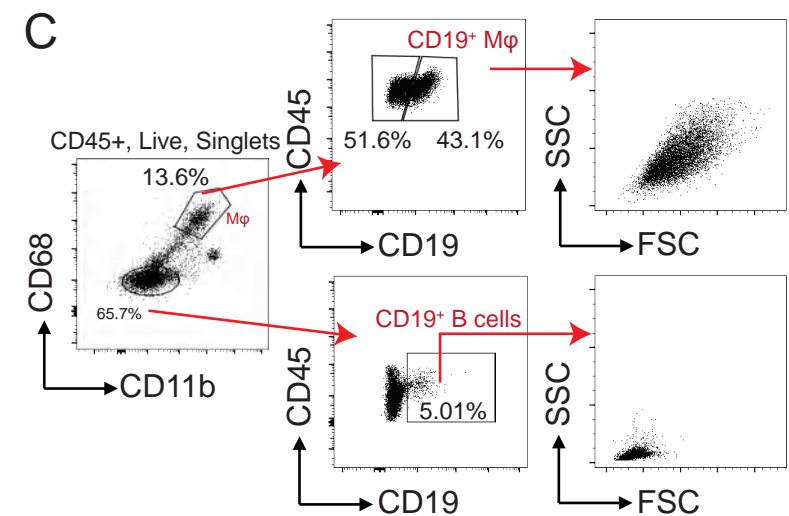
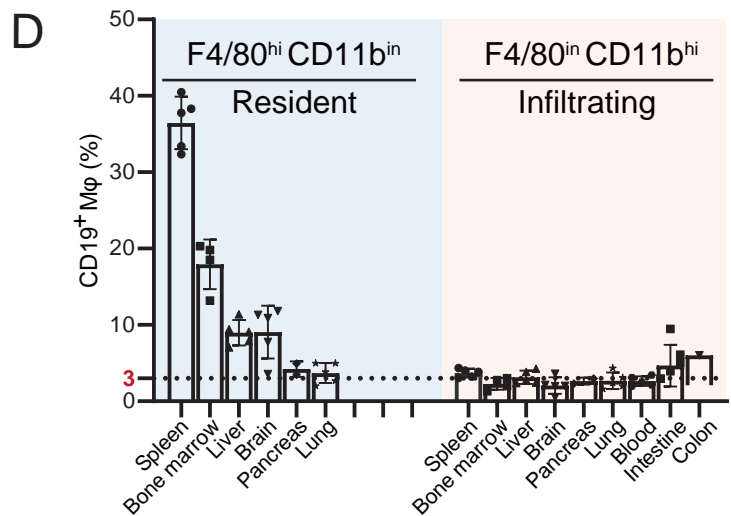
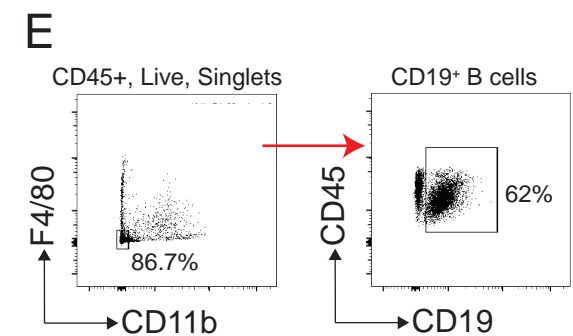
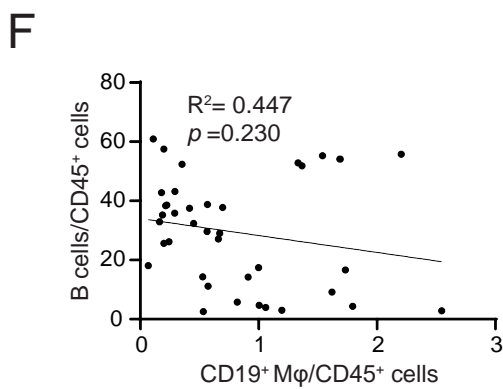

**Fig. S1. CD19$^+$ macrophages are enriched in HCC (Related to Fig. 1).**

(**A**) ImageStream gating strategy for identifying CD19$^+$ TAMs, CD19$^-$ TAMs, and B cells in HCC. CD19$^+$ TAMs are defined as CD45$^+$CD11b$^+$CD14$^+$CD19$^+$, CD19$^-$ TAMs are defined as CD45$^+$CD11b$^+$CD14$^+$CD19$^-$, and B cells are defined as CD45$^+$CD11b$^-$CD14$^-$CD19$^+$. (**B**) Flow cytometry gating strategy for identifying CD19$^+$ TAMs and CD19$^-$ TAMs in HCC. CD19$^+$ TAMs are defined as CD45$^+$CD11b$^+$CD68$^+$CD19$^+$, and CD19$^-$ TAMs are defined as CD45$^+$CD11b$^+$CD68$^+$CD19$^-$. (**C**) Back-gating flow cytometry plots of CD19$^+$ TAMs and B cells to FSC and SSC values. (**D**) Quantification of CD19$^+$ macrophages proportion in normal tissues of C57 BL/6 mice. Resident macrophages are identified as F4/80$^{high}$CD11b$^{intermediate}$, while infiltrating macrophages are identified as F4/80$^{intermediate}$CD11b$^{high}$. Data were detected by flow cytometry. n = 5. (**E**) Flow cytometry gating strategy for identifying B cells in mice (**F**) Correlation analysis between CD19$^+$ macrophages and B cells in mice tissues. n = 17. Data are presented as the mean ± SEM. Pearson correlation test was used in (**F**).

Fig. S2

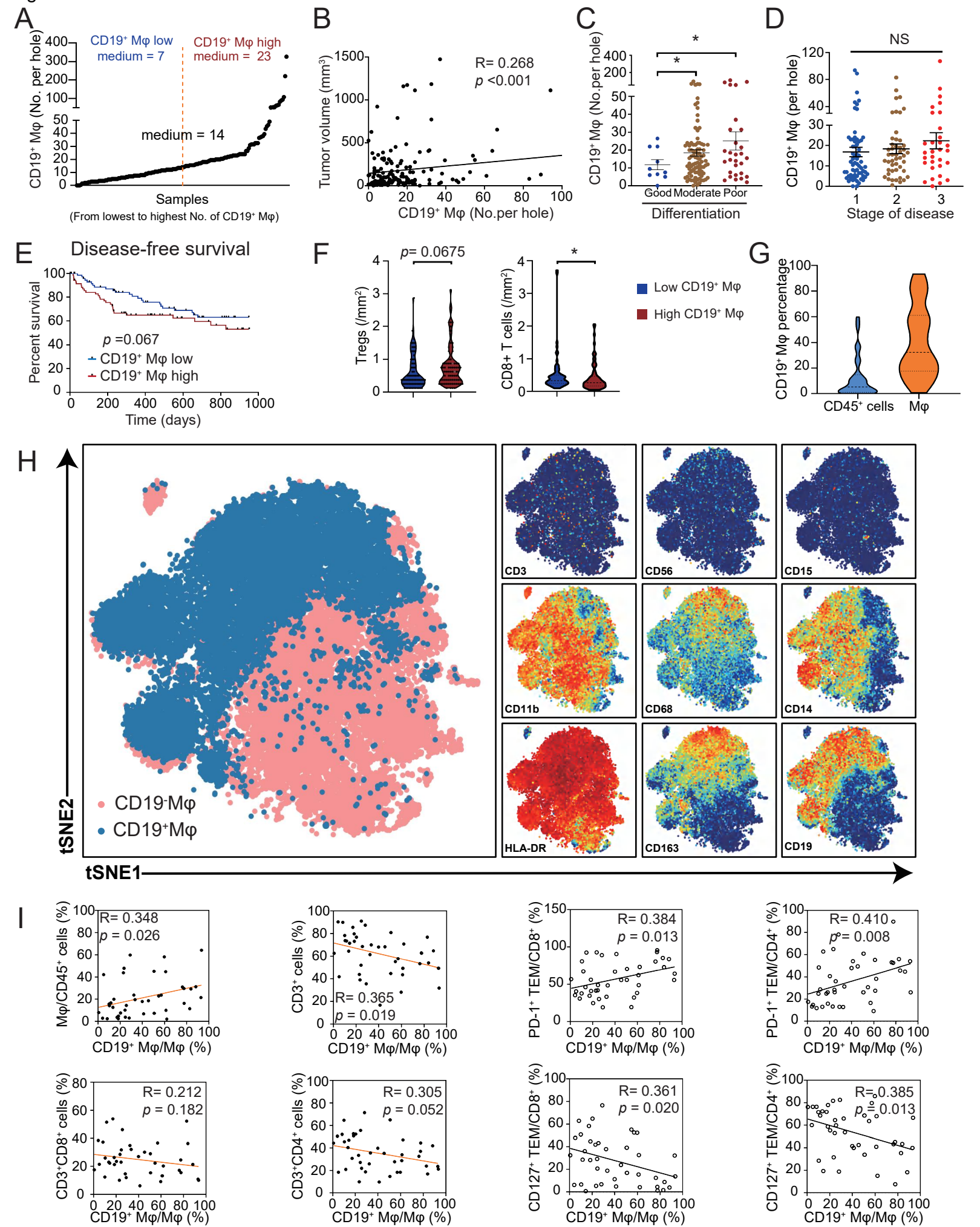

**Fig. S2. CD19$^+$ TAMs are associated with poor clinical outcome and immunotherapy response (Related to Fig. 2).**

(**A**) Densities of CD19$^+$ TAMs in HCC tissues included in an HCC tissue microarray. n = 156. (**B**) Correlation analysis between tumor volume and the density of CD19$^+$ TAMs in the tumor. Data are derived from immunofluorescence analysis of the HCC tissue microarray. n = 156. (**C**) Tumor differentiation status is correlated with the density of CD19$^+$ TAMs in the tumor. Data are derived from immunofluorescence of HCC tissue microarray. n = 145. (**D**) Disease stage of HCC is not correlated with the density of CD19$^+$ TAMs in the tumor. Data are derived from immunofluorescence analysis of the HCC tissue microarray. n = 145. (**E**) Kaplan-Meier curves showing disease-free survival (DFS) of HCC patients with a high (> median) or low (< median) ration of CD19$^+$ TAMs. Data are derived from immunofluorescence analysis of the HCC tissue microarray. n = 149. (**F**) Violin plot showing the proportions of Tregs and CD8$^+$ T cells in HCC patients with a high (> median) or low (< median) ration of CD19$^+$ TAMs. Data are derived from the human HCC tissue microarray. n = 156. (**G**) Violin plot showing the proportion of CD19$^+$ TAMs in CD45$^+$ cells or total macrophages in human HCC samples. Data are derived from CyTOF. n = 41. (**H**) Visualized *t*-SNE map showing CD19$^+$ TAMs, CD19$^-$ TAMs, and cell subpopulations with specific markers. Data are derived from CyTOF. (**I**) Correlation between percentage of CD19$^+$ TAMs and percentage of other types of immune cells as indicated, including macrophages, γδT cells, NK cells, T cells, CD8$^+$ T cells, CD4$^+$ T cells, PD-1$^+$ TEMs and CD127$^+$ TEMs. Data are derived from CyTOF. n = 41. Data are presented as the mean ± SEM. Pearson correlation test (**B, I**), unpaired one-way ANOVA with Welch's correction (**C, D**), log-rank test (**E**), or unpaired two-tailed *t*-test (**F**) was used. $^*P < 0.05$; NS, not significant.

Fig. S3

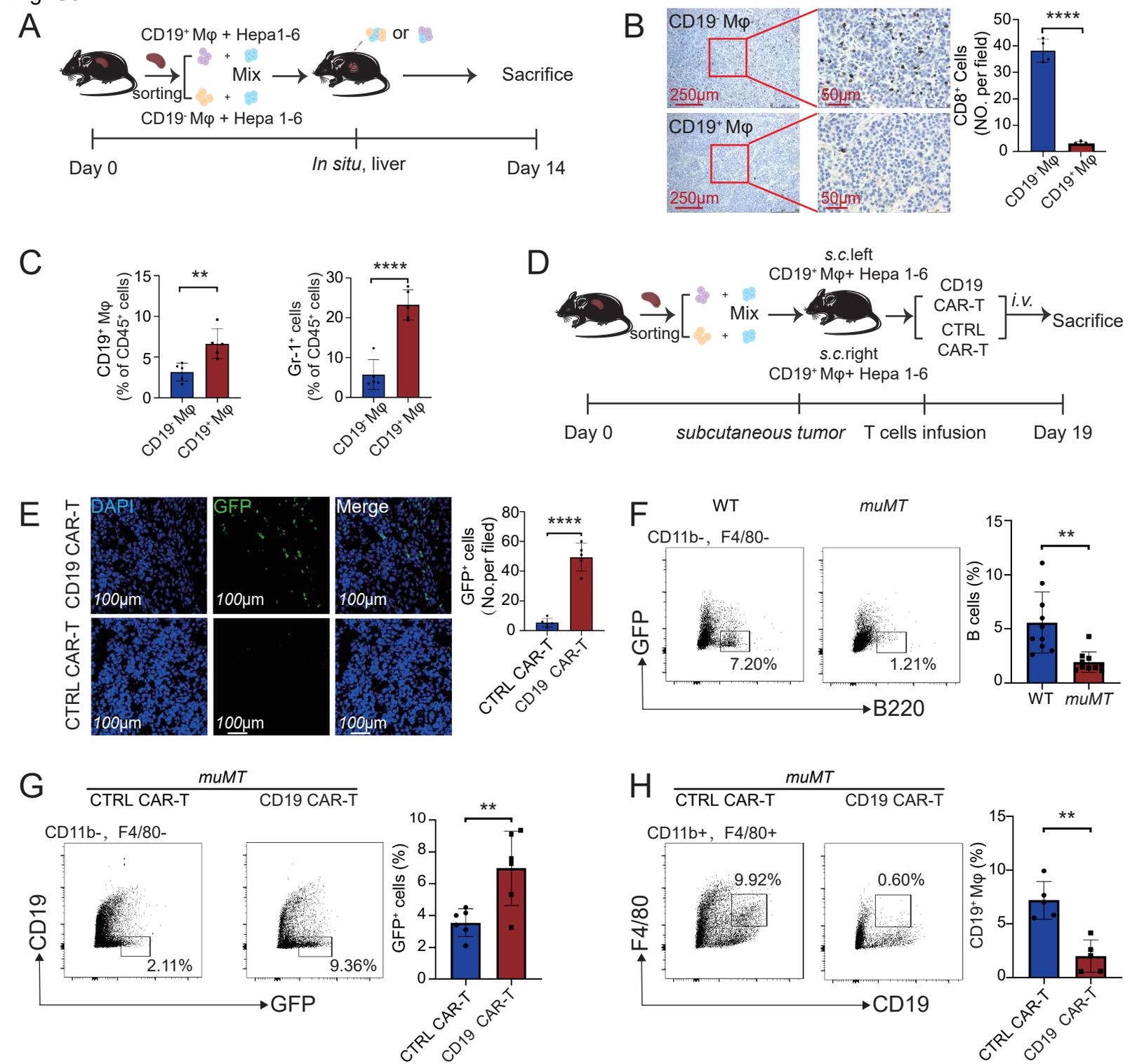

**Fig. S3. CD19⁺ TAMs promote HCC tumor progression in mice (Related to Fig. 2).**

(**A**) Schematic illustration of in vivo experiment for testing the role of CD19⁺ macrophages on tumor progression. Splenic CD19⁺ and CD19⁻ macrophages were sorted and separately mixed with 5× 10⁵ Hepa1-6 cells at a ratio of 1:5, and then injected into the liver of mice. (**B**) Flow cytometrical quantification of CD19⁺ macrophages and Gr-1⁺ cells in tumors in **Fig. 2F**. n = 5 per group. (**C**) Representative immunohistochemical images and quantification of CD8⁺ T cells in **Fig. 2F**. (**D**) Schematic illustration of anti-CD19 CAR-T cell targeted therapy on HCC. Splenic CD19⁺ and CD19⁻ macrophages were sorted and separately mixed with 5× 10⁵ Hepa1-6 cells at a ratio of 1:5, and then inoculated subcutaneously (*s.c.*) into the left and right flanks of mice. When tumors reached 35 to 45 mm², mice received intravenous injection (*i.v.*) with GFP⁺ control CAR-T cells or GFP⁺ anti-CD19 CAR-T cells (1 × 10⁷ cells in 100 μL of PBS). (**E**) Representative immunofluorescence images and quantification of control GFP⁺ CAR-T cells or GFP⁺ anti-CD19 CAR-T infiltration in tumors (**Fig. 2G**). n = 5 per group. (**F**) Representative flow cytometry plots showing B cells quantification in *muMT* and wild-type mice. n = 10 per group. (**G**) Representative flow cytometry plots and quantification of control GFP⁺ CAR-T cells and GFP⁺ anti-CD19 CAR-T infiltration in tumors from *muMT* mice. n = 5 per group. (**H**) Representative flow cytometry plots and quantification of the after percentage of CD19⁺ TAMs in *muMT* mice with control CAR-T cells or anti-CD19 CAR-T treatment. n = 5 per group. Data are presented as the mean ± SEM. An unpaired two-tailed *t*-test (**B, C, E, F, G, H**) was used. **P* < 0.01; ****P* < 0.0001.

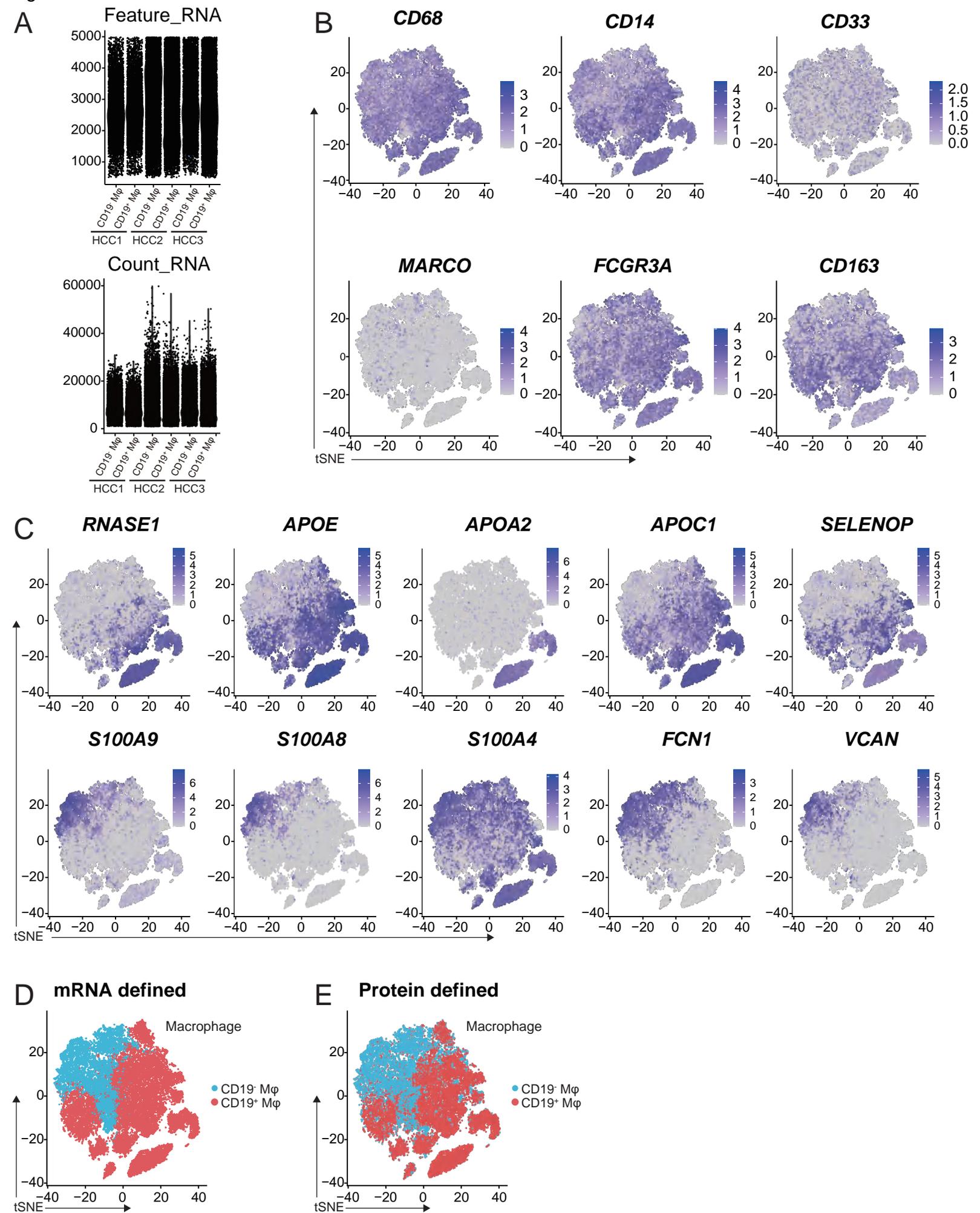

Fig. S4

**Fig. S4. CD19$^+$ TAMs exhibit a distinct gene expression profile (Related to Fig. 3).**

(**A**) VlnPlot analysis of sc-RNAs in three HCC samples after quality control. (**B**) UMAP plot showing *CD19* mRNA expression levels in CD19$^+$ TAMs, CD19$^-$ TAMs, and B cells. Macrophages were labelled with red dotted lines, B cells were labelled with orange dotted lines. (**C**) t-SNE plots showing expression levels of classical macrophage markers in CD19$^+$ TAMs and CD19$^-$ TAMs, including *CD68, CD14, CD33, MARCO, FCGR3A, CD163*. (**D**) t-SNE plots showing the expression levels of five feature genes in CD19$^+$ TAMs (*RNASE1, APOE, APOA2, APOC1, and SELENOP*) and CD19$^-$ TAMs (*S100A9, S100A8, S100A4, FCN1, VCAN*), respectively. (**E**) t-SNE plot showing CD19$^+$ TAMs and CD19$^-$ TAMs distribution based on the protein levels of the above 10 feature genes. (**F**) t-SNE plot showing CD19$^+$ TAMs and CD19$^-$ TAMs distribution based on the mRNA expression levels of the above 10 feature genes.

Fig. S5

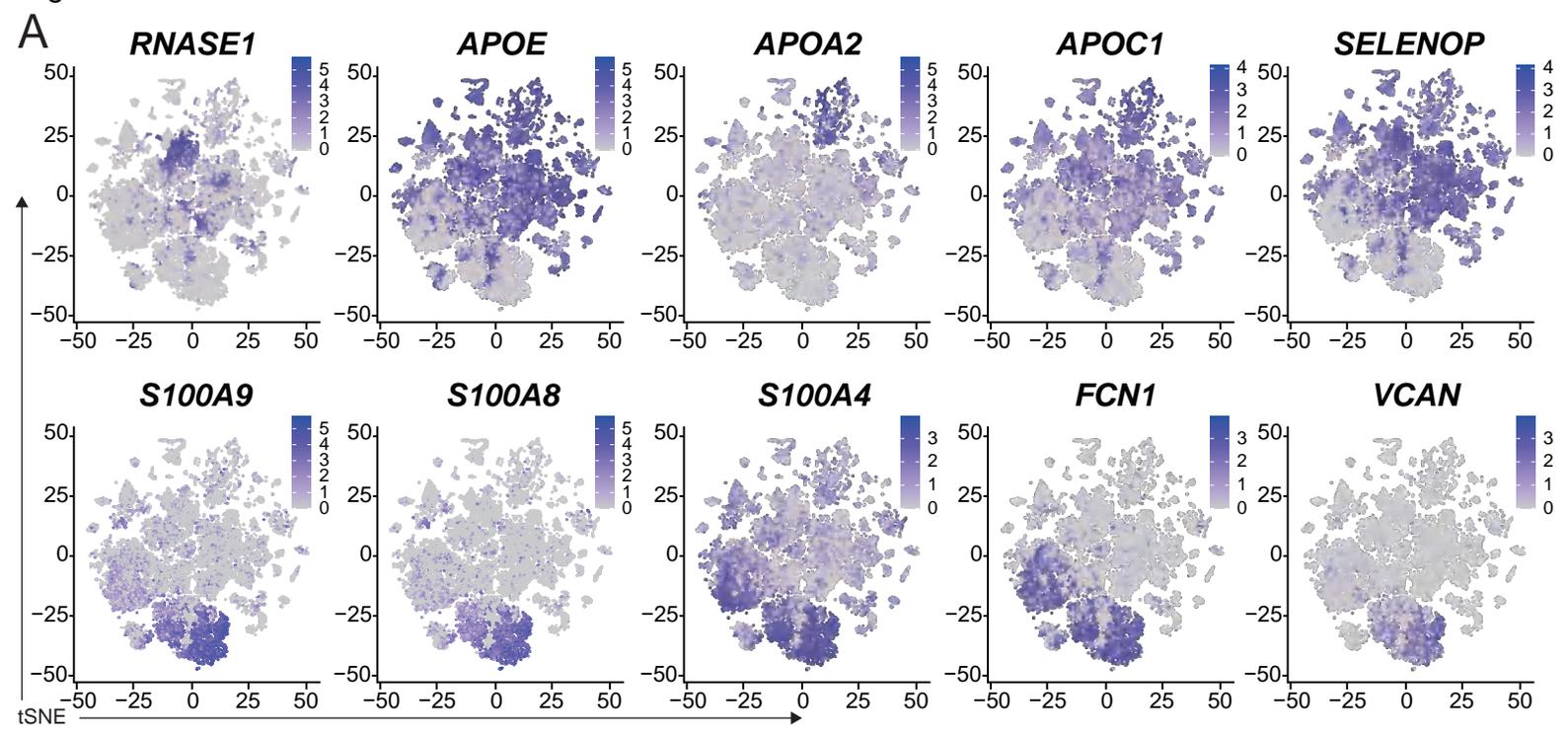
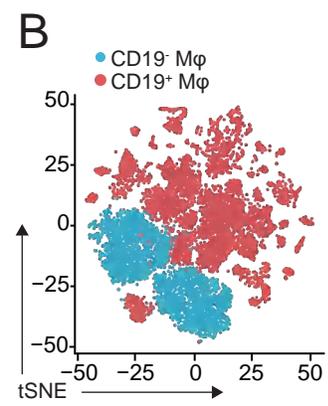
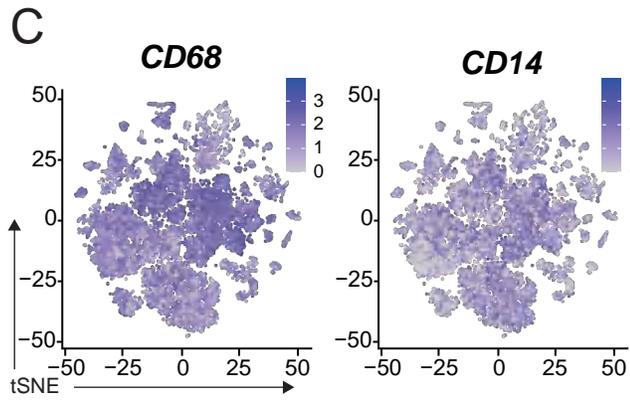
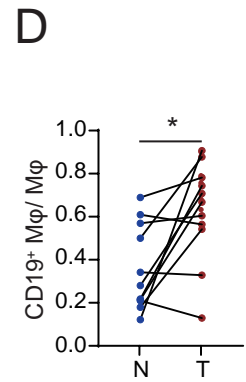
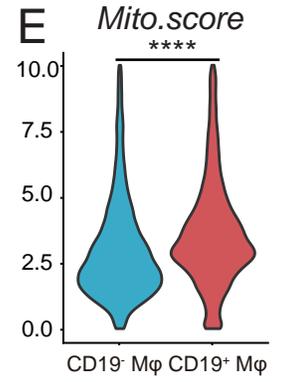

**Fig. S5. CD19⁺ TAMs exhibit a distinct gene expression profile (Public dataset: Bioproject, PRJCA010606) (Related to Fig. 3).**

(**A**) t-SNE plot showing mRNA expression levels of the 10 feature genes in CD19⁺ TAMs and CD19⁻ TAMs. (**B**) t-SNE plot showing the distribution of CD19⁺ TAMs and CD19⁻ TAMs defined by the expression of the 10 feature genes. (**C**) t-SNE plot showing mRNA expression levels of *CD68*, *CD14*, and *CD19* in CD19⁺ TAMs and CD19⁻ TAMs. (**D**) Quantification of CD19⁺ TAMs proportion in tumors and matched adjacent normal tissues. n =11. (**E**) Violin plot showing the expression levels of mitochondrion-related genes in CD19⁺ TAMs and CD19⁻ TAMs. Data are presented as the mean ± SEM. A paired two-tailed *t*-test (**D**) or an unpaired two-tailed *t*-test (**E**) was used. *$P$ < 0.05.

Fig. S6

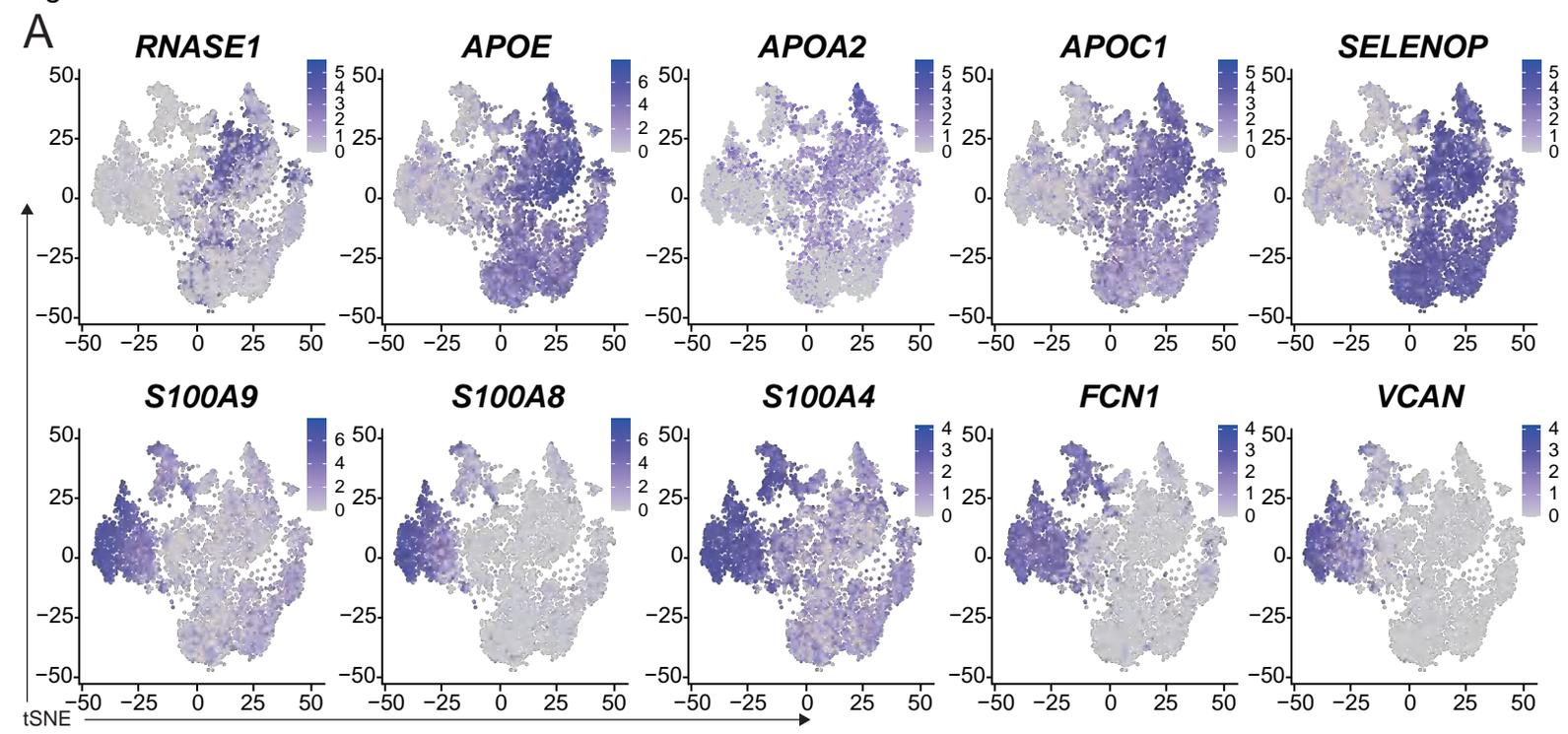
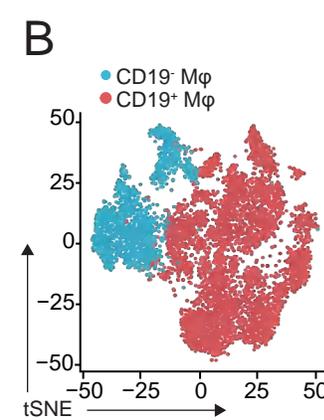
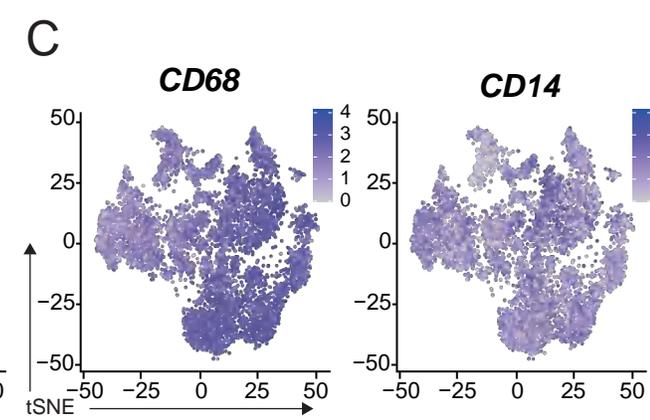
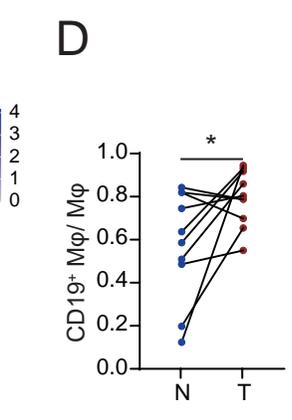
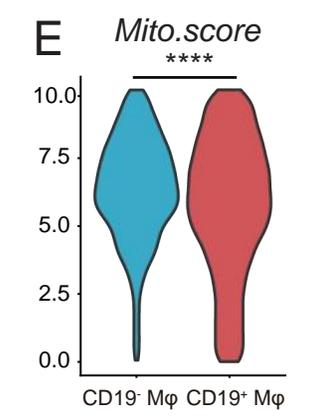

**Fig. S6. CD19⁺ TAMs exhibit a distinct gene expression profile (Public dataset: Bioproject, PRJCA007744) (Related to Fig. 3).**

(**A**) t-SNE plot showing mRNA expression levels of the 10 feature genes in CD19⁺ TAMs and CD19⁻ TAMs. (**B**) t-SNE plot showing the distribution of CD19⁺ TAMs and CD19⁻ TAMs defined by the expression of the 10 feature genes. (**C**) t-SNE plot showing mRNA expression levels of *CD68*, *CD14*, and *CD19* in CD19⁺ TAMs and CD19⁻ TAMs. (**D**) Quantification of CD19⁺ TAMs proportion in tumors and matched adjacent normal tissues. n =11. (**E**) Violin plot showing the expression levels of mitochondrion-related genes in CD19⁺ TAMs and CD19⁻ TAMs. Data are presented as the mean ± SEM. A paired two-tailed *t*-test (**D**) or an unpaired two-tailed *t*-test (**E**) was used. *$P$ < 0.05.

Fig. S7

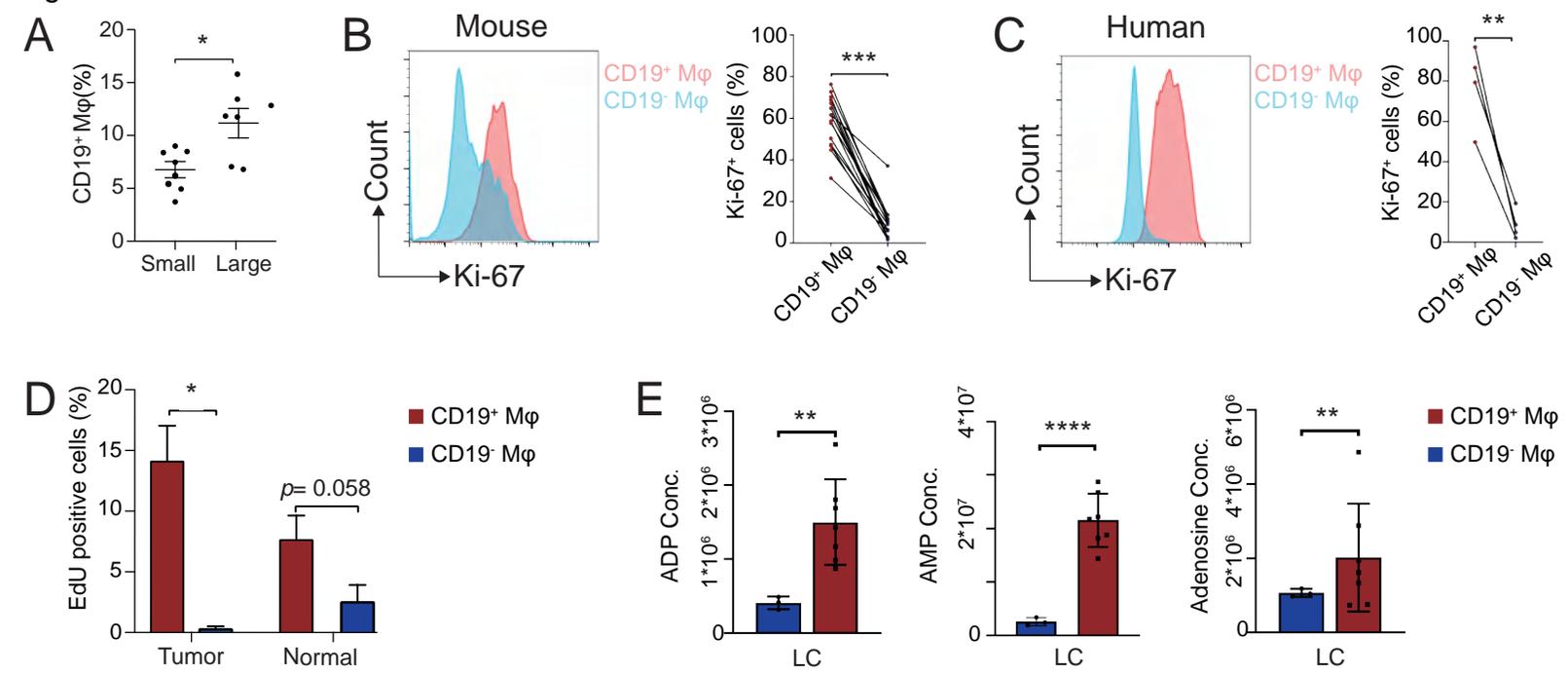

**Fig. S7. CD19⁺ TAMs are highly proliferative and immunosuppressive (Related to Fig. 4).**

(**A**) Flow cytometry quantification of CD19⁺ TAM proportions in small or large tumors of syngeneic HCC mouse model. n = 15. (**B**) Representative flow cytometry histogram and quantification of Ki-67⁺ cells in CD19⁺ TAMs and CD19⁻ TAMs from the tumors in (**A**). (**C**) Representative flow cytometry histogram and quantification of Ki-67⁺ cells in CD19⁺ TAMs and CD19⁻ TAMs from human HCC tissues. n = 4. (**D**) Quantification of Edu⁺ cells in CD19⁺ TAMs and CD19⁻ TAMs of HCC tumors in orthotopic HCC mouse model. n = 10. (**E**) High Performance Liquid Chromatography (HPLC) analysis showing ADP, AMP, and adenosine concentrations in HCC samples with high (> median) or low (< median) CD19⁺ TAM proportions. Data are presented as the mean ± SEM. Unpaired Wilcoxon test (**A**), paired two-tailed *t*-test (**B, C, D**), or unpaired two-tailed *t*-test (**E**) was used. *$P < 0.05$; **$P < 0.01$; ***$P < 0.001$; ****$P < 0.0001$.

Fig. S8

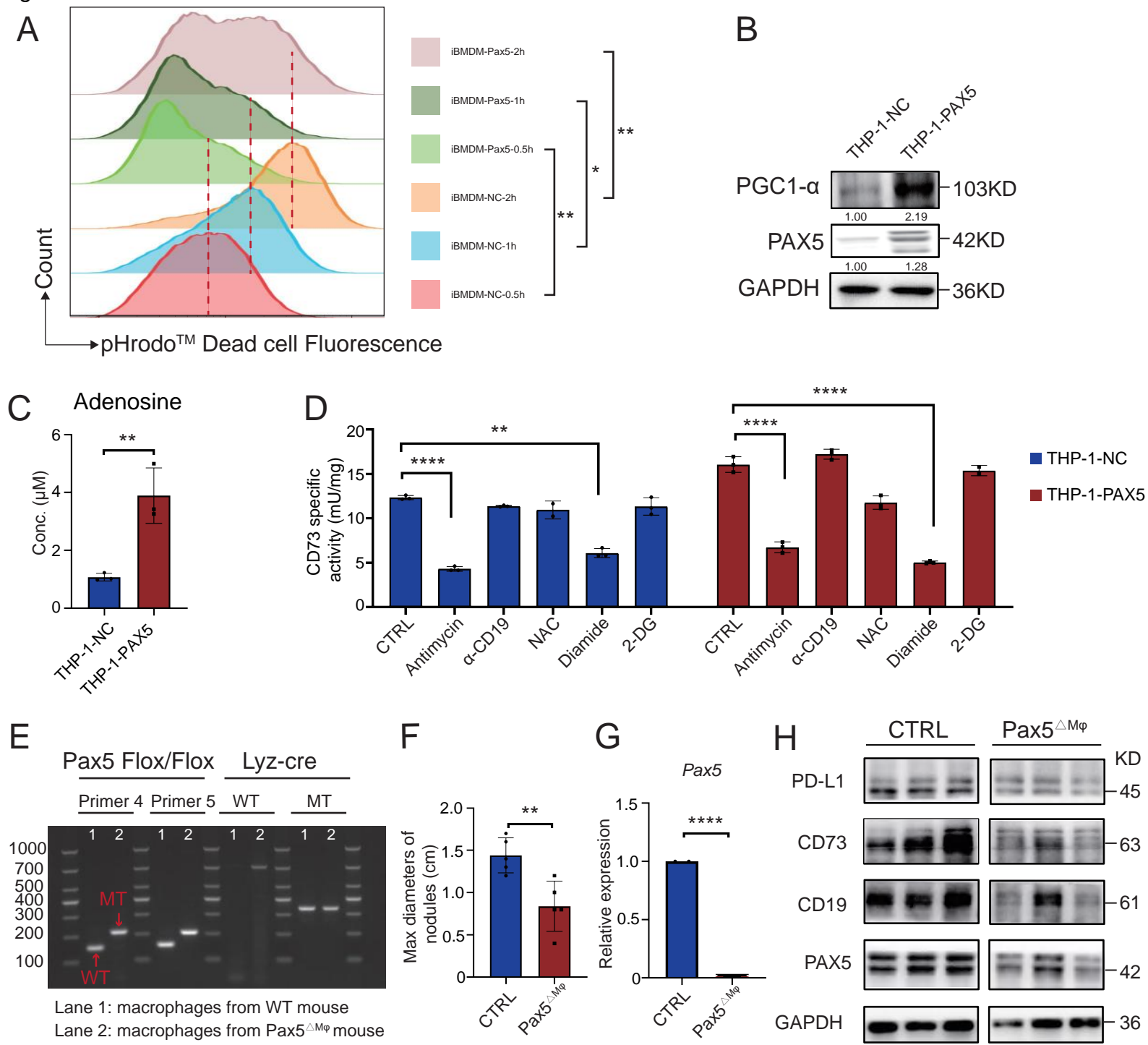

**Fig. S8. PAX5 is a master regulator for CD19+ macrophages (Related to Fig. 5).**

(**A**) Flow cytometry histogram showing the phagocytosis of pHydro<sup>TM</sup> Red SE labelled dead cells by iBMDM cells with or without *Pax5* overexpression. (**B**) Western blotting analysis of PGC1-α and PAX5 in THP-1 cells with or without *PAX5* overexpression. (**C**) Quantification of normalized adenosine concentration in conditioned media of THP-1 cells with or without *PAX5* overexpression. n = 3 independent experiments. (**D**) Quantification of normalized 5′-nucleotidase (CD73) specific activity in THP-1 cells with or without *PAX5* overexpression, or treated with antimycin (mitochondrial respiration inhibitor, 0.5μM), anti-CD19 neutralizing antibody (10g/ml), NAC (ROS inhibitor, 2mM), diamide (ROS inducer, 1mM), or 2-DG (glycolysis inhibitor, 20μM). n = 3 independent experiments. (**E**) Macrophages were sorted from the spleens of wild-type mice and $Pax5^{\triangle M\varphi}$ mice. Genomic DNA was extracted and subject to Southern blotting analysis to verify *Pax5* gene DNA deletion. Primer 4 and Primer 5 were used for *Pax5* floxed allele. WT: wild type allele; MT: floxed/Cre allele. Lane 1: macrophages from littermate control mice.; Lane 2: macrophages from $Pax5^{\triangle M\varphi}$ mice. (**F**) Maximum diameter quantification of tumors in Figure 5i. (**G**) Macrophages were sorted from the spleens of littermate control mice and $Pax5^{\triangle M\varphi}$ mice. Total RNA was extracted and subjected to RT-PCR for *Pax5* mRNA levels. (**H**) Macrophages were differentiated from the bone marrow cells of littermate control mice and $Pax5^{\triangle M\varphi}$ mice. Proteins were extracted and subjected to western blotting analysis for PAX5, CD19, CD73, and PD-L1 levels. Data are presented as the mean ± SEM. Two-way ANOVA test (**A**), unpaired two-tailed *t*-test is used (**C, D, G**), or unpaired One-way ANOVA with Welch's correction (**F**) was used. *$P < 0.05$; **$P < 0.01$; ****$P < 0.0001$.

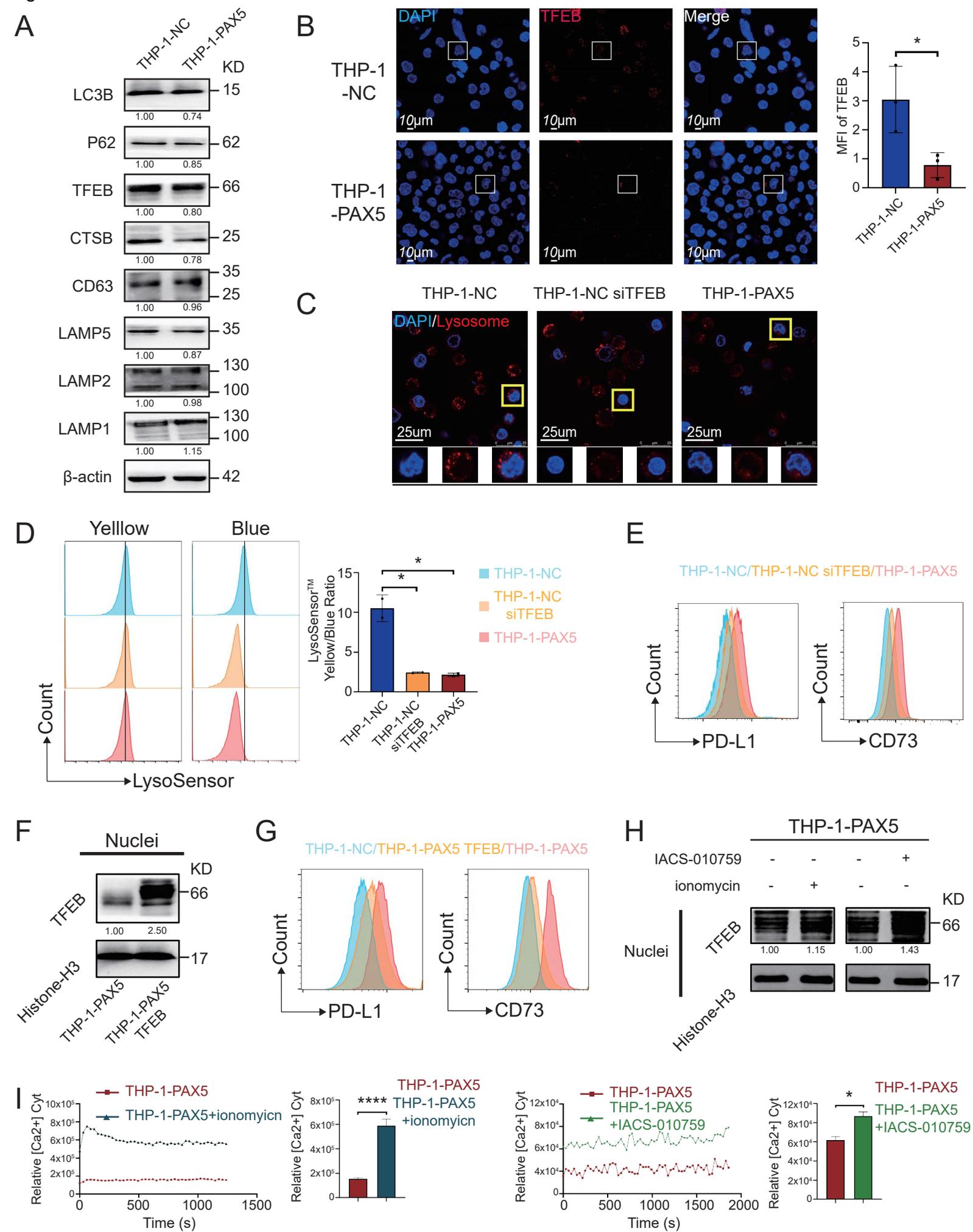

**Fig. S9. PAX5 induces PD-L1 and CD73 in a post-transcriptional manner (Related to Fig. 6).**

(**A**) Western blotting analysis of lysosome-associated proteins in THP-1 cells with or without *PAX5* overexpression. (**B**) Representative immunofluorescence images and quantification of TFEB (red) in THP-1 cells with or without *PAX5* overexpression. Blue, DAPI; scale bar, 10 µm. (**C**) Representative immunofluorescence showing Lysotracker (red) and DAPI (blue) in THP-1 cells with or without *PAX5* overexpression or *TFEB* knockdown. (**D**) Representative immunofluorescence images and quantification of Lysotracker (red) in THP-1 cells with or without *PAX5* overexpression or *TFEB* knockdown. Blue, DAPI. (**E**) Representative flow cytometry histograms showing CD19, CD73, and PD-L1 expression in THP-1 cells with or without *PAX5* overexpression or *TFEB* knockdown. (**F**) Western blotting analysis of nuclear TFEB in *PAX5*-overexpressing THP-1 cells. (**G**) Representative flow cytometry histograms showing CD19, CD73, and PD-L1 levels in *PAX5*-overexpressing THP-1 cells with or without *TFEB* overexpression. (**H**) Western blotting analysis of nuclear TFEB levels in *PAX5*-overexpressing THP-1 cells pretreated with ionomycin (a $Ca^{2+}$ ionophore, 1 µM, 24 h), or oxidative phosphorylation inhibitor IACS-010759 (100 nM, 24 h). (**I**) Fluorescent probe indicating relative $Ca^{2+}$ levels in the cytoplasm of *PAX5*-overexpressing THP-1 cells pretreated with ionomycin or IACS-010759. Data are presented as the mean ± SEM. Unpaired two-tailed *t*-test (**B, D**), or two-way ANOVA test (**I**) was used. *$P < 0.05$; ****$P < 0.0001$.

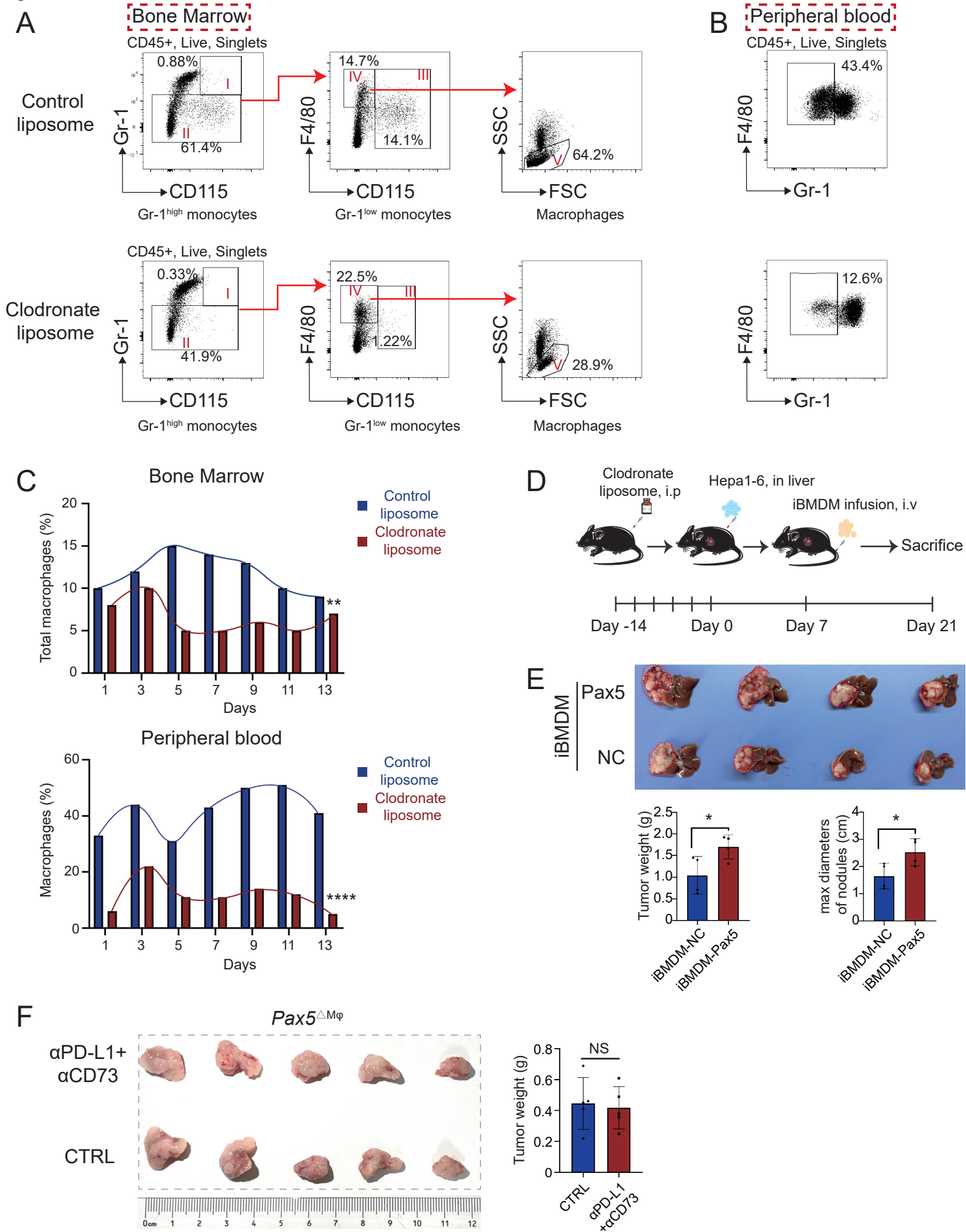

**Fig. S10. Blocking CD19⁺ TAMs enhances the efficacy of immune checkpoint blockade therapy (Related to Fig. 7).**

(**A**) Representative flow cytometry plots showing gating strategy for bone marrow mononuclear phagocytes. Part I indicates Gr-1$^{high}$ monocytes (Gr-1$^{high}$CD115$^+$), Part II indicates Gr-1$^{low}$ fraction, which was further divided into Part III (Gr-1$^{low}$ monocytes, Gr-1$^{low}$CD115$^+$) and Part IV (F4/80$^+$CD115$^-$) populations. Part IV fraction was further subdivided into Part V (SSC$^{low}$ macrophages). (**B**) Representative flow cytometry plots showing gating strategy of peripheral blood mononuclear phagocytes F4/80$^+$Gr-$^-$ population. (**C**) Quantification of macrophages depletion efficiency by clodronate liposome or control liposome. Mice received clodronate liposome or control liposome every other day for four times, then bone marrow and peripheral macrophages were subjected to flow cytometrical analysis (gating as in **A** and **B**). (**D**) Schematic illustration showing the experiment for testing CD19$^+$ TAMs function in vivo in case of macrophage elimination. C57 BL/6 mice received clodronate liposomes every other day for four times, then were inoculated with Hepa 1-6 cells *in situ*, following by intravenous injection (*i.v.*) with iBMDMs with or without *Pax5* overexpression. (**E**) Images and sizes of the tumors in (**D**). n = 4 per group. (**F**) Images and sizes of the tumors treated with a combined therapy (anti-CD73/anti-PD-L1 neutralizing antibodies) or control IgG antibodies in orthortopic HCC model using *Pax5$^{\triangle M\varphi}$* (*Pax5$^{flox/flox}$;Lyz-Cre*) mice. n = 5 per group. Data are presented as mean ± SEM; Two-way ANOVA test (**C**), or unpaired one-way ANOVA with Welch's correction (**D, E, F**) was used. $^*P < 0.05$; $^{**}P < 0.01$; $^{****}P < 0.0001$.

Fig. S11

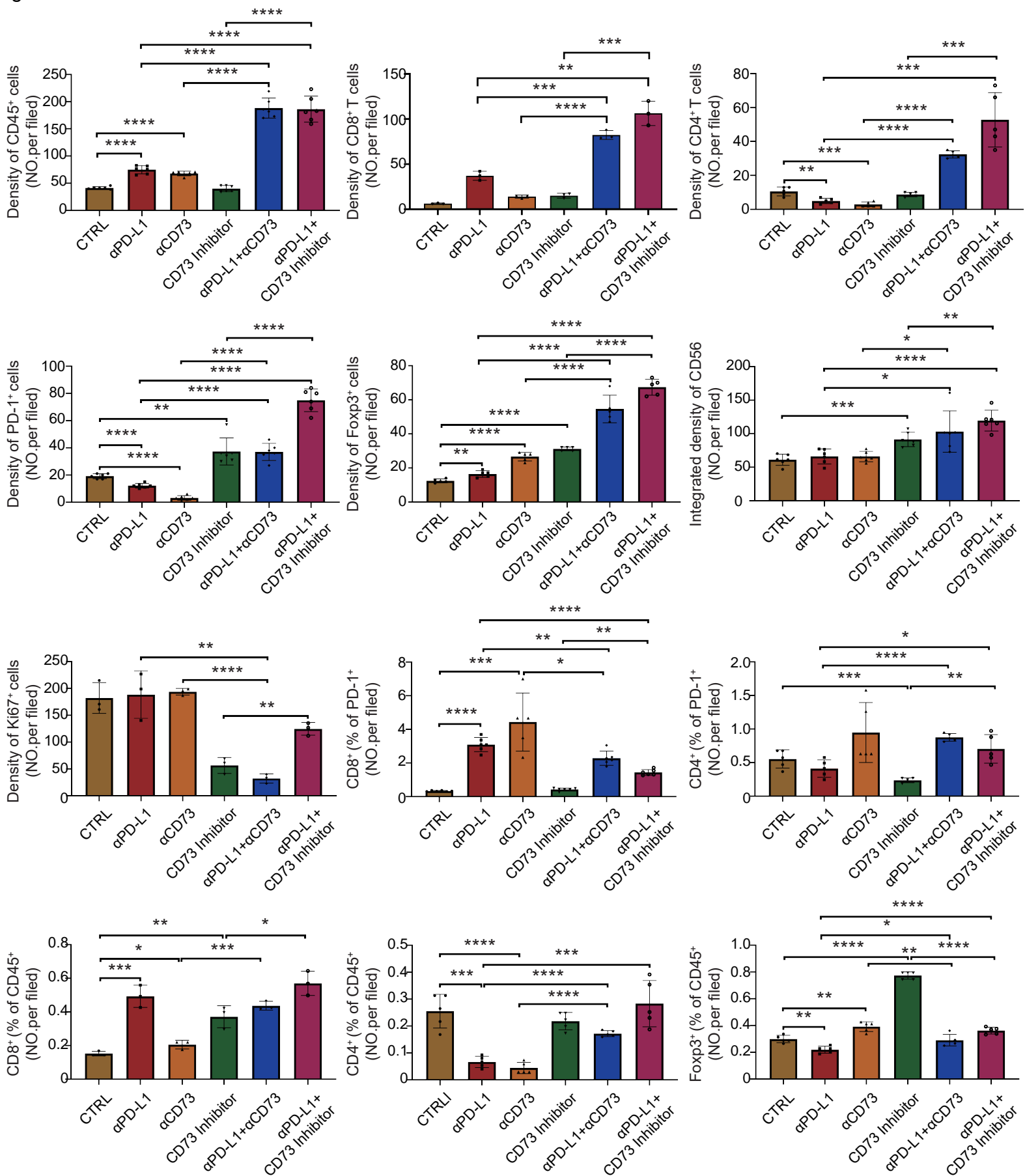

**Fig. S11. Blocking CD73 of CD19⁺ TAMs enhances the efficacy of immune checkpoint blockade therapy (Related to Fig. 7).**

Tumor Immune microenvironment analysis of the HCC tumors with targeted therapy of CD73 and PD-L1 in CD19⁺ TAM. Analyzed immune cells include $CD45^+$ cells, $CD8^+$ T cells, $CD4^+$ T cells, $PD-1^+$ cells, $Foxp3^+$ cells, $CD56^+$ cells, $Ki67^+$ cells, $CD8^+/PD-1^+$ T cells, $CD4^+/PD-1^+$ T cells, $CD8^+$ T/$CD45^+$ cells, $CD4^+$ T/$CD45^+$ cells, and $Foxp3^+$ T/$CD45^+$ cells. Statistical analyses were performed using an unpaired One-way ANOVA with Welch's correction. Data are presented as the mean ± SEM. $^*P < 0.05$; $^{**}P < 0.01$; $^{***}P < 0.001$, $^{****}P < 0.0001$.

Fig. S12

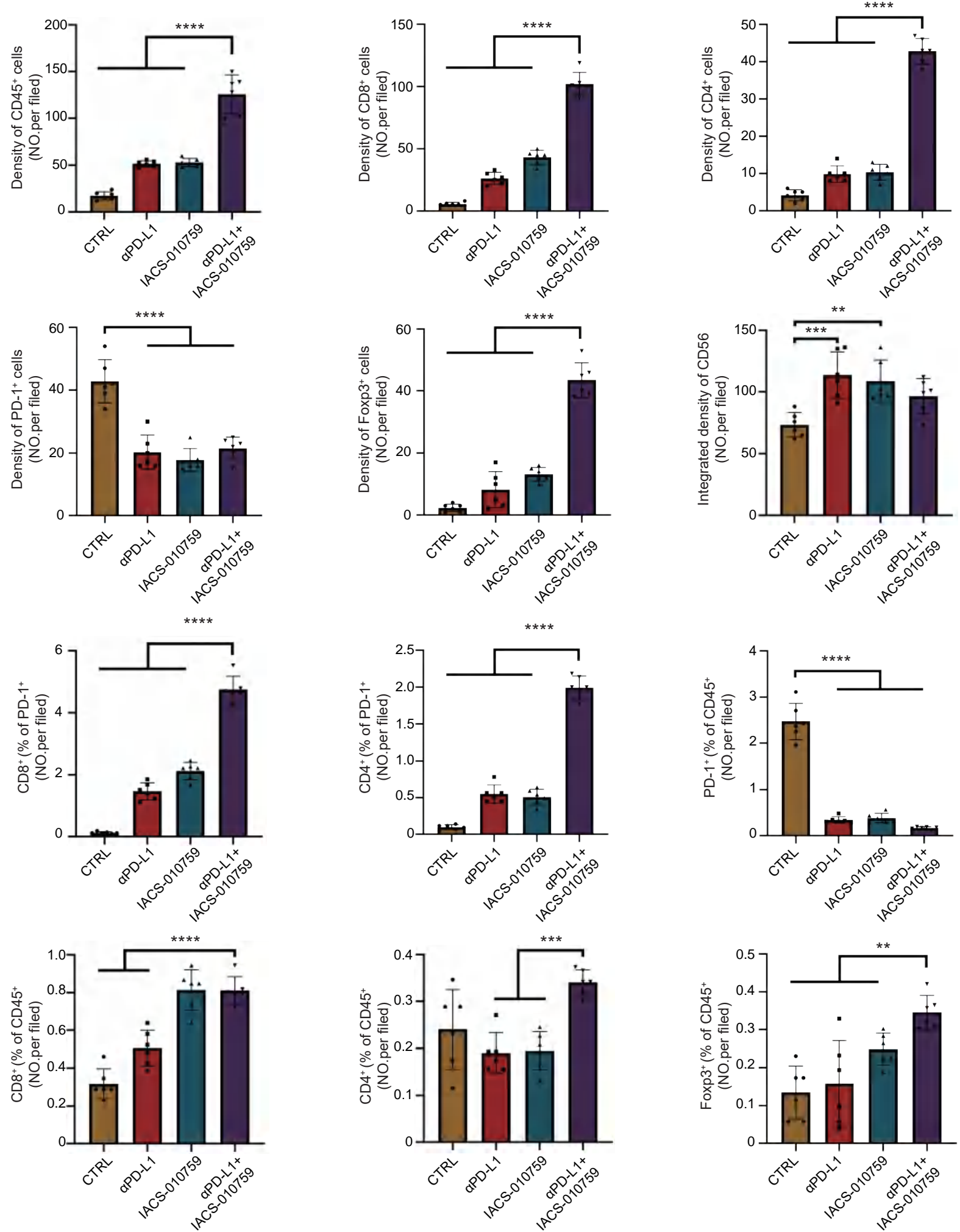

**Fig. S12. Blocking OXPHOS activity of CD19+ TAMs enhances the efficacy of immune checkpoint blockade therapy (Related to Fig. 7).**

Tumor Immune microenvironment analysis of the HCC tumors with targeted therapy of OXPHOS and PD-L1 in CD19+ TAM. Analyzed immune cells include CD45+ cells, CD8+ T cells, CD4+ T cells, PD-1+ cells, Foxp3+ cells, CD56+ cells, CD8+/PD-1+ T cells, CD4+/PD-1+ T cells, PD-1+/CD45+ cells, CD8+ T/CD45+ cells, CD4+ T/CD45+cells, and Foxp3+ T/CD45+ cells. Statistical analyses were performed using unpaired One-way ANOVA with Welch's correction. Data are presented as the mean ± SEM. *$P$ < 0.05; **$P$ < 0.01; ***$P$ < 0.001, ****$P$ < 0.0001.

**Supplementary Table 1. Patients' clinical information, related to Fig. 1, Fig. 2, and Fig. S2.**

**HCC**

| Patient | Age | Gender | Tumor size (cm) | Histopathology | Grade | TNM Stage |
|---|---|---|---|---|---|---|
| Patient 1 | 62 | Male | <5 | HCC | III | T1aN0M0 |
| Patient 2 | 67 | Male | <5 | HCC | III | T1aN0M0 |
| Patient 3 | 55 | Female | <5 | HCC | III | T1aN0M0 |
| Patient 4 | 46 | Male | <5 | HCC | I | T2N0M0 |
| Patient 5 | 57 | Male | >=5 | HCC | II | T3aN0M0 |
| Patient 6 | 52 | Male | <5 | HCC | II | T1aN0M0 |
| Patient 7 | 80 | Female | <5 | HCC | III | T2N0M0 |
| Patient 8 | 71 | Male | <5 | HCC | II | T1aN0M0 |
| Patient 9 | 52 | Male | >=5 | HCC | III | T1bN0M0 |
| Patient 10 | 53 | Male | >=5 | HCC | III | T1bN0M0 |
| Patient 11 | 68 | Female | <5 | HCC | I | T1aN0M0 |
| Patient 12 | 54 | Male | <5 | HCC | I | T1aN0M0 |
| Patient 13 | 64 | Male | >=5 | HCC | II | T1bN0M0 |
| Patient 14 | 62 | Male | <5 | HCC | II | T1aN0M0 |
| Patient 15 | 65 | Male | >=5 | HCC | II | T1bN0M0 |
| Patient 16 | 64 | Male | >=5 | HCC | III | T3N1M0 |
| Patient 17 | 66 | Female | <5 | HCC | III | T1aN0M0 |

| Patient | Age | Gender | Tumor size (cm) | Histopathology | Grade | TNM Stage |
|---|---|---|---|---|---|---|
| Patient 18 | 60 | Female | <5 | HCC | III | T1aN0M0 |
| Patient 19 | 47 | Male | <5 | HCC | II | T1aN0M0 |
| Patient 20 | 63 | Male | >=5 | HCC | II | T1bN0M0 |
| Patient 21 | 68 | Male | >=5 | HCC | II | T1bN0M0 |
| Patient 22 | 45 | Male | <5 | HCC | I | T1aN0M0 |
| Patient 23 | 74 | Male | <5 | HCC | II | T2N0M0 |
| Patient 24 | 63 | Male | <5 | HCC | II | T2N0M0 |
| Patient 25 | 71 | Female | >=5 | HCC | II | T2N0M0 |
| Patient 26 | 77 | Male | <5 | HCC | II | T1aN0M0 |
| Patient 27 | 78 | Female | >=5 | HCC | II | T2N0M0 |
| Patient 28 | 74 | Female | <5 | HCC | II | T1aN0M0 |

**PDAC**

| Patient | Age | Gender | Tumor size (cm) | Histopathology | Grade | TNM Stage |
|---|---|---|---|---|---|---|
| Patient 1 | 74 | Female | >=5 | Adenocarcinoma | Poorly-moderate | T3N0M0 |
| Patient 2 | 68 | Male | >=5 | Adenocarcinoma | Moderate | T3N0M0 |
| Patient 3 | 59 | Female | >=5 | Adenocarcinoma | Moderate | T3N0M0 |
| Patient 4 | 61 | Female | >=2, <5 | Adenocarcinoma | Poorly-moderate | T3N1M1 |
| Patient 5 | 71 | Male | <2 | Adenocarcinoma | Moderate | T1cN1M0 |
| Patient 6 | 66 | Female | >=2, <5 | Adenocarcinoma | Moderate-highly | T2N1M0 |
| Patient 7 | 55 | Male | >=2, <5 | Adenocarcinoma | Poorly-moderate | T2N0M0 |

| Patient | Age | Gender | Tumor size (cm) | Histopathology | Grade | TNM Stage |
|---|---|---|---|---|---|---|
| Patient 8 | 72 | Female | >=2, <5 | Adenocarcinoma | Moderate | T2N0M0 |
| Patient 9 | 48 | Female | >=2, <5 | Adenocarcinoma | Moderate | T2N0M0 |
| Patient 10 | 69 | Female | >=2, <5 | Adenocarcinoma | Moderate | T2N0M0 |
| Patient 11 | 55 | Male | >=2, <5 | Adenocarcinoma | Moderate | T1cN0M0 |
| Patient 12 | 67 | Female | >=2, <5 | Adenocarcinoma | Poorly-moderate | T2N1M0 |
| Patient 13 | 48 | Male | >=2, <5 | Adenocarcinoma | Poorly | T2N0M0 |

**Breast carcinoma**

| Patient | Age | Gender | Tumor size (cm) | Histopathology | Grade | TNM Stage |
|---|---|---|---|---|---|---|
| Patient 1 | 53 | Female | >=2, <5 | Invasive ductal carcinoma | III | T2N1aM0 |
| Patient 2 | 62 | Female | >=2, <5 | Invasive carcinoma (micropapillary + mucinous) | II | T2N3aM0 |
| Patient 3 | 55 | Female | >=2, <5 | Invasive ductal carcinoma | III | T2N0M0 |
| Patient 4 | 56 | Female | >=2, <5 | Invasive ductal carcinoma | III | T2N1aM0 |
| Patient 5 | 30 | Female | >=2, <5 | Invasive ductal carcinoma | III | T2N1aM0 |
| Patient 6 | 49 | Female | >=2, <5 | Invasive lobular carcinoma | I | T2N0M0 |
| Patient 7 | 45 | Female | >=2, <5 | Invasive ductal carcinoma | III | T2N0M0 |
| Patient 8 | 40 | Female | <2 | Invasive ductal carcinoma | II | T1cN0M0 |
| Patient 9 | 51 | Female | >=2, <5 | Invasive ductal carcinoma | II | T2N2M0 |
| Patient 10 | 53 | Female | <2 | Invasive ductal carcinoma | III | T1cN3M0 |
| Patient 11 | 48 | Female | >=2, <5 | Invasive ductal carcinoma | III | T1cN0M0 |

| Patient 12 | 57 | Female | <2 | Invasive ductal carcinoma | II | T1cN2M0 |

**Gastric carcinoma**

| Patient | Age | Gender | Tumor size (cm) | Histopathology | Grade | TNM Stage |
|---|---|---|---|---|---|---|
| Patient 1 | 38 | Female | <5 | Adenocarcinoma | I | T1N0M0 |
| Patient 2 | 62 | Female | <5 | Adenocarcinoma | I | T1N1M0 |
| Patient 3 | 74 | Male | >=5 | Adenocarcinoma | II | T3N0M0 |
| Patient 4 | 57 | Male | <5 | Adenocarcinoma | II | T1N0M0 |
| Patient 5 | 71 | Female | >=5 | Adenocarcinoma | I | T3N0M0 |
| Patient 6 | 63 | Female | <5 | Adenocarcinoma | II | T1N0M0 |

**Renal carcinoma**

| Patient | Age | Gender | Tumor size (cm) | Histopathology | Grade | TNM Stage |
|---|---|---|---|---|---|---|
| Patient 1 | 60 | Female | >7, <=10 | No classification | NA | T2aN0M0 |
| Patient 2 | 60 | Male | >4, <=7 | Clear cell carcinomma | NA | T1bN0M0 |
| Patient 3 | 63 | Male | <=4 | Clear cell carcinomma | NA | T1aN0M0 |
| Patient 4 | 53 | Female | >4, <=7 | Clear cell carcinomma | NA | T1bN0M0 |
| Patient 5 | 69 | Male | <=4 | Clear cell carcinomma | NA | T1aN0M0 |
| Patient 6 | 51 | Female | <=4 | No classification | NA | T1aN0M0 |
| Patient 7 | 56 | Male | <=4 | Clear cell carcinomma | NA | T1aN0M0 |
| Patient 8 | 70 | Female | >4, <=7 | Clear cell carcinomma | NA | T1bN0M0 |

| Patient | Age | Gender | Tumor size (cm) | Histopathology | Grade | TNM Stage |
|---|---|---|---|---|---|---|
| Patient 9 | 53 | Male | <=4 | Clear cell carcinomma | NA | T1aN0M0 |
| Patient 10 | 45 | Male | <=4 | Clear cell carcinomma | NA | T1aN0M0 |
| Patient 11 | 54 | Female | >4, <=7 | Clear cell carcinomma | NA | T1bN0M0 |

**Colorectal carcinoma**

| Patient | Age | Gender | Tumor size (cm) | Histopathology | Grade | TNM Stage |
|---|---|---|---|---|---|---|
| Patient 1 | 68 | Male | >=5cm | Adenocarcinoma | Poorly-moderate | T3N1M0 |
| Patient 2 | 61 | Female | <5 | Adenocarcinoma | Moderate | T1N0M0 |
| Patient 3 | 74 | Female | <5 | Adenocarcinoma | Moderate | T3N0M0 |
| Patient 4 | 67 | Male | <5 | Adenocarcinoma | Poorly | T3N1M0 |
| Patient 5 | 62 | Male | <5 | Adenocarcinoma | Moderate | T3N0M0 |
| Patient 6 | 62 | Female | >=5cm | Adenocarcinoma | Moderate | T3N1M0 |
| Patient 7 | 67 | Male | <5 | Adenocarcinoma | Poorly-moderate | T3N0M0 |
| Patient 8 | 70 | Male | >=5cm | Adenocarcinoma | Highly | T1N0M0 |
| Patient 9 | 82 | Female | >=5cm | Adenocarcinoma | Poorly-moderate | T4aN1M0 |
| Patient 10 | 68 | Female | <5 | Adenocarcinoma | Moderate | T1N1M0 |
| Patient 11 | 69 | Male | <5 | Adenocarcinoma | Moderate | T1N0M0 |

| Patient | Age | Gender | Tumor size (cm) | Histopathology | Grade | TNM Stage | OS (m) |
|---|---|---|---|---|---|---|---|
| Patient 1 | 58 | Male | <5 | HCC | II | T4N1M1 | 13 |
| Patient 2 | 65 | Male | <5 | HCC | III | T1aN0M0 | 14 |
| Patient 3 | 18 | Male | <5 | HCC | I | T2N0M0 | 15 |
| Patient 4 | 72 | Male | >=5 | HCC | II | T1bN0M0 | 15 |
| Patient 5 | 49 | Female | >=5 | HCC | II | T4N1M1 | 7 |
| Patient 6 | 61 | Male | <5 | HCC | II | T3N0M0 | 20 |
| Patient 7 | 46 | Male | >=5 | HCC | II | T3N0M0 | 24 |
| Patient 8 | 58 | Male | <5 | HCC | II | T2N0M0 | 25 |
| Patient 9 | 59 | Male | <5 | HCC | II | T1aN0M0 | 25 |
| Patient 10 | 65 | Male | <5 | HCC | III | T1aN0M0 | 26 |
| Patient 11 | 62 | Female | <5 | HCC | III | T1aN0M0 | 26 |
| Patient 12 | 59 | Male | <5 | HCC | III | T1aN0M0 | 43 |
| Patient 13 | 35 | Male | >=5 | HCC | II | T2N0M0 | 19 |
| Patient 14 | 61 | Male | >=5 | HCC | II | T3N0M0 | 32 |
| Patient 15 | 46 | Male | <5 | HCC | II | T2N0M0 | 34 |
| Patient 16 | 63 | Male | <5 | HCC | II | T2N0M0 | 35 |
| Patient 17 | 58 | Male | <5 | HCC | II | T2N0M0 | 35 |
| Patient 18 | 69 | Male | <5 | HCC | II | T2N0M0 | 47 |
| Patient 19 | 79 | Male | <5 | HCC | III | T2N0M0 | 43 |
| Patient 20 | 43 | Female | <5 | HCC | III | T1aN0M0 | 45 |
| Patient 21 | 49 | Male | <5 | HCC | II | T1aN0M0 | 45 |
| Patient 22 | 55 | Male | <5 | HCC | I | T2N0M0 | 6 |

| Patient | Age | Sex | Size | Type | Grade | TNM | Months |
|---|---|---|---|---|---|---|---|
| Patient 23 | 64 | Male | <5 | HCC | II | T2N0M0 | 46 |
| Patient 24 | 47 | Male | <5 | HCC | III | T1aN0M0 | 48 |
| Patient 25 | 81 | Male | <5 | HCC | II | T1aN0M0 | 38 |
| Patient 26 | 59 | Male | <5 | HCC | II | T1aN0M0 | 50 |
| Patient 27 | 53 | Male | <5 | HCC | II | T1aN0M0 | 51 |
| Patient 28 | 58 | Male | <5 | HCC | II | T2N0M0 | 63 |
| Patient 29 | 39 | Male | >=5 | HCC | II | T1bN0M0 | 12 |
| Patient 30 | 47 | Male | >=5 | HCC | II | T3N0M0 | 39 |
| Patient 31 | 58 | Male | <5 | HCC | II | T1aN0M0 | 51 |
| Patient 32 | 67 | Male | <5 | HCC | II | T1aN0M0 | 52 |
| Patient 33 | 45 | Female | <5 | HCC | II | T1aN0M0 | 53 |
| Patient 34 | 64 | Male | >=5 | HCC | I | T4N1M1 | 25 |
| Patient 35 | 53 | Male | >=5 | HCC | II | T2N0M0 | 26 |
| Patient 36 | 22 | Male | <5 | HCC | II | T1aN0M0 | 54 |
| Patient 37 | 64 | Male | >=5 | HCC | II | T1bN0M0 | 54 |
| Patient 38 | 73 | Male | >=5 | HCC | I | T3N0M0 | 55 |
| Patient 39 | 52 | Male | >=5 | HCC | II | T4N1M0 | 55 |
| Patient 40 | 69 | Male | <5 | HCC | II | T1aN0M0 | 55 |
| Patient 41 | 74 | Male | <5 | HCC | II | T4N1M1 | 56 |
| Patient 42 | 84 | Male | <5 | HCC | II | T1aN0M0 | 43 |
| Patient 43 | 57 | Male | >=5 | HCC | III | T1bN0M0 | 56 |
| Patient 44 | 54 | Male | >=5 | HCC | I | T1bN0M0 | 56 |
| Patient 45 | 67 | Female | <5 | HCC | II | T1aN0M0 | 57 |
| Patient 46 | 60 | Male | <5 | HCC | II | T1aN0M0 | 58 |
| Patient 47 | 66 | Male | <5 | HCC | II | T1aN0M0 | 61 |
| Patient 48 | 67 | Female | >=5 | HCC | II | T1bN0M0 | 62 |
| Patient 49 | 46 | Male | <5 | HCC | II | T1aN0M0 | 63 |
| Patient 50 | 84 | Male | <5 | HCC | III | T1aN0M0 | 27 |
| Patient 51 | 54 | Male | <5 | HCC | II | T1aN0M0 | 51 |
| Patient 52 | 60 | Female | >=5 | HCC | II | T1bN0M0 | 64 |
| Patient 53 | 63 | Male | <5 | HCC | II | T2N0M0 | 65 |
| Patient 54 | 75 | Male | <5 | HCC | II | T4N1M1 | 64 |
| Patient 55 | 58 | Female | <5 | HCC | II | T1aN0M0 | 66 |
| Patient 56 | 51 | Male | >=5 | HCC | II | T4N1M1 | 34 |
| Patient 57 | 60 | Female | <5 | HCC | III | T1aN0M0 | 66 |
| Patient 58 | 44 | Male | <5 | HCC | II | T1aN0M0 | 67 |
| Patient 59 | 41 | Female | <5 | HCC | II | T1aN0M0 | 67 |
| Patient 60 | 46 | Male | <5 | HCC | II | T1aN0M0 | 61 |
| Patient 61 | 68 | Male | >=5 | HCC | I | T3N0M0 | 17 |
| Patient 62 | 59 | Male | >=5 | HCC | II | T3N0M0 | 68 |
| Patient 63 | 59 | Female | <5 | HCC | III | T1aN0M0 | 62 |
| Patient 64 | 76 | Male | <5 | HCC | III | T1aN0M0 | 71 |
| Patient 65 | 52 | Male | <5 | HCC | I | T1aN0M0 | 64 |
| Patient 66 | 62 | Male | <5 | HCC | II | T1aN0M0 | 19 |
| Patient 67 | 79 | Male | >=5 | HCC | II | T1bN0M0 | 10 |
| Patient 68 | 65 | Male | <5 | HCC | II | T1aN0M0 | 72 |
| Patient 69 | 69 | Male | <5 | HCC | II | T1aN0M0 | 25 |
| Patient 70 | 65 | Male | <5 | HCC | II | T2N0M0 | 67 |
| Patient 71 | 48 | Male | <5 | HCC | II | T2N0M0 | 67 |
| Patient 72 | 80 | Male | <5 | HCC | II | T1aN0M0 | 35 |
| Patient 73 | 48 | Male | <5 | HCC | III | T2N0M0 | 67 |
| Patient 74 | 66 | Male | >=5 | HCC | II | T1bN0M0 | 20 |
| Patient 75 | 60 | Male | <5 | HCC | II | T4N1M0 | 66 |
| Patient 76 | 50 | Male | <5 | HCC | II | T1aN0M0 | 11 |
| Patient 77 | 43 | Male | >=5 | HCC | I | T2N0M0 | 64 |
| Patient 78 | 81 | Male | >=5 | HCC | III | T1bN0M0 | 62 |
| Patient 79 | 75 | Male | <5 | HCC | II | T2N0M0 | 17 |
| Patient 80 | 78 | Male | >=5 | HCC | II | T2N0M0 | 48 |
| Patient 81 | 54 | Male | >=5 | HCC | II | T2N0M0 | 8 |
| Patient 82 | 52 | Male | >=5 | HCC | I | T3N0M0 | 44 |
| Patient 83 | 77 | Male | <5 | HCC | II | T2N0M0 | 9 |
| Patient 84 | 61 | Male | >=5 | HCC | II | T1bN0M0 | 44 |
| Patient 85 | 73 | Female | >=5 | HCC | II | T3N0M0 | 38 |
| Patient 86 | 70 | Male | >=5 | HCC | II | T1bN0M0 | 33 |
| Patient 87 | 67 | Male | <5 | HCC | III | T2N0M0 | 12 |
| Patient 88 | 39 | Female | <5 | HCC | II | T1aN0M0 | 9 |
| Patient 89 | 62 | Male | <5 | HCC | II | T2N0M0 | 37 |
| Patient 90 | 66 | Male | <5 | HCC | II | T1aN0M0 | 10 |
| Patient 91 | 58 | Male | >=5 | HCC | II | T2N0M0 | 12 |
| Patient 92 | 55 | Female | <5 | HCC | I | T2N0M0 | 19 |
| Patient 93 | 63 | Male | <5 | HCC | I | T2N0M0 | 21 |
| Patient 94 | 65 | Male | <5 | HCC | II | T1aN0M0 | 22 |
| Patient 95 | 69 | Male | <5 | HCC | II | T1aN0M0 | 21 |
| Patient 96 | 69 | Male | >=5 | HCC | I | T2N0M0 | 19 |
| Patient 97 | 57 | Female | >=5 | HCC | II | T3N0M0 | 14 |
| Patient 98 | 73 | Male | >=5 | HCC | II | T1bN0M0 | 27 |
| Patient 99 | 58 | Male | >=5 | HCC | II | T1bN0M0 | 30 |
| Patient 100 | 58 | Male | <5 | HCC | II | T1aN0M0 | 30 |
| Patient 101 | 57 | Female | <5 | HCC | II | T1aN0M0 | 30 |
| Patient 102 | 53 | Male | <5 | HCC | I | T1aN0M0 | 32 |

| Patient | Age | Sex | Size | Type | Grade | TNM | Months |
|---|---|---|---|---|---|---|---|
| Patient 103 | 56 | Male | >=5 | HCC | II | T1aN0M0 | 33 |
| Patient 104 | 72 | Male | <5 | HCC | II | T2N0M0 | 35 |
| Patient 105 | 46 | Male | <5 | HCC | III | T1aN0M0 | 37 |
| Patient 106 | 69 | Male | <5 | HCC | I | T2N0M0 | 37 |
| Patient 107 | 55 | Male | <5 | HCC | II | T2N0M0 | 38 |
| Patient 108 | 63 | Male | <5 | HCC | II | T1aN0M0 | 41 |
| Patient 109 | 48 | Male | <5 | HCC | II | T2N0M0 | 42 |
| Patient 110 | 63 | Male | <5 | HCC | III | T2N0M0 | 43 |
| Patient 111 | 64 | Male | <5 | HCC | II | T1aN0M0 | 45 |
| Patient 112 | 74 | Male | <5 | HCC | III | T1aN0M0 | 45 |
| Patient 113 | 77 | Male | <5 | HCC | III | T1aN0M0 | 47 |
| Patient 114 | 68 | Male | <5 | HCC | III | T1aN0M0 | 48 |
| Patient 115 | 61 | Male | <5 | HCC | II | T1aN0M0 | 49 |
| Patient 116 | 57 | Male | <5 | HCC | III | T1aN0M0 | 53 |
| Patient 117 | 64 | Male | <5 | HCC | III | T1aN0M0 | 54 |
| Patient 118 | 45 | Male | <5 | HCC | II | T1aN0M0 | 52 |
| Patient 119 | 63 | Male | <5 | HCC | III | T1aN0M0 | 52 |
| Patient 120 | 61 | Male | <5 | HCC | I | T1aN0M0 | 48 |
| Patient 121 | 65 | Male | <5 | HCC | II | T1aN0M0 | 11 |
| Patient 122 | 67 | Female | <5 | HCC | II | T1aN0M0 | 44 |

| Patient | Age | Sex | Size | Type | Grade | TNM | Months |
|---|---|---|---|---|---|---|---|
| Patient 123 | 58 | Female | <5 | HCC | II | T1aN0M0 | 36 |
| Patient 124 | 36 | Male | <5 | HCC | III | T2N0M0 | 15 |
| Patient 125 | 55 | Male | >=5 | HCC | II | T2N0M0 | 15 |
| Patient 126 | 59 | Male | <5 | HCC | III | T1aN0M0 | 29 |
| Patient 127 | 54 | Male | <5 | HCC | II | T1aN0M0 | 26 |
| Patient 128 | 63 | Male | >=5 | HCC | II | T1bN0M0 | 24 |
| Patient 129 | 75 | Male | >=5 | HCC | II | T3N0M0 | 23 |
| Patient 130 | 53 | Male | <5 | HCC | II | T1aN0M0 | 23 |
| Patient 131 | 63 | Male | >=5 | HCC | II | T1bN0M0 | 54 |
| Patient 132 | 76 | Female | <5 | HCC | III | T1aN0M0 | 32 |
| Patient 133 | 62 | Male | >=5 | HCC | II | T1bN0M0 | 8 |
| Patient 134 | 55 | Male | <5 | HCC | I | T2N0M0 | 11 |
| Patient 135 | 56 | Male | >=5 | HCC | II | T1bN0M0 | 11 |
| Patient 136 | 69 | Male | <5 | HCC | III | T1aN0M0 | 11 |
| Patient 137 | 53 | Male | <5 | HCC | III | T1aN0M0 | 47 |
| Patient 138 | 55 | Female | <5 | HCC | I | T1aN0M0 | 40 |
| Patient 139 | 76 | Female | >=5 | HCC | II | T2N0M0 | 36 |
| Patient 140 | 69 | Male | >=5 | HCC | II | T1bN0M0 | 29 |
| Patient 141 | 63 | Male | >=5 | HCC | I | T2N0M0 | 25 |
| Patient 142 | 62 | Male | <5 | HCC | II | T1aN0M0 | 22 |
| Patient 143 | 67 | Male | <5 | HCC | III | T1aN0M0 | 19 |
| Patient 144 | 75 | Male | <5 | HCC | II | T2N0M0 | 19 |
| Patient 145 | 32 | Female | >=5 | HCC | II | T1aN0M0 | 17 |
| Patient 146 | 38 | Male | >=5 | HCC | II | T2N0M0 | 16 |
| Patient 147 | 55 | Male | >=5 | HCC | II | T1bN0M0 | 17 |
| Patient 148 | 66 | Male | <5 | HCC | II | T2N0M0 | 17 |
| Patient 149 | 48 | Male | <5 | HCC | I | T2N0M0 | 16 |
| Patient 150 | 59 | Female | <5 | HCC | II | T2N0M0 | 16 |
| Patient 151 | 49 | Male | >=5 | HCC | II | T2N0M0 | 17 |
| Patient 152 | 50 | Male | >=5 | HCC | I | T2N0M0 | 16 |
| Patient 153 | 66 | Male | <5 | HCC | II | T1aN0M0 | 16 |
| Patient 154 | 71 | Male | >=5 | HCC | II | T2N0M0 | 15 |
| Patient 155 | 68 | Male | <5 | HCC | II | T2N0M0 | 7 |
| Patient 156 | 65 | Male | <5 | HCC | II | T1aN0M0 | 14 |
| Patient 157 | 50 | Male | >=5 | HCC | I | T1bN0M0 | 12 |
| Patient 158 | 62 | Male | <5 | HCC | II | T1aN0M0 | 10 |
| Patient 159 | 64 | Female | <5 | HCC | II | T1aN0M0 | 11 |
| Patient 160 | 70 | Male | <5 | HCC | II | T2N0M0 | 10 |
| Patient 161 | 60 | Female | >=5 | HCC | II | T2N0M0 | 6 |
| Patient 162 | 71 | Male | >=5 | HCC | II | T1bN0M0 | 10 |
| Patient 163 | 52 | Male | >=5 | HCC | II | T2N0M0 | 41 |
| Patient 164 | 47 | Male | >=5 | HCC | II | T1bN0M0 | 22 |
| Patient 165 | 46 | Male | >=5 | HCC | II | T1bN0M0 | 12 |
| Patient 166 | 49 | Male | >=5 | HCC | II | T1bN0M0 | 35 |
| Patient 167 | 71 | Male | >=5 | HCC | I | T2N0M0 | 10 |
| Patient 168 | 56 | Male | <5 | HCC | II | T2N0M0 | 19 |
| Patient 169 | 55 | Male | <5 | HCC | II | T2N0M0 | 50 |
| Patient 170 | 69 | Male | >=5 | HCC | II | T2N0M0 | 10 |
| Patient 171 | 57 | Male | <5 | HCC | III | T2N0M0 | 15 |
| Patient 172 | 70 | Male | <5 | HCC | II | T1aN0M0 | 16 |
| Patient 173 | 80 | Male | <5 | HCC | I | T2N0M0 | 20 |
| Patient 174 | 44 | Male | >=5 | HCC | I | T2N0M0 | 20 |
| Patient 175 | 47 | Male | >=5 | HCC | II | T1bN0M0 | 20 |
| Patient 176 | 48 | Male | >=5 | HCC | II | T2N0M0 | 20 |
| Patient 177 | 52 | Male | <5 | HCC | II | T2N0M0 | 26 |
| Patient 178 | 59 | Female | <5 | HCC | II | T1aN0M0 | 28 |
| Patient 179 | 56 | Male | <5 | HCC | II | T1aN0M0 | 29 |
| Patient 180 | 72 | Male | <5 | HCC | II | T1aN0M0 | 33 |
| Patient 181 | 65 | Male | <5 | HCC | II | T1aN0M0 | 25 |
| Patient 182 | 64 | Male | <5 | HCC | III | T2N0M0 | 35 |

| Patient | Age | Sex | Size | Type | Stage | TNM | Months |
|---|---|---|---|---|---|---|---|
| Patient 183 | 67 | Male | <5 | HCC | I | T1aN0M0 | 44 |
| Patient 184 | 57 | Male | <5 | HCC | II | T1aN0M0 | 48 |
| Patient 185 | 44 | Male | >=5 | HCC | II | T2N0M0 | 48 |
| Patient 186 | 48 | Male | <5 | HCC | II | T1aN0M0 | 49 |
| Patient 187 | 76 | Male | <5 | HCC | II | T1aN0M0 | 53 |
| Patient 188 | 65 | Male | <5 | HCC | I | T4N1M1 | 55 |
| Patient 189 | 66 | Male | <5 | HCC | II | T1aN0M0 | 11 |
| Patient 190 | 75 | Male | >=5 | HCC | III | T1bN0M0 | 54 |
| Patient 191 | 69 | Male | <5 | HCC | II | T1aN0M0 | 55 |
| Patient 192 | 59 | Male | <5 | HCC | II | T2N0M0 | 10 |
| Patient 193 | 65 | Male | <5 | HCC | II | T1aN0M0 | 8 |
| Patient 194 | 71 | Male | >=5 | HCC | II | T2N0M0 | 9 |
| Patient 195 | 62 | Male | >=5 | HCC | I | T1bN0M0 | 9 |
| Patient 196 | 71 | Male | >=5 | HCC | I | T2N0M0 | 10 |
| Patient 197 | 52 | Male | >=5 | HCC | I | T2N0M0 | 10 |
| Patient 198 | 60 | Male | <5 | HCC | II | T1aN0M0 | 13 |
| Patient 199 | 63 | Male | <5 | HCC | II | T1aN0M0 | 18 |
| Patient 200 | 68 | Male | <5 | HCC | II | T2N0M0 | 22 |
| Patient 201 | 68 | Male | <5 | HCC | II | T1aN0M0 | 24 |
| Patient 202 | 45 | Male | >=5 | HCC | II | T3N0M0 | 30 |
| Patient 203 | 53 | Male | <5 | HCC | II | T2N1M0 | 30 |
| Patient 204 | 55 | Male | <5 | HCC | III | T2N1M0 | 31 |
| Patient 205 | 84 | Male | <5 | HCC | I | T1aN0M0 | 31 |
| Patient 206 | 58 | Male | >=5 | HCC | II | T2N0M0 | 32 |
| Patient 207 | 67 | Male | <5 | HCC | I | T2N0M0 | 32 |
| Patient 208 | 68 | Male | <5 | HCC | I | T2N0M0 | 95 |
| Patient 209 | 61 | Male | >=5 | HCC | II | T2N0M0 | 31 |
| Patient 210 | 64 | Male | <5 | HCC | III | T1aN0M0 | 95 |
| Patient 211 | 73 | Male | <5 | HCC | II | T1aN0M0 | 41 |
| Patient 212 | 67 | Female | >=5 | HCC | II | T2N0M0 | 16 |
| Patient 213 | 58 | Male | <5 | HCC | II | T1aN0M0 | 12 |
| Patient 214 | 65 | Male | >=5 | HCC | II | T1bN0M0 | 43 |
| Patient 215 | 78 | Male | <5 | HCC | II | T2N0M0 | 43 |
| Patient 216 | 45 | Male | >=5 | HCC | II | T2N0M0 | 43 |
| Patient 217 | 69 | Male | >=5 | HCC | III | T1bN0M0 | 44 |
| Patient 218 | 52 | Female | >=5 | HCC | I | T3N0M0 | 45 |
| Patient 219 | 66 | Male | <5 | HCC | II | T1aN0M0 | 21 |
| Patient 220 | 75 | Female | <5 | HCC | II | T3N0M0 | 60 |
| Patient 221 | 58 | Male | <5 | HCC | II | T4N1M1 | 59 |
| Patient 222 | 44 | Male | <5 | HCC | I | T1aN0M0 | 50 |

| Patient | Age | Sex | Size | Type | Grade | TNM | Months |
|---|---|---|---|---|---|---|---|
| Patient 223 | 79 | Male | <5 | HCC | II | T1aN0M0 | 56 |
| Patient 224 | 69 | Male | >=5 | HCC | II | T3N0M0 | 55 |
| Patient 225 | 58 | Male | >=5 | HCC | II | T2N0M0 | 36 |
| Patient 226 | 49 | Male | >=5 | HCC | I | T3N0M0 | 53 |
| Patient 227 | 46 | Male | >=5 | HCC | | T1bN0M0 | 52 |
| Patient 228 | 47 | Male | <5 | HCC | II | T3N0M0 | 48 |
| Patient 229 | 63 | Male | <5 | HCC | III | T1aN0M0 | 48 |
| Patient 230 | 59 | Female | >=5 | HCC | II | T1bN0M0 | 47 |
| Patient 231 | 50 | Male | >=5 | HCC | III | T2N0M0 | 45 |
| Patient 232 | 59 | Male | >=5 | HCC | II | T1bN0M0 | 43 |
| Patient 233 | 62 | Male | >=5 | HCC | II | T3N0M0 | 41 |
| Patient 234 | 67 | Male | >=5 | HCC | II | T2N0M0 | 16 |
| Patient 235 | 57 | Male | <5 | HCC | III | T2N0M0 | 40 |
| Patient 236 | 70 | Male | <5 | HCC | II | T1aN0M0 | 40 |
| Patient 237 | 52 | Female | >=5 | HCC | II | T1bN0M0 | 57 |
| Patient 238 | 77 | Male | >=5 | HCC | II | T2N0M0 | 14 |
| Patient 239 | 52 | Female | <5 | HCC | I | T3N0M0 | 28 |
| Patient 240 | 55 | Male | <5 | HCC | II | T1aN0M0 | 37 |
| Patient 241 | 66 | Male | <5 | HCC | I | T2N0M0 | 35 |
| Patient 242 | 60 | Male | <5 | HCC | II | T1aN0M0 | 34 |

| Patient 243 | 74 | Male | <5 | HCC | III | T4N1M1 | 30 |
| Patient 244 | 61 | Male | >=5 | HCC | II | T2N0M0 | 22 |
| Patient 245 | 62 | Male | >=5 | HCC | I | T2N0M0 | 28 |
| Patient 246 | 52 | Male | <5 | HCC | II | T1aN0M0 | 9 |
| Patient 247 | 77 | Male | <5 | HCC | II | T1aN0M0 | 28 |
| Patient 248 | 81 | Male | >=5 | HCC | II | T1bN0M0 | 28 |
| Patient 249 | 62 | Male | >=5 | HCC | II | T1bN0M0 | 28 |
| Patient 250 | 52 | Female | >=5 | HCC | II | T3N0M0 | 23 |
| Patient 251 | 50 | Male | >=5 | HCC | II | T1bN0M0 | 18 |
| Patient 252 | 57 | Male | <5 | HCC | I | T2N0M0 | 15 |
| Patient 253 | 58 | Female | >=5 | HCC | II | T2N0M0 | 14 |
| Patient 254 | 68 | Female | >=5 | HCC | II | T2N0M0 | 23 |
| Patient 255 | 67 | Male | <5 | HCC | II | T2N0M0 | 15 |
| Patient 256 | 70 | Male | <5 | HCC | III | T2N0M0 | 47 |
| Patient 257 | 50 | Male | <5 | HCC | II | T1aN0M0 | 45 |
| Patient 258 | 66 | Male | >=5 | HCC | II | T2N0M0 | 8 |
| Patient 259 | 71 | Male | <5 | HCC | I | T2N0M0 | 8 |
| Patient 260 | 60 | Female | >=5 | HCC | II | T3N0M0 | 7 |
| Patient 261 | 88 | Male | <5 | HCC | II | T2N0M0 | 35 |
| Patient 262 | 48 | Male | >=5 | HCC | II | T3N0M0 | 19 |

| Patient | Age | Sex | Size | Type | Stage | TNM | Months |
|---|---|---|---|---|---|---|---|
| Patient 263 | 70 | Male | <5 | HCC | I | T2N0M0 | 17 |
| Patient 264 | 55 | Male | >=5 | HCC | II | T3N0M0 | 34 |
| Patient 265 | 64 | Male | >=5 | HCC | II | T2N0M0 | 19 |
| Patient 266 | 58 | Female | >=5 | HCC | II | T4N1M1 | 31 |
| Patient 267 | 70 | Male | >=5 | HCC | II | T1bN0M0 | 29 |
| Patient 268 | 64 | Male | <5 | HCC | III | T3N0M0 | 17 |
| Patient 269 | 70 | Male | >=5 | HCC | II | T2N0M0 | 22 |
| Patient 270 | 58 | Male | >=5 | HCC | II | T3N0M0 | 22 |
| Patient 271 | 43 | Male | <5 | HCC | II | T2N0M0 | 18 |
| Patient 272 | 53 | Male | >=5 | HCC | II | T1bN0M0 | 16 |
| Patient 273 | 57 | Male | >=5 | HCC | II | T2N0M0 | 16 |
| Patient 274 | 49 | Male | <5 | HCC | I | T3N0M0 | 15 |
| Patient 275 | 69 | Male | >=5 | HCC | II | T2N0M0 | 14 |
| Patient 276 | 50 | Male | >=5 | HCC | II | T3N0M0 | 9 |
| Patient 277 | 64 | Male | >=5 | HCC | II | T2N0M0 | 14 |
| Patient 278 | 65 | Female | >=5 | HCC | II | T3N0M0 | 12 |
| Patient 279 | 63 | Male | <5 | HCC | II | T2N0M0 | 12 |
| Patient 280 | 52 | Male | <5 | HCC | I | T2N0M0 | 12 |
| Patient 281 | 66 | Male | <5 | HCC |  | T2N0M0 | 11 |
| Patient 282 | 75 | Male | >=5 | HCC | II | T3N0M0 | 10 |
| Patient 283 | 59 | Female | <5 | HCC | II | T2N0M0 | 10 |
| Patient 284 | 55 | Male | >=5 | HCC | I | T3N0M0 | 10 |
| Patient 285 | 47 | Male | >=5 | HCC | I | T3N0M0 | 9 |
| Patient 286 | 64 | Male | >=5 | HCC | II | T3N0M0 | 9 |
| Patient 287 | 63 | Male | >=5 | HCC | II | T3N0M0 | 9 |
| Patient 288 | 55 | Male | >=5 | HCC | I | T3N0M0 | 9 |
| Patient 289 | 40 | Female | >=5 | HCC | I | T2N0M0 | 9 |

**Supplementary Table 2. Sequence of primer forTransgenic mice genotyping, related to Fig. 5, and Fig. S8.**

| Primers' name | Primers' sequence | Amplified band size |
|---|---|---|
| Lyz-cre common | CTTGGGCTGCCAGAATTTCTC | Mutan allele：700bp Wildtype allele：350bp |
| Lyz-cre MT | CCCAGAAATGCCAGATTACG | |
| Lyz-cre WT | TTACAGTCGGCCAGGCTGAC | |
| Pax5 flox forward primer (primer 4) | TGCTCATGCACAGAAAAGTGAAAAG | Mutan allele：219bp Wildtype allele：152bp |
| Pax5 flox reverse primer (primer 4) | GTGGATTCGGACCAGTCTGA | |
| Pax5 flox forward primer (primer 5) | ACGTAAACGGCCACAAGTTC | Mutan allele：212bp Wildtype allele：156bp |
| Pax5 flox reverse primer (primer 5) | AGGAGAGTGAGAAAGGATGAGGTG | |
| CD19 flox forward pirmer | AAGCGGATGCCAGACATTGTGG | Mutan allele：208bp Wildtype allele：140bp |
| CD19 flox forward pirmer | TTCAGATGAGTGGGTGCAGGCTT | |

**Supplementary Table 3. CytoF antibody panel, related to Fig. 1 and Fig. S2.**

| Antibodies | Metal | Clone | Source |
|---|---|---|---|
| CD45 | 89Y | HI30 | BioLegend |
| CD3 | 115In | UCHT1 | BioXcell |
| CD68 | 139La | Y1/82A | BioLegend |
| CD56 | 141Pr | NCAM16.2 | BD Biosciences |
| CD19 | 142Nd | HIB19 | BioLegend |
| CD27 | 143Nd | O323 | BioLegend |
| CD14 | 144Nd | M5E2 | BioLegend |
| IgD | 145Nd | IA6-2 | BioLegend |
| CD123 | 146Nd | 6H6 | BioLegend |
| CD15 | 147Sm | W6D3 | BioLegend |
| CD33 | 148Nd | WM53 | BioLegend |
| CD25 | 149Sm | 24212 | R&D Systems |
| CD7 | 150Nd | CD7-6B7 | BioLegend |
| CD38 | 151Eu | HIT2 | BioLegend |
| CD103 | 152Sm | B-Ly7 | Ebioscience |
| CD274 (PD-L1) | 153Eu | 29E.2A3 | BioLegend |
| CD163 | 154Sm | GHI/61 | BioLegend |
| CD45RA | 155Gd | HI100 | BioLegend |
| CD24 | 156Gd | ML5 | BioLegend |
| CD172a/b | 157Gd | SE5A5 | BioLegend |
| CD197 (CCR7) | 158Gd | G043H7 | BioLegend |
| CD11c | 159Tb | BU15 | BioLegend |
| CD28 | 160Gd | CD28.2 | BioLegend |
| CTLA-4 | 161Dy | 14D3 | Ebioscience |
| FoxP3 | 162Dy | PCH101 | BioLegend |
| CD137 (4-1BB) | 163Dy | 4B4-1 | BioLegend |
| CD95 | 164Dy | DX2 | BioLegend |
| IL-1beita | 165Ho | 8516 | R&D Systems |
| CX3CR1 | 166Er | K0124E1 | BioLegend |
| CD278(ICOS) | 167Er | C398.4A | BioLegend |
| Tbet | 168Er | 4B10 | BioLegend |
| Ki-67 | 169Tm | SolA15 | Ebioscience |
| IL-7Ra(CD127) | 170Er | A019D5 | BioLegend |
| Eomes | 171Yb | 644730 | R&D Systems |
| Tim3 | 172Yb | F38-2E2 | BioLegend |
| CD154(CD40L) | 173Yb | 24-31 | BioLegend |
| CD279 (PD-1) | 174Yb | EH12.2H7 | BioLegend |
| CD16 | 175Lu | 3G8 | BioLegend |
| HLA-DR | 176Yb | L243 | BioLegend |
| CD4 | 197gd | RPA-T4 | BioLegend |
| CD8a | 198pt | RPA-T8 | BioLegend |
| CD11b | 209Bi | ICRF44 | BioLegend |

**Supplementary Table 4. CD19 CAR sequences, related to Fig. 2 and Fig. S3.**

**Nucleotide sequence**

ATGGCCTCACCGTTGACCCGCTTTCTGTCGCTGAACCTGCTGCTGCTGGGTGA
GTCGATTATCCTGGGGAGTGGAGAAGCTGAAGTCCAGCTGCAGCAGTCTGGG
GCTGAGCTTGTGAGACCTGGGACCTCTGTGAAGTTATCTTGCAAAGTTTCTGG
CGATACCATTACATTTTACTACATGCACTTTGTGAAGCAAAGGCCTGGACAGGG
TCTGGAATGGATAGGAAGGATTGATCCTGAGGATGAAAGTACTAAATATTCTGA
GAAGTTCAAAAACAAGGCGACACTCACTGCAGATACATCTTCCAACACAGCCTA
CCTGAAGCTCAGCAGCCTGACCTCTGAGGACACTGCAACCTATTTTGTATCTA
CGGAGGATACTACTTTGATTACTGGGGCCAAGGGGTCATGGTCACAGTCTCCT
CAGGTGGAGGTGGATCAGGTGGAGGTGGATCTGGTGGAGGTGGATCTGACAT
CCAGATGACACAGTCTCCAGCTTCCCTGTCTACATCTCTGGGAGAAACTGTCA
CCATCCAATGTCAAGCAAGTGAGGACATTTACAGTGGTTTAGCGTGGTATCAG
CAGAAGCCAGGGAAATCTCCTCAGCTCCTGATCTATGGTGCAAGTGACTTACA
AGACGGCGTCCCATCACGATTCAGTGGCAGTGGATCTGGCACACAGTATTCTC
TCAAGATCACCAGCATGCAAACTGAAGATGAAGGGGTTTATTTCTGTCAACAGG
GTTTAACGTATCCTCGGACGTTCGGTGGCGGCACCAAGCTGGAATTGAAACGG
GCGGCCGCATCTACTACTACCAAGCCAGTGCTGCGAACTCCCTCACCTGTGCA
CCCTACCGGGACATCTCAGCCCCAGAGACCAGAAGATTGTCGGCCCCGTGGC
TCAGTGAAGGGGACCGGATTGGACTTCGCCTGTGATATTTACATCTGGGCACC
CTTGGCCGGAATCTGCGTGGCCCTTCTGCTGTCCTTGATCATCACTCTCATCT
GCTACAATAGTAGAAGGAACAGACTCCTTCAAAGTGACTACATGAACATGACTC
CCCGGAGGCCTGGGCTCACTCGAAAGCCTTACCAGCCCTACGCCCCTGCCAG
AGACTTTGCAGCGTACCGCCCCAGAGCAAAATTCAGCAGGAGTGCAGAGACT
GCTGCCAACCTGCAGGACCCCAACCAGCTCTACAATGAGCTCAATCTAGGGCG
AAGAGAGGAATATGACGTCTTGGAGAAGAAGCGGGCTCGCGATCCAGAGATG
GGAGGCAAACAGCAGAGGAGGAGGAACCCCCAGGAAGGCGTATACAATGCAC
TGCAGAAAGACAAGATGGCAGAAGCCTACAGTGAGATCGGCACAAAAGGCGA
GAGGCGGAGAGGCAAGGGGCACGATGGCCTTTACCAGGGTCTCAGCACTGCC
ACCAAGGACACCTATGATGCCCTGCATATGCAGACCCTGGCCCCTCGCTAA

**Amino acid sequence**

MASPLTRFLSLNLLLLGESIILGSGEAEVQLQQSGAELVRPGTSVKLSCKVS
GDTITFYYMHFVKQRPGQGLEWIGRIDPEDESTKYSEKFKNKATLTADTSS
NTAYLKLSSLTSEDTATYFCIYGGYYFDYWGQGVMVTVSSGGGGSGGGG
SGGGGSDIQMTQSPASLSTSLGETVTIQCQASEDIYSGLAWYQQKPGKSP
QLLIYGASDLQDGVPSRFSGSGSGTQYSLKITSMQTEDEGVYFCQQGLTY
PRTFGGGTKLELKRAAASTTTKPVLRTPSPVHPTGTSQPQRPEDCRPRGS
VKGTGLDFACDIYIWAPLAGICVALLLSLIITLICYNSRRNRLLQSDYMNMTP
RRPGLTRKPYQPYAPARDFAAYRPRAKFSRSAETAANLQDPNQLYNELNL
GRREEYDVLEKKRARDPEMGGKQQRRRNPQEGVYNALQKDKMAEAYSEI
GTKGERRRGKGHDGLYQGLSTATKDTYDALHMQTLAPR

**Supplementary Table 5. qRT-PCR Primer sequences, related to Fig. 4 and Fig. 6.**

| Genes | Species | Forward primer (5'-3') | Reverse primer (5'-3') |
|---|---|---|---|
| NDUFS8 | Human | TCGACATGACCAAGTGCATC | CCCGTTGTTGAGCAACTTCT |
| NDUFA4 | Human | AGCAGCACTGTATGTGATGCGC | TGTAGTCCACATTCACAGAGTAGA |
| NDUFA9 | Human | GCCTATCGATGGGTAGCAAGA | CTTCTAAGCCAGGCAGGTGA |
| NDUFA1 | Human | TTTTGGCCTAGGTAACGGGG | TGTGGATGTACGCAGTAGCC |
| NDUFB2 | Human | AGCGGACTCATGTGGTTCTGGA | AGAGTGAGGCTGAGTCTACACC |
| SDHA | Human | GAGATGTGGTGTCTCGGTCCAT | GCTGTCTCTGAAATGCCAGGCA |
| SDHB | Human | CAATGGAGGCAACACTCTAGCTT | GGAAGAGGGTAGATTTTTGAGACCTT |
| SDHD | Human | GACTAGCGAGAGGGTTGTCAGTGT | CATCGCAGAGCAAGGATTCA |
| CYC1 | Human | CTACGGACACCTCAGGCAGT | CAGGTCACTGGCACTCACAG |
| CYCS | Human | GAGGCAAGCATAAGACTGGACC | ACTCCATCAGGGTATCCTCTCC |
| UQCRC2 | Human | GTGGCATGTAAGAACCAGCA | TCCAACACAGATGTCCAAGC |
| COX5A | Human | GGGAATTGCGTAAAGGGATAA | TCCTGCTTTGTCCTTAACAACC |
| COX5B | Human | AGTCGCCTGCTCTTCATCAG | TGGCTTCAAGGTTACTTCGC |
| COX6A1 | Human | AGCCAGTTGGAAGTGGATTC | TGTACCTGAAGTCGCACCAC |
| ATP5G1 | Human | TGAGACCAAGGGCTAAAGCTG | TCAGTCTGCACTCCTACTACCC |
| ATP5A1 | Human | AAGACACGCCCAGTTTCTTC | TTTGGGTTCATCTTTCATTGC |
| ATP5B | Human | CAAGTCATCAGCAGGCACAT | GTGGGCTATCAGCCTACCCT |
| ATP5E | Human | CTTCAGTGCATCTCTCACTGC | TACAGCATGGTGGCCTACTG |
| ND1 | Human | GGCTATATACAACTACGCAAAGGC | GGTAGATGTGGCGGGTTTTAGG |

| Gene | Species | Forward | Reverse |
|---|---|---|---|
| ND2 | Human | CTTCTGAGTCCCAGAGGTTACC | GAGAGTGAGGAGAAGGCTTACG |
| ND3 | Human | TTGATCTAGAAATTGCCCTCC | GGCAGGTTAGTTGTTTGTAGG |
| ND4 | Human | CCCTCGTAGTAACAGCCATTCTC | CGACTGTGAGTGCGTTCGTAGT |
| ND4L | Human | CTCATAACCCTCAACACCCA | AGACTAGTATGGCAATAGGCAC |
| ND6 | Human | GCGATGGCTATTGAGGAGTATCC | CACAGCACCAATCCTACCTCCA |
| ATP6 | Human | TCCCTCTACACTTATCATCTTCAC | GACAGCGATTTCTAGGATAGTC |
| ATP8 | Human | TAAATACTACCGTATGGCCCAC | GTGATGAGGAATAGTGTAAGGAG |
| ND5 | Human | TCTTAGTTACCGCTAACAACC | ATAATTCCTACGCCCTCTCAG |
| COX2 | Human | ACGCATCCTTTACATAACAGAC | GCCAATTGATTTGATGGTAAGG |
| COX1 | Human | ATATTTCACCTCCGCTACCA | TCAGCTAAATACTTTGACGCC |
| COX3 | Human | CTCTCAGCCCTCCTAATGAC | GCGTTATGGAGTGGAAGTG |
| CYTB | Human | ATCACTTTATTGACTCCTAGCC | TGGTTGTCCTCCGATTCAG |
| 12S/MT-RNR1 | Human | AAGATTACACATGCAAGCATCC | TTGATCGTGGTGATTTAGAGG |
| 16S/MT-RNR2 | Human | AACTCGGCAAATCTTACCC | AATACTGGTGATGCTAGAGGTG |
| PAX5 | Human | CTTGCTCATCAAGGTGTCAGGC | TGGCGACCTTTGGTTTGGATCC |
| CD73 | Human | AGTCCACTGGAGAGTTCCTGCA | TGAGAGGGTCATAACTGGGCAC |
| CD19 | Human | GGCTATGAGGAACCTGACAGTG | TCATCCTCAGGGTTCTCGTAGC |
| PD-L1 | Human | GCCGACTACAAGCGAATTAC | TCTCAGTGTGCTGGTCACAT |

| Gene | | Forward | Reverse |
|---|---|---|---|
| *Pax5* | Mouse | ACGTAAACGGCCACAAGTTC | AGGAGAGTGAGAAAGGATGAGGTG |
| *Cd73* | Mouse | CGCTCAGAAAGTTCGAGGTGTG | CGCAGGCACTTCTTTGGAAGGT |
| *Cd19* | Mouse | GGCACCTATTATTGTCTCCG | GGGTCAGTCATTCGCTTC |
| *Pd-l1* | Mouse | TGCGGACTACAAGCGAATCACG | CTCAGCTTCTGGATAACCCTCG |

**Supplementary Table 6. Differentially expressed genes (DEGs) in CD19+ TAMs, CD19-TAMs, and B cells (data from sc-RNA seq), related to Fig. 3.**

| Cluster | Gene | Avg_log2FC |
| --- | --- | --- |
| CD19- TAMs | S100A9 | 3.340504294 |
| CD19- TAMs | S100A8 | 2.930711861 |
| CD19- TAMs | FCN1 | 2.566707976 |
| CD19- TAMs | LYZ | 2.131023307 |
| CD19- TAMs | VCAN | 2.080575685 |
| CD19- TAMs | EREG | 1.940225065 |
| CD19- TAMs | MNDA | 1.642424955 |
| CD19- TAMs | COTL1 | 1.538471379 |
| CD19- TAMs | G0S2 | 1.466540093 |
| CD19- TAMs | S100A4 | 1.444668371 |
| CD19- TAMs | IL1B | 1.435207909 |
| CD19- TAMs | LGALS2 | 1.338150614 |
| CD19- TAMs | AREG | 1.335322968 |
| CD19- TAMs | SLC8A1 | 1.282108583 |
| CD19- TAMs | LST1 | 1.278503463 |
| CD19- TAMs | TIMP1 | 1.275011214 |
| CD19- TAMs | S100A6 | 1.251437828 |
| CD19- TAMs | AOAH | 1.212618069 |
| CD19- TAMs | DPYD | 1.194388645 |
| CD19- TAMs | BCL2A1 | 1.192443505 |
| CD19- TAMs | CD52 | 1.189102179 |
| CD19- TAMs | PLXDC2 | 1.172425514 |
| CD19- TAMs | SERPINA1 | 1.124294435 |
| CD19- TAMs | CLEC12A | 1.119820159 |
| CD19- TAMs | LYST | 1.116733768 |
| CD19- TAMs | CD44 | 1.098460979 |
| CD19- TAMs | BAG3 | 1.076846334 |
| CD19- TAMs | NEAT1 | 1.07568367 |
| CD19- TAMs | JAML | 1.072806356 |
| CD19- TAMs | RETN | 1.065992664 |
| CD19- TAMs | OLR1 | 1.057102245 |
| CD19- TAMs | HSPH1 | 1.056390765 |
| CD19- TAMs | MAML3 | 1.02137655 |
| CD19- TAMs | ZEB2 | 1.020788305 |
| CD19- TAMs | CSTA | 1.016991289 |
| CD19- TAMs | WARS | 1.001425195 |
| CD19- TAMs | SLC2A3 | 0.995337894 |
| CD19- TAMs | LRRFIP1 | 0.990029128 |
| CD19- TAMs | LCP1 | 0.985788879 |
| CD19- TAMs | DNAJA4 | 0.982852734 |
| CD19- TAMs | S100A10 | 0.979073318 |
| CD19- TAMs | C1orf162 | 0.974251251 |
| CD19- TAMs | ARHGAP26 | 0.974194192 |
| CD19- TAMs | GNAQ | 0.95056841 |
| CD19- TAMs | MAP2K1 | 0.947428329 |
| CD19- TAMs | DNAJB1 | 0.938068425 |
| CD19- TAMs | ZFAND2A | 0.92519094 |
| CD19- TAMs | H3F3A | 0.91512567 |
| CD19- TAMs | IFI30 | 0.914823948 |
| CD19- TAMs | CLEC10A | 0.912435883 |
| CD19- TAMs | RTN1 | 0.88349309 |
| CD19- TAMs | FGD4 | 0.882660782 |
| CD19- TAMs | HSPD1 | 0.87024431 |
| CD19- TAMs | C15orf48 | 0.868163747 |
| CD19- TAMs | PLCB1 | 0.866398668 |
| CD19- TAMs | PTPRE | 0.852584503 |
| CD19- TAMs | SAMHD1 | 0.844902861 |
| CD19- TAMs | HSP90AA1 | 0.844124589 |
| CD19- TAMs | SH3BGRL3 | 0.840692954 |
| CD19- TAMs | FYB1 | 0.838711669 |
| CD19- TAMs | IRAK3 | 0.837783119 |
| CD19- TAMs | SLC11A1 | 0.836157407 |
| CD19- TAMs | SNHG5 | 0.833479835 |
| CD19- TAMs | JARID2 | 0.832968847 |
| CD19- TAMs | AIF1 | 0.828193322 |
| CD19- TAMs | SOD2 | 0.826088666 |
| CD19- TAMs | ACTB | 0.818902612 |
| CD19- TAMs | MXD1 | 0.818163621 |
| CD19- TAMs | FGR | 0.8162538 |
| CD19- TAMs | PLAUR | 0.816230435 |
| CD19- TAMs | PTPRC | 0.815871218 |
| CD19- TAMs | PTGS2 | 0.815572563 |
| CD19- TAMs | PLAC8 | 0.814526183 |
| CD19- TAMs | SLC25A37 | 0.813378812 |
| CD19- TAMs | FYN | 0.812595701 |
| CD19- TAMs | JAK2 | 0.811804722 |
| CD19- TAMs | TKT | 0.810208876 |
| CD19- TAMs | UTRN | 0.809378742 |
| CD19- TAMs | GBP1 | 0.807950619 |
| CD19- TAMs | CORO1A | 0.800797112 |
| CD19- TAMs | LRRK2 | 0.794657026 |
| CD19- TAMs | TYMP | 0.791613663 |
| CD19- TAMs | TAOK3 | 0.789578161 |
| CD19- TAMs | SORL1 | 0.786968124 |
| CD19- TAMs | RPL39 | 0.783893524 |
| CD19- TAMs | CSF3R | 0.780770778 |
| CD19- TAMs | PPIF | 0.779653733 |
| CD19- TAMs | IFITM3 | 0.778147182 |
| CD19- TAMs | GPCPD1 | 0.776165165 |
| CD19- TAMs | PSTPIP2 | 0.776110737 |
| CD19- TAMs | PTMA | 0.773133914 |
| CD19- TAMs | ZDHHC20 | 0.771295364 |
| CD19- TAMs | DOCK8 | 0.769219503 |
| CD19- TAMs | SSH2 | 0.769207893 |
| CD19- TAMs | MBNL1 | 0.766246135 |
| CD19- TAMs | RPS24 | 0.764929201 |
| CD19- TAMs | AC020916.1 | 0.760095724 |
| CD19- TAMs | AP1S2 | 0.755867482 |
| CD19- TAMs | S100A12 | 0.753151341 |
| CD19- TAMs | PID1 | 0.751296337 |
| CD19- TAMs | LRMDA | 0.750946292 |
| CD19- TAMs | TMSB10 | 0.750569741 |
| CD19- TAMs | CTSS | 0.750144008 |
| CD19- TAMs | TNFSF10 | 0.749933655 |
| CD19- TAMs | SOCS3 | 0.742798512 |
| CD19- TAMs | PABPC1 | 0.741473005 |
| CD19- TAMs | GBP2 | 0.741089246 |
| CD19- TAMs | ATF3 | 0.740773983 |
| CD19- TAMs | KYNU | 0.738633814 |
| CD19- TAMs | CALHM6 | 0.733594613 |
| CD19- TAMs | TET2 | 0.732846447 |
| CD19- TAMs | MYO1F | 0.732668426 |
| CD19- TAMs | DENND1A | 0.731309165 |
| CD19- TAMs | PPA1 | 0.725591908 |
| CD19- TAMs | ANXA1 | 0.72523012 |
| CD19- TAMs | ARL5B | 0.719479517 |
| CD19- TAMs | NAP1L1 | 0.718363661 |
| CD19- TAMs | DOCK5 | 0.716053775 |
| CD19- TAMs | CFP | 0.711703049 |
| CD19- TAMs | SERPINB1 | 0.711214749 |
| CD19- TAMs | GCH1 | 0.70698975 |
| CD19- TAMs | FOSL2 | 0.70589484 |
| CD19- TAMs | FOXP1 | 0.703475313 |
| CD19- TAMs | SLCO3A1 | 0.703133546 |
| CD19- TAMs | NCF2 | 0.702331795 |
| CD19- TAMs | CRIP1 | 0.698750223 |
| CD19- TAMs | CUX1 | 0.696713297 |
| CD19- TAMs | TLR2 | 0.696208661 |
| CD19- TAMs | PRKCB | 0.696203778 |
| CD19- TAMs | SMIM25 | 0.695742541 |
| CD19- TAMs | PSME2 | 0.688585096 |
| CD19- TAMs | AGTPBP1 | 0.687591379 |
| CD19- TAMs | CD1C | 0.687535002 |
| CD19- TAMs | CHORDC1 | 0.686965047 |
| CD19- TAMs | UPP1 | 0.686538669 |
| CD19- TAMs | HSPA1B | 0.685883422 |
| CD19- TAMs | TNFRSF1B | 0.685020311 |
| CD19- TAMs | S100A11 | 0.684787866 |
| CD19- TAMs | H2AFY | 0.683742274 |
| CD19- TAMs | CFL1 | 0.682338611 |
| CD19- TAMs | HCK | 0.681422849 |
| CD19- TAMs | HSPA8 | 0.67993017 |
| CD19- TAMs | HSPE1 | 0.679461891 |
| CD19- TAMs | CD48 | 0.67933419 |
| CD19- TAMs | CD36 | 0.678109396 |
| CD19- TAMs | KLF4 | 0.673971053 |
| CD19- TAMs | TCF7L2 | 0.673456749 |
| CD19- TAMs | CPPED1 | 0.67310843 |
| CD19- TAMs | GK | 0.668431557 |
| CD19- TAMs | FAM49B | 0.667142546 |
| CD19- TAMs | TSPO | 0.667118609 |
| CD19- TAMs | NFKBIZ | 0.66626754 |
| CD19- TAMs | EFHD2 | 0.664474704 |
| CD19- TAMs | LYN | 0.662491946 |
| CD19- TAMs | PAK1 | 0.65881822 |
| CD19- TAMs | PLEK | 0.656913286 |
| CD19- TAMs | LILRB3 | 0.656548798 |
| CD19- TAMs | VMP1 | 0.656118939 |
| CD19- TAMs | TREM1 | 0.656008696 |
| CD19- TAMs | CMIP | 0.655893298 |
| CD19- TAMs | CD300E | 0.655407205 |
| CD19- TAMs | RPL23 | 0.655083512 |
| CD19- TAMs | LIMS1 | 0.654837901 |
| CD19- TAMs | PRKAG2 | 0.653754493 |
| CD19- TAMs | MAP3K2 | 0.650538112 |
| CD19- TAMs | ETV6 | 0.648961044 |
| CD19- TAMs | CARD16 | 0.646524565 |
| CD19- TAMs | NR4A1 | 0.645673748 |
| CD19- TAMs | CLEC7A | 0.645508478 |
| CD19- TAMs | SNX10 | 0.644277936 |
| CD19- TAMs | IL1RN | 0.643615361 |
| CD19- TAMs | RPL28 | 0.643471182 |
| CD19- TAMs | CYBB | 0.642319442 |
| CD19- TAMs | FPR1 | 0.639028408 |
| CD19- TAMs | APBB1IP | 0.63862511 |
| CD19- TAMs | DSE | 0.637967525 |
| CD19- TAMs | ADGRE2 | 0.636149846 |
| CD19- TAMs | PTP4A2 | 0.635029586 |
| CD19- TAMs | DNAJA1 | 0.634470659 |
| CD19- TAMs | USP15 | 0.631689736 |
| CD19- TAMs | SLC20A1 | 0.630571618 |
| CD19- TAMs | VEGFA | 0.629590236 |
| CD19- TAMs | PICALM | 0.629098201 |
| CD19- TAMs | RIPK2 | 0.625792528 |
| CD19- TAMs | RPL26 | 0.623834214 |
| CD19- TAMs | RIPOR2 | 0.623332642 |
| CD19- TAMs | AQP9 | 0.619842778 |
| CD19- TAMs | SAT1 | 0.619377841 |
| CD19- TAMs | UBE2R2 | 0.616713355 |
| CD19- TAMs | ACTR2 | 0.616058197 |
| CD19- TAMs | VSIR | 0.614779459 |
| CD19- TAMs | ANXA2 | 0.614692714 |
| CD19- TAMs | RPL37 | 0.611943772 |
| CD19- TAMs | NUMB | 0.611169836 |
| CD19- TAMs | THBS1 | 0.610680014 |
| CD19- TAMs | STXBP2 | 0.609931232 |
| CD19- TAMs | DENND5A | 0.609836446 |
| CD19- TAMs | DLEU2 | 0.609187373 |
| CD19- TAMs | RPL36A | 0.608676521 |
| CD19- TAMs | LSP1 | 0.608519808 |
| CD19- TAMs | FBP1 | 0.607722047 |
| CD19- TAMs | ATP5F1E | 0.606704249 |
| CD19- TAMs | TPM3 | 0.601919071 |
| CD19- TAMs | MEGF9 | 0.601132637 |
| CD19- TAMs | NLRP3 | 0.599716798 |
| CD19- TAMs | RPL7 | 0.59827094 |
| CD19- TAMs | CST3 | 0.598178698 |
| CD19- TAMs | ENO1 | 0.598029301 |
| CD19- TAMs | ARPC2 | 0.597732523 |
| CD19- TAMs | FKBP1A | 0.597405114 |
| CD19- TAMs | PLIN2 | 0.597165714 |
| CD19- TAMs | OSBPL8 | 0.5968208 |
| CD19- TAMs | ACTR3 | 0.596252675 |
| CD19- TAMs | ALDH2 | 0.594965865 |
| CD19- TAMs | DNAJB6 | 0.592126317 |
| CD19- TAMs | RPL37A | 0.59194264 |
| CD19- TAMs | LILRB2 | 0.589319997 |
| CD19- TAMs | CACYBP | 0.589130934 |
| CD19- TAMs | STK38L | 0.58778445 |
| CD19- TAMs | IRF1 | 0.587653568 |
| CD19- TAMs | PFN1 | 0.587067424 |
| CD19- TAMs | CDC42EP3 | 0.585443103 |
| CD19- TAMs | PRKCA | 0.585328079 |
| CD19- TAMs | CHD1 | 0.58241754 |
| CD19- TAMs | MCTP1 | 0.582302031 |
| CD19- TAMs | VIM | 0.581882718 |
| CD19- TAMs | SERPINB9 | 0.58178442 |
| CD19- TAMs | HSPA1A | 0.581631525 |
| CD19- TAMs | IL17RA | 0.578778933 |
| CD19- TAMs | PTPRJ | 0.578164508 |
| CD19- TAMs | FRY | 0.577276758 |
| CD19- TAMs | PLSCR1 | 0.576871251 |
| CD19- TAMs | EMILIN2 | 0.576621865 |
| CD19- TAMs | CELF2 | 0.575215209 |
| CD19- TAMs | XYLT1 | 0.574437035 |
| CD19- TAMs | VASP | 0.574321722 |
| CD19- TAMs | BTG1 | 0.573461281 |
| CD19- TAMs | COP1 | 0.572231694 |
| CD19- TAMs | GNAI2 | 0.570741157 |
| CD19- TAMs | HMGN2 | 0.569578118 |
| CD19- TAMs | TNFSF13B | 0.569167049 |
| CD19- TAMs | ZSWIM6 | 0.568200721 |
| CD19- TAMs | KDM6B | 0.568058466 |
| CD19- TAMs | RPS17 | 0.565380338 |
| CD19- TAMs | SBF2 | 0.565104081 |
| CD19- TAMs | DDX21 | 0.564361003 |
| CD19- TAMs | DISC1 | 0.563277819 |
| CD19- TAMs | RPL4 | 0.561916849 |
| CD19- TAMs | LINC00278 | 0.561165868 |
| CD19- TAMs | KDM7A | 0.56085421 |
| CD19- TAMs | SULF2 | 0.560213828 |
| CD19- TAMs | PTEN | 0.560086427 |
| CD19- TAMs | TIAM1 | 0.559339245 |
| CD19- TAMs | CREB5 | 0.558834397 |
| CD19- TAMs | DIAPH2 | 0.555819411 |
| CD19- TAMs | LTA4H | 0.55447753 |
| CD19- TAMs | YBX1 | 0.551458813 |
| CD19- TAMs | ARL4A | 0.551439951 |
| CD19- TAMs | PTK2B | 0.550315231 |
| CD19- TAMs | RPS8 | 0.548475562 |
| CD19- TAMs | ALOX5 | 0.547655674 |
| CD19- TAMs | RIN3 | 0.544888566 |
| CD19- TAMs | RAB11FIP1 | 0.542917962 |
| CD19- TAMs | TNFSF14 | 0.542228313 |
| CD19- TAMs | LGALS1 | 0.542158967 |
| CD19- TAMs | SPTLC2 | 0.541703254 |
| CD19- TAMs | MSN | 0.539305415 |
| CD19- TAMs | ITGAL | 0.539016593 |
| CD19- TAMs | TRPS1 | 0.538172025 |
| CD19- TAMs | PTPN12 | 0.538128355 |
| CD19- TAMs | TES | 0.536849448 |
| CD19- TAMs | LCP2 | 0.535768481 |
| CD19- TAMs | MBP | 0.535473323 |
| CD19- TAMs | EIF4A1 | 0.535410665 |
| CD19- TAMs | BAZ1A | 0.534714436 |
| CD19- TAMs | FBXL5 | 0.533690596 |
| CD19- TAMs | RPS16 | 0.533275578 |
| CD19- TAMs | DUSP6 | 0.533134469 |
| CD19- TAMs | CAST | 0.532577999 |
| CD19- TAMs | SRGN | 0.53234394 |
| CD19- TAMs | RPS9 | 0.53224439 |
| CD19- TAMs | ATG3 | 0.529616394 |
| CD19- TAMs | AHR | 0.528794822 |
| CD19- TAMs | RBM3 | 0.527920123 |
| CD19- TAMs | JMJD1C | 0.52724378 |
| CD19- TAMs | GIMAP4 | 0.526874141 |
| CD19- TAMs | ITGB2 | 0.526772605 |
| CD19- TAMs | BACH1 | 0.526334237 |
| CD19- TAMs | DOCK2 | 0.526320249 |
| CD19- TAMs | GAS7 | 0.52585517 |
| CD19- TAMs | B4GALT5 | 0.524488449 |
| CD19- TAMs | UBA52 | 0.524124138 |
| CD19- TAMs | RGS18 | 0.523958644 |
| CD19- TAMs | RNF149 | 0.523315006 |
| CD19- TAMs | ARPC1B | 0.521134344 |
| CD19- TAMs | WSB1 | 0.521104027 |
| CD19- TAMs | SPECC1 | 0.519928615 |
| CD19- TAMs | FLNA | 0.519876283 |
| CD19- TAMs | ARHGDIB | 0.519203834 |
| CD19- TAMs | CX3CR1 | 0.517694221 |
| CD19- TAMs | DIAPH1 | 0.517313266 |
| CD19- TAMs | MYOF | 0.517252489 |
| CD19- TAMs | CKLF | 0.516424304 |
| CD19- TAMs | LPCAT2 | 0.516136451 |
| CD19- TAMs | DMXL2 | 0.515721399 |
| CD19- TAMs | HNRNPA1 | 0.515064173 |
| CD19- TAMs | PELI2 | 0.513718439 |
| CD19- TAMs | CDC42 | 0.513503966 |
| CD19- TAMs | CGAS | 0.513260801 |
| CD19- TAMs | BID | 0.513078554 |
| CD19- TAMs | CEBPB | 0.512753424 |
| CD19- TAMs | SLC43A2 | 0.511895907 |
| CD19- TAMs | NOTCH2 | 0.51125959 |
| CD19- TAMs | CD1E | 0.511124549 |
| CD19- TAMs | CD55 | 0.510785979 |
| CD19- TAMs | RAB31 | 0.510649582 |
| CD19- TAMs | ACSL4 | 0.510357907 |
| CD19- TAMs | LITAF | 0.509428779 |
| CD19- TAMs | IFITM2 | 0.509310803 |
| CD19- TAMs | GDI2 | 0.509190066 |
| CD19- TAMs | TMTC2 | 0.508563115 |
| CD19- TAMs | CSF2RA | 0.506941281 |
| CD19- TAMs | CASP1 | 0.506101901 |
| CD19- TAMs | AC009093.2 | 0.505772992 |
| CD19- TAMs | STK17B | 0.504399124 |
| CD19- TAMs | HIPK3 | 0.504275475 |
| CD19- TAMs | C9orf72 | 0.504211902 |
| CD19- TAMs | JAZF1 | 0.503989591 |
| CD19- TAMs | CTNND1 | 0.503891304 |
| CD19- TAMs | HCST | 0.503431773 |
| CD19- TAMs | PRAM1 | 0.502768543 |
| CD19- TAMs | RPLP2 | 0.502456507 |
| CD19- TAMs | ATP5MC2 | 0.501871874 |
| CD19- TAMs | PRELID1 | 0.500708567 |
| CD19- TAMs | RPS13 | 0.500518235 |
| CD19- TAMs | MYADM | 0.500498868 |
| CD19- TAMs | RPS2 | 0.49949985 |
| CD19- TAMs | ACTG1 | 0.498846454 |
| CD19- TAMs | UBE2D1 | 0.498176233 |
| CD19- TAMs | RCOR1 | 0.497693068 |
| CD19- TAMs | B3GNT5 | 0.497500457 |
| CD19- TAMs | GABARAPL1 | 0.497125983 |
| CD19- TAMs | ITGAX | 0.496880942 |
| CD19- TAMs | SLC25A5 | 0.496574904 |
| CD19- TAMs | GLIPR2 | 0.496036799 |
| CD19- TAMs | KCNE1 | 0.495921253 |
| CD19- TAMs | CTSH | 0.494177615 |
| CD19- TAMs | MRPL18 | 0.492778904 |
| CD19- TAMs | USP3 | 0.491877217 |
| CD19- TAMs | USP32 | 0.491862682 |
| CD19- TAMs | STX11 | 0.491378846 |
| CD19- TAMs | HNRNPU | 0.490655727 |
| CD19- TAMs | TUT7 | 0.490549614 |
| CD19- TAMs | MED13L | 0.490243724 |
| CD19- TAMs | DOK2 | 0.488374472 |
| CD19- TAMs | ELF2 | 0.487282345 |
| CD19- TAMs | LDLR | 0.486523953 |
| CD19- TAMs | CTBP2 | 0.485869465 |
| CD19- TAMs | APOBEC3A | 0.485545692 |
| CD19- TAMs | LAP3 | 0.485270992 |
| CD19- TAMs | RPL38 | 0.485058697 |
| CD19- TAMs | STAT1 | 0.483268956 |
| CD19- TAMs | RPL34 | 0.482586873 |
| CD19- TAMs | USP25 | 0.482334879 |
| CD19- TAMs | ABR | 0.482207314 |
| CD19- TAMs | TBXAS1 | 0.481482125 |
| CD19- TAMs | SPI1 | 0.480600288 |
| CD19- TAMs | PSMA4 | 0.478856696 |
| CD19- TAMs | YBX3 | 0.477438897 |
| CD19- TAMs | TNFAIP6 | 0.475757908 |
| CD19- TAMs | LMO2 | 0.474984479 |
| CD19- TAMs | IFNGR1 | 0.47460676 |
| CD19- TAMs | RHOA | 0.47416232 |
| CD19- TAMs | RPS29 | 0.473783013 |
| CD19- TAMs | WAS | 0.47256834 |
| CD19- TAMs | ANXA5 | 0.470889337 |
| CD19- TAMs | ATP11A | 0.470885033 |
| CD19- TAMs | GPAT3 | 0.470572528 |
| CD19- TAMs | RPS11 | 0.470566689 |
| CD19- TAMs | POU2F2 | 0.470446453 |
| CD19- TAMs | RYBP | 0.470299999 |
| CD19- TAMs | MTPN | 0.469096365 |
| CD19- TAMs | ATP13A3 | 0.466979548 |
| CD19- TAMs | HRH2 | 0.466927381 |
| CD19- TAMs | NKRF | 0.466785607 |
| CD19- TAMs | COMMD6 | 0.464410243 |
| CD19- TAMs | ALAS1 | 0.464277706 |
| CD19- TAMs | SPAG9 | 0.464233179 |
| CD19- TAMs | TPI1 | 0.463933251 |
| CD19- TAMs | RPL31 | 0.462542233 |
| CD19- TAMs | RPL27 | 0.462248123 |
| CD19- TAMs | RPL12 | 0.460837501 |
| CD19- TAMs | RPL32 | 0.460597357 |
| CD19- TAMs | RILPL2 | 0.458164856 |
| CD19- TAMs | PCBP1 | 0.457723588 |
| CD19- TAMs | ARPC5 | 0.457600655 |
| CD19- TAMs | PKM | 0.457555084 |
| CD19- TAMs | GAPT | 0.456608073 |
| CD19- TAMs | LDLRAD3 | 0.456557162 |
| CD19- TAMs | EVI2B | 0.456271502 |
| CD19- TAMs | GAPDH | 0.455827192 |
| CD19- TAMs | STK10 | 0.45581856 |
| CD19- TAMs | ZNF710 | 0.455066809 |
| CD19- TAMs | COX7B | 0.455064234 |
| CD19- TAMs | HMGB2 | 0.454547357 |
| CD19- TAMs | RPS6KA3 | 0.452776197 |
| CD19- TAMs | DEK | 0.452580493 |
| CD19- TAMs | SNAP23 | 0.450902537 |
| CD19- TAMs | RPL18A | 0.450826599 |
| CD19- TAMs | ACAP2 | 0.450162233 |
| CD19- TAMs | OGFRL1 | 0.449411807 |
| CD19- TAMs | LILRB1 | 0.449309576 |
| CD19- TAMs | EIF3L | 0.448977884 |
| CD19- TAMs | PYCARD | 0.447456137 |
| CD19- TAMs | TET3 | 0.4469606 |
| CD19- TAMs | TCP1 | 0.446492966 |
| CD19- TAMs | SH3KBP1 | 0.446212239 |
| CD19- TAMs | RPL6 | 0.444162817 |
| CD19- TAMs | GPBP1 | 0.444027922 |
| CD19- TAMs | RAB10 | 0.442712418 |
| CD19- TAMs | PSME1 | 0.441156405 |
| CD19- TAMs | RPL27A | 0.440352062 |
| CD19- TAMs | MAPK14 | 0.439870402 |
| CD19- TAMs | YWHAB | 0.439786393 |
| CD19- TAMs | COX8A | 0.438660102 |
| CD19- TAMs | PGK1 | 0.438138835 |
| CD19- TAMs | SRSF3 | 0.437923589 |
| CD19- TAMs | CMTM6 | 0.437806426 |
| CD19- TAMs | KIF13A | 0.436188013 |
| CD19- TAMs | UBAC2 | 0.435319164 |
| CD19- TAMs | TMEM170B | 0.43498949 |
| CD19- TAMs | SNAI1 | 0.434326312 |
| CD19- TAMs | AC090559.1 | 0.434260514 |
| CD19- TAMs | ADGRE5 | 0.433561375 |
| CD19- TAMs | PLEKHO1 | 0.433385563 |
| CD19- TAMs | FTH1 | 0.432785723 |
| CD19- TAMs | FAM102B | 0.432125124 |
| CD19- TAMs | PHACTR2 | 0.431421212 |
| CD19- TAMs | ANKRD22 | 0.431018135 |
| CD19- TAMs | SHTN1 | 0.42976777 |
| CD19- TAMs | BCL10 | 0.428811129 |
| CD19- TAMs | PLXNC1 | 0.428705234 |
| CD19- TAMs | ARPC3 | 0.427276281 |
| CD19- TAMs | SRGAP2 | 0.427266952 |
| CD19- TAMs | GRK3 | 0.426867096 |
| CD19- TAMs | TMSB4X | 0.425483346 |
| CD19- TAMs | LIMD2 | 0.425037395 |
| CD19- TAMs | ZFAS1 | 0.424539898 |
| CD19- TAMs | MBOAT7 | 0.423792743 |
| CD19- TAMs | MAP3K20 | 0.423139005 |
| CD19- TAMs | METRNL | 0.422674055 |
| CD19- TAMs | LUCAT1 | 0.421234469 |
| CD19- TAMs | HIF1A | 0.421007539 |
| CD19- TAMs | IPCEF1 | 0.42044749 |
| CD19- TAMs | FMNL1 | 0.420128609 |
| CD19- TAMs | ATP5MPL | 0.420024453 |
| CD19- TAMs | RPL22 | 0.419805269 |
| CD19- TAMs | EMP3 | 0.418513372 |
| CD19- TAMs | SPN | 0.415697391 |
| CD19- TAMs | ROCK1 | 0.415387643 |
| CD19- TAMs | SCLT1 | 0.415306435 |
| CD19- TAMs | WAC | 0.415024171 |
| CD19- TAMs | RAP1A | 0.415005363 |
| CD19- TAMs | PILRA | 0.414955862 |
| CD19- TAMs | PSEN1 | 0.414542674 |
| CD19- TAMs | GRB2 | 0.414390945 |
| CD19- TAMs | CCNY | 0.413842109 |
| CD19- TAMs | MYO1G | 0.412755416 |
| CD19- TAMs | ADAM17 | 0.41222379 |
| CD19- TAMs | RNF213 | 0.411943335 |
| CD19- TAMs | EIF3A | 0.410959095 |
| CD19- TAMs | FGL2 | 0.410954447 |
| CD19- TAMs | ANP32B | 0.410541663 |
| CD19- TAMs | HK1 | 0.410472451 |
| CD19- TAMs | STK38 | 0.408838331 |
| CD19- TAMs | RAP1GAP2 | 0.407905149 |
| CD19- TAMs | ATAD2B | 0.40730907 |
| CD19- TAMs | RAB7A | 0.406075273 |
| CD19- TAMs | CAP1 | 0.405732041 |
| CD19- TAMs | CBL | 0.405603968 |
| CD19- TAMs | GIMAP7 | 0.405163016 |
| CD19- TAMs | IFNGR2 | 0.404782409 |
| CD19- TAMs | GLIPR1 | 0.404380843 |
| CD19- TAMs | CD86 | 0.404051435 |
| CD19- TAMs | LHFPL6 | 0.404015313 |
| CD19- TAMs | SKAP2 | 0.404012695 |
| CD19- TAMs | ZYX | 0.403657043 |
| CD19- TAMs | PACSIN2 | 0.402695878 |
| CD19- TAMs | MYD88 | 0.402624122 |
| CD19- TAMs | CD300A | 0.40258566 |
| CD19- TAMs | GPR132 | 0.402321983 |
| CD19- TAMs | PPP1R12A | 0.402001696 |
| CD19- TAMs | DPYSL2 | 0.401707229 |
| CD19- TAMs | TACC1 | 0.401542055 |
| CD19- TAMs | PSMB9 | 0.400934471 |
| CD19- TAMs | PEA15 | 0.400917827 |
| CD19- TAMs | LINC02432 | 0.399051634 |
| CD19- TAMs | PFKFB3 | 0.398132369 |
| CD19- TAMs | CERS6 | 0.397969789 |
| CD19- TAMs | PIP4K2A | 0.397911003 |
| CD19- TAMs | EIF3H | 0.3972646 |
| CD19- TAMs | DDX60L | 0.396726982 |
| CD19- TAMs | TOM1 | 0.396202995 |
| CD19- TAMs | PAK2 | 0.396069438 |
| CD19- TAMs | ANP32A | 0.395979291 |
| CD19- TAMs | KCNAB2 | 0.395884018 |
| CD19- TAMs | CSNK1A1 | 0.395718737 |
| CD19- TAMs | HNRNPA3 | 0.395208075 |
| CD19- TAMs | MX2 | 0.394926101 |
| CD19- TAMs | PIK3CB | 0.394911052 |
| CD19- TAMs | RNF19B | 0.39486896 |
| CD19- TAMs | ATP1A1 | 0.394584041 |
| CD19- TAMs | ATP6V0D1 | 0.394167174 |
| CD19- TAMs | ADD3 | 0.393558662 |
| CD19- TAMs | CLEC4A | 0.393394866 |
| CD19- TAMs | SLC25A6 | 0.392855019 |
| CD19- TAMs | ATP5MG | 0.392308617 |
| CD19- TAMs | SFPQ | 0.392046221 |
| CD19- TAMs | ITPR1 | 0.391494215 |
| CD19- TAMs | CD300LB | 0.390637382 |
| CD19- TAMs | RAC1 | 0.39049483 |
| CD19- TAMs | CSGALNACT2 | 0.390154843 |
| CD19- TAMs | RNF130 | 0.390061964 |
| CD19- TAMs | LACTB | 0.389648845 |
| CD19- TAMs | GNG5 | 0.388803053 |
| CD19- TAMs | MIR181A1HG | 0.388441255 |
| CD19- TAMs | ARL8B | 0.387921158 |
| CD19- TAMs | YWHAZ | 0.387631423 |
| CD19- TAMs | HIGD2A | 0.387287989 |
| CD19- TAMs | ARRB1 | 0.386914158 |
| CD19- TAMs | MAP2K3 | 0.386846227 |
| CD19- TAMs | ETS2 | 0.386077213 |
| CD19- TAMs | TGFBI | 0.385666306 |
| CD19- TAMs | KDM1B | 0.385073327 |
| CD19- TAMs | HNRNPC | 0.384997729 |
| CD19- TAMs | ATP6V1B2 | 0.384326996 |
| CD19- TAMs | CCDC88A | 0.384192372 |
| CD19- TAMs | SMCO4 | 0.384058137 |
| CD19- TAMs | HNRNPAB | 0.384038798 |
| CD19- TAMs | PNPLA8 | 0.383733027 |
| CD19- TAMs | MOB1A | 0.383378396 |
| CD19- TAMs | ACTN1 | 0.382948001 |
| CD19- TAMs | HSBP1 | 0.382923649 |
| CD19- TAMs | RPS15 | 0.382424687 |
| CD19- TAMs | PLBD1 | 0.382376277 |
| CD19- TAMs | PREX1 | 0.381951517 |
| CD19- TAMs | PRR13 | 0.38105791 |
| CD19- TAMs | ZNF516 | 0.380695085 |
| CD19- TAMs | XRN2 | 0.380611409 |
| CD19- TAMs | EIF3M | 0.380041562 |
| CD19- TAMs | REL | 0.378204555 |
| CD19- TAMs | SESTD1 | 0.378128438 |
| CD19- TAMs | ABRACL | 0.377973699 |
| CD19- TAMs | SLA | 0.376864402 |
| CD19- TAMs | SEC11A | 0.376261113 |
| CD19- TAMs | CAPZA1 | 0.376086687 |
| CD19- TAMs | CD1D | 0.375675691 |
| CD19- TAMs | VAMP8 | 0.374636761 |
| CD19- TAMs | GABARAP | 0.374306997 |
| CD19- TAMs | RSL1D1 | 0.373852055 |
| CD19- TAMs | RIOK3 | 0.373849471 |
| CD19- TAMs | TP53BP2 | 0.373476655 |
| CD19- TAMs | PFDN5 | 0.373373616 |
| CD19- TAMs | COX5B | 0.372422937 |
| CD19- TAMs | DNAJC7 | 0.370912597 |
| CD19- TAMs | RPL35A | 0.37078677 |
| CD19- TAMs | IL13RA1 | 0.370764423 |
| CD19- TAMs | ZBTB43 | 0.370655616 |
| CD19- TAMs | PLK3 | 0.370314133 |
| CD19- TAMs | RNF144B | 0.370220059 |
| CD19- TAMs | HCAR3 | 0.369986517 |
| CD19- TAMs | COX6B1 | 0.369607417 |
| CD19- TAMs | RHOG | 0.368639224 |
| CD19- TAMs | CPVL | 0.368527462 |
| CD19- TAMs | S100Z | 0.368379055 |
| CD19- TAMs | RBM47 | 0.3681617 |
| CD19- TAMs | RAP1B | 0.368157052 |
| CD19- TAMs | GNA15 | 0.368064849 |
| CD19- TAMs | PGLS | 0.36769789 |
| CD19- TAMs | KSR1 | 0.367418241 |
| CD19- TAMs | FAR1 | 0.367357175 |
| CD19- TAMs | PSMB10 | 0.366717621 |
| CD19- TAMs | VASH1 | 0.366607992 |
| CD19- TAMs | NACA | 0.366387755 |
| CD19- TAMs | AGO4 | 0.366322189 |
| CD19- TAMs | FAM120A | 0.36631059 |
| CD19- TAMs | OXSR1 | 0.366274115 |
| CD19- TAMs | EVI2A | 0.365951195 |
| CD19- TAMs | CRYBG1 | 0.365754239 |
| CD19- TAMs | PRRG4 | 0.365614889 |
| CD19- TAMs | APAF1 | 0.36542648 |
| CD19- TAMs | CNN2 | 0.364498651 |
| CD19- TAMs | THAP9-AS1 | 0.364366597 |
| CD19- TAMs | ZNF385A | 0.363265957 |
| CD19- TAMs | SH3BGRL | 0.363123566 |
| CD19- TAMs | ETF1 | 0.362726982 |
| CD19- TAMs | NSA2 | 0.360535649 |
| CD19- TAMs | GMFG | 0.359957628 |
| CD19- TAMs | PDLIM5 | 0.359668903 |
| CD19- TAMs | BZW1 | 0.359648651 |
| CD19- TAMs | VCL | 0.35912458 |
| CD19- TAMs | PHC2 | 0.358689371 |
| CD19- TAMs | SPOPL | 0.357037596 |
| CD19- TAMs | NAMPT | 0.356859088 |
| CD19- TAMs | CDC42EP2 | 0.354955721 |
| CD19- TAMs | MAP3K3 | 0.35486751 |
| CD19- TAMs | ZMIZ1 | 0.354271092 |
| CD19- TAMs | GSTO1 | 0.354034091 |
| CD19- TAMs | HMGA1 | 0.353946987 |
| CD19- TAMs | ASGR1 | 0.353672773 |
| CD19- TAMs | ZNF207 | 0.353622919 |
| CD19- TAMs | SNTB1 | 0.35290511 |
| CD19- TAMs | UBE2W | 0.352474922 |
| CD19- TAMs | TRERF1 | 0.351398201 |
| CD19- TAMs | SVIL | 0.351363661 |
| CD19- TAMs | CASP4 | 0.350583919 |
| CD19- TAMs | CLEC4E | 0.349596941 |
| CD19- TAMs | RNF24 | 0.348851266 |
| CD19- TAMs | RHOQ | 0.348778195 |
| CD19- TAMs | BCL6 | 0.348654988 |
| CD19- TAMs | PALLD | 0.348243357 |
| CD19- TAMs | ST3GAL6 | 0.347463573 |
| CD19- TAMs | IGF2BP2 | 0.347423295 |
| CD19- TAMs | SNRPF | 0.347342307 |
| CD19- TAMs | NCOR2 | 0.346947268 |
| CD19- TAMs | RPL35 | 0.346499526 |
| CD19- TAMs | GNB1 | 0.346297937 |
| CD19- TAMs | NUDT16 | 0.344311862 |
| CD19- TAMs | CPD | 0.343979088 |
| CD19- TAMs | SNHG16 | 0.342968699 |
| CD19- TAMs | HDAC4 | 0.342915852 |
| CD19- TAMs | ZNF267 | 0.342134108 |
| CD19- TAMs | OSCAR | 0.341748228 |
| CD19- TAMs | AP2S1 | 0.341202007 |
| CD19- TAMs | CD300LF | 0.341060829 |
| CD19- TAMs | HCLS1 | 0.341034533 |
| CD19- TAMs | LRG1 | 0.339140506 |
| CD19- TAMs | PHF20L1 | 0.33901375 |
| CD19- TAMs | GLT1D1 | 0.338905076 |
| CD19- TAMs | ANPEP | 0.338598043 |
| CD19- TAMs | HNRNPD | 0.338545039 |
| CD19- TAMs | STX6 | 0.337143374 |
| CD19- TAMs | LAT2 | 0.337074748 |
| CD19- TAMs | TRAF3IP3 | 0.336912304 |
| CD19- TAMs | SLC24A4 | 0.336535824 |
| CD19- TAMs | CAPNS1 | 0.336424717 |
| CD19- TAMs | SPATA13 | 0.335640924 |
| CD19- TAMs | CCT5 | 0.3354022 |
| CD19- TAMs | POLR1D | 0.335330432 |
| CD19- TAMs | FCGR1A | 0.333385997 |
| CD19- TAMs | LATS2 | 0.333244367 |
| CD19- TAMs | MBD2 | 0.332960025 |
| CD19- TAMs | CRISPLD2 | 0.332396002 |
| CD19- TAMs | ABI3 | 0.332333178 |
| CD19- TAMs | RELT | 0.331860145 |
| CD19- TAMs | NDRG1 | 0.331680272 |
| CD19- TAMs | ZNF106 | 0.331673278 |
| CD19- TAMs | QKI | 0.331526113 |
| CD19- TAMs | CD302 | 0.330938254 |
| CD19- TAMs | TSPAN14 | 0.330817481 |
| CD19- TAMs | HADHA | 0.33048219 |
| CD19- TAMs | IL6R | 0.329268926 |
| CD19- TAMs | ELOB | 0.32855036 |
| CD19- TAMs | NFIL3 | 0.327960999 |
| CD19- TAMs | LILRA5 | 0.327835442 |
| CD19- TAMs | CDA | 0.327665299 |
| CD19- TAMs | ITGA5 | 0.325537255 |
| CD19- TAMs | SCO2 | 0.325493764 |
| CD19- TAMs | CCT6A | 0.324854975 |
| CD19- TAMs | HK2 | 0.324685996 |
| CD19- TAMs | BIN2 | 0.323669387 |
| CD19- TAMs | DAZAP2 | 0.323556132 |
| CD19- TAMs | NMI | 0.323496903 |
| CD19- TAMs | BZW2 | 0.32327998 |
| CD19- TAMs | RCBTB2 | 0.322632135 |
| CD19- TAMs | RTN4 | 0.322196463 |
| CD19- TAMs | MCUB | 0.321382548 |
| CD19- TAMs | RASSF5 | 0.320953173 |
| CD19- TAMs | PARVG | 0.320245207 |
| CD19- TAMs | AC015871.7 | 0.319487191 |
| CD19- TAMs | CHMP4B | 0.319125954 |
| CD19- TAMs | PLA2G4A | 0.318821961 |
| CD19- TAMs | DYNC1LI1 | 0.318166922 |
| CD19- TAMs | C19orf38 | 0.317573944 |
| CD19- TAMs | MAPK1 | 0.316380002 |
| CD19- TAMs | PCSK5 | 0.316312176 |
| CD19- TAMs | NDUFB1 | 0.315770375 |
| CD19- TAMs | CYB5R4 | 0.315723646 |
| CD19- TAMs | ERICH1 | 0.315470198 |
| CD19- TAMs | ZC3H12A | 0.315077666 |
| CD19- TAMs | AL138899.1 | 0.314842413 |
| CD19- TAMs | VPS35 | 0.314565447 |
| CD19- TAMs | OAZ1 | 0.314126126 |
| CD19- TAMs | TMEM71 | 0.312360584 |
| CD19- TAMs | LAMTOR4 | 0.311281019 |
| CD19- TAMs | SERF2 | 0.310678883 |
| CD19- TAMs | CCDC26 | 0.310163181 |
| CD19- TAMs | MEFV | 0.309374257 |
| CD19- TAMs | ACAA1 | 0.309047484 |
| CD19- TAMs | TMA7 | 0.308306722 |
| CD19- TAMs | KIAA0513 | 0.307936172 |
| CD19- TAMs | SIRPB1 | 0.307498953 |
| CD19- TAMs | RPS28 | 0.307465037 |
| CD19- TAMs | NUP214 | 0.306733043 |
| CD19- TAMs | POLE4 | 0.30383542 |
| CD19- TAMs | SLC2A6 | 0.303104943 |
| CD19- TAMs | MIS18BP1 | 0.302888012 |
| CD19- TAMs | ACOT9 | 0.302696979 |
| CD19- TAMs | ECHDC1 | 0.301029725 |
| CD19- TAMs | SH3BP2 | 0.300700626 |
| CD19- TAMs | RAB32 | 0.299506711 |
| CD19- TAMs | GCA | 0.298262411 |
| CD19- TAMs | RASSF2 | 0.298052399 |
| CD19- TAMs | PTGIR | 0.297861641 |
| CD19- TAMs | SGMS2 | 0.297393317 |
| CD19- TAMs | SNX27 | 0.295057232 |
| CD19- TAMs | SECTM1 | 0.294935179 |
| CD19- TAMs | LINC00877 | 0.294526319 |
| CD19- TAMs | STK26 | 0.290191062 |
| CD19- TAMs | STMP1 | 0.288166695 |
| CD19- TAMs | TTC39C | 0.286829307 |
| CD19- TAMs | DENND10 | 0.28591409 |
| CD19- TAMs | AL034397.3 | 0.285394597 |
| CD19- TAMs | HDGF | 0.28494395 |
| CD19- TAMs | GRAMD1B | 0.28353345 |
| CD19- TAMs | LYSMD2 | 0.282283901 |
| CD19- TAMs | VDR | 0.279549551 |
| CD19- TAMs | LFNG | 0.279376687 |
| CD19- TAMs | STAT6 | 0.277004236 |
| CD19- TAMs | ZDHHC7 | 0.276434444 |
| CD19- TAMs | AGTRAP | 0.276350086 |
| CD19- TAMs | PKP2 | 0.276105087 |
| CD19- TAMs | MPP7 | 0.275726332 |
| CD19- TAMs | LILRA1 | 0.273949584 |
| CD19- TAMs | PSTPIP1 | 0.271772001 |
| CD19- TAMs | PGD | 0.271145573 |
| CD19- TAMs | ARRB2 | 0.26653568 |
| CD19- TAMs | FFAR2 | 0.266522059 |
| CD19- TAMs | MARCO | 0.266161035 |
| CD19- TAMs | CAMKK2 | 0.265401797 |
| CD19- TAMs | IL27RA | 0.262745112 |
| CD19- TAMs | NRGN | 0.261376001 |
| CD19- TAMs | CYSLTR2 | 0.260715762 |
| CD19- TAMs | PRKACA | 0.258131576 |
| CD19- TAMs | BST1 | 0.254079936 |
| CD19- TAMs | CCR2 | 0.252923175 |
| CD19- TAMs | RPS12 | 0.397686334 |
| CD19- TAMs | SAP30 | 0.455389975 |
| CD19- TAMs | STIP1 | 0.340635656 |
| CD19- TAMs | CSK | 0.289364389 |
| CD19- TAMs | C20orf27 | 0.265428633 |
| CD19- TAMs | PADI2 | 0.360576263 |
| CD19- TAMs | UBAP1 | 0.409590931 |
| CD19- TAMs | RPL29 | 0.397254841 |
| CD19- TAMs | SYAP1 | 0.387948543 |
| CD19- TAMs | VAMP5 | 0.44504612 |
| CD19- TAMs | TMPO | 0.334774986 |
| CD19- TAMs | KHDRBS1 | 0.329242701 |
| CD19- TAMs | RPS15A | 0.354701142 |
| CD19- TAMs | FOXN2 | 0.335499975 |
| CD19- TAMs | ITGA4 | 0.407363175 |
| CD19- TAMs | CAPZA2 | 0.290610637 |
| CD19- TAMs | MYL12B | 0.319679451 |
| CD19- TAMs | TMEM165 | 0.344725027 |
| CD19- TAMs | AL138720.1 | 0.358288522 |
| CD19- TAMs | CLK1 | 0.409563733 |
| CD19- TAMs | SLC7A7 | 0.300227447 |
| CD19- TAMs | RBPJ | 0.324156048 |
| CD19- TAMs | TMOD3 | 0.311615531 |
| CD19- TAMs | PRKAR2B | 0.356722451 |
| CD19- TAMs | AZI2 | 0.284401899 |
| CD19- TAMs | SLC6A6 | 0.274340678 |
| CD19- TAMs | PSMA6 | 0.344134708 |
| CD19- TAMs | XPO1 | 0.341002128 |
| CD19- TAMs | IER5 | 0.624692148 |
| CD19- TAMs | EIF4G2 | 0.375254233 |
| CD19- TAMs | CIITA | 0.439433492 |
| CD19- TAMs | BRI3 | 0.2515886 |
| CD19- TAMs | ARHGAP31 | 0.410379503 |
| CD19- TAMs | UVRAG | 0.424579509 |
| CD19- TAMs | ENGASE | 0.412974203 |
| CD19- TAMs | PTPN2 | 0.316964217 |
| CD19- TAMs | RPS14 | 0.370894473 |
| CD19- TAMs | TBC1D8 | 0.335121554 |
| CD19- TAMs | ARPC4 | 0.318745627 |
| CD19- TAMs | PTGES3 | 0.445246691 |
| CD19- TAMs | SIGLEC10 | 0.279824328 |
| CD19- TAMs | TNFAIP3 | 0.392692554 |
| CD19- TAMs | LSM6 | 0.272977842 |
| CD19- TAMs | FLVCR2 | 0.274112849 |
| CD19- TAMs | MOB3A | 0.310220301 |
| CD19- TAMs | FAM117B | 0.292712517 |
| CD19- TAMs | MARCKSL1 | 0.280701646 |
| CD19- TAMs | MYO9B | 0.372199021 |
| CD19- TAMs | ABHD5 | 0.321449027 |
| CD19- TAMs | C17orf49 | 0.324184396 |
| CD19- TAMs | 1-Mar | 0.341480148 |
| CD19- TAMs | SIPA1L1 | 0.672282129 |
| CD19- TAMs | IRS2 | 0.351839858 |
| CD19- TAMs | FKBP4 | 0.522971003 |
| CD19- TAMs | RHOT1 | 0.279231521 |
| CD19- TAMs | COX7C | 0.31513449 |
| CD19- TAMs | MTMR3 | 0.334562664 |
| CD19- TAMs | RPS27 | 0.378680518 |
| CD19- TAMs | PTPN18 | 0.263551762 |
| CD19- TAMs | HSPA6 | 0.803196892 |
| CD19- TAMs | TRA2B | 0.398892699 |
| CD19- TAMs | ASAP1 | 0.488892965 |
| CD19- TAMs | CUL1 | 0.369681537 |
| CD19- TAMs | ZFAND3 | 0.576354311 |
| CD19- TAMs | TBC1D2 | 0.266724822 |
| CD19- TAMs | PPIA | 0.405988077 |
| CD19- TAMs | YWHAG | 0.339412838 |
| CD19- TAMs | NFE2L2 | 0.351629882 |
| CD19- TAMs | SLC8B1 | 0.366804211 |
| CD19- TAMs | IQGAP1 | 0.264952262 |
| CD19- TAMs | WNK1 | 0.33933174 |
| CD19- TAMs | JOSD1 | 0.336035138 |
| CD19- TAMs | CLEC5A | 0.309082359 |
| CD19- TAMs | ADA2 | 0.309178171 |
| CD19- TAMs | RGS2 | 0.391424451 |
| CD19- TAMs | BTF3 | 0.383026245 |
| CD19- TAMs | RPS20 | 0.382876939 |
| CD19- TAMs | DAPK1 | 0.302950022 |
| CD19- TAMs | CAMK1 | 0.282488165 |
| CD19- TAMs | DENND3 | 0.342551335 |
| CD19- TAMs | SP110 | 0.294898612 |
| CD19- TAMs | LRRC25 | 0.278776366 |
| CD19- TAMs | XAF1 | 0.30347961 |
| CD19- TAMs | LGALS9 | 0.284754517 |
| CD19- TAMs | CYTH1 | 0.423102725 |
| CD19- TAMs | RPL30 | 0.367453217 |
| CD19- TAMs | TPM4 | 0.355293758 |
| CD19- TAMs | TAX1BP1 | 0.301378183 |
| CD19- TAMs | MAP3K5 | 0.409008166 |
| CD19- TAMs | HNRNPK | 0.334386477 |
| CD19- TAMs | SYF2 | 0.299938584 |
| CD19- TAMs | TMEM167A | 0.28119852 |
| CD19- TAMs | NFAT5 | 0.57398145 |
| CD19- TAMs | APOL3 | 0.259926147 |
| CD19- TAMs | TCIRG1 | 0.260243514 |
| CD19- TAMs | PSMB3 | 0.30826652 |
| CD19- TAMs | HES4 | 0.458668898 |
| CD19- TAMs | TBC1D1 | 0.340526636 |
| CD19- TAMs | MICU1 | 0.377828504 |
| CD19- TAMs | ADSS | 0.253093186 |
| CD19- TAMs | HBEGF | 0.55542621 |
| CD19- TAMs | IL15 | 0.307099188 |
| CD19- TAMs | MRTFA | 0.373190633 |
| CD19- TAMs | PGAM1 | 0.300290681 |
| CD19- TAMs | CSRNP1 | 0.305797002 |
| CD19- TAMs | COX4I1 | 0.325504508 |
| CD19- TAMs | TRIM25 | 0.265371018 |
| CD19- TAMs | DOCK10 | 0.58552555 |
| CD19- TAMs | ACER3 | 0.355188345 |
| CD19- TAMs | LRRFIP2 | 0.312234998 |
| CD19- TAMs | KMT2C | 0.444454417 |
| CD19- TAMs | SCIMP | 0.26806365 |
| CD19- TAMs | ARID1A | 0.298422171 |
| CD19- TAMs | MARK2 | 0.251087324 |
| CD19- TAMs | TNFRSF10B | 0.409730007 |
| CD19- TAMs | TMEM131L | 0.455874311 |
| CD19- TAMs | MT2A | 0.262447408 |
| CD19- TAMs | IER3 | 0.427349991 |
| CD19- TAMs | SLC25A25 | 0.337261133 |
| CD19- TAMs | SEMA4D | 0.295275362 |
| CD19- TAMs | VDAC1 | 0.341128579 |
| CD19- TAMs | RP2 | 0.25561748 |
| CD19- TAMs | SET | 0.371299997 |
| CD19- TAMs | EIF3K | 0.365064233 |
| CD19- TAMs | CPQ | 0.337756017 |
| CD19- TAMs | ENY2 | 0.312881971 |
| CD19- TAMs | ADAM10 | 0.353180115 |
| CD19- TAMs | BCL11A | 0.280757355 |
| CD19- TAMs | TFCP2 | 0.299437599 |
| CD19- TAMs | MTHFD2 | 0.326918 |
| CD19- TAMs | YY1 | 0.303139174 |
| CD19- TAMs | HIVEP3 | 0.338637694 |
| CD19- TAMs | SLC25A13 | 0.31400404 |
| CD19- TAMs | ACTN4 | 0.259114835 |
| CD19- TAMs | SKIL | 0.400355519 |
| CD19- TAMs | NFKBIE | 0.287719 |
| CD19- TAMs | PPP1R15A | 0.52838111 |
| CD19- TAMs | CXCL8 | 0.827951501 |
| CD19- TAMs | GRASP | 0.343187755 |
| CD19- TAMs | LIMK2 | 0.268814438 |
| CD19- TAMs | SCAF11 | 0.295087921 |
| CD19- TAMs | PRKCE | 0.517725477 |
| CD19- TAMs | FAM49A | 0.336624116 |
| CD19- TAMs | SMAD3 | 0.300359333 |
| CD19- TAMs | FLOT1 | 0.297919002 |
| CD19- TAMs | LDHA | 0.420468894 |
| CD19- TAMs | LDLRAD4 | 0.341114431 |
| CD19- TAMs | BAZ2B | 0.358606494 |
| CD19- TAMs | DUSP1 | 0.592620584 |
| CD19- TAMs | SKI | 0.311673381 |
| CD19- TAMs | PSMA7 | 0.365719785 |
| CD19- TAMs | RPS26 | 0.434303506 |
| CD19- TAMs | MIDN | 0.353936933 |
| CD19- TAMs | AGPAT3 | 0.264595454 |
| CD19- TAMs | CBX3 | 0.28424595 |
| CD19- TAMs | HSPA1L | 0.282717756 |
| CD19- TAMs | GBP4 | 0.339504559 |
| CD19- TAMs | PLEKHB2 | 0.287225058 |
| CD19- TAMs | TAF15 | 0.271900555 |
| CD19- TAMs | HNRNPA2B1 | 0.299503076 |
| CD19- TAMs | LINC01578 | 0.317928666 |
| CD19- TAMs | RASGEF1B | 0.407712283 |
| CD19- TAMs | DYNLT1 | 0.258378976 |
| CD19- TAMs | USP10 | 0.251120717 |
| CD19- TAMs | RPL36 | 0.258751714 |
| CD19- TAMs | DEDD2 | 0.294537542 |
| CD19- TAMs | STK3 | 0.35916867 |
| CD19- TAMs | HSD17B11 | 0.260097461 |
| CD19- TAMs | SRRM1 | 0.303001858 |
| CD19- TAMs | LRRC8D | 0.299988846 |
| CD19- TAMs | PITPNA | 0.297467371 |
| CD19- TAMs | EIF3E | 0.342517879 |
| CD19- TAMs | GAB2 | 0.390726518 |
| CD19- TAMs | NIPBL | 0.349880606 |
| CD19- TAMs | BANF1 | 0.286159594 |
| CD19- TAMs | FNIP2 | 0.28462891 |
| CD19- TAMs | RPL24 | 0.303738382 |
| CD19- TAMs | ZC3H15 | 0.291215229 |
| CD19- TAMs | CAMK2D | 0.67143355 |
| CD19- TAMs | PAG1 | 0.350173571 |
| CD19- TAMs | DUSP10 | 0.293753927 |
| CD19- TAMs | ST6GALNAC3 | 0.397448673 |
| CD19- TAMs | ZBTB7A | 0.298111642 |
| CD19- TAMs | RASSF3 | 0.356232497 |
| CD19- TAMs | CHP1 | 0.252419868 |
| CD19- TAMs | LAPTM5 | 0.257408315 |
| CD19- TAMs | SERBP1 | 0.340617923 |
| CD19- TAMs | NAA38 | 0.252223408 |
| CD19- TAMs | ZBTB16 | 0.655194002 |
| CD19- TAMs | WTAP | 0.287026252 |
| CD19- TAMs | FKBP15 | 0.253024783 |
| CD19- TAMs | UQCRB | 0.293893967 |
| CD19- TAMs | GCNT2 | 0.278855007 |
| CD19- TAMs | TRIM38 | 0.314057703 |
| CD19- TAMs | ATP5PD | 0.287617004 |
| CD19- TAMs | NAAA | 0.266159128 |
| CD19- TAMs | PSMB8 | 0.326597465 |
| CD19- TAMs | DNMT1 | 0.284474995 |
| CD19- TAMs | HNRNPF | 0.286132825 |
| CD19- TAMs | MYH9 | 0.293356568 |
| CD19- TAMs | EPB41L3 | 0.264361524 |
| CD19- TAMs | PRPF40A | 0.325468282 |
| CD19- TAMs | LPGAT1 | 0.282692157 |
| CD19- TAMs | EIF3F | 0.318647798 |
| CD19- TAMs | JPT1 | 0.289645932 |
| CD19- TAMs | MICOS10 | 0.306520326 |
| CD19- TAMs | AMD1 | 0.334316409 |
| CD19- TAMs | RPS6KC1 | 0.309022162 |
| CD19- TAMs | PPP2R2A | 0.280373475 |
| CD19- TAMs | PHACTR1 | 0.413881489 |
| CD19- TAMs | DDX5 | 0.319197093 |
| CD19- TAMs | TSC22D2 | 0.326925536 |
| CD19- TAMs | SYK | 0.349151851 |
| CD19- TAMs | IL10RB | 0.254700955 |
| CD19- TAMs | CERT1 | 0.298619934 |
| CD19- TAMs | EHD1 | 0.27497228 |
| CD19- TAMs | CALCOCO2 | 0.25530156 |
| CD19- TAMs | AKAP13 | 0.29952939 |
| CD19- TAMs | RPS21 | 0.292192776 |
| CD19- TAMs | MID1IP1 | 0.315318537 |
| CD19- TAMs | PECAM1 | 0.344962088 |
| CD19- TAMs | MORF4L1 | 0.284863952 |
| CD19- TAMs | PHF21A | 0.4261191 |
| CD19- TAMs | SNRPE | 0.267534642 |
| CD19- TAMs | MIER1 | 0.260813915 |
| CD19- TAMs | EPSTI1 | 0.350844347 |
| CD19- TAMs | RPS19 | 0.304375411 |
| CD19- TAMs | TWISTNB | 0.680905968 |
| CD19- TAMs | APP | 0.315582578 |
| CD19- TAMs | ATP5PB | 0.300016171 |
| CD19- TAMs | LAMTOR2 | 0.303393062 |
| CD19- TAMs | RGS19 | 0.265591059 |
| CD19- TAMs | AC016831.5 | 0.392771359 |
| CD19- TAMs | PPT1 | 0.264024641 |
| CD19- TAMs | ATP5F1C | 0.285720397 |
| CD19- TAMs | CEBPD | 0.253504108 |
| CD19- TAMs | RPLP1 | 0.251482713 |
| CD19- TAMs | TAOK1 | 0.300824651 |
| CD19- TAMs | ZNF217 | 0.269316316 |
| CD19- TAMs | RIN2 | 0.279584821 |
| CD19- TAMs | HSPB1 | 0.485929063 |
| CD19- TAMs | DRAP1 | 0.292796249 |
| CD19- TAMs | MARK3 | 0.311889779 |
| CD19- TAMs | NDUFB2 | 0.304398408 |
| CD19- TAMs | GPSM3 | 0.251750171 |
| CD19- TAMs | EEF1A1 | 0.369544485 |
| CD19- TAMs | SOAT1 | 0.295654371 |
| CD19- TAMs | NBN | 0.273682324 |
| CD19- TAMs | ATP5MF | 0.261857084 |
| CD19- TAMs | TENT5A | 0.365805093 |
| CD19- TAMs | SLC7A5 | 0.311744177 |
| CD19- TAMs | PITPNC1 | 0.343677577 |
| CD19- TAMs | CORO1C | 0.285987884 |
| CD19- TAMs | RPL11 | 0.314788028 |
| CD19- TAMs | PBX3 | 0.394771408 |
| CD19- TAMs | RPS3A | 0.380020955 |
| CD19- TAMs | CCNL1 | 0.385354265 |
| CD19- TAMs | FAM241A | 0.285859835 |
| CD19- TAMs | FAM110A | 0.250876144 |
| CD19- TAMs | RUNX3 | 0.371390826 |
| CD19- TAMs | IPMK | 0.253483797 |
| CD19- TAMs | GPR65 | 0.275143086 |
| CD19- TAMs | MCL1 | 0.3078787 |
| CD19- TAMs | TLN1 | 0.290190117 |
| CD19- TAMs | PCBP2 | 0.275906137 |
| CD19- TAMs | KIF5B | 0.250580451 |
| CD19- TAMs | SMURF1 | 0.276129033 |
| CD19- TAMs | CREBBP | 0.299079289 |
| CD19- TAMs | CTNNB1 | 0.416637378 |
| CD19- TAMs | APEX1 | 0.306450512 |
| CD19- TAMs | RPS6 | 0.361140109 |
| CD19- TAMs | PTPN6 | 0.271522107 |
| CD19- TAMs | SGK3 | 0.297623104 |
| CD19- TAMs | CDK19 | 0.304525372 |
| CD19- TAMs | SOX4 | 0.405988448 |
| CD19- TAMs | PMAIP1 | 0.526707167 |
| CD19- TAMs | ARID4B | 0.328340694 |
| CD19- TAMs | CD58 | 0.278152697 |
| CD19- TAMs | BRK1 | 0.265084996 |
| CD19- TAMs | PDE3B | 0.330376655 |
| CD19- TAMs | MAP3K1 | 0.393589882 |
| CD19- TAMs | MAPKAP1 | 0.337228412 |
| CD19- TAMs | CIB1 | 0.31501276 |
| CD19- TAMs | OSM | 0.3209141 |
| CD19- TAMs | KLF10 | 0.333301755 |
| CD19- TAMs | SLC25A33 | 0.303850197 |
| CD19- TAMs | RACK1 | 0.347090075 |
| CD19- TAMs | PPP1CA | 0.28194012 |
| CD19- TAMs | SNHG6 | 0.274915039 |
| CD19- TAMs | AGFG1 | 0.266886887 |
| CD19- TAMs | ATP2B1 | 0.42774248 |
| CD19- TAMs | PDZD8 | 0.263827738 |
| CD19- TAMs | ZNF281 | 0.266257366 |
| CD19- TAMs | FKBP5 | 0.418198053 |
| CD19- TAMs | ZC3HAV1 | 0.463419469 |
| CD19- TAMs | PNRC2 | 0.25885721 |
| CD19- TAMs | GATAD2A | 0.259302816 |
| CD19- TAMs | MYL6 | 0.259028486 |
| CD19- TAMs | AZIN1 | 0.275190183 |
| CD19- TAMs | ZNF804A | 0.340296163 |
| CD19- TAMs | CCRL2 | 0.282810268 |
| CD19- TAMs | HOOK3 | 0.282700055 |
| CD19- TAMs | KDM2A | 0.389576192 |
| CD19- TAMs | PRRC2C | 0.275978429 |
| CD19- TAMs | HSP90AB1 | 0.420510259 |
| CD19- TAMs | HMGB1 | 0.423726875 |
| CD19- TAMs | WIPF1 | 0.265669333 |
| CD19- TAMs | THBD | 0.383327417 |
| CD19- TAMs | RBX1 | 0.260325896 |
| CD19- TAMs | APRT | 0.30323289 |
| CD19- TAMs | PIK3AP1 | 0.288087935 |
| CD19- TAMs | ARHGEF3 | 0.276448093 |
| CD19- TAMs | EIF4G3 | 0.397668422 |
| CD19- TAMs | CCNI | 0.283957858 |
| CD19- TAMs | HAUS2 | 0.371699932 |
| CD19- TAMs | NLRP1 | 0.310769212 |
| CD19- TAMs | WWP2 | 0.272565963 |
| CD19- TAMs | WDFY2 | 0.342620253 |
| CD19- TAMs | NIN | 0.264213597 |
| CD19- TAMs | TRIP12 | 0.279572949 |
| CD19- TAMs | LSM7 | 0.252489902 |
| CD19- TAMs | UBE2E2 | 0.461980431 |
| CD19- TAMs | ZNF131 | 0.316540131 |
| CD19- TAMs | ATP5F1B | 0.299934249 |
| CD19- TAMs | SLTM | 0.266660594 |
| CD19- TAMs | RC3H1 | 0.303531414 |
| CD19- TAMs | YPEL2 | 0.259757114 |
| CD19- TAMs | NUDC | 0.296447803 |
| CD19- TAMs | TENT4B | 0.302536694 |
| CD19- TAMs | NEDD9 | 0.267998946 |
| CD19- TAMs | SEPTIN9 | 0.259920178 |
| CD19- TAMs | BHLHE40 | 0.339089773 |
| CD19- TAMs | EP300 | 0.278117107 |
| CD19- TAMs | CDC37 | 0.26189178 |
| CD19- TAMs | CPNE8 | 0.26878241 |
| CD19- TAMs | PSMA1 | 0.268496498 |
| CD19- TAMs | RAB8B | 0.250443773 |
| CD19- TAMs | GBP5 | 0.425261694 |
| CD19- TAMs | ATG7 | 0.394009693 |
| CD19- TAMs | HIP1 | 0.438977947 |
| CD19- TAMs | SNHG15 | 0.250734541 |
| CD19- TAMs | RPL21 | 0.341295963 |
| CD19- TAMs | RPL41 | 0.251318871 |
| CD19- TAMs | RPL10A | 0.392282584 |
| CD19- TAMs | PA2G4 | 0.286873071 |
| CD19- TAMs | RAB1A | 0.292559754 |
| CD19- TAMs | MYL12A | 0.278728231 |
| CD19- TAMs | PPP2R5A | 0.253130503 |
| CD19- TAMs | RPL15 | 0.383202977 |
| CD19- TAMs | SNRPB | 0.285921639 |
| CD19- TAMs | NXF1 | 0.284187642 |
| CD19- TAMs | PKIB | 0.283485225 |
| CD19- TAMs | SETX | 0.322852962 |
| CD19- TAMs | MYLIP | 0.353129583 |
| CD19- TAMs | RPS25 | 0.262867756 |
| CD19- TAMs | TAGLN2 | 0.300830386 |
| CD19- TAMs | GLUD1 | 0.260032585 |
| CD19- TAMs | CRLF3 | 0.254824672 |
| CD19- TAMs | HNRNPM | 0.284714068 |
| CD19- TAMs | NOCT | 0.285838855 |
| CD19- TAMs | PHLPP1 | 0.361166832 |
| CD19- TAMs | PPFIA1 | 0.25740294 |
| CD19- TAMs | CD109 | 0.312629935 |
| CD19- TAMs | MGAT1 | 0.266259876 |
| CD19- TAMs | STAT3 | 0.317028824 |
| CD19- TAMs | NR4A3 | 0.382922349 |
| CD19- TAMs | RBM25 | 0.269319666 |
| CD19- TAMs | SYNCRIP | 0.255759333 |
| CD19- TAMs | KAT6A | 0.293722099 |
| CD19- TAMs | ZNF438 | 0.283328024 |
| CD19- TAMs | FNIP1 | 0.313920961 |
| CD19- TAMs | PPP2CA | 0.264023616 |
| CD19- TAMs | PIAS1 | 0.326378401 |
| CD19- TAMs | AL163541.1 | 0.260447971 |
| CD19- TAMs | C5AR1 | 0.264290454 |
| CD19- TAMs | SPIDR | 0.407341357 |
| CD19- TAMs | NUFIP2 | 0.294699539 |
| CD19- TAMs | FRYL | 0.274147601 |
| CD19- TAMs | FAR2 | 0.256466301 |
| CD19- TAMs | PPM1B | 0.278574043 |
| CD19- TAMs | UBE2K | 0.256247437 |
| CD19- TAMs | RPS7 | 0.310850795 |
| CD19- TAMs | ZFR | 0.253879122 |
| CD19- TAMs | MOB3B | 0.304294641 |
| CD19- TAMs | XRCC5 | 0.263357785 |
| CD19- TAMs | PTGER4 | 0.295128019 |
| CD19- TAMs | SH3PXD2B | 0.251723088 |
| CD19- TAMs | P2RX7 | 0.291049986 |
| CD19- TAMs | PNRC1 | 0.308675897 |
| CD19- TAMs | TTC7A | 0.271204173 |
| CD19- TAMs | RPL9 | 0.306528084 |
| CD19- TAMs | TOP1 | 0.325746346 |
| CD19- TAMs | IL18 | 0.294879603 |
| CD19- TAMs | UBE2L3 | 0.250416798 |
| CD19- TAMs | TBL1X | 0.283375981 |
| CD19- TAMs | SPG11 | 0.271499519 |
| CD19- TAMs | IL1R2 | 0.385799945 |
| CD19- TAMs | LCOR | 0.286747921 |
| CD19- TAMs | NABP1 | 0.468523821 |
| CD19- TAMs | CA2 | 0.300193174 |
| CD19- TAMs | DNAJC8 | 0.250225961 |
| CD19- TAMs | UBE2E1 | 0.270858891 |
| CD19- TAMs | NDUFA12 | 0.270544866 |
| CD19- TAMs | DOCK11 | 0.261672526 |
| CD19- TAMs | ARFGEF1 | 0.264505904 |
| CD19- TAMs | YWHAH | 0.518002513 |
| CD19- TAMs | TLE4 | 0.265840254 |
| CD19- TAMs | HECA | 0.266230143 |
| CD19- TAMs | PCNX1 | 0.323218139 |
| CD19- TAMs | KDM4B | 0.278285746 |
| CD19- TAMs | RGS10 | 0.271512656 |
| CD19- TAMs | FOSB | 0.505990847 |
| CD19- TAMs | AP3B1 | 0.29100408 |
| CD19- TAMs | ATP2B1-AS1 | 0.319887342 |
| CD19- TAMs | EVI5 | 0.274982813 |
| CD19- TAMs | ZNF609 | 0.323903767 |
| CD19- TAMs | OTULINL | 0.251121053 |
| CD19- TAMs | NCOA1 | 0.351325715 |
| CD19- TAMs | HSPA9 | 0.266110973 |
| CD19- TAMs | ATP2C1 | 0.403579713 |
| CD19- TAMs | MED13 | 0.28324618 |
| CD19- TAMs | SLC39A11 | 0.375842412 |
| CD19- TAMs | ZFAND6 | 0.333266706 |
| CD19- TAMs | RPS23 | 0.288527725 |
| CD19- TAMs | ENTPD1 | 0.299468315 |
| CD19- TAMs | INSR | 0.369018308 |
| CD19- TAMs | CLIC1 | 0.250423988 |
| CD19- TAMs | TAB2 | 0.284507475 |
| CD19- TAMs | RPS18 | 0.316259315 |
| CD19- TAMs | KDM6A | 0.29371813 |
| CD19- TAMs | YWHAE | 0.252541396 |
| CD19- TAMs | VAV3 | 0.284644293 |
| CD19- TAMs | SLC25A3 | 0.292818948 |
| CD19- TAMs | ATP6V1G1 | 0.302192248 |
| CD19- TAMs | ELOVL5 | 0.331628518 |
| CD19- TAMs | WDFY3 | 0.303797632 |
| CD19- TAMs | R3HDM2 | 0.268382217 |
| CD19- TAMs | VPS13C | 0.251769532 |
| CD19- TAMs | GBP3 | 0.273679993 |
| CD19- TAMs | CALM1 | 0.342012957 |
| CD19- TAMs | TAGAP | 0.314889283 |
| CD19- TAMs | RPSA | 0.374062757 |
| CD19- TAMs | KIAA1109 | 0.260127081 |
| CD19- TAMs | LINC00243 | 0.278541317 |
| CD19- TAMs | NR3C1 | 0.251422443 |
| CD19- TAMs | LPIN2 | 0.25175383 |
| CD19- TAMs | FOXN3 | 0.441571448 |
| CD19- TAMs | RAP1GDS1 | 0.298586531 |
| CD19- TAMs | RALA | 0.313815548 |
| CD19- TAMs | FAM133B | 0.264807351 |
| CD19- TAMs | RPL19 | 0.258041797 |
| CD19- TAMs | FLT1 | 0.319447734 |
| CD19- TAMs | UBASH3B | 0.26013414 |
| CD19- TAMs | PDIA3 | 0.289274427 |
| CD19- TAMs | ISG15 | 0.499774971 |
| CD19- TAMs | ACSL3 | 0.32503697 |
| CD19- TAMs | VTI1A | 0.312196421 |
| CD19- TAMs | SERTAD2 | 0.2657022 |
| CD19- TAMs | GSK3B | 0.308178682 |
| CD19- TAMs | CHST11 | 0.343314649 |
| CD19- TAMs | WDR70 | 0.271056343 |
| CD19- TAMs | SENP6 | 0.258379116 |
| CD19- TAMs | RPL17 | 0.286530984 |
| CD19- TAMs | PPP4R3A | 0.266776093 |
| CD19- TAMs | SIK3 | 0.514426523 |
| CD19- TAMs | LY6E | 0.25121836 |
| CD19- TAMs | TTC19 | 0.272938541 |
| CD19- TAMs | TFEC | 0.251293256 |
| CD19- TAMs | TOB1 | 0.26370941 |
| CD19- TAMs | RHOB | 0.284228074 |
| CD19- TAMs | FAM168A | 0.269184831 |
| CD19- TAMs | PPP1R15B | 0.272290034 |
| CD19- TAMs | SLC5A3 | 0.768750137 |
| CD19- TAMs | ZFP36L2 | 0.274310797 |
| CD19- TAMs | ANKRD44 | 0.258707478 |
| CD19- TAMs | PLCL2 | 0.287662784 |
| CD19- TAMs | EIF4E | 0.273705362 |
| CD19- TAMs | DYRK1A | 0.29646825 |
| CD19- TAMs | HERC1 | 0.330225391 |
| CD19- TAMs | MYCBP2 | 0.25984441 |
| CD19- TAMs | NCL | 0.252232407 |
| CD19- TAMs | TUBA1A | 0.327903579 |
| CD19- TAMs | FLI1 | 0.254789033 |
| CD19- TAMs | SNHG29 | 0.270693961 |
| CD19- TAMs | EXT1 | 0.342821608 |
| CD19- TAMs | ATP5MC3 | 0.272031779 |
| CD19- TAMs | SUMF1 | 0.295761909 |
| CD19- TAMs | MAP4K3 | 0.313254507 |
| CD19- TAMs | LPP | 0.305971824 |
| CD19- TAMs | TBC1D22A | 0.363602606 |
| CD19- TAMs | COPA | 0.270705165 |
| CD19- TAMs | JUND | 0.377186154 |
| CD19- TAMs | PIK3R5 | 0.250311159 |
| CD19- TAMs | FBXO11 | 0.280832867 |
| CD19- TAMs | CAMK1D | 0.390746575 |
| CD19- TAMs | GPRIN3 | 0.28166197 |
| CD19- TAMs | TNFAIP8 | 0.299499468 |
| CD19- TAMs | LRRC23 | 0.659542435 |
| CD19- TAMs | EEF1G | 0.294559917 |
| CD19- TAMs | FOS | 0.305854951 |
| CD19- TAMs | DDHD1 | 0.284995981 |
| CD19- TAMs | USPL1 | 0.351540073 |
| CD19- TAMs | RAN | 0.266295312 |
| CD19- TAMs | RPL13A | 0.252098731 |
| CD19- TAMs | CDKN1A | 0.305206912 |
| CD19- TAMs | MLKL | 0.340494994 |
| CD19- TAMs | BIRC6 | 0.29367826 |
| CD19- TAMs | NR4A2 | 0.329308544 |
| CD19- TAMs | PPM1L | 0.314687987 |
| CD19- TAMs | CHMP1B | 0.276260107 |
| CD19- TAMs | PAN3 | 0.301661021 |
| CD19- TAMs | GPR183 | 0.502768071 |
| CD19- TAMs | GSN | 0.307016553 |
| CD19- TAMs | S100B | 0.273125822 |
| CD19- TAMs | ARIH1 | 0.281547993 |
| CD19- TAMs | NCOA2 | 0.299731335 |
| CD19- TAMs | TUBB | 0.255922341 |
| CD19- TAMs | DNAJB4 | 0.390835778 |
| CD19- TAMs | FBXL14 | 0.355406693 |
| CD19- TAMs | H3F3B | 0.257316218 |
| CD19- TAMs | MRPS6 | 0.280920356 |

| Cell type | Gene | Value |
|---|---|---|
| CD19- TAMs | SSBP2 | 0.405519465 |
| CD19- TAMs | BIRC3 | 0.563287058 |
| CD19- TAMs | ANKRD17 | 0.268299881 |
| CD19- TAMs | CYLD | 0.258147354 |
| CD19- TAMs | ARL15 | 0.287736895 |
| CD19- TAMs | PAPSS2 | 0.372422982 |
| CD19- TAMs | AC025164.1 | 0.315453188 |
| CD19- TAMs | RUNX1 | 0.357661826 |
| CD19- TAMs | CPEB2 | 0.286432108 |
| CD19- TAMs | HDAC9 | 0.526147964 |
| CD19- TAMs | VPS13B | 0.295759219 |
| CD19- TAMs | ABHD3 | 0.275318506 |
| CD19- TAMs | FBXO34 | 0.293077266 |
| CD19- TAMs | INTS6 | 0.323638505 |
| CD19- TAMs | MEF2A | 0.481725451 |
| CD19- TAMs | PHF20 | 0.270269648 |
| CD19- TAMs | MAP4K4 | 0.272395085 |
| CD19- TAMs | ATP8B4 | 0.286838813 |
| CD19- TAMs | EML4 | 0.250041445 |
| CD19- TAMs | DLEU1 | 0.340026928 |
| CD19- TAMs | CKS2 | 0.289726529 |
| CD19- TAMs | EIF4A3 | 0.254937567 |
| CD19- TAMs | CXCL9 | 0.375363255 |
| CD19- TAMs | INSIG1 | 0.272401522 |
| CD19- TAMs | ALOX5AP | 0.475470011 |
| CD19- TAMs | MAML2 | 0.327287132 |
| CD19- TAMs | CXCL10 | 0.455922955 |
| CD19- TAMs | BTBD9 | 0.260978021 |
| CD19- TAMs | TNF | 0.356402433 |
| CD19- TAMs | CRADD | 0.328739298 |
| CD19- TAMs | IL10 | 0.284406858 |
| CD19- TAMs | FCGR2B | 0.259824914 |
| B cells | IGHG3 | 5.864420508 |
| B cells | IGHG1 | 5.548688779 |
| B cells | IGLC2 | 4.394092929 |
| B cells | IGKC | 4.348637947 |
| B cells | JCHAIN | 3.980681405 |
| B cells | IGHA1 | 3.661309444 |
| B cells | IGHM | 3.296325392 |
| B cells | IGLC1 | 3.211330562 |
| B cells | IGKV1-5 | 3.119942249 |

| Cell type | Gene | Value |
|---|---|---|
| B cells | IGHG4 | 3.115409814 |
| B cells | MZB1 | 3.052821042 |
| B cells | IGHV3-23 | 2.99101681 |
| B cells | ANKRD28 | 2.624137194 |
| B cells | IFNG-AS1 | 2.508638845 |
| B cells | IGKV3-11 | 2.502220338 |
| B cells | FKBP11 | 2.442082241 |
| B cells | IGHG2 | 2.179163164 |
| B cells | TXNDC5 | 2.160220924 |
| B cells | SEC11C | 2.045149811 |
| B cells | CD79A | 1.992744763 |
| B cells | BANK1 | 1.927636179 |
| B cells | IGHV1-24 | 1.88675895 |
| B cells | RALGPS2 | 1.880410439 |
| B cells | DERL3 | 1.81572439 |
| B cells | XBP1 | 1.785942815 |
| B cells | TENT5C | 1.760493999 |
| B cells | PRDX4 | 1.629772788 |
| B cells | ITM2C | 1.627508491 |
| B cells | ISG20 | 1.560012468 |
| B cells | TPD52 | 1.554588814 |
| B cells | RHOH | 1.502526591 |
| B cells | FUT8 | 1.500677675 |
| B cells | CDK14 | 1.475387451 |
| B cells | PDE4D | 1.462643597 |
| B cells | COBLL1 | 1.396933767 |
| B cells | POU2AF1 | 1.390483027 |
| B cells | ST6GAL1 | 1.382446383 |
| B cells | SOX5 | 1.347594759 |
| B cells | BACH2 | 1.343561641 |
| B cells | AC008014.1 | 1.342839521 |
| B cells | MYO1D | 1.314309442 |
| B cells | CD69 | 1.303452708 |
| B cells | CEP128 | 1.282464483 |
| B cells | MS4A1 | 1.281096234 |
| B cells | SPCS2 | 1.265116587 |
| B cells | TMEM156 | 1.262287477 |
| B cells | CREB3L2 | 1.23522551 |
| B cells | LARGE1 | 1.233189104 |
| B cells | CD27 | 1.201682082 |
| B cells | MSI2 | 1.197588406 |
| B cells | LTB | 1.177969016 |
| B cells | DENND5B | 1.17675929 |
| B cells | RAB30 | 1.175820057 |
| B cells | OSBPL10 | 1.169235833 |
| B cells | HIPK2 | 1.154195476 |
| B cells | FCRL5 | 1.149919528 |
| B cells | SEL1L3 | 1.144578227 |
| B cells | LINC02362 | 1.143414842 |
| B cells | PLCG2 | 1.132062498 |
| B cells | FBXW7 | 1.127553207 |
| B cells | NDUFAF6 | 1.094255585 |
| B cells | EBF1 | 1.08761323 |
| B cells | IKZF3 | 1.069731437 |
| B cells | CADPS2 | 1.054974114 |
| B cells | TP53INP1 | 1.051806057 |
| B cells | TNFRSF17 | 1.017929704 |
| B cells | RABGAP1L | 1.007743383 |
| B cells | CARMIL1 | 1.004392123 |
| B cells | NCOA3 | 1.003118192 |
| B cells | AFF3 | 1.001042507 |
| B cells | NUCB2 | 1.000448322 |
| B cells | HSH2D | 0.977898741 |
| B cells | LIME1 | 0.969308648 |
| B cells | PDK1 | 0.966057209 |
| B cells | GLCCI1 | 0.965741411 |
| B cells | CD79B | 0.963286784 |
| B cells | EIF2AK3 | 0.956823419 |
| B cells | STK17A | 0.950092066 |
| B cells | RHEX | 0.945186669 |
| B cells | SLC38A1 | 0.939935425 |
| B cells | ERLEC1 | 0.913550736 |
| B cells | SYNE2 | 0.908391838 |
| B cells | TNFRSF13C | 0.907671987 |
| B cells | LMAN1 | 0.902949015 |
| B cells | GAB1 | 0.902262256 |
| B cells | FNDC3A | 0.902116827 |
| B cells | STRBP | 0.899843635 |
| B cells | GNG7 | 0.898054255 |
| B cells | PIM2 | 0.893010069 |
| B cells | BCL2 | 0.892432665 |
| B cells | ZEB1 | 0.891399537 |

| Cell type | Gene | Value |
|---|---|---|
| B cells | CYTIP | 0.88146198 |
| B cells | PRDM2 | 0.880858033 |
| B cells | AP001011.1 | 0.865882745 |
| B cells | JSRP1 | 0.865567564 |
| B cells | DIPK1A | 0.848478581 |
| B cells | LINC02384 | 0.847403921 |
| B cells | SPATS2 | 0.835250111 |
| B cells | SP140 | 0.832785467 |
| B cells | BTD | 0.828020756 |
| B cells | ARID5B | 0.826643241 |
| B cells | FAM214A | 0.825261945 |
| B cells | AC120193.1 | 0.824560622 |
| B cells | C11orf80 | 0.816397121 |
| B cells | AC007569.1 | 0.807526255 |
| B cells | RASGRP3 | 0.80593982 |
| B cells | PLPP5 | 0.804699947 |
| B cells | UBE2J1 | 0.804521065 |
| B cells | AC078883.1 | 0.800956529 |
| B cells | SELENOM | 0.792819766 |
| B cells | DNAJC1 | 0.788950304 |
| B cells | FAM30A | 0.788445449 |
| B cells | ADAM28 | 0.783769621 |
| B cells | PVT1 | 0.781659439 |
| B cells | PPP1R16B | 0.780742051 |
| B cells | AC023424.3 | 0.773393097 |
| B cells | ZBTB20 | 0.77238884 |
| B cells | SSPN | 0.77158193 |
| B cells | SPAG4 | 0.768118854 |
| B cells | AL589693.1 | 0.757639645 |
| B cells | OSBPL3 | 0.756714794 |
| B cells | ADAM19 | 0.752683273 |
| B cells | CCDC88C | 0.751786787 |
| B cells | ICAM3 | 0.748883624 |
| B cells | RRAS2 | 0.747304343 |
| B cells | GMDS | 0.744676569 |
| B cells | STAP1 | 0.739993885 |
| B cells | FAAH2 | 0.738922194 |
| B cells | IRF4 | 0.734816096 |
| B cells | PIP5K1B | 0.728113753 |
| B cells | SETBP1 | 0.726161644 |
| B cells | SMCHD1 | 0.724435255 |
| B cells | PLEKHG1 | 0.719937116 |
| B cells | UBE2H | 0.716270132 |
| B cells | TLE1 | 0.709275705 |
| B cells | MEI1 | 0.703814761 |
| B cells | FAM107B | 0.700340851 |
| B cells | BICD1 | 0.695547627 |
| B cells | FCRL1 | 0.693176255 |
| B cells | LY9 | 0.683926753 |
| B cells | ZCCHC7 | 0.67029537 |
| B cells | BLK | 0.669476068 |
| B cells | PPP3CC | 0.667475461 |
| B cells | DENND4A | 0.666193593 |
| B cells | HIST1H1D | 0.665493415 |
| B cells | DNAJB9 | 0.663439154 |
| B cells | PDE7A | 0.663335248 |
| B cells | SNX25 | 0.657565313 |
| B cells | WWOX | 0.65109014 |
| B cells | SDC1 | 0.645218073 |
| B cells | AL050309.1 | 0.644951525 |
| B cells | PAX5 | 0.644877019 |
| B cells | MEF2C-AS1 | 0.634726651 |
| B cells | LINC00926 | 0.624897733 |
| B cells | ITGA8 | 0.624749627 |
| B cells | PALM2-AKAP2 | 0.622065479 |
| B cells | AC007384.1 | 0.60578921 |
| B cells | RHBDD1 | 0.604874683 |
| B cells | LRBA | 0.603074241 |
| B cells | MBNL2 | 0.603062979 |
| B cells | CRELD2 | 0.601982515 |
| B cells | KLF13 | 0.591789883 |
| B cells | SKAP1 | 0.585888364 |
| B cells | IGLV3-1 | 0.585187649 |
| B cells | FCHSD2 | 0.585093582 |
| B cells | SEC24A | 0.584467826 |
| B cells | TUBA4A | 0.583214259 |
| B cells | RASSF6 | 0.577984152 |
| B cells | SMYD3 | 0.573063689 |
| B cells | TMEM243 | 0.569002109 |
| B cells | ZBP1 | 0.567608537 |
| B cells | TOX | 0.565258167 |
| B cells | RAB11A | 0.564861881 |
| B cells | HIVEP1 | 0.564028935 |
| B cells | LPIN1 | 0.558088709 |
| B cells | NEDD9 | 0.55807375 |
| B cells | MCTP2 | 0.55558914 |
| B cells | DANCR | 0.554307278 |
| B cells | GMDS-DT | 0.548826322 |
| B cells | TAF4B | 0.545912559 |
| B cells | ZHX2 | 0.544869131 |
| B cells | BLNK | 0.544355555 |
| B cells | SLAMF7 | 0.544314627 |
| B cells | ACOXL | 0.538400814 |
| B cells | TRAM2 | 0.534804002 |
| B cells | TNFRSF13B | 0.531939187 |
| B cells | AL591518.1 | 0.528662736 |
| B cells | LINC01934 | 0.526552252 |
| B cells | CD70 | 0.525606527 |
| B cells | ATP8A1 | 0.52422888 |
| B cells | TSHZ2 | 0.524138864 |
| B cells | ANGPTL1 | 0.523126582 |
| B cells | TMC3-AS1 | 0.521153051 |
| B cells | CHODL | 0.518282325 |
| B cells | CAMKMT | 0.516524716 |
| B cells | P2RY10 | 0.510108144 |
| B cells | ANKRD44 | 0.509705299 |
| B cells | KCNQ5 | 0.509672477 |
| B cells | KCNN3 | 0.503919643 |
| B cells | KRT10 | 0.501708485 |
| B cells | GPHN | 0.501048706 |
| B cells | LAX1 | 0.5008046 |
| B cells | PARP15 | 0.500723132 |
| B cells | IQCB1 | 0.499527007 |
| B cells | RORA | 0.497963645 |
| B cells | TUT4 | 0.497822054 |
| B cells | CARD11 | 0.497816023 |
| B cells | EPB41 | 0.490786324 |
| B cells | FAM3C | 0.488905093 |
| B cells | AP002075.1 | 0.484952708 |
| B cells | COL19A1 | 0.48346739 |
| B cells | MANEA | 0.481557583 |
| B cells | SIDT1 | 0.478858186 |
| B cells | SVIP | 0.478125898 |
| B cells | TNRC6B | 0.477514303 |
| B cells | IGHD | 0.470628754 |
| B cells | HMCES | 0.470122467 |
| B cells | AC009522.1 | 0.469911992 |
| B cells | STK4 | 0.469610004 |
| B cells | AC079793.1 | 0.464910917 |
| B cells | TVP23C | 0.46459306 |
| B cells | CPNE5 | 0.458928672 |
| B cells | MICAL3 | 0.456275802 |
| B cells | VPREB3 | 0.456140842 |
| B cells | ETS1 | 0.4556552 |
| B cells | ESR2 | 0.455414083 |
| B cells | ITGA6 | 0.454398646 |
| B cells | PNOC | 0.453699526 |
| B cells | PDLIM1 | 0.451762842 |
| B cells | FCMR | 0.448238415 |
| B cells | HIST1H2BD | 0.447815019 |
| B cells | ODC1 | 0.446659162 |
| B cells | BMP6 | 0.444418189 |
| B cells | CDKAL1 | 0.443400854 |
| B cells | COL4A4 | 0.441673301 |
| B cells | FCRL2 | 0.441525287 |
| B cells | IFT57 | 0.439286359 |
| B cells | GBF1 | 0.436226197 |
| B cells | MIATNB | 0.43440699 |
| B cells | CD24 | 0.426274511 |
| B cells | TLK1 | 0.425102221 |
| B cells | SYNGR1 | 0.420710091 |
| B cells | JADE3 | 0.420127876 |
| B cells | CDC42SE2 | 0.41988387 |
| B cells | AC012236.1 | 0.41977732 |
| B cells | RIPOR2 | 0.418284681 |
| B cells | PLP2 | 0.417556894 |
| B cells | SIPA1L3 | 0.417041306 |
| B cells | KLF12 | 0.416379481 |
| B cells | PPM1K | 0.412966923 |
| B cells | P2RX5 | 0.412191398 |
| B cells | NUGGC | 0.411122181 |
| B cells | AC016074.2 | 0.410781345 |
| B cells | AC022182.1 | 0.408433774 |
| B cells | ATP2A3 | 0.405040094 |
| B cells | TMEM238 | 0.403842478 |
| B cells | P2RY8 | 0.403739641 |
| B cells | LINC02397 | 0.402047996 |
| B cells | ORAI2 | 0.401630707 |
| B cells | SRPRB | 0.400405586 |
| B cells | KHDRBS2 | 0.397200213 |
| B cells | TMEM263 | 0.39707678 |
| B cells | LINC01004 | 0.396301136 |
| B cells | XKR6 | 0.395020508 |
| B cells | TMEM154 | 0.394017987 |
| B cells | ASXL1 | 0.393123976 |
| B cells | SCMH1 | 0.393091065 |
| B cells | TSHR | 0.392368881 |
| B cells | ANXA6 | 0.386659364 |
| B cells | C16orf74 | 0.385585548 |
| B cells | EPB41L4A | 0.384515453 |
| B cells | CYFIP2 | 0.383919734 |
| B cells | SINHCAF | 0.38317008 |
| B cells | ANK3 | 0.382199388 |
| B cells | GRK5 | 0.381844915 |
| B cells | OCIAD2 | 0.38080901 |
| B cells | NIBAN3 | 0.376709936 |
| B cells | EIF1AY | 0.374417918 |
| B cells | CCDC32 | 0.374238581 |
| B cells | PKIG | 0.371735617 |
| B cells | MACROD2 | 0.370543467 |
| B cells | TEX9 | 0.367711941 |
| B cells | E2F5 | 0.367400677 |
| B cells | PBX4 | 0.361946138 |
| B cells | TSBP1-AS1 | 0.359879261 |
| B cells | LINC02576 | 0.358095558 |
| B cells | ABCA5 | 0.357911642 |
| B cells | PMM2 | 0.354203727 |
| B cells | CAMK4 | 0.353761769 |
| B cells | P2RX1 | 0.353507855 |
| B cells | SCNN1B | 0.351421637 |
| B cells | LINC01473 | 0.350437842 |
| B cells | PARM1 | 0.348518833 |
| B cells | ST6GALNAC4 | 0.348513432 |
| B cells | AC021678.2 | 0.347488861 |
| B cells | OPTN | 0.345218172 |
| B cells | TSTD1 | 0.344188682 |
| B cells | ARHGAP15 | 0.343742037 |
| B cells | CAV1 | 0.341494856 |
| B cells | TMEM117 | 0.34072029 |
| B cells | HIST1H1E | 0.339754859 |
| B cells | AL355076.2 | 0.339349523 |
| B cells | CHPF | 0.33831405 |
| B cells | ZNF107 | 0.33763595 |
| B cells | LINC02541 | 0.334717823 |
| B cells | HLA-DOB | 0.334160058 |
| B cells | DPEP1 | 0.333637931 |
| B cells | ANKRD36B | 0.332399552 |
| B cells | TARSL2 | 0.332367302 |
| B cells | SLC25A4 | 0.33223125 |
| B cells | NXPE3 | 0.331676084 |
| B cells | RASGRP1 | 0.331426314 |
| B cells | RIC3 | 0.330337743 |
| B cells | KIAA0040 | 0.329693125 |
| B cells | SPAG1 | 0.329556291 |
| B cells | FAM117A | 0.328684439 |
| B cells | ZNF827 | 0.328132776 |
| B cells | ITGB7 | 0.32694634 |
| B cells | SYTL1 | 0.326771623 |
| B cells | BFSP2 | 0.326234394 |
| B cells | LINC00571 | 0.325501262 |
| B cells | CCDC69 | 0.324396713 |
| B cells | VCPKMT | 0.322910146 |
| B cells | RORA-AS1 | 0.322236362 |
| B cells | ACAP1 | 0.318247535 |
| B cells | AC079781.5 | 0.31522726 |
| B cells | RAPGEF5 | 0.313398857 |
| B cells | MAP3K9 | 0.312497802 |
| B cells | TRAF3 | 0.312299967 |
| B cells | SLAMF6 | 0.310231929 |
| B cells | ZFAT | 0.308704179 |
| B cells | C11orf24 | 0.305875658 |
| B cells | 8-Mar | 0.302889612 |
| B cells | C12orf74 | 0.302130827 |
| B cells | PAWR | 0.301378174 |
| B cells | AFF2 | 0.299739919 |
| B cells | IL21R | 0.298444252 |
| B cells | LINC01484 | 0.298429119 |
| B cells | MCEE | 0.297678054 |
| B cells | LBH | 0.296088016 |
| B cells | ZNF215 | 0.295429557 |
| B cells | DNAH8 | 0.294023684 |
| B cells | DENND2C | 0.289853208 |
| B cells | MPP6 | 0.288380089 |
| B cells | BCAS4 | 0.288354796 |
| B cells | PPP1R9A | 0.280881549 |
| B cells | PATJ | 0.2765179 |
| B cells | ITM2A | 0.276248532 |
| B cells | PKHD1L1 | 0.275877285 |
| B cells | COL4A3 | 0.272789848 |
| B cells | YES1 | 0.269875317 |
| B cells | C12orf65 | 0.268364033 |
| B cells | AQP3 | 0.267989971 |
| B cells | SYVN1 | 0.267628624 |
| B cells | FAM102A | 0.266158624 |
| B cells | KNTC1 | 0.265242895 |
| B cells | AIM2 | 0.265174664 |
| B cells | CLNK | 0.264971065 |
| B cells | GNB5 | 0.263690603 |
| B cells | KIZ | 0.263600926 |
| B cells | ALDH18A1 | 0.260778996 |
| B cells | AC243960.1 | 0.258577194 |
| B cells | FCRLA | 0.257912072 |
| B cells | SLAMF1 | 0.251769228 |
| B cells | LINC01781 | 0.251713677 |
| B cells | CFAP54 | 0.250946806 |
| B cells | GRAMD1C | 0.250840334 |
| B cells | PTPN4 | 0.273702604 |
| B cells | PCED1B | 0.310286874 |
| B cells | USP48 | 0.535177292 |
| B cells | HSPA13 | 0.320611012 |
| B cells | TMEM131 | 0.374400964 |
| B cells | PTPN1 | 0.260064369 |
| B cells | TXNDC15 | 0.478587923 |
| B cells | EHMT1 | 0.680361503 |
| B cells | TXNDC11 | 0.621183636 |
| B cells | EAF2 | 0.505337472 |
| B cells | AL162253.2 | 0.345993067 |

| Cell type | Gene | Value |
|---|---|---|
| B cells | IGHV3-66 | 1.646626967 |
| B cells | MIR4435-2HG | 0.605822204 |
| B cells | AHI1 | 0.279126907 |
| B cells | STEAP1B | 0.377844451 |
| B cells | ZBTB38 | 0.642553297 |
| B cells | U62317.4 | 0.33286814 |
| B cells | XRRA1 | 0.26031321 |
| B cells | TMEM131L | 0.261028051 |
| B cells | AC239799.2 | 0.391500984 |
| B cells | SGO1-AS1 | 0.253937144 |
| B cells | PGM3 | 0.258626501 |
| B cells | STAG1 | 0.374404167 |
| B cells | PCED1B-AS1 | 0.282580435 |
| B cells | LIN52 | 0.264424938 |
| B cells | FBH1 | 0.330168884 |
| B cells | UMAD1 | 0.345367512 |
| B cells | AC092546.1 | 0.27772988 |
| B cells | CD38 | 0.631640487 |
| B cells | NSD3 | 0.381188454 |
| B cells | AL592429.2 | 0.32585317 |
| B cells | ERC1 | 0.428206094 |
| B cells | LRRK1 | 0.445259586 |
| B cells | SEC24D | 0.548945958 |
| B cells | ELL2 | 0.345713289 |
| B cells | HIST1H2AC | 0.557568337 |
| B cells | CLPTM1L | 0.418627539 |
| B cells | BET1 | 0.252503218 |
| B cells | SNRPN | 0.292871002 |
| B cells | SREBF2 | 0.299731345 |
| B cells | ATF7IP | 0.474538457 |
| B cells | DAPP1 | 0.337263743 |
| B cells | HIST1H1C | 0.599361052 |
| B cells | SCAPER | 0.377980792 |
| B cells | CYTOR | 1.095413166 |
| B cells | SEL1L | 0.758977979 |
| B cells | ZNF791 | 0.276967402 |
| B cells | LARP1B | 0.334805896 |
| B cells | GORASP2 | 0.302535624 |
| B cells | IL2RG | 0.269753014 |
| B cells | TIFA | 0.26349383 |
| B cells | CHD7 | 0.28021416 |
| B cells | ESR1 | 0.5001838 |
| B cells | ATP11B | 0.288456152 |
| B cells | RFX3 | 0.313149221 |
| B cells | HIST1H4C | 0.302375723 |
| B cells | RPS6KA5 | 0.32156828 |
| B cells | RPL3 | 0.523271676 |
| B cells | SELENOK | 0.624164758 |
| B cells | XRN1 | 0.441925169 |
| B cells | UBE2G1 | 0.304133612 |
| B cells | RNF19A | 0.364238748 |
| B cells | VPS37B | 0.392665553 |
| B cells | SMAP2 | 0.373062084 |
| B cells | HIST1H2BJ | 0.273397603 |
| B cells | NFX1 | 0.284270145 |
| B cells | AUTS2 | 0.434451529 |
| B cells | PRDX2 | 0.381234126 |
| B cells | LINC00910 | 0.468807541 |
| B cells | DUSP5 | 0.377144312 |
| B cells | RPS5 | 0.401133383 |
| B cells | OGT | 0.378654632 |
| B cells | UAP1 | 0.251034342 |
| B cells | VWA8 | 0.268681359 |
| B cells | SPCS1 | 1.079745346 |
| B cells | MRPS31 | 0.321760497 |
| B cells | EZR | 0.274632611 |
| B cells | SRPRA | 0.331336063 |
| B cells | IMMP2L | 0.33392401 |
| B cells | AC012447.1 | 0.382495437 |
| B cells | ALG5 | 0.375495902 |
| B cells | UBA6-AS1 | 0.267621338 |
| B cells | SCFD1 | 0.39115707 |
| B cells | BCL11A | 0.253262452 |
| B cells | SFMBT1 | 0.267733202 |
| B cells | RNGTT | 0.264815209 |
| B cells | Z93241.1 | 0.275890451 |
| B cells | SDF2L1 | 0.734842973 |
| B cells | MAN1A2 | 0.351507386 |
| B cells | MTMR12 | 0.263444361 |
| B cells | C12orf57 | 0.275346793 |
| B cells | ANKRD12 | 0.321100848 |
| B cells | GABPB1 | 0.306584047 |
| B cells | ASCC3 | 0.289159348 |
| B cells | RASA2 | 0.298822562 |
| B cells | MANF | 0.707982299 |
| B cells | BRAF | 0.254908714 |
| B cells | RIC1 | 0.280010601 |
| B cells | SLC44A1 | 0.478629376 |
| B cells | VOPP1 | 0.452140278 |
| B cells | RBM6 | 0.339752188 |
| B cells | BCAR3 | 0.273466978 |
| B cells | CCNC | 0.250815921 |
| B cells | HIST1H2BG | 0.293489885 |
| B cells | TTC28 | 0.310012999 |
| B cells | CD37 | 0.545181626 |
| B cells | AC016831.7 | 0.371825947 |
| B cells | SEC62 | 0.541828531 |
| B cells | SRP54 | 0.311364093 |
| B cells | STIM2 | 0.299686412 |
| B cells | MAP4K4 | 0.256883258 |
| B cells | PRKD3 | 0.255809117 |
| B cells | ICAM2 | 0.256568073 |
| B cells | ERN1 | 0.539194417 |
| B cells | SMDT1 | 0.332375627 |
| B cells | LINC00513 | 0.296130788 |
| B cells | EEF1B2 | 0.396695468 |
| B cells | PDCD4 | 0.287031427 |
| B cells | CCPG1 | 0.394580364 |
| B cells | CADM1 | 0.561402543 |
| B cells | ARID2 | 0.310672938 |
| B cells | 6-Mar | 0.259550164 |
| B cells | CASK | 0.256751215 |
| B cells | EXOC4 | 0.26787937 |
| B cells | CCDC50 | 0.294648635 |
| B cells | ERGIC2 | 0.287630973 |
| B cells | OSBPL9 | 0.330615108 |
| B cells | ANAPC16 | 0.25544124 |
| B cells | CLEC2D | 0.288723842 |
| B cells | SELENOS | 0.679162471 |
| B cells | TCEA1 | 0.279005772 |
| B cells | ADK | 0.280341648 |
| B cells | SUB1 | 0.74161544 |
| B cells | RABAC1 | 0.648416575 |
| B cells | SND1 | 0.391757939 |
| B cells | RPL22L1 | 0.274359154 |
| B cells | DERL1 | 0.297981735 |
| B cells | MON2 | 0.258781365 |
| B cells | RUNX2 | 0.400353108 |
| B cells | EIF2AK4 | 0.399391519 |
| B cells | SP100 | 0.254428742 |
| B cells | CCND3 | 0.275391076 |
| B cells | R3HDM1 | 0.29038092 |
| B cells | RPS4X | 0.322210081 |
| B cells | BTG2 | 0.707784591 |
| B cells | SRM | 0.257590924 |
| B cells | SSR4 | 2.564688964 |
| B cells | CLIC4 | 0.318468964 |
| B cells | TSC22D3 | 0.59434998 |
| B cells | BCL2L11 | 0.387572635 |
| B cells | PELI1 | 0.616937665 |
| B cells | GALNT2 | 0.340710138 |
| B cells | TEX14 | 1.108163698 |
| B cells | DENND1B | 0.33446084 |
| B cells | MEF2C | 0.262485891 |
| B cells | CASP3 | 0.271522464 |
| B cells | TMEM140 | 0.263700312 |
| B cells | ATXN1 | 0.490116508 |
| B cells | NSMCE2 | 0.444950899 |
| B cells | SPCS3 | 0.685545432 |
| B cells | IGHV3-53 | 2.648495087 |
| B cells | ARF4 | 0.297404136 |
| B cells | ANKRD37 | 0.318687035 |
| B cells | HSP90B1 | 1.293182021 |
| B cells | SEC61B | 0.768729097 |
| B cells | HERPUD1 | 0.688951947 |
| B cells | YPEL5 | 0.259356891 |
| B cells | ARHGAP24 | 0.362607674 |
| B cells | CUTA | 0.323190493 |
| B cells | MRPS24 | 0.303587404 |
| B cells | SSR2 | 0.364395483 |
| B cells | RPLP0 | 0.251199004 |
| B cells | FNDC3B | 0.948392126 |
| B cells | HDLBP | 0.422308802 |
| B cells | KDELR1 | 0.43600709 |

| Cell type | Gene | Value |
|---|---|---|
| B cells | TMEM107 | 0.268106924 |
| B cells | PRDM1 | 0.3287783 |
| B cells | SERP1 | 0.365162211 |
| B cells | RPL36AL | 0.33582272 |
| B cells | CCSER1 | 0.252744915 |
| B cells | IGHA2 | 3.302809124 |
| B cells | CEMIP2 | 0.368056069 |
| B cells | H1FX | 0.93680862 |
| B cells | IGLC3 | 4.325137273 |
| B cells | TMEM59 | 0.317812084 |
| B cells | SSR3 | 0.570497307 |
| B cells | PPIB | 0.375597437 |
| B cells | RRBP1 | 0.408682184 |
| B cells | MYDGF | 0.349770271 |
| B cells | MDM2 | 0.42105483 |
| CD19+ TAMs | RNASE1 | 3.290767755 |
| CD19+ TAMs | SELENOP | 3.014380344 |
| CD19+ TAMs | APOE | 2.752114467 |
| CD19+ TAMs | SLC40A1 | 2.746750281 |
| CD19+ TAMs | APOA2 | 2.538999278 |
| CD19+ TAMs | LGMN | 2.534461043 |
| CD19+ TAMs | FOLR2 | 2.293578007 |
| CD19+ TAMs | APOC1 | 2.133389332 |
| CD19+ TAMs | CTSD | 2.09878348 |
| CD19+ TAMs | MAF | 1.938261725 |
| CD19+ TAMs | DAB2 | 1.916545241 |
| CD19+ TAMs | C1QB | 1.905356348 |
| CD19+ TAMs | A2M | 1.890830463 |
| CD19+ TAMs | C1QA | 1.877578695 |
| CD19+ TAMs | GPNMB | 1.875451582 |
| CD19+ TAMs | SPP1 | 1.824372018 |
| CD19+ TAMs | C1QC | 1.788469074 |
| CD19+ TAMs | PLD3 | 1.752274902 |
| CD19+ TAMs | FCGRT | 1.660393642 |
| CD19+ TAMs | LIPA | 1.649375503 |
| CD19+ TAMs | CTSB | 1.64625461 |
| CD19+ TAMs | MS4A7 | 1.643340472 |
| CD19+ TAMs | MS4A4A | 1.617640642 |
| CD19+ TAMs | MRC1 | 1.431393169 |
| CD19+ TAMs | GYPC | 1.429518996 |
| CD19+ TAMs | STAB1 | 1.414347414 |
| CD19+ TAMs | VCAM1 | 1.412999712 |
| CD19+ TAMs | NRP1 | 1.371969915 |
| CD19+ TAMs | NUPR1 | 1.362024623 |
| CD19+ TAMs | LILRB5 | 1.345166122 |
| CD19+ TAMs | CTSL | 1.326783229 |
| CD19+ TAMs | MSR1 | 1.310124562 |
| CD19+ TAMs | CCL4L2 | 1.306589432 |
| CD19+ TAMs | CD163 | 1.302638802 |
| CD19+ TAMs | APOC2 | 1.28733032 |
| CD19+ TAMs | ARL4C | 1.285350792 |
| CD19+ TAMs | CD14 | 1.275070485 |
| CD19+ TAMs | IGF1 | 1.271125713 |
| CD19+ TAMs | CD63 | 1.267297754 |
| CD19+ TAMs | GPR34 | 1.264436758 |
| CD19+ TAMs | SLCO2B1 | 1.257764854 |
| CD19+ TAMs | CD68 | 1.256820771 |
| CD19+ TAMs | FTL | 1.249004126 |
| CD19+ TAMs | TREM2 | 1.223720602 |
| CD19+ TAMs | ALB | 1.189504873 |
| CD19+ TAMs | CREG1 | 1.167159838 |
| CD19+ TAMs | GRN | 1.153102121 |
| CD19+ TAMs | KCNMA1 | 1.152699997 |
| CD19+ TAMs | WWP1 | 1.148290348 |
| CD19+ TAMs | BLVRB | 1.14794117 |
| CD19+ TAMs | CD59 | 1.124151861 |
| CD19+ TAMs | CTSZ | 1.095230377 |
| CD19+ TAMs | CD209 | 1.088308627 |
| CD19+ TAMs | FABP5 | 1.070719295 |
| CD19+ TAMs | PLTP | 1.061173978 |
| CD19+ TAMs | HLA-DRB5 | 1.058467687 |
| CD19+ TAMs | ITM2B | 1.057971675 |
| CD19+ TAMs | CXCL12 | 1.052072944 |
| CD19+ TAMs | TIMP2 | 1.048022427 |
| CD19+ TAMs | ADAP2 | 1.045768663 |
| CD19+ TAMs | ITSN1 | 1.024221283 |
| CD19+ TAMs | CFD | 1.018891411 |
| CD19+ TAMs | APOC3 | 1.016629748 |
| CD19+ TAMs | MT1G | 1.014037822 |
| CD19+ TAMs | CD81 | 1.013603252 |
| CD19+ TAMs | C2 | 1.011732212 |
| CD19+ TAMs | ASAH1 | 1.002145925 |
| CD19+ TAMs | CD163L1 | 0.999159851 |
| CD19+ TAMs | FCHO2 | 0.99185275 |
| CD19+ TAMs | FCGR2A | 0.990384377 |
| CD19+ TAMs | PSAP | 0.981315709 |
| CD19+ TAMs | FUCA1 | 0.980722253 |
| CD19+ TAMs | FRMD4A | 0.977086425 |
| CD19+ TAMs | TYROBP | 0.976660709 |
| CD19+ TAMs | SLC1A3 | 0.97637299 |
| CD19+ TAMs | MARCKS | 0.976051117 |
| CD19+ TAMs | CD74 | 0.975351682 |
| CD19+ TAMs | MS4A6A | 0.974098367 |
| CD19+ TAMs | CTSC | 0.969983964 |
| CD19+ TAMs | C3AR1 | 0.966034752 |
| CD19+ TAMs | FCGR3A | 0.962631441 |
| CD19+ TAMs | TSPAN4 | 0.960906511 |
| CD19+ TAMs | SDCBP | 0.959644238 |
| CD19+ TAMs | TMEM176B | 0.952373202 |
| CD19+ TAMs | RAP2B | 0.948169295 |
| CD19+ TAMs | MCOLN1 | 0.93687894 |
| CD19+ TAMs | HLA-DPA1 | 0.936304497 |
| CD19+ TAMs | MAFB | 0.931052559 |
| CD19+ TAMs | FMN1 | 0.929970554 |
| CD19+ TAMs | LGALS3 | 0.92952993 |
| CD19+ TAMs | HLA-DRB1 | 0.927925977 |
| CD19+ TAMs | CXCL16 | 0.924595278 |
| CD19+ TAMs | CCL4 | 0.921221594 |
| CD19+ TAMs | NRP2 | 0.905230308 |
| CD19+ TAMs | MPEG1 | 0.903183663 |
| CD19+ TAMs | FRMD4B | 0.900965645 |
| CD19+ TAMs | SDC3 | 0.891648592 |
| CD19+ TAMs | HLA-DQA1 | 0.890821314 |
| CD19+ TAMs | MERTK | 0.882626922 |
| CD19+ TAMs | NPL | 0.867173587 |
| CD19+ TAMs | CD9 | 0.866771193 |
| CD19+ TAMs | NPC2 | 0.864936097 |
| CD19+ TAMs | ATP1B1 | 0.860131069 |
| CD19+ TAMs | HLA-DQB1 | 0.857854345 |
| CD19+ TAMs | SGK1 | 0.857258929 |
| CD19+ TAMs | HLA-DMA | 0.857151456 |
| CD19+ TAMs | F13A1 | 0.848268886 |
| CD19+ TAMs | MT-CYB | 0.848081003 |
| CD19+ TAMs | FMNL2 | 0.844102119 |
| CD19+ TAMs | HLA-DPB1 | 0.842823479 |
| CD19+ TAMs | GADD45B | 0.842394894 |
| CD19+ TAMs | HLA-DRA | 0.841268734 |
| CD19+ TAMs | ACP5 | 0.840054679 |
| CD19+ TAMs | GATM | 0.836059142 |
| CD19+ TAMs | VSIG4 | 0.835933348 |
| CD19+ TAMs | CD84 | 0.831983086 |
| CD19+ TAMs | GLUL | 0.829044432 |
| CD19+ TAMs | ABCA1 | 0.823144146 |
| CD19+ TAMs | TSC22D1 | 0.822188731 |
| CD19+ TAMs | MT-CO3 | 0.81889247 |
| CD19+ TAMs | MFSD1 | 0.81654977 |
| CD19+ TAMs | LAMP1 | 0.810510483 |
| CD19+ TAMs | STARD13 | 0.807509497 |
| CD19+ TAMs | FCER1G | 0.805547037 |
| CD19+ TAMs | CSF1R | 0.805467582 |
| CD19+ TAMs | PDK4 | 0.798812645 |
| CD19+ TAMs | FAM20A | 0.798313113 |
| CD19+ TAMs | ME1 | 0.795497294 |
| CD19+ TAMs | PLA2G7 | 0.793633975 |
| CD19+ TAMs | CSTB | 0.790283581 |
| CD19+ TAMs | IGSF6 | 0.789698503 |
| CD19+ TAMs | LRP1 | 0.78757899 |
| CD19+ TAMs | CCL3 | 0.7865266 |
| CD19+ TAMs | AP2A2 | 0.781350427 |
| CD19+ TAMs | VMO1 | 0.776072212 |
| CD19+ TAMs | SERPING1 | 0.775380531 |
| CD19+ TAMs | CPM | 0.772600116 |
| CD19+ TAMs | AXL | 0.769075293 |
| CD19+ TAMs | PTMS | 0.765771449 |
| CD19+ TAMs | MAN2B1 | 0.762932031 |
| CD19+ TAMs | HLA-E | 0.757744812 |
| CD19+ TAMs | PDGFC | 0.756371773 |
| CD19+ TAMs | IQGAP2 | 0.752871766 |
| CD19+ TAMs | DOCK4 | 0.749978301 |
| CD19+ TAMs | DST | 0.748537955 |
| CD19+ TAMs | TTTY14 | 0.745930139 |
| CD19+ TAMs | CXCL3 | 0.741054831 |
| CD19+ TAMs | CEBPD | 0.733215868 |
| CD19+ TAMs | BNC2 | 0.73316218 |
| CD19+ TAMs | ZFHX3 | 0.732910918 |
| CD19+ TAMs | LTC4S | 0.731958507 |
| CD19+ TAMs | FARP1 | 0.731254586 |
| CD19+ TAMs | HLA-A | 0.729263977 |
| CD19+ TAMs | TUBA1B | 0.726243984 |
| CD19+ TAMs | HMOX1 | 0.725768476 |
| CD19+ TAMs | TPP1 | 0.723871572 |
| CD19+ TAMs | GAA | 0.722979168 |
| CD19+ TAMs | ZFP36L1 | 0.722684756 |
| CD19+ TAMs | ID2 | 0.719794237 |
| CD19+ TAMs | SLC16A10 | 0.71970179 |
| CD19+ TAMs | AKR1B1 | 0.715350842 |
| CD19+ TAMs | RNASET2 | 0.715172697 |
| CD19+ TAMs | TLR4 | 0.706653748 |
| CD19+ TAMs | TMEM37 | 0.700802942 |
| CD19+ TAMs | CCL3L1 | 0.698715882 |
| CD19+ TAMs | HEXA | 0.691562367 |
| CD19+ TAMs | HLA-DQA2 | 0.69000523 |
| CD19+ TAMs | ABCC5 | 0.684361206 |
| CD19+ TAMs | FPR3 | 0.683972278 |
| CD19+ TAMs | BST2 | 0.682058401 |
| CD19+ TAMs | SNX6 | 0.676141904 |
| CD19+ TAMs | MT1X | 0.674262691 |
| CD19+ TAMs | HLA-DMB | 0.674079143 |
| CD19+ TAMs | MT-ATP6 | 0.674048199 |
| CD19+ TAMs | CD4 | 0.672760849 |
| CD19+ TAMs | LGALS3BP | 0.670317989 |
| CD19+ TAMs | MT-ND1 | 0.6697393 |
| CD19+ TAMs | LYVE1 | 0.666189889 |
| CD19+ TAMs | EPB41L2 | 0.660774145 |
| CD19+ TAMs | ANKH | 0.658527589 |
| CD19+ TAMs | BCAT1 | 0.657748791 |
| CD19+ TAMs | CTSA | 0.65727844 |
| CD19+ TAMs | CALM3 | 0.653081743 |
| CD19+ TAMs | EGR1 | 0.646880354 |
| CD19+ TAMs | PLXND1 | 0.645838506 |
| CD19+ TAMs | MGST3 | 0.645410782 |
| CD19+ TAMs | LAIR1 | 0.645153024 |
| CD19+ TAMs | CCL8 | 0.643237399 |
| CD19+ TAMs | SPATS2L | 0.641534832 |
| CD19+ TAMs | ITGB5 | 0.641264696 |
| CD19+ TAMs | STMN1 | 0.64120384 |
| CD19+ TAMs | LAMP2 | 0.639532706 |
| CD19+ TAMs | SASH1 | 0.63879473 |
| CD19+ TAMs | SIGLEC1 | 0.638498419 |
| CD19+ TAMs | GADD45G | 0.637280392 |
| CD19+ TAMs | KCTD12 | 0.635035053 |
| CD19+ TAMs | IGSF21 | 0.634186973 |
| CD19+ TAMs | SCN1B | 0.630100748 |
| CD19+ TAMs | SLC7A8 | 0.628934217 |
| CD19+ TAMs | EBI3 | 0.627933482 |
| CD19+ TAMs | SESN1 | 0.625801382 |
| CD19+ TAMs | DPP7 | 0.621323143 |
| CD19+ TAMs | TTYH3 | 0.61834114 |
| CD19+ TAMs | NFIA | 0.611838054 |
| CD19+ TAMs | GAS2L3 | 0.608172468 |
| CD19+ TAMs | PRDX1 | 0.607435989 |
| CD19+ TAMs | EPHX1 | 0.607261682 |
| CD19+ TAMs | PRNP | 0.606388213 |
| CD19+ TAMs | KIF1B | 0.603915258 |
| CD19+ TAMs | APLP2 | 0.601799827 |
| CD19+ TAMs | OTUD1 | 0.601725072 |
| CD19+ TAMs | RHOB | 0.598659898 |
| CD19+ TAMs | CLTC | 0.597304486 |
| CD19+ TAMs | TNFAIP2 | 0.592875223 |
| CD19+ TAMs | IFI27 | 0.591852282 |
| CD19+ TAMs | TIMP3 | 0.591371961 |
| CD19+ TAMs | NCOA4 | 0.590847285 |
| CD19+ TAMs | GM2A | 0.58793349 |
| CD19+ TAMs | HRH1 | 0.585495142 |
| CD19+ TAMs | MT-ND4 | 0.583783574 |
| CD19+ TAMs | RAB20 | 0.581458573 |
| CD19+ TAMs | FTH1 | 0.581424034 |
| CD19+ TAMs | BRI3 | 0.579809278 |
| CD19+ TAMs | GPX4 | 0.579010271 |
| CD19+ TAMs | GPR155 | 0.578192942 |
| CD19+ TAMs | GOLIM4 | 0.577349988 |
| CD19+ TAMs | VEGFB | 0.576975238 |
| CD19+ TAMs | ATP6V0B | 0.575762265 |
| CD19+ TAMs | COLEC12 | 0.574539178 |
| CD19+ TAMs | TMIGD3 | 0.572376407 |
| CD19+ TAMs | UNC93B1 | 0.57192875 |
| CD19+ TAMs | C5AR1 | 0.570939032 |
| CD19+ TAMs | RNASE6 | 0.568185688 |
| CD19+ TAMs | LPAR6 | 0.566094118 |
| CD19+ TAMs | OTOA | 0.564843083 |
| CD19+ TAMs | AMBP | 0.564636322 |
| CD19+ TAMs | ADAM9 | 0.562997844 |
| CD19+ TAMs | MT-ND3 | 0.562584439 |
| CD19+ TAMs | FCGBP | 0.562338885 |
| CD19+ TAMs | C6orf62 | 0.560791918 |
| CD19+ TAMs | SERPINB6 | 0.560506549 |
| CD19+ TAMs | HNMT | 0.56004068 |
| CD19+ TAMs | SMS | 0.559905365 |
| CD19+ TAMs | ALDH1A1 | 0.559504127 |
| CD19+ TAMs | MPP1 | 0.557834118 |
| CD19+ TAMs | MYO1E | 0.557647796 |
| CD19+ TAMs | FGL2 | 0.556237028 |
| CD19+ TAMs | GPX3 | 0.554916053 |
| CD19+ TAMs | B2M | 0.554885482 |
| CD19+ TAMs | DBI | 0.553457797 |
| CD19+ TAMs | C1orf54 | 0.553262889 |
| CD19+ TAMs | PSD3 | 0.5531246 |
| CD19+ TAMs | TMBIM6 | 0.552796296 |
| CD19+ TAMs | ATP6AP1 | 0.552674228 |
| CD19+ TAMs | MITF | 0.549643922 |
| CD19+ TAMs | GNPDA1 | 0.548163513 |
| CD19+ TAMs | PIK3IP1 | 0.547813397 |
| CD19+ TAMs | VAT1 | 0.54527336 |
| CD19+ TAMs | MT-ND2 | 0.542729555 |
| CD19+ TAMs | CPVL | 0.542497415 |
| CD19+ TAMs | CLTA | 0.542262055 |
| CD19+ TAMs | DHRS3 | 0.541287277 |
| CD19+ TAMs | SRGAP1 | 0.541239022 |
| CD19+ TAMs | RGS10 | 0.540398197 |
| CD19+ TAMs | ARHGAP18 | 0.540328033 |
| CD19+ TAMs | SLC22A23 | 0.535746207 |
| CD19+ TAMs | FEZ2 | 0.533535289 |
| CD19+ TAMs | SLC15A3 | 0.53245417 |
| CD19+ TAMs | P4HA1 | 0.531413923 |
| CD19+ TAMs | LAPTM4A | 0.530832243 |
| CD19+ TAMs | IFRD1 | 0.530760646 |
| CD19+ TAMs | PAX8-AS1 | 0.530078399 |
| CD19+ TAMs | OSBPL1A | 0.529925493 |
| CD19+ TAMs | BCAP31 | 0.529583219 |
| CD19+ TAMs | ATP6V0C | 0.52950461 |
| CD19+ TAMs | AP1B1 | 0.527787768 |
| CD19+ TAMs | CYFIP1 | 0.526216006 |
| CD19+ TAMs | EGFL7 | 0.524849822 |
| CD19+ TAMs | CMKLR1 | 0.524115824 |
| CD19+ TAMs | CD28 | 0.523839681 |
| CD19+ TAMs | NAIP | 0.523198451 |
| CD19+ TAMs | GCLC | 0.522639302 |
| CD19+ TAMs | TTR | 0.519699967 |
| CD19+ TAMs | CALM2 | 0.519091474 |
| CD19+ TAMs | HLA-DOA | 0.518327517 |
| CD19+ TAMs | PEBP1 | 0.516823318 |
| CD19+ TAMs | PRXL2A | 0.516539171 |
| CD19+ TAMs | CD93 | 0.514620673 |
| CD19+ TAMs | GPR137B | 0.513986789 |
| CD19+ TAMs | AC100849.1 | 0.513568397 |
| CD19+ TAMs | IDH1 | 0.513447214 |
| CD19+ TAMs | PLEKHA1 | 0.511682263 |
| CD19+ TAMs | UGCG | 0.508669765 |
| CD19+ TAMs | CST3 | 0.506567085 |
| CD19+ TAMs | PHLDA1 | 0.505423329 |
| CD19+ TAMs | CCL13 | 0.505060038 |
| CD19+ TAMs | SPRED1 | 0.504780986 |
| CD19+ TAMs | LILRB4 | 0.504530479 |
| CD19+ TAMs | RND3 | 0.503896429 |
| CD19+ TAMs | DDAH2 | 0.502962457 |
| CD19+ TAMs | IER3 | 0.500999686 |
| CD19+ TAMs | GAS6 | 0.500146301 |
| CD19+ TAMs | NINJ1 | 0.49965259 |
| CD19+ TAMs | ARL6IP1 | 0.49784466 |
| CD19+ TAMs | SLC18B1 | 0.496300795 |
| CD19+ TAMs | MT-CO1 | 0.496063256 |
| CD19+ TAMs | GDF15 | 0.494949931 |
| CD19+ TAMs | TMEM176A | 0.494723424 |
| CD19+ TAMs | ITGAV | 0.494164606 |
| CD19+ TAMs | RGS1 | 0.493705495 |
| CD19+ TAMs | WLS | 0.492041472 |
| CD19+ TAMs | RGL1 | 0.491116441 |
| CD19+ TAMs | MKNK1 | 0.490661248 |
| CD19+ TAMs | PLAU | 0.486009063 |
| CD19+ TAMs | SDS | 0.485482735 |
| CD19+ TAMs | LY96 | 0.485350962 |
| CD19+ TAMs | CAPZB | 0.484947691 |
| CD19+ TAMs | SERINC1 | 0.484926096 |
| CD19+ TAMs | ATP6V0E1 | 0.484152772 |
| CD19+ TAMs | EVA1B | 0.481480407 |
| CD19+ TAMs | EMP2 | 0.481142881 |
| CD19+ TAMs | PDGFB | 0.4795825 |
| CD19+ TAMs | CD164 | 0.479377379 |
| CD19+ TAMs | SCARB2 | 0.477831503 |
| CD19+ TAMs | ATP6AP2 | 0.477687131 |
| CD19+ TAMs | CCL2 | 0.476086576 |
| CD19+ TAMs | LAPTM5 | 0.474925903 |
| CD19+ TAMs | CYTH4 | 0.474894364 |
| CD19+ TAMs | SCD | 0.473944505 |
| CD19+ TAMs | ATP6V1F | 0.47254628 |
| CD19+ TAMs | GFRA2 | 0.471349643 |
| CD19+ TAMs | RGS2 | 0.466640528 |
| CD19+ TAMs | TNFRSF1A | 0.463931665 |
| CD19+ TAMs | MGAT1 | 0.461345987 |
| CD19+ TAMs | GIMAP5 | 0.459220646 |
| CD19+ TAMs | CITED2 | 0.458858496 |
| CD19+ TAMs | DNASE1L3 | 0.458655455 |
| CD19+ TAMs | GSN | 0.457484552 |
| CD19+ TAMs | HOMER3 | 0.457141641 |
| CD19+ TAMs | ETV5 | 0.455647167 |
| CD19+ TAMs | CYBA | 0.453997134 |
| CD19+ TAMs | PMP22 | 0.452783883 |
| CD19+ TAMs | IL2RA | 0.451631544 |
| CD19+ TAMs | MGLL | 0.451017187 |
| CD19+ TAMs | KCNJ5 | 0.448885818 |
| CD19+ TAMs | DSC2 | 0.448557228 |
| CD19+ TAMs | DIRC3 | 0.447660184 |
| CD19+ TAMs | SCAMP2 | 0.445559625 |
| CD19+ TAMs | FSCN1 | 0.444531684 |
| CD19+ TAMs | DNPH1 | 0.441807705 |
| CD19+ TAMs | QKI | 0.441511586 |
| CD19+ TAMs | PLA2G15 | 0.44042319 |
| CD19+ TAMs | TNFSF13 | 0.440195401 |
| CD19+ TAMs | SLC38A6 | 0.438309672 |
| CD19+ TAMs | TENT5A | 0.438184552 |
| CD19+ TAMs | NFIC | 0.4379106 |
| CD19+ TAMs | DENND4C | 0.435293593 |
| CD19+ TAMs | CD302 | 0.434995615 |
| CD19+ TAMs | GNB4 | 0.434857961 |
| CD19+ TAMs | CRYL1 | 0.434349667 |
| CD19+ TAMs | MGAT4A | 0.432886006 |
| CD19+ TAMs | MXD4 | 0.431818007 |
| CD19+ TAMs | SLAMF8 | 0.431244517 |
| CD19+ TAMs | COMT | 0.430931787 |
| CD19+ TAMs | RAC1 | 0.429941899 |
| CD19+ TAMs | ADGRG6 | 0.429154703 |
| CD19+ TAMs | CCDC152 | 0.428803805 |
| CD19+ TAMs | MFHAS1 | 0.428050322 |
| CD19+ TAMs | EDA | 0.427309389 |
| CD19+ TAMs | M6PR | 0.426287867 |
| CD19+ TAMs | BLVRA | 0.425514617 |
| CD19+ TAMs | RNF130 | 0.425498732 |
| CD19+ TAMs | COLGALT1 | 0.425445017 |
| CD19+ TAMs | CCL18 | 0.424699696 |
| CD19+ TAMs | TCEAL9 | 0.422816713 |
| CD19+ TAMs | ARHGAP5 | 0.422497069 |
| CD19+ TAMs | SPATA7 | 0.421596546 |
| CD19+ TAMs | RAPH1 | 0.420631761 |
| CD19+ TAMs | TGOLN2 | 0.419794843 |
| CD19+ TAMs | BEX4 | 0.419482686 |
| CD19+ TAMs | SPIRE1 | 0.419364704 |
| CD19+ TAMs | SGPL1 | 0.416661046 |
| CD19+ TAMs | AAK1 | 0.413709237 |
| CD19+ TAMs | EPB41L3 | 0.413667167 |
| CD19+ TAMs | MYL6 | 0.413388415 |
| CD19+ TAMs | BNIP3L | 0.412203964 |
| CD19+ TAMs | NCF4 | 0.411937265 |
| CD19+ TAMs | PPT1 | 0.410302513 |
| CD19+ TAMs | BMP2K | 0.410097088 |
| CD19+ TAMs | IER5L | 0.410005461 |
| CD19+ TAMs | GNS | 0.41000349 |
| CD19+ TAMs | VAMP8 | 0.409574741 |
| CD19+ TAMs | NENF | 0.407614251 |
| CD19+ TAMs | RHOBTB3 | 0.407605319 |
| CD19+ TAMs | CD99 | 0.407555351 |
| CD19+ TAMs | TMSB4X | 0.406315346 |
| CD19+ TAMs | CCND1 | 0.405969838 |
| CD19+ TAMs | RAB5C | 0.404741363 |
| CD19+ TAMs | NEU1 | 0.40403695 |
| CD19+ TAMs | RASSF4 | 0.403423168 |
| CD19+ TAMs | HES1 | 0.403262725 |
| CD19+ TAMs | FNIP2 | 0.401537083 |
| CD19+ TAMs | UACA | 0.401402462 |
| CD19+ TAMs | SIRPA | 0.40036941 |
| CD19+ TAMs | SCPEP1 | 0.399864918 |
| CD19+ TAMs | TGFBR1 | 0.399694513 |
| CD19+ TAMs | ARHGAP10 | 0.39885887 |
| CD19+ TAMs | TRIM14 | 0.398614782 |
| CD19+ TAMs | HGF | 0.396819234 |
| CD19+ TAMs | IL18 | 0.396453867 |
| CD19+ TAMs | RTN4 | 0.396083065 |
| CD19+ TAMs | CREBL2 | 0.395991628 |
| CD19+ TAMs | EPS8 | 0.395658955 |
| CD19+ TAMs | RNASEK | 0.395147674 |
| CD19+ TAMs | NTAN1 | 0.393716056 |
| CD19+ TAMs | PEPD | 0.393471807 |
| CD19+ TAMs | IL4I1 | 0.393210125 |
| CD19+ TAMs | TANC2 | 0.39257715 |
| CD19+ TAMs | HPGDS | 0.392533825 |
| CD19+ TAMs | FABP3 | 0.392277687 |
| CD19+ TAMs | SMPDL3A | 0.392233552 |
| CD19+ TAMs | SULT1A1 | 0.39207856 |
| CD19+ TAMs | GLMP | 0.391829619 |
| CD19+ TAMs | PIK3R1 | 0.3893586 |
| CD19+ TAMs | SESN3 | 0.388783091 |
| CD19+ TAMs | MYO5A | 0.386937194 |
| CD19+ TAMs | SAT1 | 0.38579691 |
| CD19+ TAMs | CLIC2 | 0.384914044 |
| CD19+ TAMs | TGFBI | 0.384237329 |
| CD19+ TAMs | RBM47 | 0.384009795 |
| CD19+ TAMs | APH1A | 0.383791864 |
| CD19+ TAMs | ITGA9 | 0.383650598 |
| CD19+ TAMs | STOM | 0.383278642 |
| CD19+ TAMs | CD151 | 0.382197102 |
| CD19+ TAMs | ZFYVE16 | 0.382052575 |
| CD19+ TAMs | MFSD12 | 0.381919917 |
| CD19+ TAMs | CLEC2B | 0.380921705 |
| CD19+ TAMs | GLDN | 0.38042056 |
| CD19+ TAMs | RNF13 | 0.380172525 |
| CD19+ TAMs | HSD17B14 | 0.379661166 |
| CD19+ TAMs | AKR1A1 | 0.378763374 |
| CD19+ TAMs | ZDHHC14 | 0.377913923 |
| CD19+ TAMs | H2AFJ | 0.377514285 |
| CD19+ TAMs | TMEM219 | 0.377031323 |
| CD19+ TAMs | MMD | 0.376752027 |
| CD19+ TAMs | PLEKHO2 | 0.376451235 |
| CD19+ TAMs | SEPTIN11 | 0.374638242 |
| CD19+ TAMs | DRAM2 | 0.3741726 |
| CD19+ TAMs | SNX5 | 0.373724273 |
| CD19+ TAMs | CALR | 0.372564171 |
| CD19+ TAMs | NRIP1 | 0.372326589 |
| CD19+ TAMs | TNFSF12 | 0.371095221 |
| CD19+ TAMs | PRKACB | 0.37067084 |
| CD19+ TAMs | CANX | 0.369474642 |
| CD19+ TAMs | LHFPL2 | 0.368673724 |
| CD19+ TAMs | ARRB2 | 0.368297826 |
| CD19+ TAMs | CORO1B | 0.367847137 |
| CD19+ TAMs | HLA-B | 0.366532142 |
| CD19+ TAMs | ARHGAP21 | 0.366095962 |
| CD19+ TAMs | OLMALINC | 0.364727431 |
| CD19+ TAMs | DIP2A | 0.3642343 |
| CD19+ TAMs | APOH | 0.364129644 |
| CD19+ TAMs | ADM | 0.363937671 |
| CD19+ TAMs | PEAK1 | 0.363176913 |
| CD19+ TAMs | PXDC1 | 0.362900233 |
| CD19+ TAMs | CMTM3 | 0.362650768 |
| CD19+ TAMs | IGFLR1 | 0.362045685 |
| CD19+ TAMs | CCR1 | 0.361013523 |
| CD19+ TAMs | SLC35F6 | 0.36093349 |
| CD19+ TAMs | MT-CO2 | 0.360843515 |
| CD19+ TAMs | PER3 | 0.359332175 |
| CD19+ TAMs | TMEM14C | 0.359250626 |
| CD19+ TAMs | TNFRSF21 | 0.359195506 |
| CD19+ TAMs | CEBPA | 0.358722238 |
| CD19+ TAMs | RAB13 | 0.358457988 |
| CD19+ TAMs | GUSB | 0.358136438 |
| CD19+ TAMs | SCP2 | 0.355256203 |
| CD19+ TAMs | SCN9A | 0.35444073 |
| CD19+ TAMs | ANXA4 | 0.354024535 |
| CD19+ TAMs | STX4 | 0.352165623 |
| CD19+ TAMs | ACP2 | 0.351952949 |
| CD19+ TAMs | NCKAP1L | 0.351814482 |
| CD19+ TAMs | TMEM35B | 0.351282411 |
| CD19+ TAMs | GABARAP | 0.350714223 |
| CD19+ TAMs | HEXB | 0.349307015 |
| CD19+ TAMs | FERMT3 | 0.348736788 |
| CD19+ TAMs | ARRDC2 | 0.347606753 |
| CD19+ TAMs | C1orf56 | 0.347439612 |
| CD19+ TAMs | NAGK | 0.347264548 |
| CD19+ TAMs | AMDHD2 | 0.347130656 |
| CD19+ TAMs | HS3ST2 | 0.346660833 |
| CD19+ TAMs | NCKAP5 | 0.346482106 |
| CD19+ TAMs | GNAS | 0.346325528 |
| CD19+ TAMs | LMNA | 0.346113998 |
| CD19+ TAMs | MT-ND5 | 0.345984196 |
| CD19+ TAMs | INSIG1 | 0.344245104 |
| CD19+ TAMs | RNH1 | 0.341748705 |
| CD19+ TAMs | BIN1 | 0.341644625 |
| CD19+ TAMs | PLAAT3 | 0.341597428 |
| CD19+ TAMs | GASK1B | 0.340819831 |
| CD19+ TAMs | TUBB | 0.340312123 |
| CD19+ TAMs | SNX18 | 0.340154912 |
| CD19+ TAMs | ORM1 | 0.339890039 |
| CD19+ TAMs | PCBD1 | 0.339385347 |
| CD19+ TAMs | ABHD12 | 0.339375323 |
| CD19+ TAMs | FUOM | 0.339208928 |
| CD19+ TAMs | HAVCR2 | 0.338792011 |
| CD19+ TAMs | SH3PXD2A | 0.338422989 |
| CD19+ TAMs | PIK3R3 | 0.338256924 |
| CD19+ TAMs | FAM20C | 0.338251131 |
| CD19+ TAMs | RBMS1 | 0.337978578 |
| CD19+ TAMs | SHMT1 | 0.337020693 |
| CD19+ TAMs | LGALS9 | 0.336697641 |
| CD19+ TAMs | POMP | 0.335872312 |
| CD19+ TAMs | IRF2BP2 | 0.335538934 |
| CD19+ TAMs | AIG1 | 0.334657784 |
| CD19+ TAMs | PLXNB2 | 0.334180715 |
| CD19+ TAMs | TLN1 | 0.333966578 |
| CD19+ TAMs | CTTNBP2NL | 0.333334536 |
| CD19+ TAMs | LPL | 0.333284622 |
| CD19+ TAMs | ITGB2 | 0.332417702 |
| CD19+ TAMs | MGST2 | 0.332301047 |
| CD19+ TAMs | ARL6IP5 | 0.331672798 |
| CD19+ TAMs | SORBS3 | 0.331187633 |
| CD19+ TAMs | CRHBP | 0.331150378 |
| CD19+ TAMs | ARPIN | 0.32979864 |
| CD19+ TAMs | CHID1 | 0.329311408 |
| CD19+ TAMs | AP2S1 | 0.328956825 |
| CD19+ TAMs | SLC46A3 | 0.328545464 |
| CD19+ TAMs | CNRIP1 | 0.328535544 |
| CD19+ TAMs | RHOC | 0.327066713 |
| CD19+ TAMs | S100A13 | 0.325916604 |
| CD19+ TAMs | OSTF1 | 0.325302513 |
| CD19+ TAMs | SPPL2A | 0.322506604 |
| CD19+ TAMs | TPT1 | 0.321897569 |
| CD19+ TAMs | RAB3IL1 | 0.321674683 |
| CD19+ TAMs | UCP2 | 0.321455074 |
| CD19+ TAMs | MDH1 | 0.321209863 |
| CD19+ TAMs | TWF2 | 0.320738005 |
| CD19+ TAMs | QPRT | 0.32069077 |
| CD19+ TAMs | RENBP | 0.320318975 |
| CD19+ TAMs | FEM1B | 0.320198121 |
| CD19+ TAMs | ARHGAP4 | 0.319762742 |
| CD19+ TAMs | MAT2A | 0.319478184 |
| CD19+ TAMs | SLC11A2 | 0.31939895 |
| CD19+ TAMs | LITAF | 0.318302914 |
| CD19+ TAMs | SYNGR2 | 0.31824409 |
| CD19+ TAMs | CYBRD1 | 0.317889515 |
| CD19+ TAMs | CCDC107 | 0.317686552 |
| CD19+ TAMs | SPINK1 | 0.316898362 |
| CD19+ TAMs | MAMDC2 | 0.316336846 |
| CD19+ TAMs | C20orf194 | 0.313737089 |
| CD19+ TAMs | TNFRSF14 | 0.313634117 |
| CD19+ TAMs | ATOX1 | 0.313606192 |
| CD19+ TAMs | USF2 | 0.313035643 |
| CD19+ TAMs | TMEM70 | 0.311943742 |
| CD19+ TAMs | DTNA | 0.311766738 |
| CD19+ TAMs | BLOC1S1 | 0.311307758 |
| CD19+ TAMs | APPL2 | 0.311292653 |
| CD19+ TAMs | SNHG12 | 0.31107359 |
| CD19+ TAMs | SUCNR1 | 0.309916662 |
| CD19+ TAMs | MSRB2 | 0.309883626 |
| CD19+ TAMs | ACTG1 | 0.309304178 |
| CD19+ TAMs | GNG10 | 0.308918436 |
| CD19+ TAMs | CTSF | 0.308859646 |
| CD19+ TAMs | PI4K2A | 0.308668507 |
| CD19+ TAMs | MTSS1 | 0.30852664 |
| CD19+ TAMs | ENG | 0.307238925 |
| CD19+ TAMs | OTULINL | 0.306954308 |
| CD19+ TAMs | LAMTOR1 | 0.306502571 |
| CD19+ TAMs | IGFBP4 | 0.306203851 |
| CD19+ TAMs | CYB5A | 0.30599866 |
| CD19+ TAMs | NAGLU | 0.305229554 |
| CD19+ TAMs | ARPC5 | 0.304400256 |
| CD19+ TAMs | ITPRIPL2 | 0.303210979 |
| CD19+ TAMs | HNRNPH1 | 0.303164175 |
| CD19+ TAMs | TCN2 | 0.303127105 |
| CD19+ TAMs | ARHGEF12 | 0.302791895 |
| CD19+ TAMs | PITHD1 | 0.302293779 |
| CD19+ TAMs | DEGS1 | 0.302125148 |
| CD19+ TAMs | SH2B3 | 0.301096126 |
| CD19+ TAMs | DHRS7 | 0.300920889 |
| CD19+ TAMs | CYB5R1 | 0.299463674 |
| CD19+ TAMs | SDC2 | 0.299383175 |
| CD19+ TAMs | PRKN | 0.298867966 |
| CD19+ TAMs | OAZ2 | 0.298507868 |
| CD19+ TAMs | SERPINF1 | 0.297697296 |
| CD19+ TAMs | ETS2 | 0.296595577 |
| CD19+ TAMs | OGFRL1 | 0.2962876 |
| CD19+ TAMs | SPIN1 | 0.296243908 |
| CD19+ TAMs | H1F0 | 0.296010607 |
| CD19+ TAMs | BAIAP2 | 0.295947913 |
| CD19+ TAMs | GRINA | 0.295365551 |
| CD19+ TAMs | ADORA3 | 0.293358172 |
| CD19+ TAMs | FAM91A1 | 0.293222889 |
| CD19+ TAMs | CEP170 | 0.293006973 |
| CD19+ TAMs | OLFML2B | 0.292822243 |
| CD19+ TAMs | LACC1 | 0.292324877 |
| CD19+ TAMs | SMIM30 | 0.291208054 |
| CD19+ TAMs | LY86 | 0.29113305 |
| CD19+ TAMs | HCST | 0.29049392 |
| CD19+ TAMs | CCDC47 | 0.290368314 |
| CD19+ TAMs | TM2D2 | 0.289741432 |
| CD19+ TAMs | SUMF2 | 0.289524752 |
| CD19+ TAMs | EVL | 0.28752045 |
| CD19+ TAMs | HEBP1 | 0.287489239 |
| CD19+ TAMs | HECTD2 | 0.286917784 |
| CD19+ TAMs | STON2 | 0.286868972 |
| CD19+ TAMs | RAB32 | 0.286525645 |
| CD19+ TAMs | FGFR1 | 0.286104794 |
| CD19+ TAMs | APMAP | 0.285700079 |
| CD19+ TAMs | HSPB1 | 0.284471212 |
| CD19+ TAMs | IFNGR1 | 0.284211167 |
| CD19+ TAMs | RRAGD | 0.28284892 |
| CD19+ TAMs | P4HB | 0.282620765 |
| CD19+ TAMs | EPAS1 | 0.282598723 |
| CD19+ TAMs | SLC39A1 | 0.281545348 |
| CD19+ TAMs | GAL3ST4 | 0.281520096 |
| CD19+ TAMs | DDRGK1 | 0.280298089 |
| CD19+ TAMs | MMP14 | 0.28016654 |
| CD19+ TAMs | VKORC1 | 0.279566424 |
| CD19+ TAMs | NUCB1 | 0.279461037 |
| CD19+ TAMs | TCEAL4 | 0.279171378 |
| CD19+ TAMs | IL18BP | 0.278449584 |
| CD19+ TAMs | GBA | 0.278288458 |
| CD19+ TAMs | DBNDD2 | 0.278112793 |
| CD19+ TAMs | TM6SF1 | 0.277956693 |
| CD19+ TAMs | ANXA5 | 0.276704513 |
| CD19+ TAMs | BCL2L1 | 0.276534898 |
| CD19+ TAMs | SAMD4A | 0.275115027 |
| CD19+ TAMs | CLN8 | 0.274305752 |
| CD19+ TAMs | EYA2 | 0.274274837 |
| CD19+ TAMs | SGPP1 | 0.274272691 |
| CD19+ TAMs | TMBIM1 | 0.274128399 |
| CD19+ TAMs | LEPROT | 0.270714885 |
| CD19+ TAMs | TFPT | 0.268896661 |
| CD19+ TAMs | TRIM47 | 0.267541119 |
| CD19+ TAMs | ARPC1B | 0.267210855 |
| CD19+ TAMs | PFKL | 0.267074347 |
| CD19+ TAMs | CXCL9 | 0.267069716 |
| CD19+ TAMs | CCDC88A | 0.26684799 |
| CD19+ TAMs | RASAL2 | 0.264803726 |
| CD19+ TAMs | ADIPOR1 | 0.264672032 |
| CD19+ TAMs | TMEM86A | 0.264364849 |
| CD19+ TAMs | CD33 | 0.263944475 |
| CD19+ TAMs | SLC43A3 | 0.263815625 |
| CD19+ TAMs | NMRK1 | 0.261425121 |
| CD19+ TAMs | MRC2 | 0.261262455 |
| CD19+ TAMs | MPPED2 | 0.260691865 |
| CD19+ TAMs | CNTLN | 0.259860893 |
| CD19+ TAMs | LINC00472 | 0.259013258 |
| CD19+ TAMs | ZNF358 | 0.258965105 |
| CD19+ TAMs | ALDH9A1 | 0.258328056 |
| CD19+ TAMs | TMEM106A | 0.25736182 |
| CD19+ TAMs | PTTG1IP | 0.256774603 |
| CD19+ TAMs | BEX3 | 0.256641248 |
| CD19+ TAMs | RPP25 | 0.256339568 |
| CD19+ TAMs | CHCHD6 | 0.255934224 |
| CD19+ TAMs | SNX24 | 0.255672286 |
| CD19+ TAMs | PPIA | 0.255242782 |
| CD19+ TAMs | CLN6 | 0.254815399 |
| CD19+ TAMs | PYCARD | 0.254068411 |
| CD19+ TAMs | SLC22A18 | 0.253709924 |
| CD19+ TAMs | TNFRSF11A | 0.253632553 |
| CD19+ TAMs | NAGA | 0.253509241 |
| CD19+ TAMs | SLC12A5 | 0.252864231 |
| CD19+ TAMs | MATK | 0.252213507 |
| CD19+ TAMs | IQGAP1 | 0.326989065 |
| CD19+ TAMs | MCRIP1 | 0.254963685 |
| CD19+ TAMs | DNASE2 | 0.286698441 |
| CD19+ TAMs | HLA-C | 0.319711046 |
| CD19+ TAMs | GNB2 | 0.260372932 |
| CD19+ TAMs | AP2M1 | 0.265414953 |
| CD19+ TAMs | TLE5 | 0.280884795 |
| CD19+ TAMs | CHMP1B | 0.368499412 |
| CD19+ TAMs | NCEH1 | 0.256038457 |
| CD19+ TAMs | ARHGDIA | 0.267508278 |
| CD19+ TAMs | RDX | 0.406554251 |
| CD19+ TAMs | CXCL2 | 0.594433695 |
| CD19+ TAMs | SNX9 | 0.840236172 |
| CD19+ TAMs | ZNF331 | 0.665368328 |
| CD19+ TAMs | PAPSS2 | 0.353866162 |
| CD19+ TAMs | APH1B | 0.292380591 |
| CD19+ TAMs | DAPK1 | 0.371818982 |
| CD19+ TAMs | NDFIP1 | 0.267708232 |
| CD19+ TAMs | DYNLL1 | 0.293370535 |
| CD19+ TAMs | WASHC4 | 0.293826674 |
| CD19+ TAMs | NECAP2 | 0.265141668 |
| CD19+ TAMs | TRAPPC5 | 0.288154976 |
| CD19+ TAMs | TMEM50A | 0.297983814 |
| CD19+ TAMs | CLIC1 | 0.298461669 |
| CD19+ TAMs | ATP6V0A1 | 0.296190671 |
| CD19+ TAMs | ATP6V1C1 | 0.271285278 |
| CD19+ TAMs | FAM13A | 0.297043123 |
| CD19+ TAMs | PLCL1 | 0.348295871 |
| CD19+ TAMs | TMBIM4 | 0.279379538 |
| CD19+ TAMs | CREM | 0.493443103 |
| CD19+ TAMs | VPS28 | 0.283356068 |
| CD19+ TAMs | BSG | 0.267449509 |
| CD19+ TAMs | PPP1CB | 0.423424556 |
| CD19+ TAMs | SELENOW | 0.290196203 |
| CD19+ TAMs | AHNAK | 0.306597679 |
| CD19+ TAMs | CBR1 | 0.251980541 |
| CD19+ TAMs | ANXA11 | 0.277392082 |
| CD19+ TAMs | TCF12 | 0.379447448 |
| CD19+ TAMs | RANBP2 | 0.590398943 |
| CD19+ TAMs | C9orf16 | 0.397797095 |
| CD19+ TAMs | DHRS9 | 0.25415643 |
| CD19+ TAMs | IL10RA | 0.25385869 |
| CD19+ TAMs | OS9 | 0.251077242 |
| CD19+ TAMs | GSTO1 | 0.283397017 |
| CD19+ TAMs | PTBP3 | 0.319087438 |
| CD19+ TAMs | FXYD5 | 0.312805492 |
| CD19+ TAMs | GATAD1 | 0.26193292 |
| CD19+ TAMs | NHSL1 | 0.516097467 |
| CD19+ TAMs | CEPT1 | 0.252082189 |
| CD19+ TAMs | GNA13 | 0.460537329 |
| CD19+ TAMs | NORAD | 0.27642695 |
| CD19+ TAMs | MT2A | 0.718106849 |
| CD19+ TAMs | EIF4A3 | 0.36016246 |
| CD19+ TAMs | CAPG | 0.468787029 |
| CD19+ TAMs | WDR61 | 0.252641227 |
| CD19+ TAMs | S100A11 | 0.301787644 |
| CD19+ TAMs | DYNC1H1 | 0.257189207 |
| CD19+ TAMs | EIF3D | 0.2572143 |
| CD19+ TAMs | SMURF2 | 0.262328079 |
| CD19+ TAMs | SNX2 | 0.253595376 |
| CD19+ TAMs | MMP9 | 0.383324355 |
| CD19+ TAMs | RPS27L | 0.260339926 |
| CD19+ TAMs | MPC2 | 0.256777412 |
| CD19+ TAMs | NAMPT | 0.588383771 |
| CD19+ TAMs | LNCAROD | 0.306515595 |
| CD19+ TAMs | CD72 | 0.277438875 |
| CD19+ TAMs | PDK3 | 0.25595727 |
| CD19+ TAMs | SLC3A2 | 0.277468102 |
| CD19+ TAMs | DYNLRB1 | 0.251092286 |
| CD19+ TAMs | TNS3 | 0.266800944 |
| CD19+ TAMs | WASF2 | 0.304755717 |
| CD19+ TAMs | RBPJ | 0.250123342 |
| CD19+ TAMs | ATP2A2 | 0.520450232 |
| CD19+ TAMs | PLA2G4C | 0.275493308 |
| CD19+ TAMs | TCF4 | 0.317094185 |
| CD19+ TAMs | P2RY13 | 0.293798333 |
| CD19+ TAMs | CD40 | 0.258794107 |
| CD19+ TAMs | REX1BD | 0.259631775 |
| CD19+ TAMs | SRP14 | 0.256094577 |
| CD19+ TAMs | FOS | 0.317078391 |
| CD19+ TAMs | ZFAND5 | 0.631400544 |
| CD19+ TAMs | AC084871.1 | 0.36671899 |
| CD19+ TAMs | B4GALT1 | 0.371239844 |
| CD19+ TAMs | FUS | 0.278089373 |
| CD19+ TAMs | FILIP1L | 0.454646837 |
| CD19+ TAMs | CEBPB | 0.309371434 |
| CD19+ TAMs | CHCHD10 | 0.26075932 |
| CD19+ TAMs | BAX | 0.258785692 |
| CD19+ TAMs | ARRDC3 | 0.301206968 |
| CD19+ TAMs | NFE2L2 | 0.255995942 |
| CD19+ TAMs | TALDO1 | 0.283405892 |
| CD19+ TAMs | KLF6 | 0.483810975 |
| CD19+ TAMs | CREBRF | 0.280580233 |
| CD19+ TAMs | ID3 | 0.263107752 |
| CD19+ TAMs | C4orf48 | 0.254673661 |
| CD19+ TAMs | UQCR11 | 0.265126354 |
| CD19+ TAMs | SLC31A2 | 0.300059555 |
| CD19+ TAMs | EIF4A2 | 0.2699903 |
| CD19+ TAMs | FHIT | 0.326673349 |
| CD19+ TAMs | DDX3X | 0.298504381 |
| CD19+ TAMs | CSGALNACT1 | 0.339469978 |
| CD19+ TAMs | RNF144B | 0.389020973 |
| CD19+ TAMs | FOXO3 | 0.304155882 |
| CD19+ TAMs | SAMSN1 | 0.401026514 |
| CD19+ TAMs | LDLRAD4 | 0.356235071 |
| CD19+ TAMs | GLA | 0.256065957 |
| CD19+ TAMs | CD83 | 0.379473933 |
| CD19+ TAMs | MMP19 | 0.4025614 |
| CD19+ TAMs | CHKA | 0.384909705 |
| CD19+ TAMs | ABCG1 | 0.265622469 |
| CD19+ TAMs | SERINC5 | 0.269337157 |
| CD19+ TAMs | PDE8A | 0.543291961 |
| CD19+ TAMs | WDR45B | 0.383786764 |
| CD19+ TAMs | PRMT9 | 0.273199651 |
| CD19+ TAMs | AVPI1 | 0.255958851 |
| CD19+ TAMs | ALPK3 | 0.313509921 |
| CD19+ TAMs | SLC19A2 | 0.25673093 |
| CD19+ TAMs | ADAMDEC1 | 0.279564906 |
| CD19+ TAMs | ACSL1 | 0.615487287 |
| CD19+ TAMs | JUNB | 0.409139662 |
| CD19+ TAMs | AC007319.1 | 0.29308236 |
| CD19+ TAMs | RAB31 | 0.301375106 |
| CD19+ TAMs | SH3BP5 | 0.328097456 |
| CD19+ TAMs | PFKFB3 | 0.270662321 |
| CD19+ TAMs | ICAM1 | 0.262391819 |
| CD19+ TAMs | MT1E | 0.420698365 |
| CD19+ TAMs | FOXK2 | 0.317446339 |
| CD19+ TAMs | HIF1A | 0.259362529 |
| CD19+ TAMs | RAB7A | 0.371992401 |
| CD19+ TAMs | MAP1LC3B | 0.282452846 |
| CD19+ TAMs | UBE2D3 | 0.31220936 |
| CD19+ TAMs | MANBA | 0.32837071 |
| CD19+ TAMs | ATP1B3 | 0.632237887 |
| CD19+ TAMs | NR4A2 | 0.38153332 |
| CD19+ TAMs | IFIT2 | 0.258227696 |
| CD19+ TAMs | ZFP36 | 0.493950691 |
| CD19+ TAMs | SRGN | 0.323366016 |
| CD19+ TAMs | DDX3Y | 0.294700194 |
| CD19+ TAMs | LGALS1 | 0.295823092 |
| CD19+ TAMs | TFRC | 0.384779558 |
| CD19+ TAMs | UST | 0.580468203 |
| CD19+ TAMs | MAFF | 0.28910063 |
| CD19+ TAMs | TTN | 0.329275431 |
| CD19+ TAMs | SIAH2 | 0.293876934 |
| CD19+ TAMs | NFKB1 | 0.589416132 |
| CD19+ TAMs | KLF2 | 0.304061094 |
| CD19+ TAMs | ELL2 | 0.271398093 |
| CD19+ TAMs | CCNH | 0.391547718 |
| CD19+ TAMs | ATF6 | 0.456523194 |

**Supplementary Table 7. Classical M1 and M2 macrophage marker genes expression in CD19+ TAMs and CD19- TAMs, related to Fig. 3.**

| Gene | avg.exp | pct.exp | Cluster | avg.exp.scaled |
|---|---|---|---|---|
| MARCO | 0.31911087 | 15.50013554 | CD19- TAMs | 0.707106781 |
| IRF1 | 2.369314177 | 66.16427216 | CD19- TAMs | 0.707106781 |
| CD80 | 0.215234481 | 14.37245866 | CD19- TAMs | -0.707106781 |
| CD86 | 2.061100569 | 73.04960694 | CD19- TAMs | 0.707106781 |
| CD40 | 0.380807807 | 25.56248306 | CD19- TAMs | -0.707106781 |
| IL1B | 7.452805638 | 53.60260233 | CD19- TAMs | 0.707106781 |
| TNF | 0.87975061 | 24.96069396 | CD19- TAMs | 0.707106781 |
| IL1A | 0.10750881 | 4.597451884 | CD19- TAMs | 0.707106781 |
| NOS2 | 0.000271076 | 0.027107617 | CD19- TAMs | -0.707106781 |
| IL6 | 0.022933044 | 1.192735159 | CD19- TAMs | -0.707106781 |
| CXCL9 | 1.298021144 | 12.42071022 | CD19- TAMs | 0.707106781 |
| CXCL10 | 1.753537544 | 15.20737327 | CD19- TAMs | 0.707106781 |
| CXCL11 | 0.163513147 | 4.066142586 | CD19- TAMs | 0.707106781 |
| CCR7 | 0.079587964 | 4.212523719 | CD19- TAMs | 0.707106781 |
| IDO1 | 0.170344267 | 5.578747628 | CD19- TAMs | 0.707106781 |
| VEGFA | 1.279425319 | 36.06397398 | CD19- TAMs | 0.707106781 |
| VEGFB | 0.507834101 | 32.67552182 | CD19- TAMs | -0.707106781 |
| VEGFC | 0.000325291 | 0.032529141 | CD19- TAMs | 0.707106781 |
| VEGFD | 0.001680672 | 0.135538086 | CD19- TAMs | -0.707106781 |
| IRF4 | 0.141772838 | 8.262401735 | CD19- TAMs | 0.707106781 |
| FN1 | 1.894985091 | 17.57657902 | CD19- TAMs | 0.707106781 |
| CLEC7A | 1.875142315 | 72.97370561 | CD19- TAMs | 0.707106781 |
| TGFB1 | 1.330008132 | 62.87340743 | CD19- TAMs | 0.707106781 |
| TGFB2 | 0.078991597 | 4.440227704 | CD19- TAMs | -0.707106781 |
| IL1RN | 1.061100569 | 25.87693142 | CD19- TAMs | 0.707106781 |
| IL1R2 | 0.594524261 | 16.41095148 | CD19- TAMs | 0.707106781 |
| IL4R | 0.393819463 | 28.77202494 | CD19- TAMs | 0.707106781 |
| IL10 | 0.76416373 | 17.34345351 | CD19- TAMs | 0.707106781 |
| ARG2 | 0.264028192 | 12.27975061 | CD19- TAMs | 0.707106781 |
| CCL2 | 0.138465709 | 3.822174031 | CD19- TAMs | -0.707106781 |
| CCL4 | 8.174247764 | 47.33532123 | CD19- TAMs | -0.707106781 |
| CCL5 | 0.27975061 | 10.74545947 | CD19- TAMs | -0.707106781 |
| CCL18 | 0.095310382 | 1.718622933 | CD19- TAMs | -0.707106781 |
| CCL22 | 0.01854161 | 1.051775549 | CD19- TAMs | 0.707106781 |
| CCL24 | 0.023963134 | 1.832474925 | CD19- TAMs | -0.707106781 |
| CD274 | 0.091190024 | 6.408240716 | CD19- TAMs | -0.707106781 |
| PDCD1 | 0.005529954 | 0.525887774 | CD19- TAMs | -0.707106781 |
| CD276 | 0.148495527 | 11.67796151 | CD19- TAMs | -0.707106781 |
| CD200R1 | 0.107454595 | 8.181078883 | CD19- TAMs | -0.707106781 |
| TNFSF12 | 0.350121984 | 26.99376525 | CD19- TAMs | -0.707106781 |
| MMP9 | 0.505773922 | 14.74112226 | CD19- TAMs | -0.707106781 |
| MMP14 | 0.13754405 | 11.37435619 | CD19- TAMs | -0.707106781 |
| MMP19 | 0.282895094 | 11.13038764 | CD19- TAMs | -0.707106781 |
| CSF1R | 2.514990512 | 83.38303063 | CD19- TAMs | -0.707106781 |
| CD163 | 2.48669016 | 66.98834372 | CD19- TAMs | -0.707106781 |
| MSR1 | 2.515803741 | 68.22445107 | CD19- TAMs | -0.707106781 |
| MARCO1 | 0.130183934 | 4.428226244 | CD19+ TAMs | -0.707106781 |
| IRF11 | 1.227279452 | 49.41565166 | CD19+ TAMs | -0.707106781 |
| CD801 | 0.248039844 | 17.40716998 | CD19+ TAMs | 0.707106781 |
| CD861 | 1.803984417 | 71.22639183 | CD19+ TAMs | -0.707106781 |
| CD401 | 0.618768184 | 38.59164653 | CD19+ TAMs | 0.707106781 |
| IL1B1 | 2.487400759 | 35.8597564 | CD19+ TAMs | -0.707106781 |
| TNF1 | 0.556930815 | 26.47566448 | CD19+ TAMs | -0.707106781 |
| IL1A1 | 0.051925637 | 2.830514325 | CD19+ TAMs | -0.707106781 |
| NOS21 | 0.000394497 | 0.039449677 | CD19+ TAMs | 0.707106781 |
| IL61 | 0.024360176 | 1.439913211 | CD19+ TAMs | 0.707106781 |
| CXCL91 | 1.166872134 | 24.22703289 | CD19+ TAMs | -0.707106781 |
| CXCL101 | 1.448296267 | 27.24986439 | CD19+ TAMs | -0.707106781 |
| CXCL111 | 0.114058879 | 5.493367523 | CD19+ TAMs | -0.707106781 |
| CCR71 | 0.054835051 | 2.934069727 | CD19+ TAMs | -0.707106781 |
| IDO11 | 0.071403915 | 2.791064648 | CD19+ TAMs | -0.707106781 |
| VEGFA1 | 0.628482667 | 27.90571527 | CD19+ TAMs | -0.707106781 |
| VEGFB1 | 1.107451058 | 60.2988313 | CD19+ TAMs | 0.707106781 |
| VEGFC1 | 0.000147936 | 0.00493121 | CD19+ TAMs | -0.707106781 |
| VEGFD1 | 0.00493121 | 0.438877657 | CD19+ TAMs | 0.707106781 |
| IRF41 | 0.074116081 | 5.365156073 | CD19+ TAMs | -0.707106781 |
| FN11 | 0.899206075 | 24.99630159 | CD19+ TAMs | -0.707106781 |
| CLEC7A1 | 1.189999507 | 54.28275556 | CD19+ TAMs | -0.707106781 |
| TGFB11 | 1.088268652 | 57.22668771 | CD19+ TAMs | -0.707106781 |
| TGFB21 | 0.141081907 | 9.172049904 | CD19+ TAMs | 0.707106781 |
| IL1RN1 | 0.391784605 | 14.73938557 | CD19+ TAMs | -0.707106781 |
| IL1R21 | 0.314265989 | 11.33191972 | CD19+ TAMs | -0.707106781 |
| IL4R1 | 0.307362296 | 24.55249273 | CD19+ TAMs | -0.707106781 |
| IL101 | 0.631934514 | 19.49307165 | CD19+ TAMs | -0.707106781 |
| ARG21 | 0.186005227 | 11.74121012 | CD19+ TAMs | -0.707106781 |
| CCL21 | 0.530598156 | 18.46738005 | CD19+ TAMs | 0.707106781 |
| CCL41 | 9.277232605 | 64.32269836 | CD19+ TAMs | 0.707106781 |
| CCL51 | 0.350609004 | 16.76611273 | CD19+ TAMs | 0.707106781 |
| CCL181 | 0.376399231 | 15.73055871 | CD19+ TAMs | 0.707106781 |
| CCL221 | 0.012328024 | 0.542433059 | CD19+ TAMs | -0.707106781 |
| CCL241 | 0.033236353 | 2.722027713 | CD19+ TAMs | 0.707106781 |
| CD2741 | 0.113910942 | 9.172049904 | CD19+ TAMs | 0.707106781 |
| PDCD11 | 0.006903693 | 0.631194832 | CD19+ TAMs | 0.707106781 |
| CD2761 | 0.279994083 | 22.31372356 | CD19+ TAMs | 0.707106781 |
| CD200R11 | 0.281720006 | 21.30282558 | CD19+ TAMs | 0.707106781 |
| TNFSF121 | 0.584693525 | 40.7811036 | CD19+ TAMs | 0.707106781 |
| MMP91 | 0.679964495 | 19.90729326 | CD19+ TAMs | 0.707106781 |
| MMP141 | 0.325509147 | 25.48449135 | CD19+ TAMs | 0.707106781 |
| MMP191 | 0.482124365 | 14.78376646 | CD19+ TAMs | 0.707106781 |
| CSF1R1 | 3.984811874 | 93.46121604 | CD19+ TAMs | 0.707106781 |
| CD1631 | 5.85369101 | 90.90684945 | CD19+ TAMs | 0.707106781 |
| MSR11 | 5.663395631 | 95.09837763 | CD19+ TAMs | 0.707106781 |

# Supplementary Table 8. Key Resources Table

| REAGENT or RESOURCE | SOURCE | IDENTIFIER |
|---|---|---|
| **Antibodies** | | |
| *In vivo* mAb anti-mouse CD73 | bioxcell | Cat#BE0209 |
| *In vivo* Plus anti-mouse PD-L1 (B7-H1) | bioxcell | Cat#BP0101 |
| *In vivo* mAb anti-human CD19 | bioxcell | Cat#BE0281 |
| FC: anti-Human CD45 | Biolegend | Cat#368516 |
| FC: anti-Human CD14 | Biolegend | Cat#325603 |
| FC: anti-Human CD11b | Biolegend | Cat#301322 |
| FC: anti-Human CD19 (For CD19$^+$ TAMs sorting and ImageStream) | Biolegend | Cat#302234 |
| FC: anti-Human CD68 | Biolegend | Cat#333808 |
| FC: anti-Human CD73 | Biolegend | Cat#344013 |
| FC: anti-Human CD19 (For flow cytometry analysis) | Biolegend | Cat#302206 |
| FC: anti-Human CD274 | Biolegend | Cat#329740 |
| FC: anti-Human CD64 | Biolegend | Cat#305027 |
| FC: anti-Human PD-1 | Biolegend | Cat#367405 |
| FC: anti-Human CD80 | Biolegend | Cat#305217 |
| FC: anti-Human CCR2 | Biolegend | Cat#357203 |
| FC: anti-Human ARG-1 | Biolegend | Cat#369705 |
| FC: anti-Human CD163 | Biolegend | Cat#333610 |
| FC: anti-Human CD206 | Biolegend | Cat#321103 |
| FC: anti-Human CD19 (For flow cytometry in Fig 5f) | BD Bioscience | Cat#557921 |
| FC: anti-Human PAX5 | BD Bioscience | Cat#562814 |
| FC: anti-Human Ki-67 | BD Bioscience | Cat#563757d |
| FC: anti-Human CD86 | BD Bioscience | Cat#562432 |
| FC: anti-Human HLA_DR | BD Bioscience | Cat#564231 |
| FC: anti-Human CD326 | BD Bioscience | Cat#745841 |
| FC: anti-Human Fc Block | BD Bioscience | Cat#564219 |
| FC: anti-Mouse F4/80 (For CD19$^+$ macrophages sorting) | Biolegend | Cat#123110 |
| FC: anti-Mouse F4/80 (For flow cytometry analysis) | Biolegend | Cat#123114 |
| FC: anti-Mouse Ly6C | Biolegend | Cat#128036 |
| FC: anti-Mouse CD274 | Biolegend | Cat#124312 |
| FC: anti-Mouse CD279 | Biolegend | Cat#135218 |
| FC: anti-Mouse CD45.1 | Biolegend | Cat#110723 |
| FC: anti-Mouse CD45.2 | Biolegend | Cat#109813 |
| FC: anti-Mouse Ki-67 | Biolegend | Cat#652410 |
| FC: anti-Mouse Ly6G | Biolegend | Cat#127605 |
| FC: anti-Mouse CD3 | Biolegend | Cat#100203 |
| FC: anti-Mouse CD8a | Biolegend | Cat#100722 |
| FC: anti-Mouse PD-1 | Biolegend | Cat#135218 |
| FC: anti-Mouse CD19 (For CD19+ macrophages sorting) | Biolegend | Cat#152410 |
| FC: anti-Mouse CD169 | Biolegend | Cat#142406 |
| FC: anti-Mouse CD209b | Thermo Fisher Scientific | Cat# 17-2093-82 |
| FC: anti-Mouse CLEC4F | Biolegend | Cat#156804 |
| FC: anti-Mouse CD11c | Biolegend | Cat#117306 |
| FC: anti-Mouse CD170 | Biolegend | Cat#155508 |
| FC: anti-Mouse CD115 | Biolegend | Cat#135510 |
| FC: anti-Mouse Ly-6G/Ly6c (Gr-1) | Biolegend | Cat#108438 |
| FC: anti-Mouse CD45 | BD Bioscience | Cat#563053 |
| FC: anti-Mouse CD11b | BD Bioscience | Cat#562287 |
| FC: anti-Mouse MHCII (I-A/I-E) | BD Bioscience | Cat#562564 |
| FC: anti-Mouse CD19 (For flow cytometry analysis) | BD Bioscience | Cat#557921 |
| FC: anti-Mouse CD11c | BD Bioscience | Cat#560584 |
| FC: anti-Mouse CD4 | BD Bioscience | Cat#552051 |
| FC: anti-Mouse CD45R/B220 | BD Bioscience | Cat#563893 |
| FC: anti-Mouse Fc Block | BD Bioscience | Cat#553142 |
| FC: 7-AAD | BD Bioscience | Cat#559925 |
| FC: Fixable Viability Stain 780 | BD Bioscience | Cat#565388 |
| FC: Fixable viability stain 700 | BD Bioscience | Cat#564997 |
| Chip, Wb: Rabbit IgG, monoclonal-Isotype Control | abcam | Cat#ab172730 |
| Chip, Wb: Anti-PAX5 antibody | abcam | Cat#ab227635 |
| Wb: Anti-PGC1 alpha antibody | abcam | Cat#ab106814 |
| Wb: Anti-LAMP1 antibody | abcam | Cat#ab24170 |
| Wb: Anti-CD63 antibody | abcam | Cat#ab134045 |
| Wb: LC3B (E7X4S) XP® Rabbit mAb | Cell Signaling Technology | Cat#43566S |
| Wb: SQSTM1/p62 (D5L7G) Mouse mAb | Cell Signaling Technology | Cat#88588S |
| Wb: NT5E antibody | Affinity | Cat#DF6763 |
| Wb: Cathepsin B (D1C7Y) XP® Rabbit mAb | Cell Signaling Technology | Cat#31718T |
| Wb: Cathepsin H Antibody - Internal | Affinity | Cat#DF8514 |
| Wb: Phospho-TFEB (Ser142) antibody | Affinity | Cat#AF3845 |
| Wb: β-Actin Mouse Monoclonal Antibody | Beyotime | Cat#AF5009 |
| Wb: GAPDH Mouse Monoclonal Antibody | Beyotime | Cat#AF5001 |
| Wb: Histone H3 Mouse Monoclonal Antibody | Beyotime | Cat#AF0009 |
| Wb: HRP-labeled Goat Anti-Mouse IgG(H+L) | Beyotime | Cat#A0216 |
| Wb: HRP-labeled Goat Anti-Rabbit IgG(H+L) | Beyotime | Cat#A0208 |
| Wb: LAMP-5 antibody | santa cruz | Cat#sc-398190 |
| Wb: LAMP2 Monoclonal antibody | proteintech | Cat#66301-1-Ig |
| Wb, IF: TFEB Polyclonal antibody | proteintech | Cat#13372-1-AP |
| Wb, IF: PD-L1/CD274 monoclonal antibody | proteintech | Cat#2B11D11 |

| Wb, IF: CD19 antibody | Affinity | Cat#DF7030 |
|---|---|---|
| IF: CD68 | abcam | Cat#ab213363 |
| IF: PD-L1 | abcam | Cat#ab213524 |
| IF: F4/80 | abcam | Cat#ab6640 |
| IF: Alexa Fluor® 488 Anti-F4/80 | Abcam | Cat#ab204266 |
| IF: Alexa Fluor® 647 Anti-PD-L1 antibody | Abcam | Cat#ab224030 |
| IF: CD19 | Cell Signaling Technology | Cat#3574S |
| IF: Anti-rabbit IgG (H+L), F(ab')$_2$ Fragment (Alexa Fluor® 555 Conjugate) | Cell Signaling Technology | Cat#4413S |
| IF:Anti-mouse IgG (H+L), F(ab')$_2$ Fragment (Alexa Fluor® 488 Conjugate) | Cell Signaling Technology | Cat#4408S |
| IF:Anti-mouse IgG (H+L), F(ab')$_2$ Fragment (Alexa Fluor® 647 Conjugate) | Cell Signaling Technology | Cat#4410S |
| IHC: Anti-CD45 antibody | abcam | Cat#ab40763 |
| IHC: Ki-67 (D2H10) Rabbit mAb (IHC Specific) | Cell Signaling Technology | Cat#9027S |
| IHC: CD8α (D4W2Z) XP® Rabbit mAb (Mouse Specific) | Cell Signaling Technology | Cat#98941S |
| IHC: CD4 (D7D2Z) Rabbit mAb | Cell Signaling Technology | Cat#25229S |
| IHC: NCAM1 (CD56) (E7X9M) XP® Rabbit mAb | Cell Signaling Technology | Cat#99746S |
| IHC: Ki-67 (8D5) Mouse mAb | Cell Signaling Technology | Cat#94497 |
| IHC: PD-1 (D7D5W) XP® Rabbit mAb (Mouse Specific) | Cell Signaling Technology | Cat#84651s |
| IHC: CD45 (D3F8Q) Rabbit mAb | Cell Signaling Technology | Cat#70257S |
| IHC: FoxP3 (D6O8R) Rabbit mAb | Cell Signaling Technology | Cat#12653S |
| IHC: FoxP3 (D2W8E™) Rabbit mAb (IHC Specific) | Cell Signaling Technology | Cat#98377S |
| IHC: CD8α (D8A8Y) Rabbit mAb | Cell Signaling Technology | Cat#85336S |
| IHC: CD31 (PECAM-1) (89C2) Mouse mAb | Cell Signaling Technology | Cat#3528S |
| IHC: CD4 (D7D2Z) Rabbit mAb (BSA and Azide Free) | Cell Signaling Technology | Cat#27520SF |
| IHC: PD-1 (D4W2J) XP® Rabbit mAb | Cell Signaling Technology | Cat#86163S |
| mIHC: CD14 | Abcam | Cat#ab245235 |
| mIHC: CD68 | Abcam | Cat#ab192847 |
| mIHC: CD20 | Cell Signaling Technology | Cat#48750S |
| mIHC: CD19 | Abcam | Cat#ab134114 |
| **Bacterial and virus strains** | | |
| pMSCV-IRES-GFP II | Addgene | Cat#52107 |
| Lentivirus | Jikai | N/A |
| **Biological samples** | | |

| Tumor samples | The First Affiliated Hospital of Zhejiang University | NA |
|---|---|---|
| Adjacent normal tissues | The First Affiliated Hospital of Zhejiang University | NA |
| PBMC from HCC patients | The First Affiliated Hospital of Zhejiang University | NA |
| **Critical commercial assays** | | |
| 5'-Nucleotidase (CD73) Activity Assay kit | Abcam | Cat#ab235945 |
| cytofix/Cytoperm Fixation/ Permeabilization Kit | BD Bioscience | Cat#554714 |
| Nuclear and Cytoplasmic Protein Extraction Kit | Beyotime | Cat#P0028 |
| Adenosine Assay | Cell Biolabs | Cat#MET-5090 |
| Chromium Next GEM Single Cell 3' GEM, Library & Gel Bead Kit v3.1 | 10X Genomics | Cat#1000121 |
| Chromium Next GEM Chip G Single Cell Kit, | 10X Genomics | Cat#1000120 |
| Opal 6-plex manual detection kit | Akoya Biosciences | Cat#NEL861001KT |
| **Experimental models: Cell lines** | | |
| THP-1 | ATCC | N/A |
| Hepa1-6 | ATCC | N/A |
| Plat-E | ATCC | N/A |
| iBMDM | Zhejiang University | N/A |
| **Experimental models: Organisms/strains** | | |
| C57BL/6 (males, 6-12 weeks old) | Nanjing University | N/A |
| *Pax5$^{flox/flox}$* | Cyagen Biosciences | N/A |
| *CD19$^{flox/flox}$* | Cyagen Biosciences | N/A |
| *muMT* | Zhejiang University | N/A |
| *Lyz-Cre* | Jackson Laboratory | Cat#018956 |
| **Oligonucleotides** | | |
| Primers, see **Table S5** | This manuscript | N/A |
| **Software and algorithms** | | |
| Venny（2.1)) | https://bioinfogp.cnb.csic.es/tools/venny/index.html | https://bioinfogp.cnb.csic.es/tools/venny/ |
| WebGestalt | http://www.webgestalt.org/ | https://doi.org/10.1093/nar/gkz401 |
| MAST | https://doi.org/10.1186/s13059-015-0844-5 | https://www.bioconductor.org/packages/release/bioc/html/MAST.html |
| STAR | doi:10.1093/bioinformatics/bts635 | http://code.google.com/p/rna-star/ |
| GSVA (v1.36.0） | https://doi.org/10.1186/1471-2105-14-7 | http://www.bioconductor.org/packages/release/bioc/html/GSVA.html |
| Monocle2 (v2.10.1) | https://doi.org/10.1038/nmeth.4402 | https://github.com/cole-trapnelllab/monocle-release |
| Seurat | doi: 10.1038/nbt.4096. | https://github.com/satijalab/seurat |
| Cell Ranger (v3.0.1) | 10x Genomics, https://doi.org/10.1038/ncomms14049 | https://10xgenomics.com |
| GSEA (v4.1.0) | https://doi.org/10.1073/pnas.0506580102 | https://www. gsea-msigdb.org/ gsea/index.jsp |
| WebGestalt (v2019) |  | https://www.webgestalt.org/ |
| Wave desktop (v2.6.1） | Agilent | https://www.agilent.com.cn/zh-cn/product/cell-analysis/real-time-cell-metabolic-analysis/xf-software/seahorse-wave-desktop-software-740897#relatedproducts |
| ACCENSE | doi: 10.1073/pnas.1321405111. | http://cellaccense.com/ |
| Bowtie 2 (v2.3.4.3) | doi: 10.1038/nmeth.1923 | http://bowtie-bio.sourceforge.net/bowtie2/index.shtml |
| Deeptools (3.02) | doi: 10.1093/nar/gku365 | https://deeptools.readthedocs.io/en/develop/ |
| ChIPseeker (1.12.1) | https://doi.org/10.1093/bioinformatics/btv145 | https://bioconductor.org/packages/release/bioc/html/ChIPseeker.html |
| MEME (5.1) | doi: 10.1093/nar/gkp335. | https://meme-suite.org/meme/ |
| Prism (v8.0) | Graph Pad | https://www.graphpad.com/ |
| Flow Jo (v10.5.3) | Treestar | https://www.flowjo.com/ |
| Image J | Freeware/NIH | https://imagej.nih.gov/ij/ |
| MicroVigene | Vigene Tech | http://www.vigenetech.com/MicroVigene.htm |
| DAVID | doi: 10.1038/nprot.2008.211 | https://david.ncifcrf.gov/summary.jsp |
| R 4.0.5 | R Project | https://www.r-project.org |
| **Deposited data** | | |
| Sc-RNA-seq | This paper | N/A |
| Sc-RNA-seq (published) | Bioproject | PRJCA010606 |
| Sc-RNA-seq (published) | Bioproject | PRJCA007744 |
| Cut&Tag | This paper | ftp://ftp.ncbi.nlm.nih.gov/genomes/all/GCF/000/0 |
| RNA-seq | This paper | N/A |
| **Other** | | |
| Protein Block | Abcam | Cat#ab64226 |
| Seahorse XF 1.0 M glucose solution | Agilent | Cat#103577-100 |
| Seahorse XF 100 mM pyruvate solution | Agilent | Cat#103579-100 |
| Seahorse XF 200 mM glutamine solution | Agilent | Cat#103578-100 |
| Seahorse XF RPMI Medium | Agilent | Cat#103576-100 |
| Seahorse XFe24 FluxPak | Agilent | 102340-100 |
| Leukocyte Activation Cocktail, with BD GolgiPlug™ | BD Bioscience | Cat#550583 |
| Lyso-Tracker Red | Beyotime | Cat#C1046 |
| Cell lysis buffer for Western and IP | Beyotime | Cat#p0013 |
| EdU | Beyotime | Cat#ST067 |
| FCCP | Cayman | Cat#15218 |
| Antimycin A1 | Cayman | Cat#18433 |
| oligomycin A | Cayman | Cat#11342 |
| Clodronate Liposomes & Control Liposomes | Clodronate Liposomes | Cat#CP-005-005 |
| Corning® Cell-Tak™ Cell and Tissue Adhesive | Corning | Cat#354240 |
| Dipyridamole | Glpbio | Cat#GC17673 |
| EHNA hydrochloride | Glpbio | Cat#GC10935 |
| NBMPR | Glpbio | Cat#GC10835 |
| MitoTracker™ Deep Red FM | Invitrogen | Cat#M22426 |
| pHrodo™ Red, SE | Invitrogen | Cat#P36600 |
| MitoSOX™ Red | Invitrogen | Cat#M36008 |
| LysoSensor™ Yellow/Blue DND-160 | Invitrogen | Cat#L7545 |
| CellLight™ Mitochondria-GFP | Invitrogen | Cat#C1050 |
| staurosporine | Merck | Cat#569397-100ug |
| NEOFECT™ DNA transfection reagent | Neofect | Cat#TF201201 |
| Animal-free recombinant murine M-CSF | Peprotech | Cat#AF-315-02 |
| *TFEB* siRNA (h) | santa cruz | Cat#sc-38509 |
| IACS-010759 | Selleck | Cat#1570496-34-2 |
| Adenosine 5'-monophosphate monohydrate | Sigma-Aldrich | Cat#A2252 |

| Adensine 5'-pyrophosphoric acid （5'-ADP） | Sigma-Aldrich | Cat#3768-14-7 |
| --- | --- | --- |
| CD73 inhibitor-AMP-CP | Sigma-Aldrich | Cat#3768-14-7 |
| Hyaluronidase | Worthington | Cat#LS005477 |
| ProLong™ Gold and Diamond Antifade Mountants | Invitrogen | Cat#S36939 |
| Rhod-2 dextran conjugate | ABD Bioquest | Cat#20451 |
| Calbryte™ 630 AM | ABD Bioquest | Cat#20720 |
| Goat serum | Zsbio | Cat#ZLI-9056 |
| Anti-CD3ε | Biolegend | Cat#100302 |
| Anti-CD28 | Biolegend | Cat#102116 |
| MojoSort™ Streptavidin Nanobeads | Biolegend | Cat#480016 |
| MojoSort™ Magnet | Biolegend | Cat#480019 |
| FBS | NEWZERUM | Cat#FBS-E500 |
| IL-2 | Novoprotein | Cat#C013 |
| 2-mercaptoethanol | Gibco | Cat#21985023 |
| 40,6-diamidino-2-phenylindole (DAPI) | Sigma-Aldrich | Cat#D9542 |